\documentclass[INR,biber]{nowfnt} 

\usepackage[utf8]{inputenc}
\usepackage{tabularx}
\usepackage{booktabs}
\usepackage{amsmath}
\usepackage{amssymb}
\usepackage{dsfont}
\usepackage{amsthm}
\usepackage{multirow}
\usepackage{url}
\usepackage{subfigure}
\usepackage{tabularx}
\usepackage{mathrsfs, amsfonts}
\usepackage{url,graphicx,times}
\usepackage{balance}
\usepackage[normalem]{ulem}
\usepackage{mathcomp}
\usepackage{setspace}
\usepackage{ltxtable}
\usepackage[skip=0pt]{caption}
\usepackage{acronym}
\usepackage{paralist}
\usepackage[inline,ignoredisplayed]{enumitem}
\usepackage{tikz}
\usetikzlibrary{positioning}
\usepackage{CJKutf8}
\usepackage{makecell}

\usepackage{array}
\usepackage{pifont}
\usepackage[english]{babel}
\usepackage{rotating}
\usepackage[noend]{algpseudocode}
\usepackage{algorithmicx,algorithm}

\usepackage{mwe}
\DeclareUnicodeCharacter{0301}{\'{e}}

\newcommand{\changed}[1]{\textcolor{blue}{#1}}

\newcommand{\mdr}[1]{\textcolor{orange}{\textbf{[#1]}}}

\newcommand{\hide}[1]{}
\newcommand{\OurParagraph}[1]{\smallskip\noindent\textbf{#1}}

\acrodef{CIS}{conversational information seeking}
\acrodef{CR}{conversational recommendation}
\acrodef{CRS}{conversational recommender system}
\acrodef{CVR}{conversion rate}
\acrodef{FM}{factorization machine}
\acrodef{KGC}{knowledge-grounded conversation}
\acrodef{KS}{knowledge selection}
\acrodef{NBR}{next basket recommendation}
\acrodef{QA}{question answering}
\acrodef{TDS}{task-oriented dialogue system}

\maintitleauthorlist{
Zhaochun Ren \\
Leiden University \\
z.ren@liacs.leidenuniv.nl
\and
Xiangnan He \\
University of Science and Technology of China \\
xiangnanhe@gmail.com
\and
Dawei Yin \\
Baidu Inc. \\
yindawei@acm.com
\and
Maarten de Rijke \\
University of Amsterdam \\
m.derijke@uva.nl
}

\author[1]{Ren, Zhaochun}
\author[2]{He, Xiangnan}
\author[3]{Yin, Dawei}
\author[4]{de Rijke, Maarten}

\affil[1]{Leiden University; z.ren@liacs.leidenuniv.nl}
\affil[2]{University of Science and Technology of China; xiangnanhe@gmail.com}
\affil[3]{Baidu Inc.; yindawei@acm.com}
\affil[4]{University of Amsterdam; m.derijke@uva.nl}

\issuesetup
{%
 copyrightowner={Z.~Ren, X.~He, D.~Yin, and M.~de Rijke},
 volume        = xx,
 issue         = xx,
 pubyear       = 2025,
 isbn          = xxx-x-xxxxx-xxx-x,
 eisbn         = xxx-x-xxxxx-xxx-x,
 doi           = 10.1561/XXXXXXXXX,
 firstpage     = 1, 
 lastpage      = 999
 }

\title{Information Discovery in E-commerce}

\articledatabox{\nowfntstandardcitation}

\begin{document}

\makeabstracttitle

\begin{abstract}
Electronic commerce, or e-commerce, is the buying and selling of goods and services, or the transmitting of funds or data online. 
E-commerce platforms come in many kinds, with global players such as Amazon, Airbnb, Alibaba, Booking.com, eBay, JD.com and platforms targeting specific geographic regions such as Bol.com and Flipkart.com.
Information retrieval has a natural role to play in e-commerce, especially in connecting people to goods and services.
Information discovery in e-commerce concerns different types of search (e.g., exploratory search vs. lookup tasks), recommender systems, and natural language processing in e-commerce portals. 
The rise in popularity of e-commerce sites has made research on information discovery in e-commerce an increasingly active research area. 
This is witnessed by an increase in publications and dedicated workshops in this space. 
Methods for information discovery in e-commerce largely focus on improving the effectiveness of e-commerce search and recommender systems, on enriching and using knowledge graphs to support e-commerce, and on developing innovative question answering and bot-based solutions that help to connect people to goods and services. 
In this survey, an overview is given of the fundamental infrastructure, algorithms, and technical solutions for information discovery in e-commerce.
The topics covered include user behavior and profiling, search, recommendation, and language technology in e-commerce.
\end{abstract}


\chapter{Introduction}

\section{Motivation}
\label{sec:motivation}

Over the past 20 years, we have seen an explosive growth of e-commerce portals, such as Alibaba, Amazon, eBay, and JD.com. These developments have reshaped people's shopping habits.
An increasing number of customers now prefer to spend more time shopping online, generating billions of user requests per day.
As part of the process of serving customer requests, large volumes of multi-modal data, including user search logs, clicks, orders, reviews, images, and chat logs, etc., are being generated. 
From an information retrieval point of view, discovering and employing pertinent information from the sheer volume of e-commerce data so as to enhance the performance of e-commerce services presents interesting challenges, both for academic and industrial researchers. 
In this survey we describe those challenges and the solutions that the community has so far proposed.

The topics of information discovery in e-commerce can be divided into several main directions: 
\begin{itemize}[leftmargin=*,nosep]
\item e-commerce presentation and users;
\item user behavior and profiling;
\item search in e-commerce;
\item recommender systems in e-commerce; and
\item question answering and dialogue systems in e-commerce.
\end{itemize}

\noindent%
Each of these areas comes with its own set of research challenges.
For example, in e-commerce search there may be no hypertext links between products, thus excluding an important type of ranking signal that is often used in the setting of web search. 
But with click streams and order streams we have two parallel sources of ranking signal, a characteristic e-commerce feature that is absent from more traditional search scenarios.

E-commerce information discovery problems are wide in scope as the underlying discovery tasks concern a broad range of interaction modalities.
There is a growing body of established methods in the e-commerce, aimed at developing algorithms for analyzing user behavior\hide{~\citep{yin2011exploiting,yi2014beyond,InfNeed4RS,zhouwsdm2018,kim2005development}}, for product search\hide{~\citep{duan2013supporting,duan2013supporting,ai2017learning}}, for recommender systems\hide{~\citep{schafer1999recommender,jiang2015life,he2016fast,kim2007impact,RS4EC,purchaserate1,LiRCRLM17}},  and for question answering and dialogue systems. 
These areas, and the methods developed, form the core around which most ongoing research efforts concerning information discovery for e-commerce are organized.
The time is right to organize this material and to present it to a broad audience of interested information retrieval researchers, whether junior or senior, whether academic or industrial~\citep{tsagkias-2020-challenges}.

\section{Aims of this survey}
A key aim of this survey is to bring together, and offer a unified perspective on, the large number of methods for e-commerce information discovery available today.
To achieve this, we describe the basic architecture used for information discovery in e-commerce, algorithms for e-commerce information discovery, and evaluation principles. 
We supplement this with an account of available datasets and software based on these. 
We also introduce e-commerce applications accompanied by examples.

The survey targets practitioners and researchers from academia and industry and aims to present them with the challenges, state-of-the-art approaches, and the most urgent open questions in information discovery for e-commerce. 
Specifically, in terms of content, the objectives of the survey are as follows:

\begin{itemize}[leftmargin=*,nosep]
\item To introduce tasks that constitute the information discovery problem in e-commerce, and to explain the difference between e-commerce information discovery and related work in other domains;
\item To describe e-commerce information discovery algorithms in a unified way, i.e., using common notation and terminology, so that different models can easily be related to each other;
\item To explain how to analyze the performance of e-commerce information discovery algorithms and why it is worth the effort;
\item To present appropriate experimental and evaluation methodologies for e-commerce information discovery in both synthetic and real world settings; and 
\item To discuss future directions of research in e-commerce information discovery.
\end{itemize}

\section{Outline}
Information discovery aims to distill pertinent information from datasets with various modalities; it plays a role in many areas, ranging from web search to academic search and medical search. 
What is different about the e-commerce setting is that many traditional ranking features are either not present or present in a different form~\citep{degenhardt-ecom-2017}.
Instead, discovery processes need to be supported based on structured information, semi-structured information, or information that might have facets such as price, ratings, title, description, seller location, etc.

\subsection{Topics covered}
We break the e-commerce information discovery problem down into five research directions:
\begin{enumerate*}[label=(\roman*)]
\item e-commerce information presentation and users, 
\item user behavior and profiling in e-commerce, 
\item search in e-com\-merce, 
\item recommendation in e-commerce, and 
\item question answering and dialogue systems in e-commerce.
\end{enumerate*}
Below, we briefly describe each of these five directions.
 
The first direction concerns preliminaries about e-commerce information presentation and users. E-commerce portals provide various modalities of information to users, e.g., rankings of products, product titles, descriptions, tips, and user reviews, etc.  
Multiple genres and types of text analysis can be employed to enhance e-commerce services, e.g., review filtering, review analysis, and normalization of production descriptions. User characteristics in e-commerce, e.g., browsing modules, clicks, purchases, and dwell time, generate multiple patterns for e-commerce scenarios. These two factors play fundamental roles in e-commerce information discovery.
In this survey, we summarize recent work on both e-commerce information presentation and user characteristics.

The second direction concerns user behavior modeling and user profiling. 
Tracking and profiling users' behavior on e-commerce portals are important prerequisites for many e-commerce services, such as recommender systems, search, and online advertising. 
In this survey, we summarize recent work on user behavior modeling in e-commerce and introduce solutions to profiling users of e-commerce services. 

The third direction of this survey concerns search in e-commerce, which examines approaches for product search scenarios on e-commerce portals. 
Just like, e.g., traditional web search, the target of this task is to satisfy users' needs. 
However, product search in e-commerce sites should be realized with different types of features than, e.g., web search, with the availability of a large number of product, query attributes, and engagement features. 
Moreover, calculating relevance in product search faces challenges regarding gaps between users and products. The target corpora can be structured, semi-structured, or unstructured, or a mixture of these; semantic search against such diverse sources raises interesting research challenges. 

The fourth direction concerns recommendations in e-commerce. 
In contrast to traditional research on recommender systems that focuses on rating prediction, e-commerce recommender systems aim to tackle three challenges: the huge volume of products, sparsity, and data richness. 
Due to the existence of a very large number of candidate items in e-commerce portals, of which only a small fraction will attract a user's attention, e-commerce recommendation methods usually follow a two-stage recommendation framework with 
\begin{enumerate*}[label=(\roman*)]
\item candidate retrieval, and 
\item candidate ranking.
\end{enumerate*}
The first phase of candidate retrieval goes through the whole product catalog, and selects a small set of products that might match the information need. 
The second phase of candidate ranking ranks the candidates to present the final top-$K$ products to the user. 
Given structured user behavior logs and semi-structured data about product features, e-commerce knowledge bases can be created to assist the candidate generation step. 
And the candidate ranking procedure ranks the retrieved candidate items for a better conversion rate or click-through rate, based on various machine learning models.\hide{~\citep{he2014practical,chen2016xgboost,he2016fast,He2017NCF}.}

The fifth and final direction of this survey concerns question answering and dialogue systems in e-commerce.
We survey recent work on e-commerce question answering and dialogue systems that have attracted increased attention. 
For dialogue systems, we describe both task-oriented dialogue systems, aimed at helping users complete a task in an e-commerce setting, and non-task-oriented dialogue systems aimed at generating fluent and engaging responses.

For the directions listed above, our ambition has been to cover related work up to the spring of 2023.

\subsection{Topics not covered}
E-commerce impacts large parts of our economy and society, including markets and retailers, supply chain management, and employment. 
With the development of data science, business intelligence studies on e-commerce marketing, e.g., sales volume forecasting and time series analysis, are receiving an increasing amount of attention.
All of these areas are important, scientifically challenging, and deserving of attention from the information retrieval community.
However, our focus will be limited to information discovery within the context of e-commerce. Specifically, we will not address topics such as computational advertising approaches that are irrelevant to search and recommendation, marketing strategies, forecasting, or information management in e-commerce.

\subsection{Structure of the survey}
The remainder of this survey is organized as follows. 
Section~\ref{chapter:terms} provides key definitions and background related to e-commerce information discovery, drawing from user modeling, search, recommender systems, question-answering, and dialogue systems.
Section~\ref{chapter:basic} describes preliminaries of e-commerce presentations as well as e-commerce users, including user behavior characteristics, and relevant language technologies and their use in e-commerce applications.
Section~\ref{chapter:user} details user behavior modeling and user profiling approaches in e-commerce, including click behavior tracking, post-click tracking, purchase behavior modeling, and user profiling in e-commerce.
Section~\ref{chapter:search} describes recent approaches proposed for e-commerce search, which we organize along two lines: research about the matching problem in e-commerce search, and about ranking strategies for e-commerce search.
Section~\ref{chp:rec} presents algorithms and solutions for recommender systems in e-commerce. After introducing the two-stage recommendation framework in e-commerce portals, we organize the e-commerce recommendation studies into two groups: candidate retrieval models and candidate ranking models.
We survey e-commerce question answering and dialogue systems in Section~\ref{chapter:qa}, where we introduce recent studies on e-commerce question answering and dialogue systems, respectively. 
In Section~\ref{chapter:conclusion} we conclude this survey and identify emerging research directions and issues for future work.

\section{Our readers}
We expect this survey to be useful to both academic and industrial researchers who either want to develop e-commerce information discovery methods, use them in their own research, or apply the methods described in the survey to improve product performance in e-commerce services.
The intention is to help our audience acquire domain knowledge and to promote information discovery research activities in e-commerce.

To be able to benefit from this survey, we expect the reader to have a background in information retrieval, natural language processing, or machine learning.
We recommend that readers read the material that we offer from start to finish, in the order that we offer it. However, readers who have a specific interest in search, or in recommender systems, or in conversational technology in e-commerce should read Section~\ref{chapter:basic} and~\ref{chapter:user} first before skipping ahead to Section~\ref{chapter:search}, \ref{chp:rec}, or~\ref{chapter:qa}, respectively.

\chapter{Definitions and background}
\label{chapter:terms}

The section presents definitions and background applied to e-commerce information discovery studies from the perspectives of research communities on user modeling, search, recommender systems, quesiton-answering, and dialogue systems. 
We first introduce relevant concepts about user modeling, information retrieval, recommender systems, and conversational AI. 
We then introduce definitions and notations associated with e-commerce information discovery. Next, we explore fundamental concepts in e-commerce information discovery, including e-commerce information presentation, e-commerce search, e-commerce recommendation, and e-commerce conversational AI systems. The glossary and notations attached to these concepts are introduced in the last part of this section.

\section{Background}

E-commerce has revolutionized how consumers interact with products and services, fundamentally altering the landscape of information discovery. There are plenty of relevant research perspectives on information discovery in e-commerce. Unlike traditional retail settings, where physical exploration and interaction drive decision-making, e-commerce relies on digital mechanisms to guide users through vast and often overwhelming amounts of information. As a result, the effectiveness of e-commerce platforms hinges on their ability to deliver personalized, relevant, and timely information to users.

Various research disciplines -- such as user modeling, information retrieval, recommender systems, question-answering, and conversational AI -- contribute significantly to enhancing the e-commerce experience. Each of these fields offers unique insights and methodologies for addressing key challenges in e-commerce, such as understanding user intent, predicting user preferences, addressing user concerns, satisfying user needs, and facilitating seamless product discovery. We list the fundamental concepts and research perspectives behind these areas, laying the groundwork for understanding how e-commerce platforms enable efficient and effective information discovery for users.

\begin{header}{Information discovery in e-commerce}
In the context of e-commerce, information discovery refers to the process by which users engage with relevant products or services based on their specific needs, regardless of the format or presentation of that information. This process encompasses a variety of functions, including search, recommendation, and personalized content delivery and presentation. At its core, information discovery in e-commerce involves not only retrieving relevant products but also understanding user intent and preferences to provide the most suitable results. It relies on algorithmic solutions to identify, search, recommend, and display information that aligns with user requirements. Whether through search queries, personalized recommendations, or curated content, information discovery systems enable users to efficiently navigate large product catalogs and find what they need in a seamless and engaging manner.
\end{header}

\section{User modeling}

User modeling refers to the process of creating a representation of a user's characteristics, behaviors, preferences, and goals in order to personalize user-system interactions or appropriate content for that user. It is widely used in fields like information retrieval, recommender systems, and conversational AI. User modeling is a critical component of personalizing e-commerce experiences. It involves the construction of user profiles based on behavioral data, preferences, demographics, and interactions within the system. These profiles help systems adapt content, recommendations, and interactions to suit individual needs.
Meanwhile, e-commerce platforms present information in various formats, such as lists, grids, or interactive elements, which can significantly affect user engagement and conversion rates. Understanding user behavior and preferences by optimizing these presentations is important for enhancing the user experience and user satisfaction.

In e-commerce, user modeling can use implicit feedback (e.g., clicks, purchases, carting, and user engagement) and explicit feedback (e.g., ratings and reviews) to predict a user's future behavior, preferences, and profiling. Techniques like collaborative filtering, content-based filtering, and hybrid models are frequently employed in user modeling during early studies on this topic. In recent years, deep neural networks and pre-trained language models have been successfully applied to user modeling in e-commerce portals. User modeling spans across domains such as cognitive science, machine learning, and human-computer interaction, contributing to the development of systems that continuously refine the understanding of users as they interact with the platform.

\section{Information retrieval in e-commerce}
Information retrieval (IR) has been playing a critical role in e-commerce services. Search and recommendation functionalities have been applied to e-commerce portals almost since the beginning of e-commerce. 
Beyond traditional search functionalities, IR techniques have evolved to support a wide range of features, including search and recommendation, making them essential for delivering a seamless user engagement. In e-commerce platforms, IR techniques are responsible for retrieving relevant items from a vast product catalog based on a user's query or search intent. This process involves not only matching query terms to product descriptions but also understanding the broader context behind the query, such as user preferences, purchase history, and real-time behaviors.

\begin{header}{E-commerce search}
E-commerce search involves techniques and algorithms used to allow users to efficiently find products or services within an online store. This includes the use of keywords, filters, and advanced semantic search technologies. E-commerce search differs from traditional information retrieval because it often focuses on product features, pricing, availability, and user preferences. Research in this area spans areas like query understanding, ranking algorithms, and the integration of multimodal data (e.g., images and reviews).
\end{header}

The effectiveness of an e-commerce search engine depends heavily on how well it can interpret user queries and match them with appropriate products or services. With natural language processing methods, query understanding and expansion techniques are playing an important role to bridge the semantic gap between queries and product information during this procedure, allowing the system to understand complex, ambiguous, or conversational queries from users. For example, users might search for ``affordable running shoes for winter,'' which requires the platform to parse the query, infer user intent (i.e., shoes for running in cold weather), and prioritize products based on pricing and seasonal relevance. Semantic search in e-commerce techniques go beyond keyword matching by understanding the user query and correlations between key entities, enabling more context-aware results.
Search engines in e-commerce also rely on machine learning models that take into account user intent, contextual data, and preferences to deliver highly relevant search results.
Additionally, search engines in e-commerce must address unique challenges like scalability and diversity. With product catalogs growing rapidly, search engines must efficiently process and rank millions of items in real-time. Advanced ranking algorithms, often powered by machine learning, play a vital role in this process, optimizing for both relevance and user engagement metrics such as click-through rates or conversion rates.

\begin{header}{E-commerce recommendations}
IR in e-commerce is increasingly intertwined with recommender systems. 
Recommender systems are a cornerstone of e-commerce platforms, helping users discover products they might not have explicitly searched for but are likely to find appealing. These systems predict user preferences using collaborative filtering, content-based filtering, or hybrid methods that combine both approaches.
While search engines retrieve items explicitly requested by the user (i.e., through queries), recommendations anticipate user needs by suggesting products the user may not have thought to search for. 
These systems analyze user data, such as browsing history, purchase patterns, and interactions, to suggest relevant items. They typically use a combination of techniques, including collaborative filtering, which recommends products based on the behaviors of similar users, and content-based filtering, which suggests items with attributes similar to those the user has shown interest in. Many modern systems employ hybrid models that integrate both types of method, sometimes enhanced with techniques like deep learning, to improve accuracy and diversity. 
By delivering personalized recommendations, these systems not only enhance the user experience but also drive business goals by increasing engagement, conversion rates, and customer satisfaction, all while introducing users to new products that may surprise or delight them.
\end{header}

Recommendation systems in e-commerce analyze user data and behavioral patterns to suggest products or services that users are likely to be interested in. These systems use various algorithms, including collaborative filtering, content-based filtering, and hybrid approaches.
In the context of e-commerce, recommender systems must balance relevance, diversity, novelty, and serendipity to enhance user engagement and satisfaction. By using a division into candidate retrieval and reranking stages, these systems often operate in two modes: personalized recommendations (based on individual profiles and history) and non-personalized recommendations (based on overall product popularity or trends). Research in recommender systems for e-commerce involves improving recommendation algorithms, addressing challenges like cold-start users, and optimizing recommendations for business goals such as conversion rates and customer retention.

In summary, information retrieval is foundational to the search and recommendation functions within e-commerce platforms. 
The convergence between search and recommendation highlights the importance of IR techniques that can balance precision (retrieving highly relevant products) with recall (offering a broader set of options that might interest the user). Hybrid models that combine collaborative filtering, content-based filtering, and neural IR approaches are commonly employed to address this dual need.
The integration of advanced IR techniques, such as search and recommendation models, allows e-commerce platforms to deliver highly personalized, efficient, and contextually relevant user experiences, ensuring that users find the products they want -- and even those they did not know they wanted.

\section{Conversational AI}

Conversational artificial intelligence (AI) techniques refer to technologies that enable natural, human-like interactions between users and machines through conversational communication. These interactions can be categorized into single-turn and multi-turn scenarios, corresponding to question-answering systems and dialogue systems, respectively.
During the interactions, conversational AI aims to understand user input, process context, and generate meaningful, human-like responses. Conversational AI is used in various applications, such as virtual assistants, customer support, and personal productivity tools. These systems can handle simple queries as well as complex, multi-turn conversations, adapting to user needs and improving over time through continuous learning. By mimicking human conversation patterns, conversational AI allows for more intuitive and accessible interactions, making it a valuable tool for enhancing communication between users and machines.

In e-commerce, chatbots and QA services powered by conversational AI can help users find relevant products, provide recommendations, and even complete transactions seamlessly. By offering a more engaging and interactive way for customers to interact with e-commerce platforms, conversational AI improves user satisfaction, increases engagement, and reduces the friction often associated with traditional search and navigation methods.

\begin{header}{E-commerce question-answering}
Question-answering (QA) systems are designed to deliver direct and precise responses to user questions, improving both user satisfaction and decision-making efficiency. 
Recently, e-commerce platforms have started to provide question-answering services. E-commerce QA systems help to enhance user experiences by enabling customers to obtain relevant, concise, and accurate answers to their product-related queries. These systems typically understand user intent and either retrieve answers from product descriptions, reviews, and FAQs or generate responses dynamically using advanced models.
\end{header}
E-commerce question-answering refers to the process in which users ask product-related questions on an e-commerce platform, and the system provides answers either from knowledge bases, reviews, or user-generated content. E-commerce QA systems use both retrieval-based and generative models to match or generate appropriate answers. This helps users make informed decisions based on product descriptions, user reviews, and frequently asked questions (FAQs). The goal is to reduce information overload and improve the user experience by providing relevant and concise answers.

E-commerce QA systems must handle a wide variety of queries ranging from simple fact-based questions (e.g., ``What is the price of this product?'') to more complex inquiries about product specifications, reviews, or usage (e.g., ``Is this laptop suitable for gaming?''). To address this diversity, QA systems often incorporate a mix of retrieval-based approaches, which search for relevant information in structured data or knowledge bases, and generative approaches, which generate answers when information is sparse or not directly available. Additionally, many e-commerce platforms enable community-based QA, where previous buyers or users of a product can contribute answers, further enriching the system’s knowledge base.
The integration of QA systems into e-commerce portals helps reduce the friction often associated with product discovery and decision-making. By offering immediate answers to user queries, these systems improve the overall shopping experience, increase user engagement, and can positively impact conversion rates. 
E-commerce QA is a rapidly evolving field with ongoing research aimed at improving the accuracy, efficiency, and personalization of responses.

\begin{header}{E-commerce automatic dialogue systems}
Automatic dialogue systems aim to engage in natural, human-like conversations with users and are widely used in applications such as customer support, virtual assistants, and e-commerce. They provide personalized, efficient, and engaging interactions, enhancing the overall user experience. In e-commerce, automatic dialogue systems, often in the form of chatbots or voice assistants, enable natural language interactions between users and platforms. 
They can support multi-turn interactions, where users ask follow-up questions, refine their preferences, or seek assistance, creating a more engaging and personalized shopping experience.
These systems assist users in discovering information, finding products, completing purchases, and sharing their opinions, all through conversational interfaces.
Systematically, automatic dialogue systems in e-commerce refers to the use of chatbots and virtual assistants that can simulate a human conversation to assist users in finding products, answering inquiries, and facilitating transactions. These systems use natural language processing and machine learning to provide timely and relevant assistance.
\end{header}

Research in conversational AI for e-commerce focuses on improving dialogue understanding, response generation, context retention across sessions, and user satisfaction. Additionally, conversational AI systems must adapt to diverse user needs and accommodate various languages and cultural contexts, making this an evolving area of study.

\chapter{E-commerce presentations and users}
\label{chapter:basic}

E-commerce presentations are composed of a series of user-facing components in e-commerce portals, e.g., various pages about items and categories, titles of items, user comments on item pages, search bars, and recommendation list. 
Such functions provided by e-commerce portals are meant to enable interactions with users on e-commerce platforms.
E-commerce users possess unique characteristics. 
There are multiple types of user behavior and feedback on an e-commerce platform, e.g., search, clicks, add-to-carts, purchases, returns, comments, and discussions with retailers. These unique characteristics of e-commerce users provide a rich source of information about the successes and failures of e-commerce platforms in helping users discover the items they need. 

We divide this section into two parts: e-commerce presentations (Section~\ref{ch2:sec1}) and e-commerce users (Section~\ref{ch2:sec2}). In Section~\ref{ch2:sec1}, we first introduce basic concepts of e-commerce interfaces; then we detail studies that analyze different aspects of e-commerce presentations, i.e., title analysis, item information analysis, and review analysis. 
In Section~\ref{ch2:sec2}, we list characteristics of e-commerce users, and examine user behavior on e-commerce portals, i.e., macro behavior, micro behavior and cross-platform behavior.

\section{E-commerce presentations}
\label{ch2:sec1}

In this section, we cover two aspects of e-commerce presenations: 
\begin{enumerate*}[label=(\roman*)]
\item basic concepts and types of e-commerce interface (Section~\ref{subsec:2.1.1}), and 
\item e-commerce presentation analysis (Section~\ref{subsec:2.1.2}). 
\end{enumerate*}

\subsection{Basic concepts}
\label{subsec:2.1.1}

\begin{figure}[!t]
	\centering
	\includegraphics[width=\columnwidth]{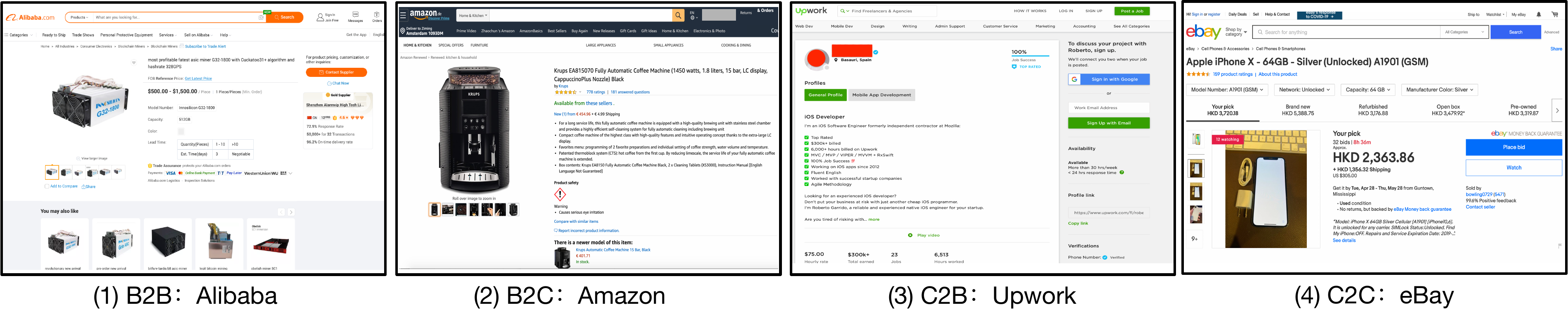}
	\caption{Four types of e-commerce businesses examples. Image sources: Alibaba.com, Amazon.com, UpWork.com, and EBay.com.}
	\vspace{-3mm}
	\label{fig_ch2sec21_typeecom}
\end{figure}

An \emph{interface} refers to an interactive component of a webpage or an application~\citep{hearst-2009-search}.
Referring to all interactive components on e-commerce portals (e.g., search bars, navigation panels, lists of recommended items, item titles, and user reviews), e-commerce interfaces play an invaluable role for the e-commerce user experience.
E-commerce interfaces dramatically impact the performance of an e-commerce platform.
Depending on the nature of the stakeholders involved, most e-commerce sites can be divided into four types of business: \emph{business-to-business}, \emph{business-to-consumer}, \emph{consumer-to-consumer}, and \emph{consumer-to-business}~\citep{nemat2011taking}:

\begin{description}[leftmargin=\parindent,nosep]
\item[B2B: Business to Business] This type of e-commerce business focuses on electronic transactions of goods or services between two corporations, i.e., one company uses the e-commerce site to sell items to another company. Fig.~\ref{fig_ch2sec21_typeecom}(1) shows an example of the interface used in a B2B setting.
\item[B2C: Business to Consumer] B2C refers to scenarios where businesses directly sell items to consumers. Most online shopping platforms, such as Amazon, Booking.com, and JD.com belong to this type of business. Fig.~\ref{fig_ch2sec21_typeecom}(2) shows an item page from Amazon as an example of the interface used by B2C businesses.
\item[C2B: Consumer to Business] Instead of a business retailing items to consumers, C2B sites such as UpWork\footnote{\url{https://www.upwork.com}} cater for a scenario where consumers provide services to businesses. Fig.~\ref{fig_ch2sec21_typeecom}(3) provides a screenshot from Upwork as an example of a C2B interface. 
\item[C2C: Consumer to Consumer] C2C refers to a type of e-commerce business where both retailers and buyers are consumers, while the C2C site itself benefits from commission fees that are normally paid by the seller. eBay is a well-known example of C2C business. Fig.~\ref{fig_ch2sec21_typeecom}(4) lists a screenshot from eBay as an example of a C2C business interface.
\end{description}
Like other web interfaces, e-commerce interfaces are evaluated in terms of user satisfaction~\citep{vergo2002commerce}. Thus, different communities of users are usually catered for with different interfaces that are designed to accommodate for their tastes and shopping interests. 
Three ingredients are shared by virtually all e-commerce interface designs:
\begin{itemize}[leftmargin=*,nosep]
\item \textbf{navigation options}, which refer to elements that help users reach a certain part of the e-commerce platform, e.g., search bars, paginations, and universal menus;
\item \textbf{input options}, which are elements of an e-commerce platform for which the user provides input from their end, e.g., search bars, checkboxes, dropdown lists, dropdown buttons, toggles, and other text fields; and 
\item \textbf{information components}, which are composed of various types of information about the products or services listed on e-commerce platforms, such as search results, recommendation results, item titles, images, item information, question answer pairs, user reviews, and tooltips. 
\end{itemize}

\noindent%
As we dive into the problem of information discovery in e-commerce, information components are our main focus in this section. 
We find that almost every e-commerce site provides six information components: \emph{search results}, \emph{recommendation results}, \emph{item titles},  \emph{item features and descriptions}, \emph{question answer pairs}, and \emph{user reviews of the item}. 
In Section~\ref{subsec:2.1.2}, we describe studies on information interface analysis of these six components. 

\subsection{Analyzing information components}
\label{subsec:2.1.2}

Many studies have been devoted to analyzing the effect of different information components. In this section, we summarize studies that focus on search results, recommendation results, titles, item descriptions, question answering, and reviews, respectively.

\subsubsection{Search results in e-commerce}

\begin{figure}[!t]
	\centering
	\includegraphics[width=0.95\columnwidth]{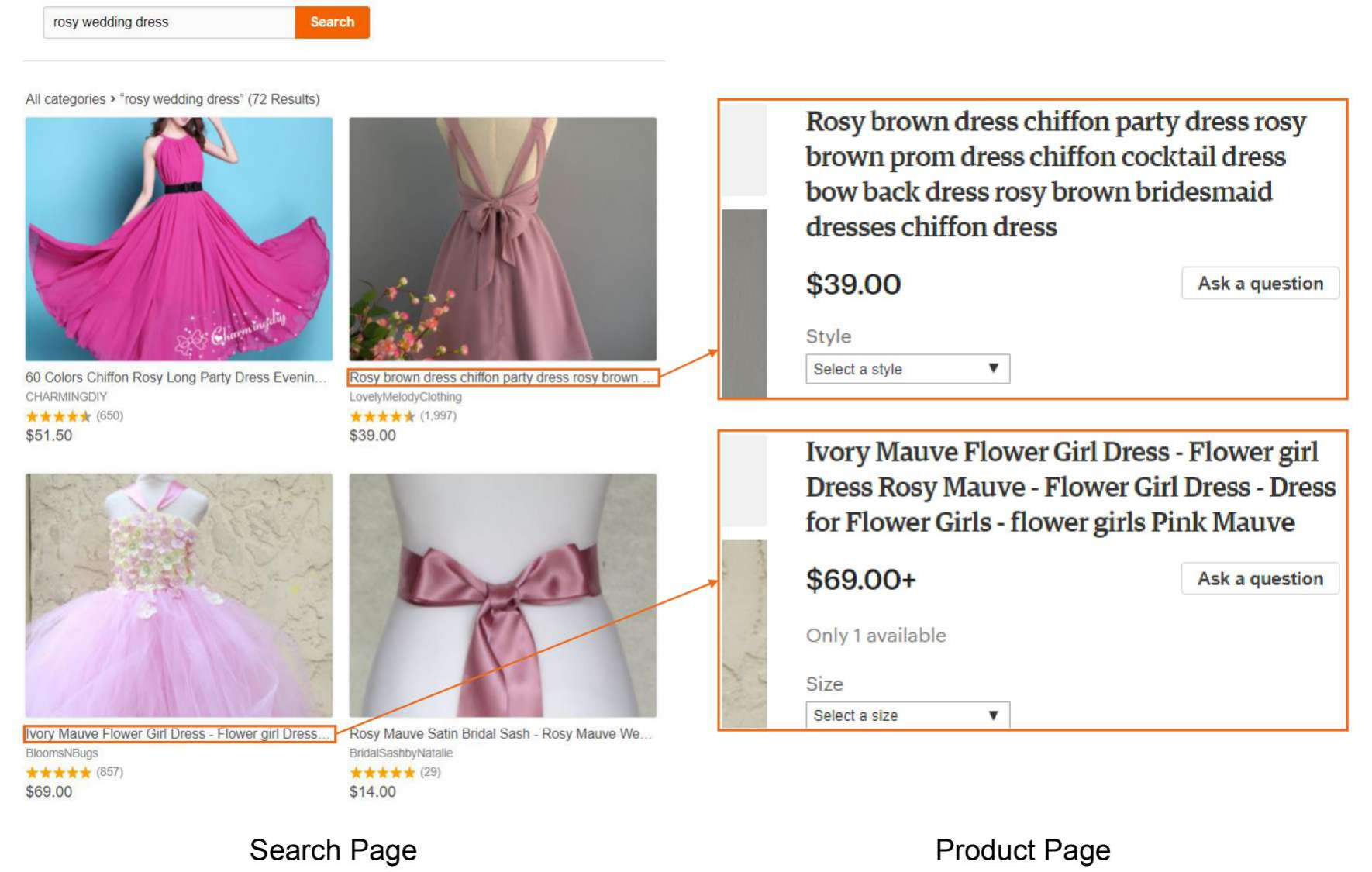}
	\caption{Illustration of a sample search session in an e-commerce platform. The query is ``rosy wedding dress,'' and the search result page is shown on the left and a portion of the item page for two items is shown on the right. This search session consists of two stages: 
	(i) selecting an item to click from a ranked list, and 
	(ii) deciding whether to purchase the item by reading its detailed description.
	Image source: \citep{wu2018turning}. }
	\label{fig_ch2sec21_etsydemo}
\end{figure}

For all e-commerce information components, search and recommendation are the two main tasks in most of e-commerce platforms. 
E-commerce search engines are often the starting points for many online consumers~\citep{wu2018turning}. 
E-commerce sites typically feature two-stage search interfaces.
As shown in Fig.~\ref{fig_ch2sec21_etsydemo}, in an e-commerce search session,\footnote{As defined in web-based search engines, a search session refers to all queries made by a user in a particular time period with a consistent underlying user need~\citep{eickhoff2014lessons}.} a consumer first searches using a query, leading to a result page, and then selects an item to click on the result page; after that, the user decides whether or not to purchase the item by examining its detailed description on the so-called item page.

E-commerce search engines provide category options with the search bar.
During the early development of e-commerce search, interfaces of different types have been considered, e.g., devoted type, divided type, co-existing type, and multi-page type~\citep{lu2006clustering}.
But with the development of e-commerce search, these types of interfaces have been blended by e-commerce platforms.
Currently, a typical e-commerce search system includes three main components: query processing, candidate retrieval and ranking~\citep{zhang2020towards}.
In query processing, the search engine rewrites a query from the user into a term-based representation that can be processed by downstream components.
In the candidate retrieval stage, the system uses the inverted index to retrieve candidate products to match queries. 
Finally, the ranking component orders the retrieved candidates based on factors such as relevance, and predicted conversion ratio.
We will discuss research into the principles and strategies of all three components in Section~\ref{chapter:search} in more detail.

\subsubsection{Recommendation results in e-commerce platforms}
\label{subsubsection:3122}
\begin{figure}[!t]
	\centering
	\includegraphics[width=0.9\columnwidth]{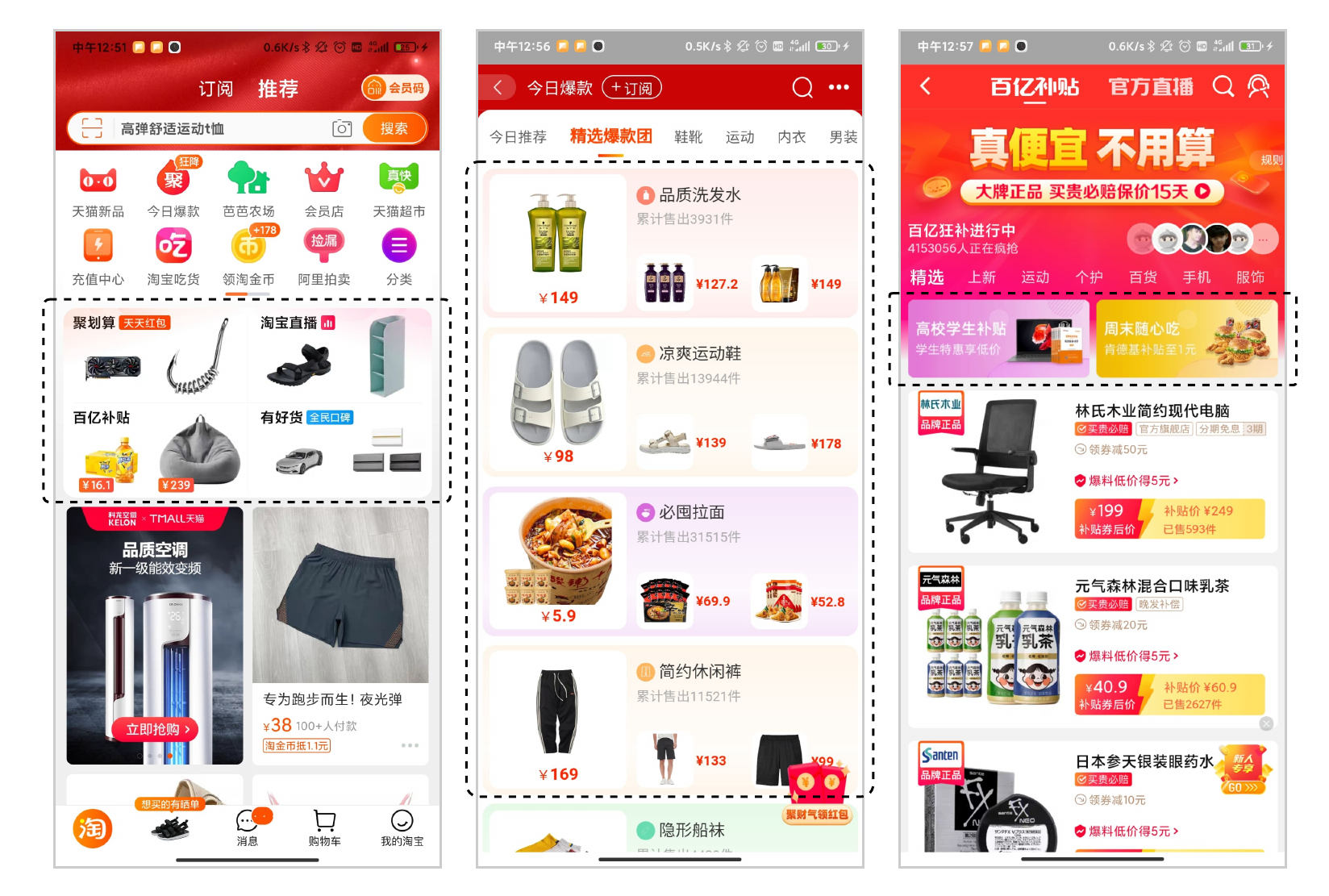}
	\caption{E-commerce recommendation scenarios in Taobao. The areas highlighted with dashed rectangles are personalized for users. Images and textual descriptions are also generated for better user experience. Image source: \citep{wang2018billion}. }
	\label{fig_ch2sec21_210704}
\end{figure}

For many e-commerce platforms, recommendations have become the most important service to help users find their needed items.
E.g., recommendations have been reported to contribute to the majority of both revenue and traffic in Taobao~\citep{wang2018billion}, where one billion users can be connected to two billion items.
To this end, the homepage on the mobile Taobao app is generated based on consumers' past behavior via recommendation algorithms, as illustrated in Fig.~\ref{fig_ch2sec21_210704}.
Fig.~\ref{fig_ch2sec21_210704} shows three recommendation areas displayed on the home page: a list of recommendation interfaces, a ``popular products'' list, and a promotion list, respectively. 
Each recommendation area is provided based on users' past behaviors with recommendation strategies.
As user behavior varies between scenarios, the recommendation strategy also needs to consider specific patterns and user preferences specific for each recommendation scenario. 
For example, on the item page, the recommendation strategy needs to provide either relevant or similar items to the item that the user is focusing on, whereas the recommendation list on the home page shows the recommendation results considering the user's personalized preferences~\citep{zhouwsdm2018}. 

\begin{figure}[!t]
  \centering
  \subfigure[JD.com]{
    \label{fig_ch2sec21_recres1}
    \includegraphics[width=0.32\columnwidth]{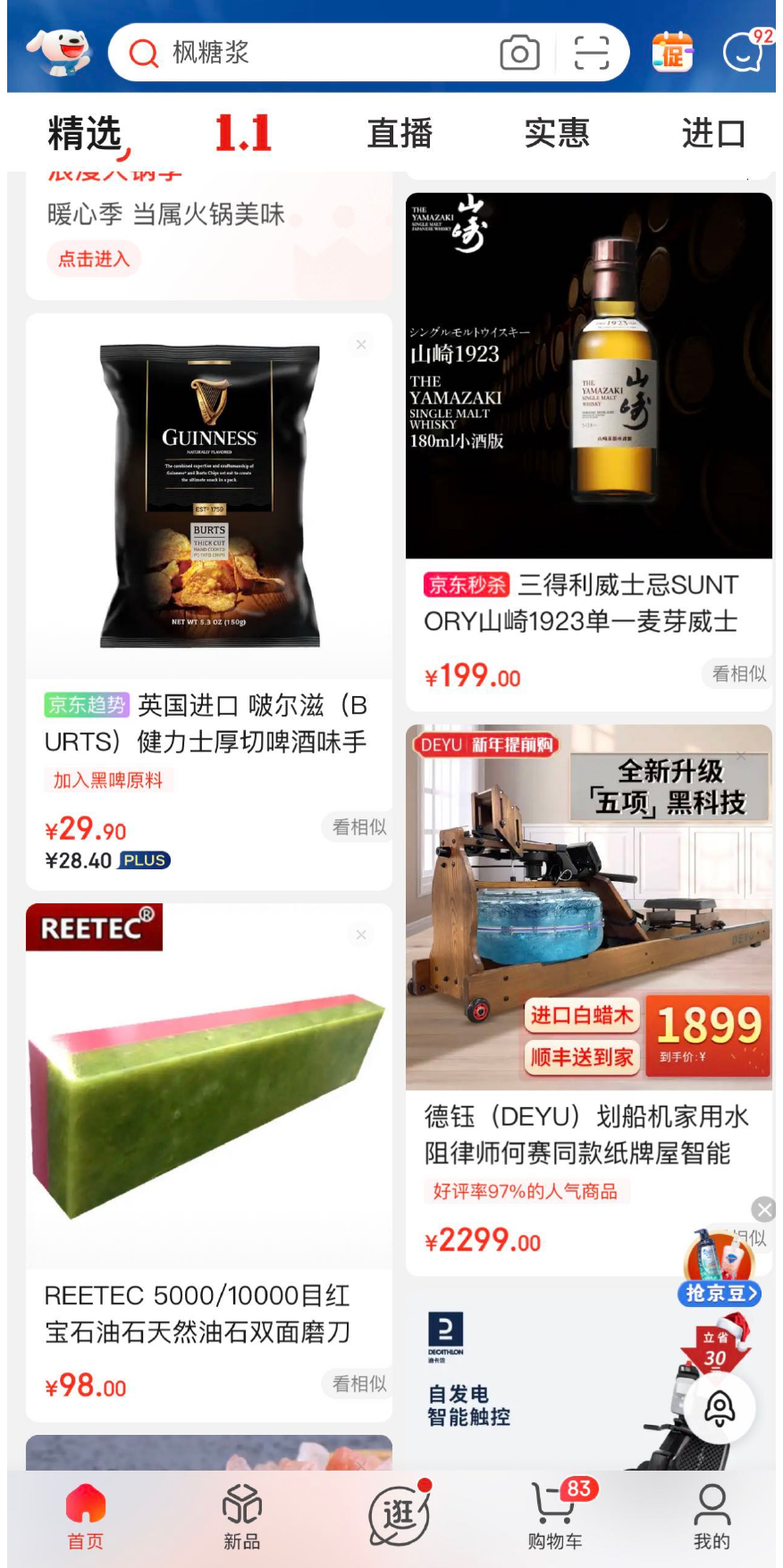}}
  \subfigure[Amazon]{
    \label{fig_ch2sec21_recres2}
    \includegraphics[width=0.32\columnwidth]{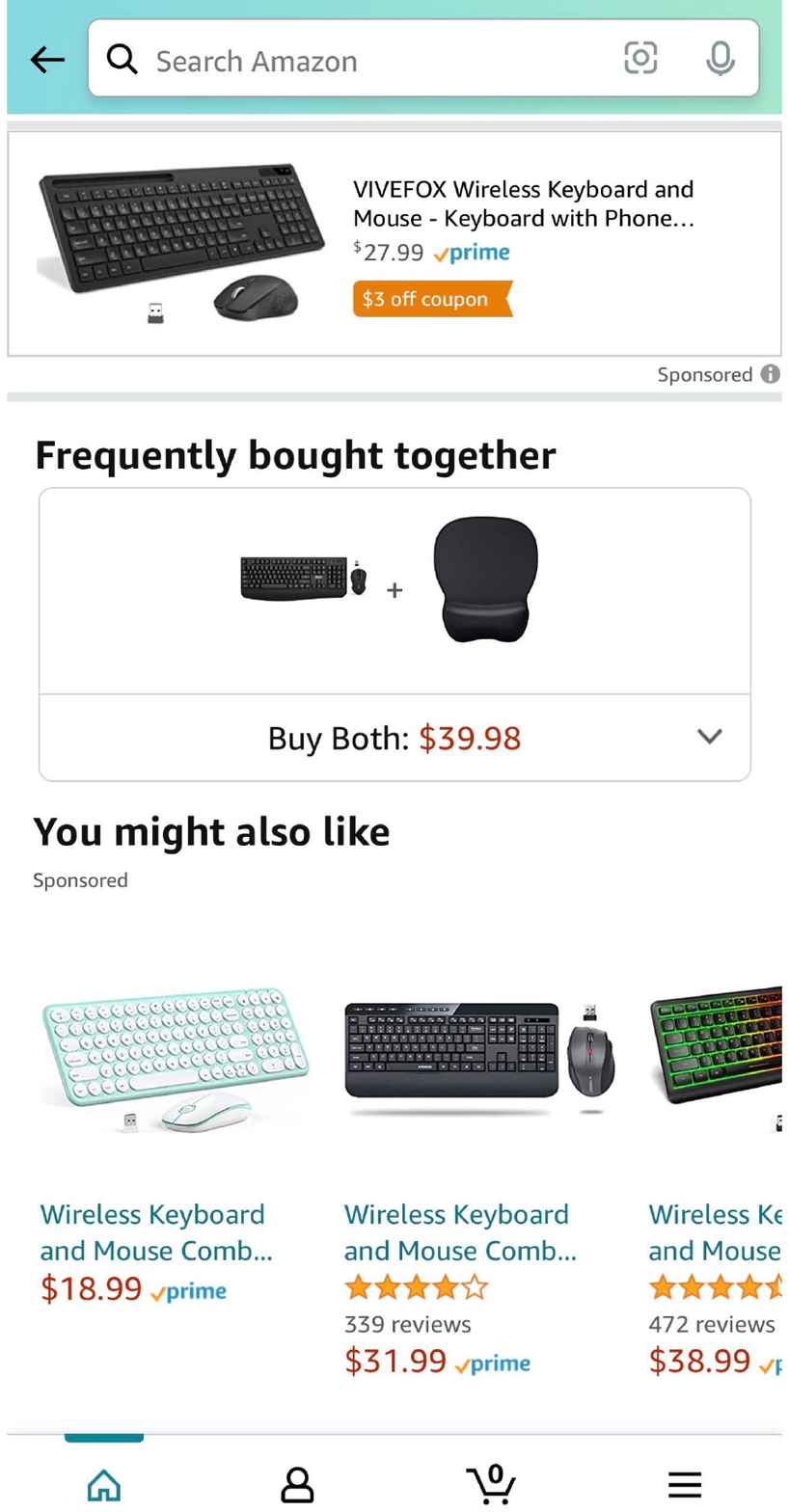}}
   \subfigure[Tmall.com]{
    \label{fig_ch2sec21_recres3}
    \includegraphics[width=0.32\columnwidth]{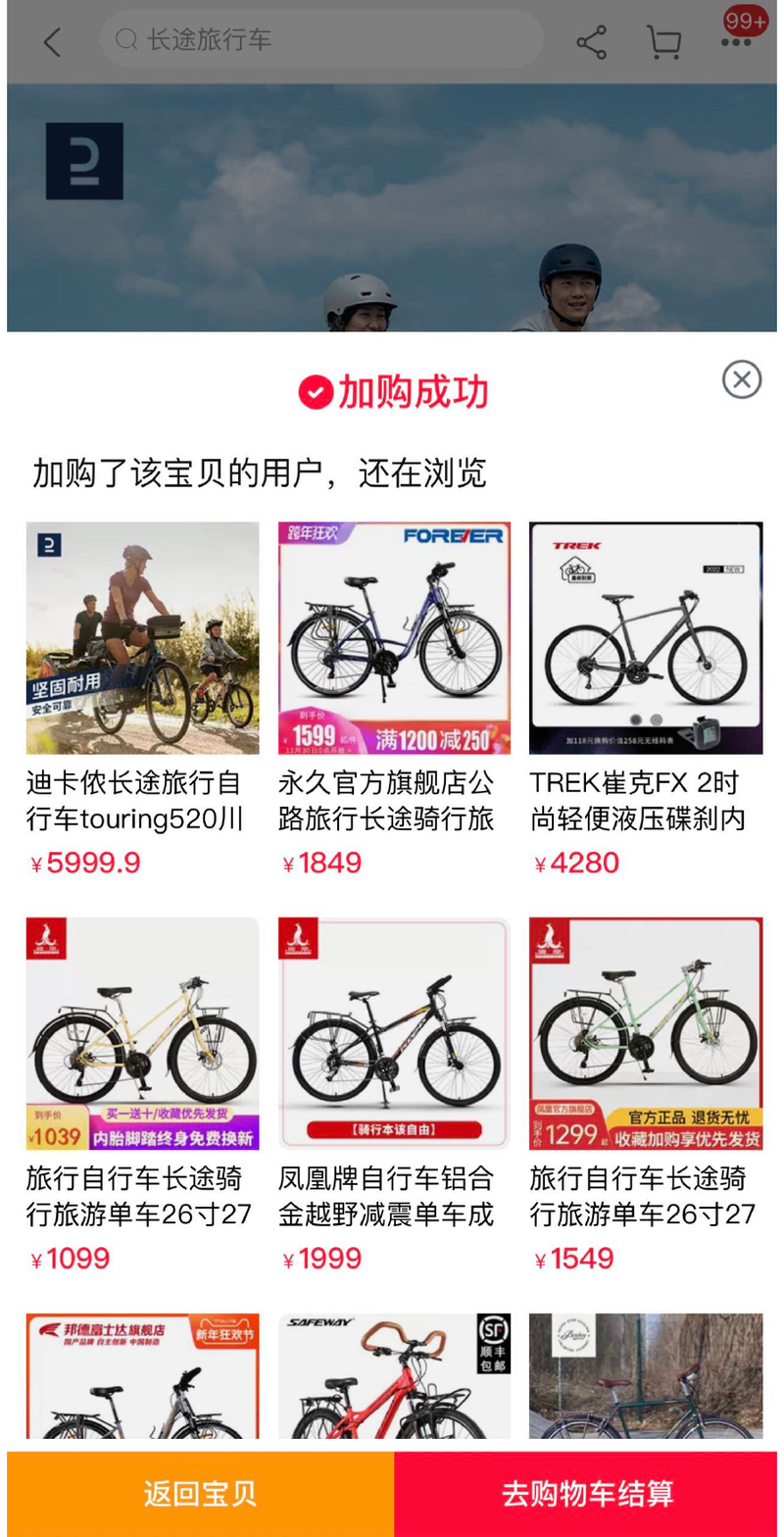}}
       \caption{Recommendation results exposed to users in three e-commerce platforms. Image sources: \url{JD.com}, \url{Amazon.com}, and \url{Tmall.com}.}
  \label{fig_ch2sec21_recres}
\end{figure}

Different types of recommendation results may be shown at different stages of a customer's.
Examples include ``substitutes'' (see Fig.~\ref{fig_ch2sec21_recres1} and ``complementary items'' before and after the user adds a product to their cart (see  Fig.~\ref{fig_ch2sec21_recres2} and~\ref{fig_ch2sec21_recres3}, respectively).
Once the consumer clicks a recommended product, the system will automatically jump to the product detail page, which includes product titles, product descriptions, categories, ratings, and reviews.
We will discuss more details about strategies and technologies of e-commerce recommendation in Section~\ref{chp:rec}.

\subsubsection{Product titles in e-commerce platforms}

\begin{figure}[!t]
  \centering
  \subfigure[Search result page]{
    \label{fig_ch2sec21_titlecomp1}
    \includegraphics[width=0.48\columnwidth]{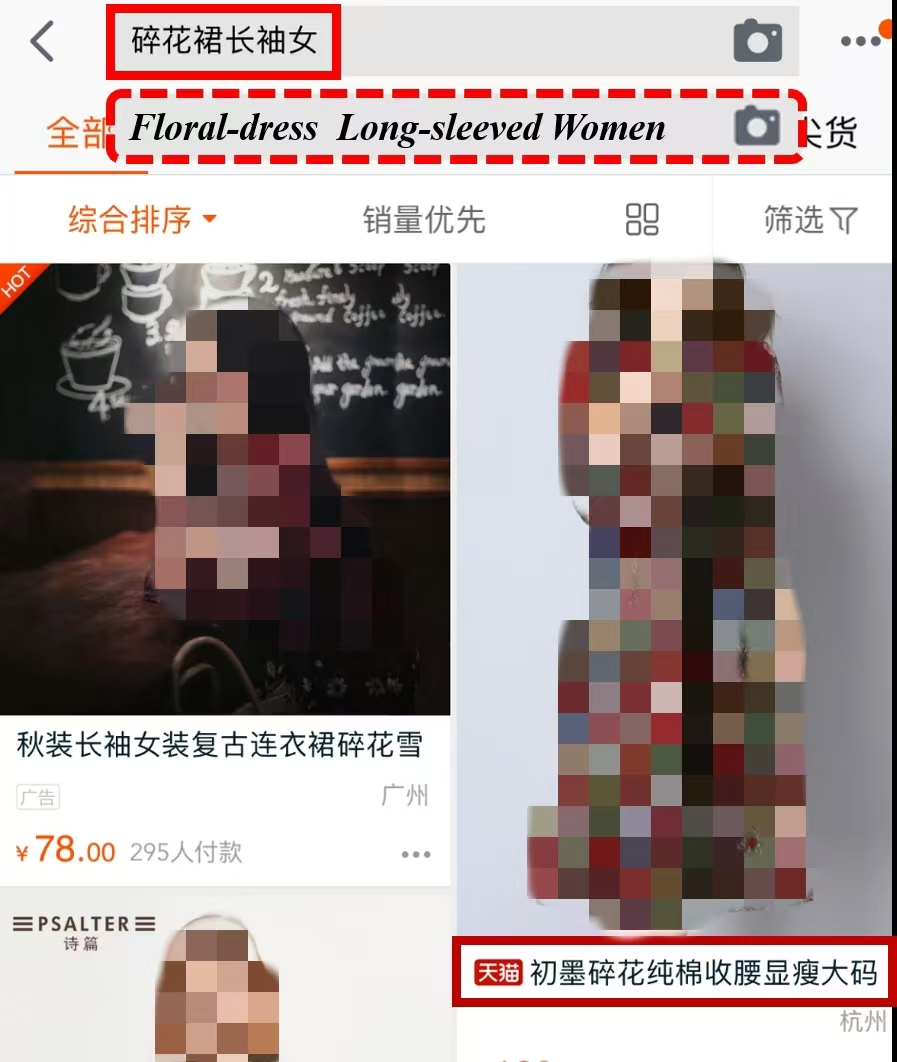}}
  \subfigure[Item detail page]{
    \label{fig_ch2sec21_titlecomp2}
    \includegraphics[width=0.48\columnwidth]{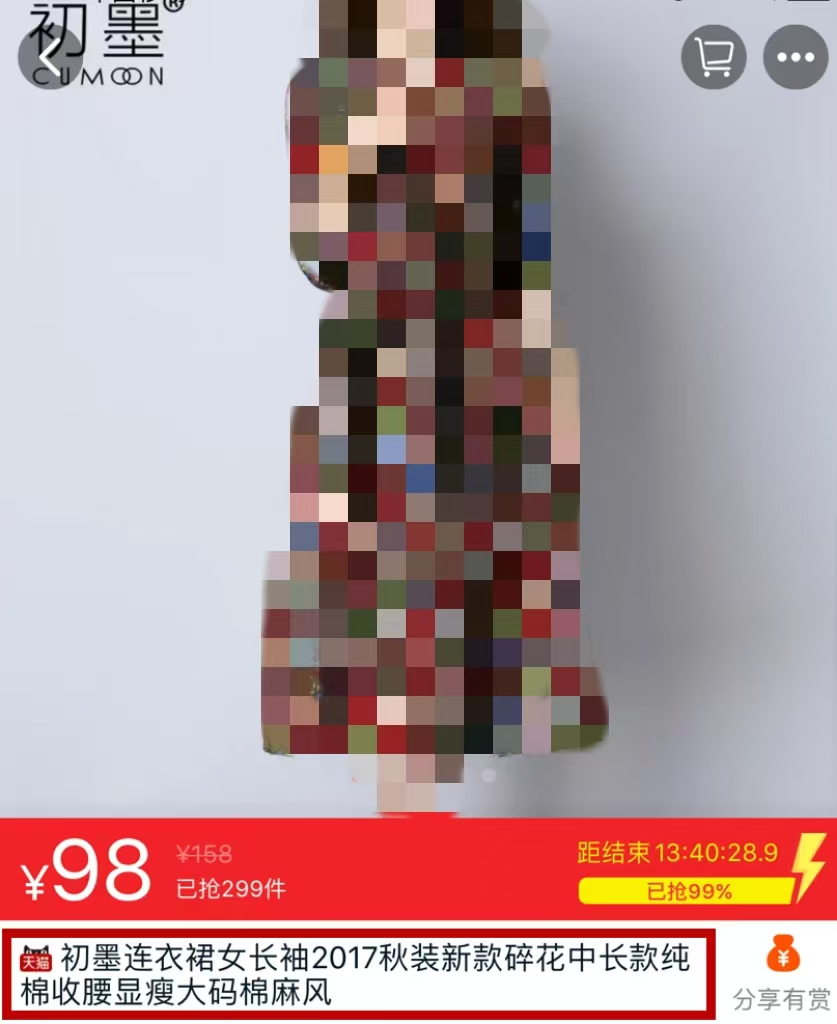}}
       \caption{Given a query ``floral-dress long sleeve women'' on Tmall, the complete title cannot be displayed in the search result page unless the user proceeds to the detail page further. Image source: \citep{wang2018multi}.}
  \label{fig_ch2sec21_titlecomp}
\end{figure}

Product titles and their images are uploaded by suppliers to showcase their items. 
As most e-commerce platforms at least provide search and recommendation services based on information in the titles, retailers have applied many search engine optimization strategies to titles~\citep{ledford2015search}.
As a result, lots of item titles are lengthy, over-informative, and sometimes incorrect.
Fig.~\ref{fig_ch2sec21_titlecomp}(b) provides an example from Tmall, the largest B2C online shopping platform in China, where the item title is composed of more than 30 Chinese words.
But when a customer browses an item on Tmall Apps, fewer than 10 Chinese words can be displayed due to screen size limitations (Fig.~\ref{fig_ch2sec21_titlecomp}(a)).
Thus, lengthy and verbose titles are inconvenient for mobile e-commerce users to search items on e-commerce platforms. 
Similarly, it has been reported that item titles with less than 80 characters improve the shopping experience on Amazon, because these shorter titles make it easier for customers to find products.\footnote{\url{https://sellercentral.amazon.com/forums/message.jspa?messageID=2921001}} 
Accordingly, research on e-commerce title analysis mainly focuses on obtaining effective compression or summaries of lengthy item titles for e-commerce search.

Item title compression, also called short title extraction~\citep{gong2019automatic}, is meant to extract sufficient words from lengthy and verbose titles to produce a succinct new title to improve the user experience on mobile devices~\citep{wang2018multi}. 
%
%
Inspired by neural extractive document summarization methods~\citep{ren2017leveraging}, item title compression methods apply neural networks to weight the importance of each word in the item title.
\citet{gong2019automatic} introduce a feature-enriched neural extractive model to extract short titles. 
Specifically, the authors apply a recurrent neural network as a sequential classifier with three types of features: content, attention, and semantics respectively. 
By using user search logs as external knowledge, \citet{wang2018multi} construct a multi-task learning approach for improving item title compression. The proposed method is composed of two seq2seq components which share an identical encoder. 
The authors combine these two components with an overall pointer neural network~\citep{vinyals2015pointer} to automatically select the most informative words from the given item title.

\begin{figure}[!t]
	\centering
	\includegraphics[width=0.95\columnwidth]{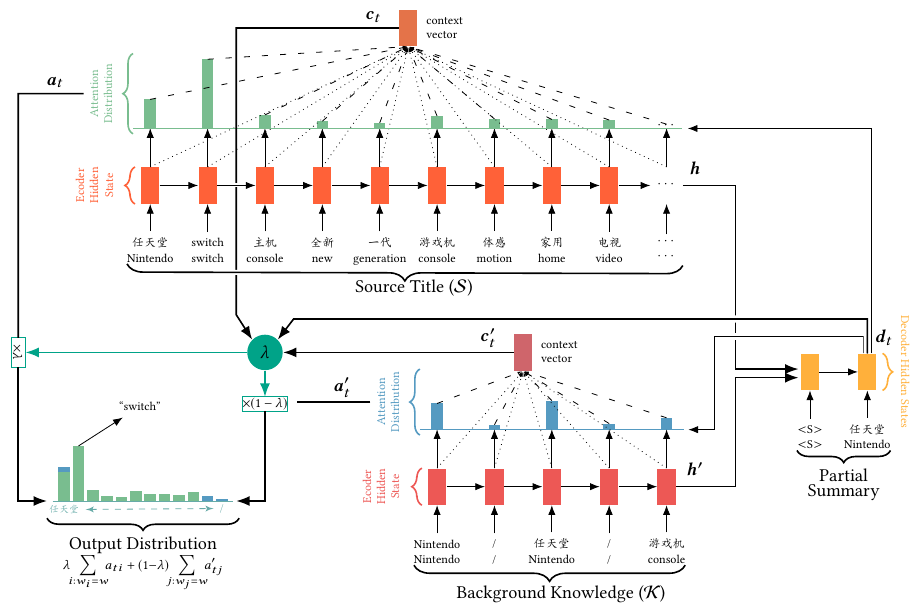}
	\caption{Multi-source pointer network (MS-Pointer) with two encoders for item title compression. MS-Pointer copies words from two encoders. At each decoding time step, a soft gating weight $\lambda \in [0,1]$ is calculated to weight the probability of words from the source title, versus words from the background knowledge. The final output distribution is the weighted sum of attention distributions $a_{t}$ and $a'_{t}$. Image source: \citep{sun2018multi}.}
	\label{fig_ch2sec21_sunfeicikm}
\end{figure}

Pointer neural networks easily omit key information. 
To tackle this problem, \citet{sun2018multi} introduce a multi-source pointer network model, named the multi-source pointer network (MS-Pointer), by considering two extra constraints: 
\begin{enumerate*}[label=(\roman*)]
\item irrelevant information reduction; and 
\item the key information retainment. 
\end{enumerate*}
Fig.~\ref{fig_ch2sec21_sunfeicikm} provides an overview of MS-Pointer, with two encoders. 
In MS-Pointer, in addition to the encoder for the source title, the authors add another knowledge encoder that uses an LSTM to embed the brand name and the commodity name. 
As shown in Fig.~\ref{fig_ch2sec21_sunfeicikm}, MS-Pointer combines the original title ``Nintendo switch console\ldots'' and background knowledge ``brand name: Nintendo'', and then it generates the short title about the item ``Nintendo switch''.
More recently, \citet{DBLP:conf/emnlp/FetahuCRM23} have proposed an instruction fine-tuning strategy to summarize product titles according to various criteria such as the number of words in a summary or the inclusion of specific phrases.

\begin{figure}[!t]
	\centering
	\includegraphics[width=0.85\columnwidth]{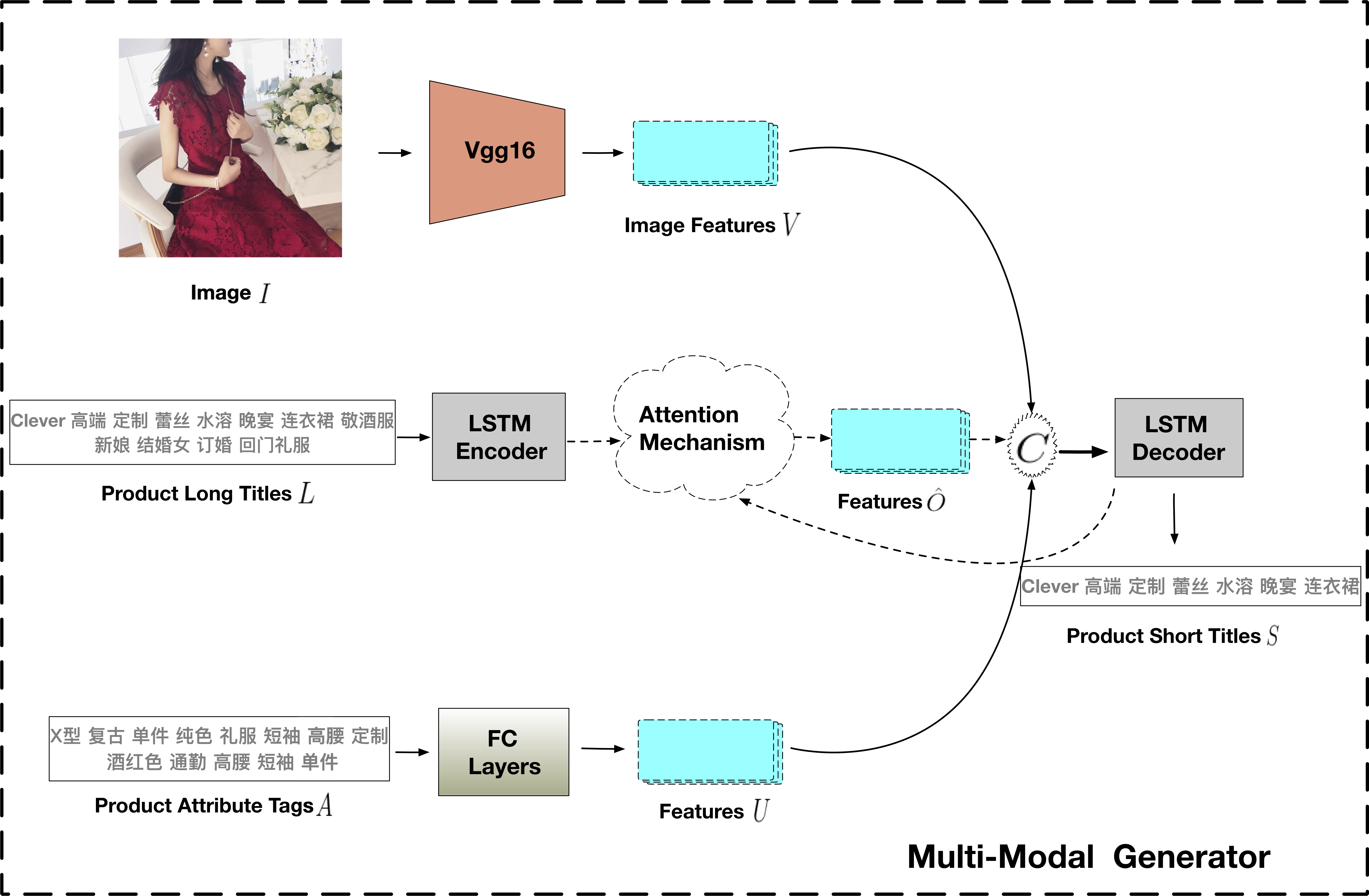}
	\vspace*{2mm}
	\caption{Overall framework of the MM-GAN model for short item title generation. Image source: \citep{zhang2019multi}.}
	\label{fig_ch2sec21_jianguofig2}
\end{figure}

The task of title generation has been proposed to extend the task of title compression into a text generation problem. 
Unlike title compression, which only extracts words from item titles, the task of title generation is to \emph{generate} a short item title so as to address the problem of inaccurate item titles in e-commerce~\citep{zhang2019multi}. 
To generate a succinct and accurate short title from a long source title, \citet{zhang2019multi} offer a multi-modal generative adversarial network, named MM-GAN, which addresses the title generation task as a reinforcement learning problem.
MM-GAN is composed of two main components, a title generator and a discriminator (Fig.~\ref{fig_ch2sec21_jianguofig2}). 
Given the source title and its corresponding tags or features, the generator applies an LSTM-based network to generate a short item title.
The discriminator, i.e., a binary classifier, distinguishes whether the generated short titles are human-generated or machine-generated. 
Thus, an adversarial learning procedure is constructed, in which the quality of the short title depends on its ability to fool the discriminator into believing it is a human-generated one, and the output of the discriminator is a reward for the generator to improve the generation performance.
Recently, scene marketing has become a new marketing mode for product promotion where scene scenarios are created to demonstrate product functions~\citep{zhao2020data}.
To help the e-commerce system find scene topics, \citet{lin2022automatic} propose a topic generation method to generate scene-based titles in e-commerce.

\subsubsection{Product descriptions in e-commerce platforms}

\begin{figure}[!t]
  \centering
  \subfigure[Recommendation page on Taobao]{
    \label{fig_ch2sec21_tozhangf1a}
    \includegraphics[width=0.48\columnwidth]{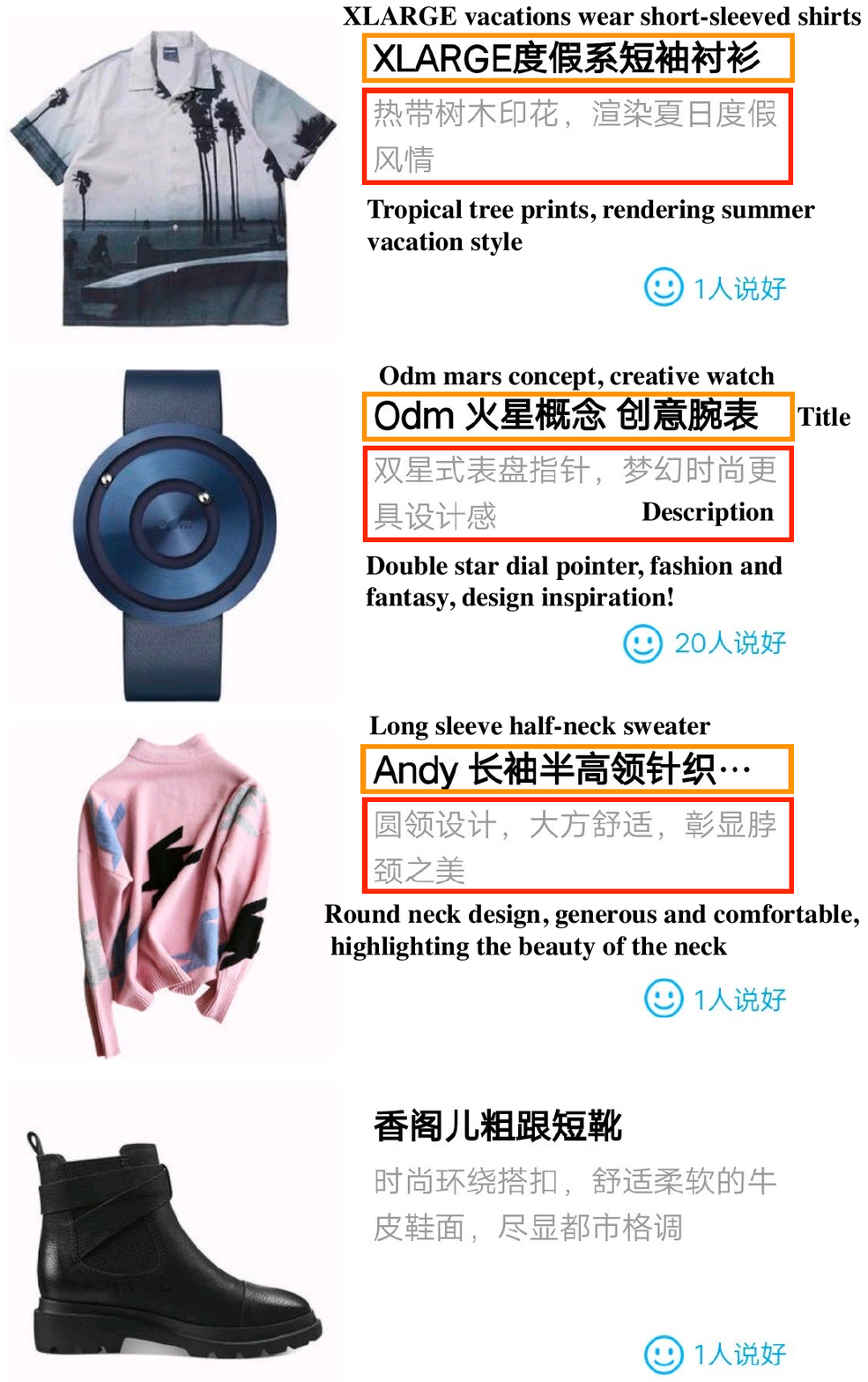}}
  \subfigure[Item detail page on Amazon]{
    \label{fig_ch2sec21_tozhangf1b}
    \includegraphics[width=0.48\columnwidth]{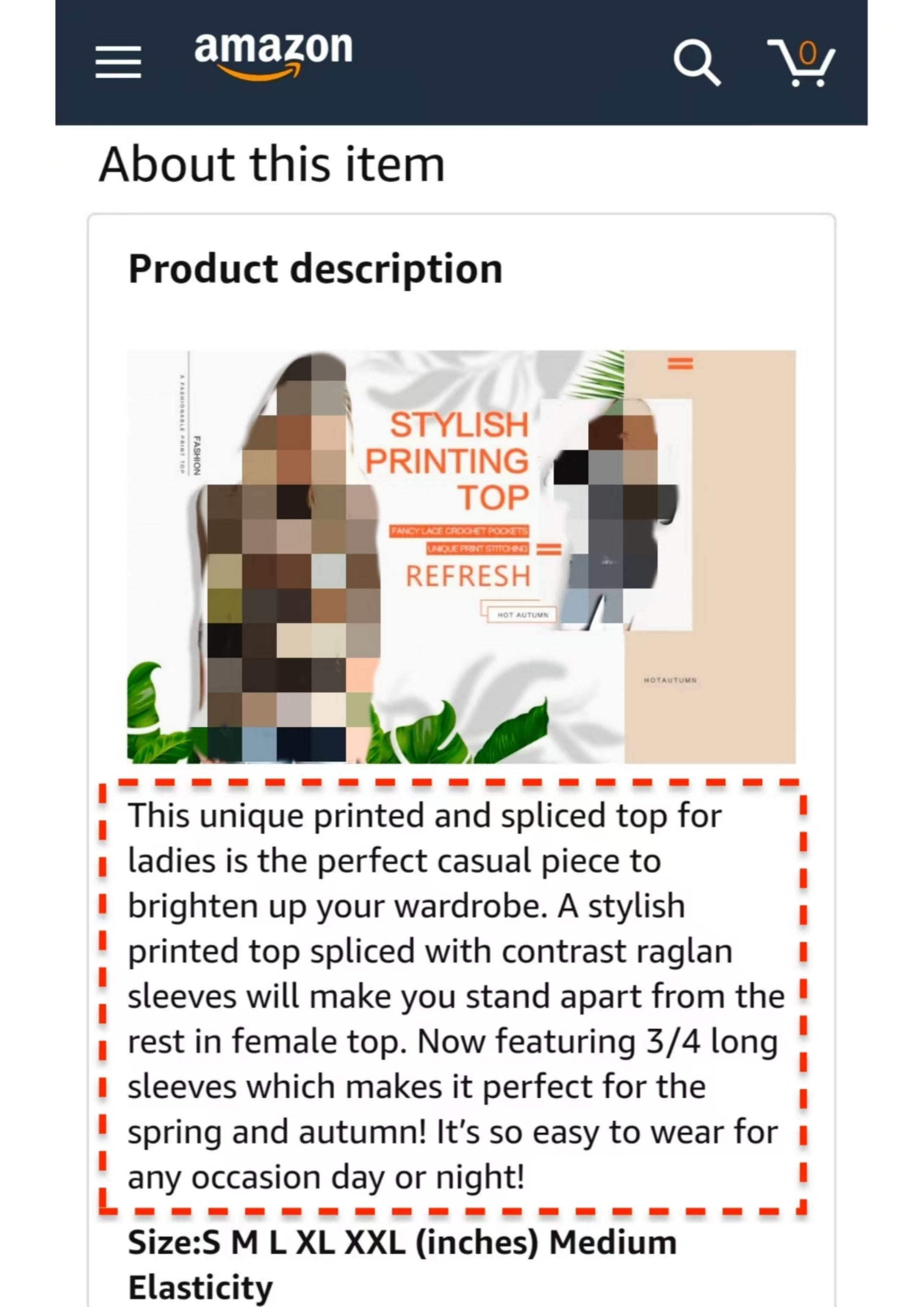}}
       \caption{Item descriptions are widely used in e-commerce platforms, e.g., (a) Taobao and (b) Amazon. Image source: \citep{zhang2019automatic}.}
  \label{fig_ch2sec21_tozhangf1}
\end{figure}

\begin{figure}[!t]
	\centering
	\includegraphics[width=0.85\columnwidth]{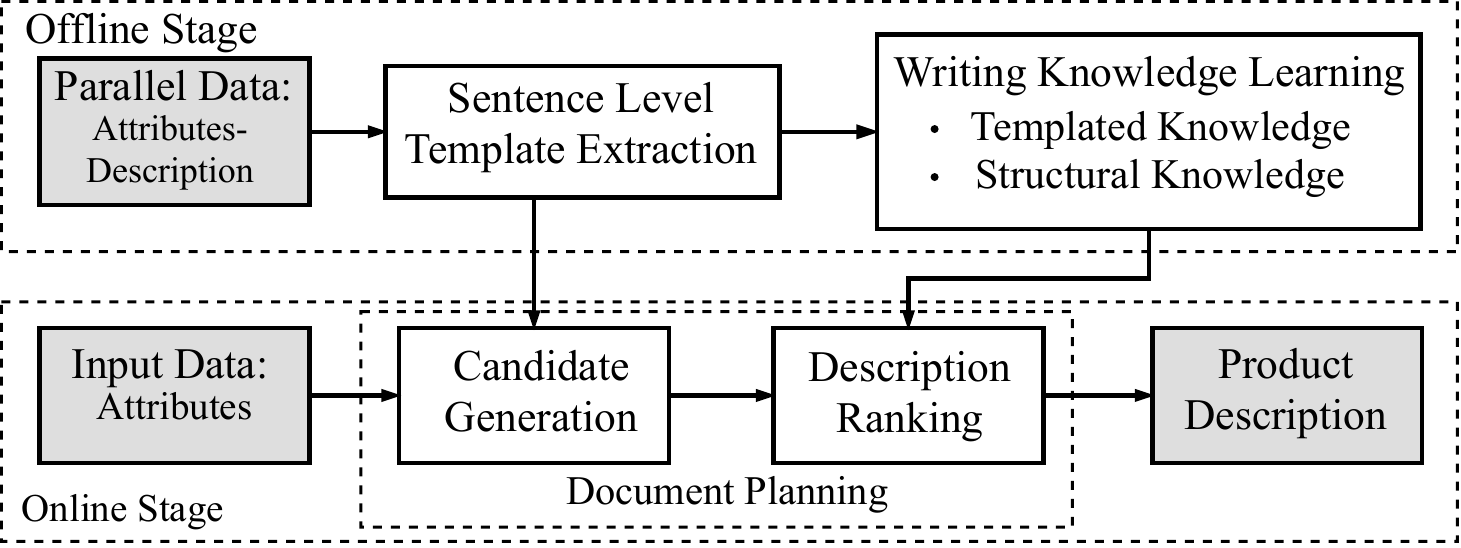}
	\caption{Overall framework for item description generation with pretrained writing knowledge. Image source: \citep{wang2017statistical}.}
	\label{fig_ch2sec21_jinpengfig2}
\end{figure}

As shown in Fig.~\ref{fig_ch2sec21_tozhangf1}, many e-commerce platforms provide a short description for each item so as to showcase the features of the item. 
As an important factor in content marketing, the item description is key for increasing consumer engagement.
During the early years of e-commerce, item descriptions were usually written or edited by human copywriters. 
However, the availability of an increasing number of items in e-commerce makes this manual process too costly.
Moreover, with the development of virtual assistants in e-commerce, such as Alexa and Tmall Genie, there is a growing demand for automatically generating a short description given item attributes.
To address this demand, the task of item description generation has been proposed. 
Item description generation needs to generate an item's description from a series of complicated attributes. \citet{wang2017statistical} detail a statistical framework to weight the relative importance of the attributes of an item and to maintain accuracy at the same time. 
In Fig.~\ref{fig_ch2sec21_jinpengfig2}  we specify the framework of the proposed item description model.
By combining sentence-level templates extracted from the input data with  knowledge from a pre-trained dataset, the authors generate and rank candidate item descriptions through an online \emph{document planning} stage.

Unlike early studies that focused on generating item descriptions purely from the item's attributes, \citet{zhang2019automatic} generate a pattern-controlled item description from multiple features, e.g., titles and item categories. 
Based on the copy mechanism~\citep{gu2016incorporating}, the authors propose a pattern-controlled pointer-generator network (PGPCN) to generate the description. 
In PGPCN, a transformer is applied in the encoder component, whereas the decoder is used to control the pattern (e.g., the category, the length, and the style of the description) of the item. 

\begin{figure}[!t]
	\centering
	\includegraphics[width=0.95\columnwidth]{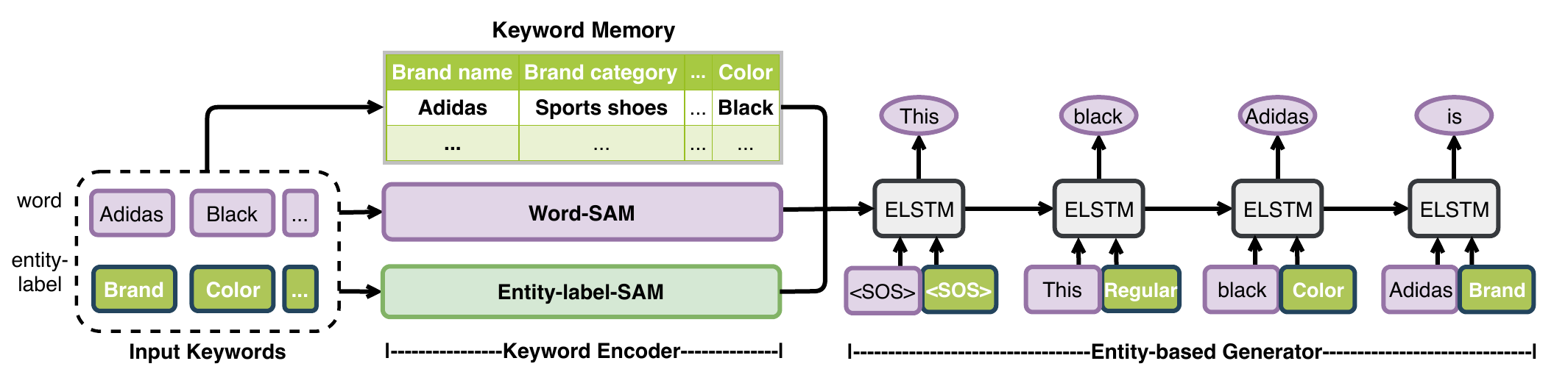}
	\caption{Overview of the fidelity-oriented item description generator. The whole model is divided into two components: (i) a keyword encoder, and (ii) an entity-based generator. Image source: \citep{chanstick2019}.}
	\label{fig_ch2sec21_xiuyingf1}
\end{figure}

It is important that the descriptions generated for item description are grounded in facts.
To generate a fact-based description, \citet{chanstick2019} offer an encoder-decoder framework, called the fidelity-oriented product description generator (FPDG), by searching key information from keyword labels. The authors establish semantic connections between item keywords and the generated product description.
As shown in Fig.~\ref{fig_ch2sec21_xiuyingf1}, FPDG has two main components: 
\begin{enumerate*}[label=(\roman*)]
\item a keyword encoder that stores the word and its entity label in the token memory and self-attention modules, and 
\item an entity-based generator that generates an item description based on the memory and self-attention modules.
\end{enumerate*}

\begin{figure}[!t]
	\centering
	\includegraphics[width=0.95\columnwidth]{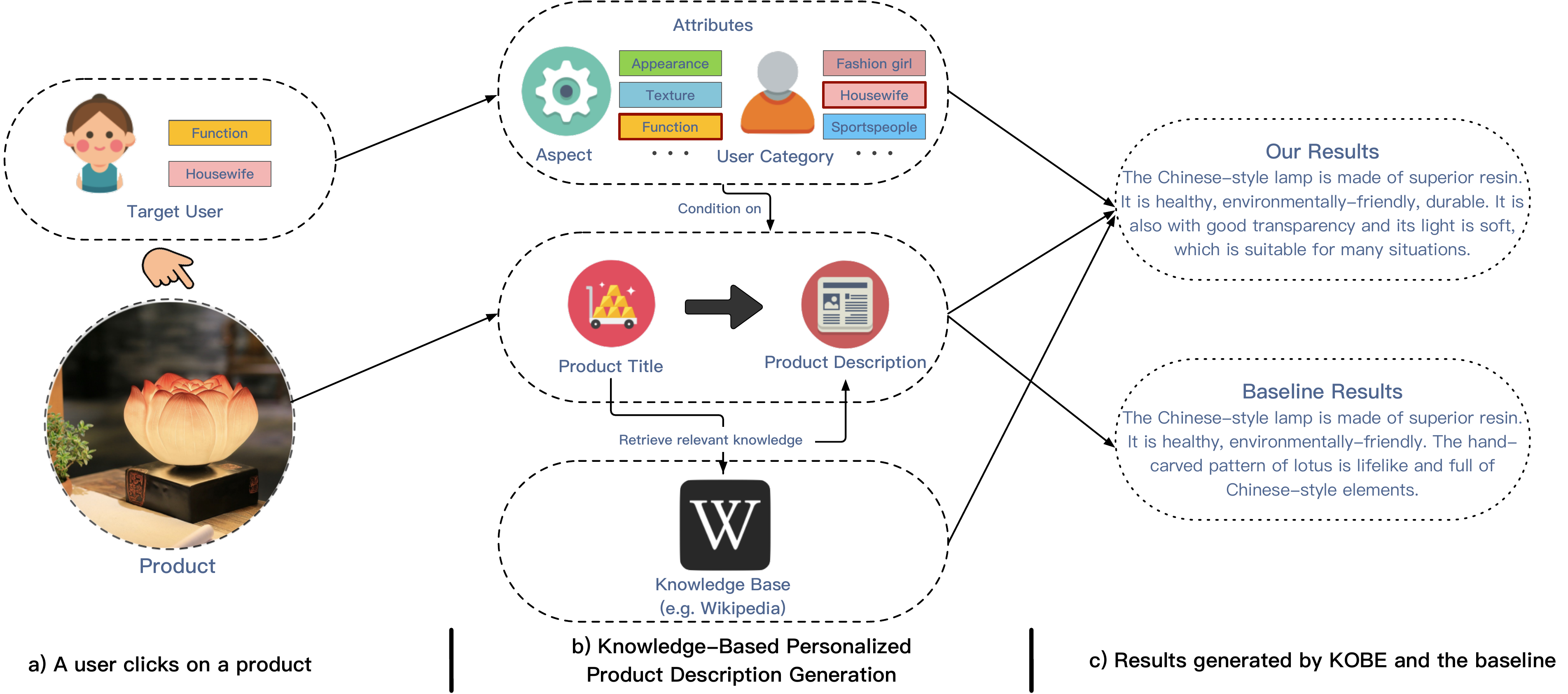}
	\caption{An example of knowledge-based personalized item description generation. The example is divided into three parts: (a) a user clicks an on item; (b) knowledge-based personalized item description generation; and (c) results generated by \citet{chen2019towards}'s proposed method and a baseline. Image source: \citep{chen2019towards}.}
	\label{fig_ch2sec21_qibin2}
\end{figure}

Personal interest is neglected by all of the above approaches that generate descriptions given attributes or keywords. 
To address this shortcoming, \citet{chen2019towards} propose a knowledge-based personalized item description generation strategy. 
The authors extend the encoder-decoder framework \citep{sutskever2014sequence} to a sequence modeling formulation using a self-attention mechanism. 
A large variety of item attributes, including the target user's personalized preference features, are combined in an attribute fusion component through multi-layer attention mechanisms; retrieved external knowledge is incorporated in a \emph{knowledge incorporation} component. In Fig.~\ref{fig_ch2sec21_qibin2}, we provide an example of the knowledge-based personalized item description generation. 

\subsubsection{Question answering in e-commerce}
\label{ch2:sec2:qa}

\begin{figure}[!t]
  \centering
  \subfigure[Question answering service on Amazon]{
    \label{fig_ch2sec21_qab}
    \includegraphics[height=4cm]{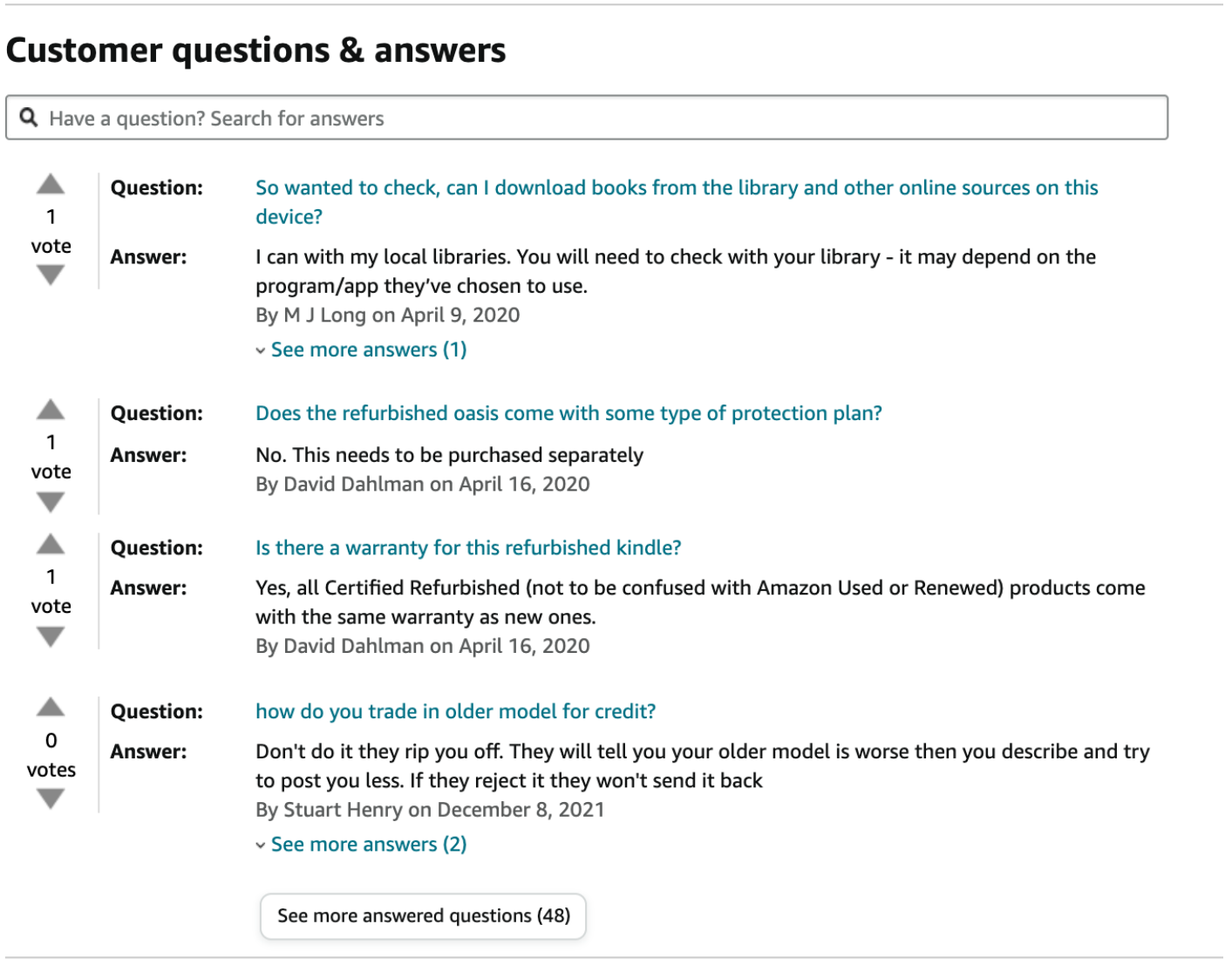}}
  \subfigure[Question answering service on JD.com]{
    \label{fig_ch2sec21_qaa}
    \includegraphics[height=4cm]{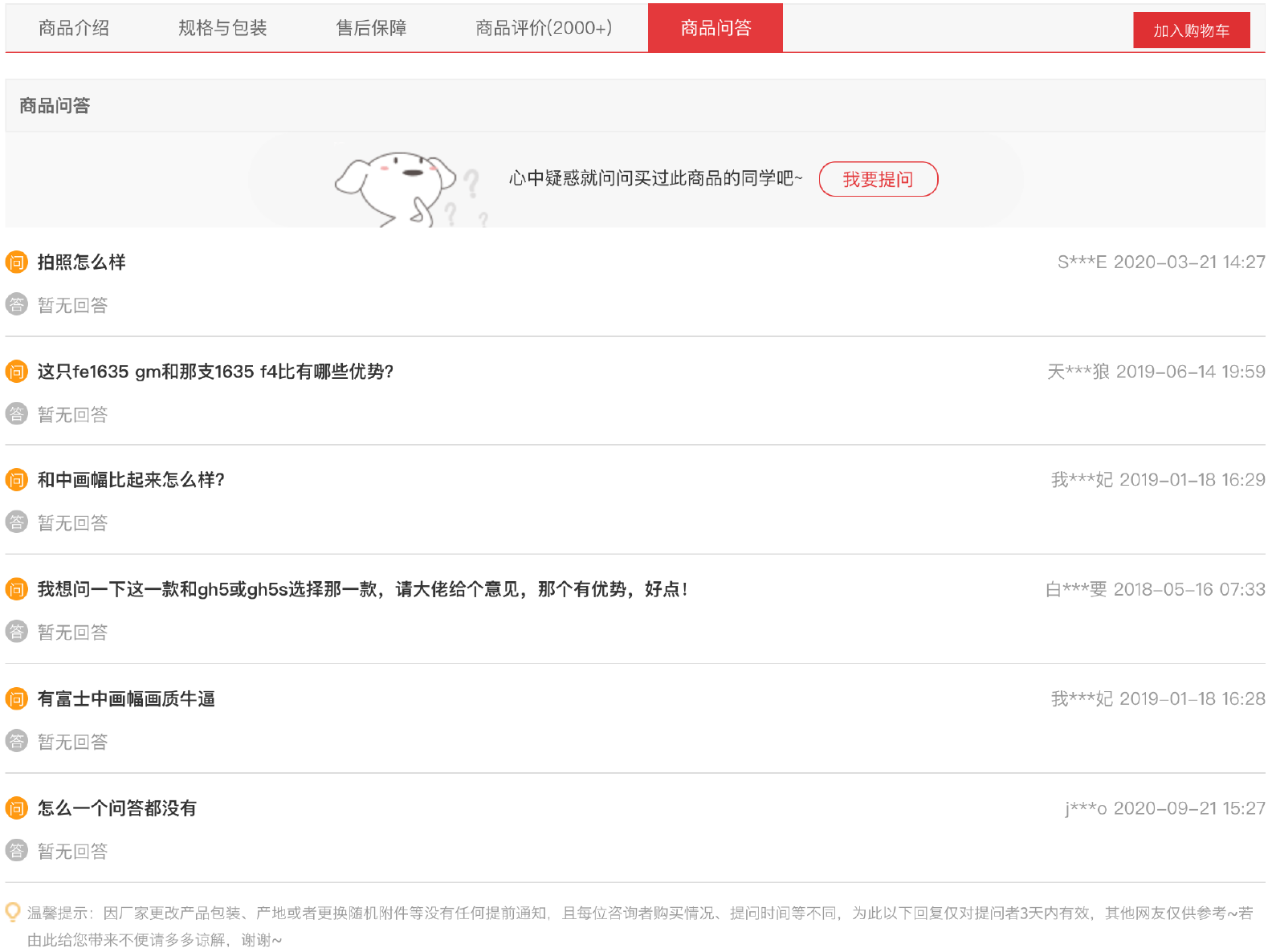}}
       \caption{Question answering services are widely applied in e-commerce platforms, (a) Amazon and (b) JD.com. Image sources: \url{Amazon.com} and \url{JD.com}.}
  \label{fig_ch2sec21_qa}
\end{figure}

Question-answering (QA) systems are designed to provide direct and concise answers to user queries by understanding the intent behind a question and retrieving or generating the most relevant information. QA systems aim to deliver specific answers from structured or unstructured data sources, such as web documents or knowledge bases. QA systems are essential across a variety of domains, including web search, customer support, where users seek quick, accurate, and contextually relevant information~\citep{radev2002evaluating,tapeh2008knowledge,yin2015neural}.
To increase the number of sales, most e-commerce portals provide a QA service to facilitate the customers' shopping procedure by answering their questions about products~\citep{gao2019product,feng2021multi}.
Currently, on many e-commerce sites, a user can ask a question about a product, and the QA system allows some users (e.g., customers who bought this product) to provide answers~\citep{gao2019product}.
In Fig.~\ref{fig_ch2sec21_qa}, we show examples of question-answering services at Amazon and JD.com. 
More detailed discussions of e-commerce QA are provided in Section~\ref{ch6:sec1}.

\subsubsection{User reviews in e-commerce}
\label{ch5:sec2:ar}

User reviews serve as a type of reliable information about the quality of items on e-commerce platforms. 
User reviews have been shown to play an essential role in determining user preference~\citep{liang2015consumer,huebner2018people,li2018document}. 
In this section, we introduce methods for review analysis in e-commerce. 
Research on review analysis can be organized into four key components:  
\begin{enumerate*}[label=(\roman*)]
\item sentiment classification, 
\item helpfulness prediction,
\item review summarization, and 
\item review generation.
\end{enumerate*}

\paragraph{(i) Sentiment classification in reviews.}
The \emph{sentiment classification} task is to label a given text with a specific opinion label.
It has received lots of attention during the past two decades~\citep{pang2002thumbs,go2009twitter,pan2010cross,kamal2012mining,tang2014learning,pontiki2015semeval,tsytsarau2016managing,sun2017review}. 
Work on sentiment classification in an e-commerce context  has attempted to capture a user's opinion about a specific item from reviews on an e-commerce platform~\citep{sun2017review,tang2015learning,xia2015dual,tripathy2016classification,li2018effect}. 
Traditional approaches to sentiment classification focus on the classification problem given textual attributes of an item~\citep{pang2002thumbs}, while largely ignoring the relation between users and the item.
To address this problem, \citet{tang2015learning} introduce a neural network, the user-product neural network (UPNN), to incorporate user and item information into a document-level sentiment classification procedure.
In particular, the authors jointly embed the user preference information, i.e., ratings and reviews, and item attributes. 
Then, a convolutional neural network is used to predict the sentiment label of the target review. 

Sentiment classification using the relation between users and products or services faces two important challenges: 
\begin{enumerate*}[label=(\roman*)]
\item the sparseness of user-item interactions, and 
\item the information in user embedding methods.
\end{enumerate*}
\citet{chen2016neural} present a fine-grained hierarchical neural network model to  incorporate global user and item information into sentiment classification. Unlike many  sentiment classifiers that use convolutional neural networks, the authors apply a hierarchical LSTM~\citep{gers1999learning} to jointly generate sentence-level representations and document-level representations. Then, user-item interaction information is applied as attention over various regions of a document to enhance the sentiment classification.
\citet{wu2018improving} distinguish between different roles of words and sentences in user reviews: to describe the user's preferences or to describe an item's characteristics. To distinguish between these roles, the authors put forward an attention-based hierarchical neural network model to embed user and item information to generate two text representations with user attention or item attention, respectively. 
\citet{Hao2021wwwsentiment} use fine-grained latent opinion knowledge into the sentiment classification process by using a variational reasoning method.

\paragraph{(ii) Helpfulness prediction.}
Given the fact that an item can be commented on by hundreds of thousands of consumers, the quality of reviews in e-commerce varies considerably and not all reviews are helpful.
To gain insights from helpful reviews, the task of review helpfulness prediction has attracted attention from both academia and industry~\citep{mcauley2016addressing}.
Early studies on review helpfulness prediction employ feature-aware methods, where multiple types of features, such as structural features~\citep{kim2006automatically,susan2010makes,xiong2011automatically}, emotional features~\citep{martin2014prediction}, semantic features~
\citep{yang2015semantic}, argument features~\citep{liu2017using}, and lexical features~\citep{xiong2014empirical}, are successfully applied.

Motivated by the progress of deep neural networks, \citet{fan2018multi} introduce a multi-task neural learning (MTNL) architecture for identifying helpful reviews. \citet{chen2018cross} propose a CNN-based neural network with multi-granularity (i.e., character-level, word-level, and topic-level) features for helpfulness prediction.
\citet{fan2019product} suggest an end-to-end deep neural architecture to capture the intrinsic relationship between the meta-data of an item and its numerous comments that could be beneficial to discover the helpful reviews.
Multi-modal data has become increasingly popular in online reviews. To analyze multi-modal reviews,~\citet{liumulti2021} introduce a multi-modal review helpfulness prediction task that is aimed at exploring multi-modal clues for review helpfulness prediction. The authors describe an item-review coherent reasoning module to capture the intra- and inter-modal coherence between the target item and the review.
\citet{han2022sancl} put forward a selective attention approach, including probe mask generation and mask-based attention computation, for the multi-modal review helpfulness prediction problem.
To mine the mutual information of cross-modal relations in the input, \citet{nguyen2022adaptive} propose an adaptive cross-modal contrastive learning mechanism, with a multi-modal interaction module to correlate modalities' features.

\paragraph{(iii) Review summarization.}
Given a set of user reviews, the task of \emph{review summarization} is to extract the main information from the reviews.
Similar to multi-document summarization, review summarization summarizes a set of item reviews for a single item.
Approaches to review summarization can be divided into two: feature-aware methods and aspect-aware methods.

Feature-aware methods are inspired by previous document summarization methods: \citet{yang2010feature} detail a feature-based item review summarization method to satisfy the detailed information needs of customers.
To jointly summarize reviews and predict ratings in a mobile environment, \citet{liu2012movie} offer a latent semantic indexing based approach to extract features and attributes from user reviews and ratings.
For new items without reviews, a probabilistic retrieval method is proposed to extract relevant opinion features from other items to describe the item information~\citep{park2015retrieval}.  
Another important part of review summarization concerns aspect extraction from user reviews~\citep{chen2014aspect,angelidis2018summarizing,bravzinskas2020few}, where target entities and aspects need to be extracted from opinionated text.
\citet{chen2014aspect} extract prior knowledge automatically from user reviews and propose a fault-tolerant model to extract aspects guided by the knowledge. 
Category hierarchy information is combined with a topic model to improve the performance of aspect extraction~\citep{yang2017aspect}. 
By jointly considering fine-grained aspect-topic-sentiment connections, \citet{tan2017sentence}  propose a generative topic aspect sentiment model.

With the development of deep learning, item review summarization has been tackled from a range of perspectives~\citep{ly2011product,wu2016aspect}. 
For instance, to tackle the weakness of the ``bag of words'' assumption, \citet{he2017unsupervised} propose an unsupervised neural network model. Considering dependencies between adjacent words, the authors used an embedding method with attention mechanism to de-emphasize saliency and extract aspects.
\citet{angelidis2018summarizing} describe a weakly supervised neural framework for the identification and extraction of salient customer opinions that combines aspect and sentiment information.
Using a small number of annotated instances with a large-scale unlabeled corpus, \citet{bravzinskas2020few} suggest a few-shot learning framework for generating an abstractive summary. 
In recent years, pre-trained language models~\citep{vaswani2017attention,kenton2019bert} have been shown to be effective in document summarization~\citep{liu2019hierarchical}. 
A domain-specific generative pre-training method, PEGASUS, has been proposed to address the e-commerce review summarization problem~\citep{3404835zhang3463037}.
Inspired by Vector-Quantized Variational Auto-encoders (VQ-VAE)~\citep{van2017neural}, \citet{angelidis2021extractive} explain an unsupervised neural model, Quantized Transformer (QT), that uses a clustering interpretation of the quantized space to discover popular opinions among hundreds of reviews.
To tackle challenges such as a lack of cross item diversity and consistency, \citet{ovedpass21} offer a method that uses strong pre-trained language models. 
 
\paragraph{(iv) Review generation.}
The task of \emph{review generation} has been proposed to understand how a specific user provides comments on items~\citep{dong2017learning}.
Unlike review summarization, where one extracts salient sentences or generates abstractive summaries, the task of \emph{review generation} is to generate sentences as user reviews to represent users' intention.
Sequence-to-sequence (seq2seq) neural networks have been applied to automatically generate text~\citep{sutskever2014sequence}.
However, it is difficult to directly use traditional seq2seq models to generate reviews because of the following challenges: 
\begin{enumerate*}[label=(\roman*)]
\item the presence of unknown factors renders the generation process non-deterministic, and 
\item both implicit and explicit information need to be handled, which makes it difficult to decode reviews.
\end{enumerate*}

\begin{figure}[!t]
	\centering
	\includegraphics[width=0.65\columnwidth]{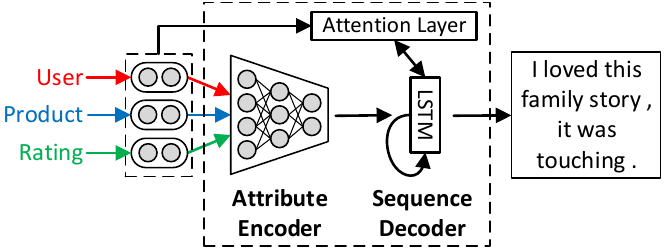}
	\caption{Overview of the attention-enhanced attribute-to-sequence model for review generation. Image source: \citep{dong2017learning}.}
	\label{fig_ch2sec21_overviewreviewgene}
\end{figure}

To address these problems, \citet{dong2017learning} propose an attention-enhanced attribute-to-sequence model to generate item reviews for given attribute information.
The authors introduce an attention-enhanced attribute-to-sequence model that learns to encode attributes into vectors and then uses a recurrent neural networks based on LSTM units to generate reviews by conditioning on the encoding vectors; see Fig.~\ref{fig_ch2sec21_overviewreviewgene}. The model can be divided into three components: an attribute encoder, a sequence decoder, and an attention mechanism. The authors use a dataset collected from Amazon to verify the effectiveness of the proposed method, especially the attention mechanism in the review generation procedure. 

There is increasing attention for combining preference prediction with review generation.
As sentiment classification plays an important role in e-commerce review analysis, \citet{radford2017learning} describe a representation learning strategy to detect opinions while generating reviews, where a generic sentiment tree bank was applied to represent the sentiment label in user reviews~\citep{socher2013recursive}.
Many e-commerce sites provide structured information, such as aspect-sentiment scores, i.e., each review text contains sentences describing a number of aspects of the item.
Focusing on generating long Chinese reviews from aspect-sentiment scores, \citet{zang2017towards} offer end-to-end sequential review generation models (SRGMs). Unlike traditional seq2seq models, SRGMs encode inputs of aspect-sentiment scores using multi-layer perceptrons. 
\citet{sharma2018cyclegen} propose an LSTM-based neural network to generate personalized reviews from multi-faceted factors, i.e., user profiles and item attributes, where an additional loss term is used to ensure consistency of the sentiment rating in the generated review.

\citet{ni2017estimating} put forward a collaborative-filtering generative concatenative network to jointly optimize item recommendation and generate personalized reviews.
To generate personalized high-fidelity reviews, \citet{ni2018personalized} come up with an encoder-decoder model to use both user and item information as well as auxiliary, textual input and aspect-aware knowledge, where an attention fusion layer is introduced to control the influence of various encoders. 

\begin{figure}[!t]
	\centering
	\includegraphics[width=0.9\columnwidth]{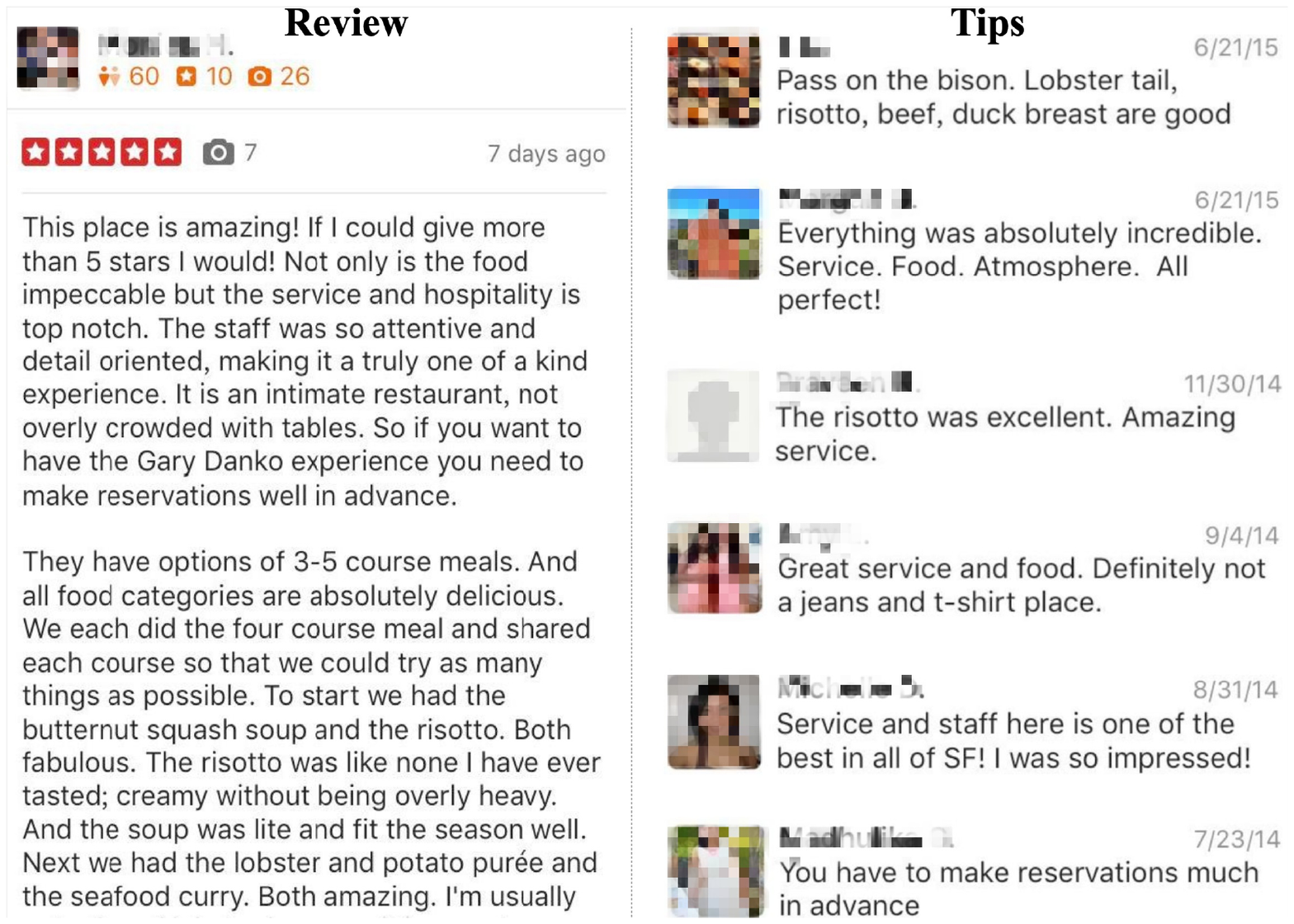}
	\caption{Examples of reviews and tips selected from the restaurant ``Gary Danko'' on Yelp. Users will get conclusions about this restaurant immediately after scanning the tips with their mobile phones. Image source: \citep{Li2017}.}
	\label{fig_ch2sec21_tipsgenerate}
\end{figure}

Some e-commerce sites have launched an interaction box called \emph{tips} on their mobile platforms.
Fig.~\ref{fig_ch2sec21_tipsgenerate} shows examples of reviews and tips on Yelp. The left column is the review from the user ``Monica H.'', and tips from several other users are shown in the right column.
Tips are more concise than reviews and can reveal user experience, feelings, and suggestions with only a few words. To generate concise tips from reviews, \citet{Li2017} suggest a multi-task learning framework for tip generation and rating prediction. 
For abstractive tip generation, gated recurrent neural networks are employed to decode user and item latent factors, whereas for rating regression, a multilayer perceptron network is employed to project user latent factors and item latent factors into ratings.
A persona-aware tip generation framework has been put forward for \emph{personalized} tip generation through adversarial variational auto-encoders (aVAE)~\citep{li2019persona}. 

\begin{figure}[!t]
	\centering
	\includegraphics[width=0.9\columnwidth]{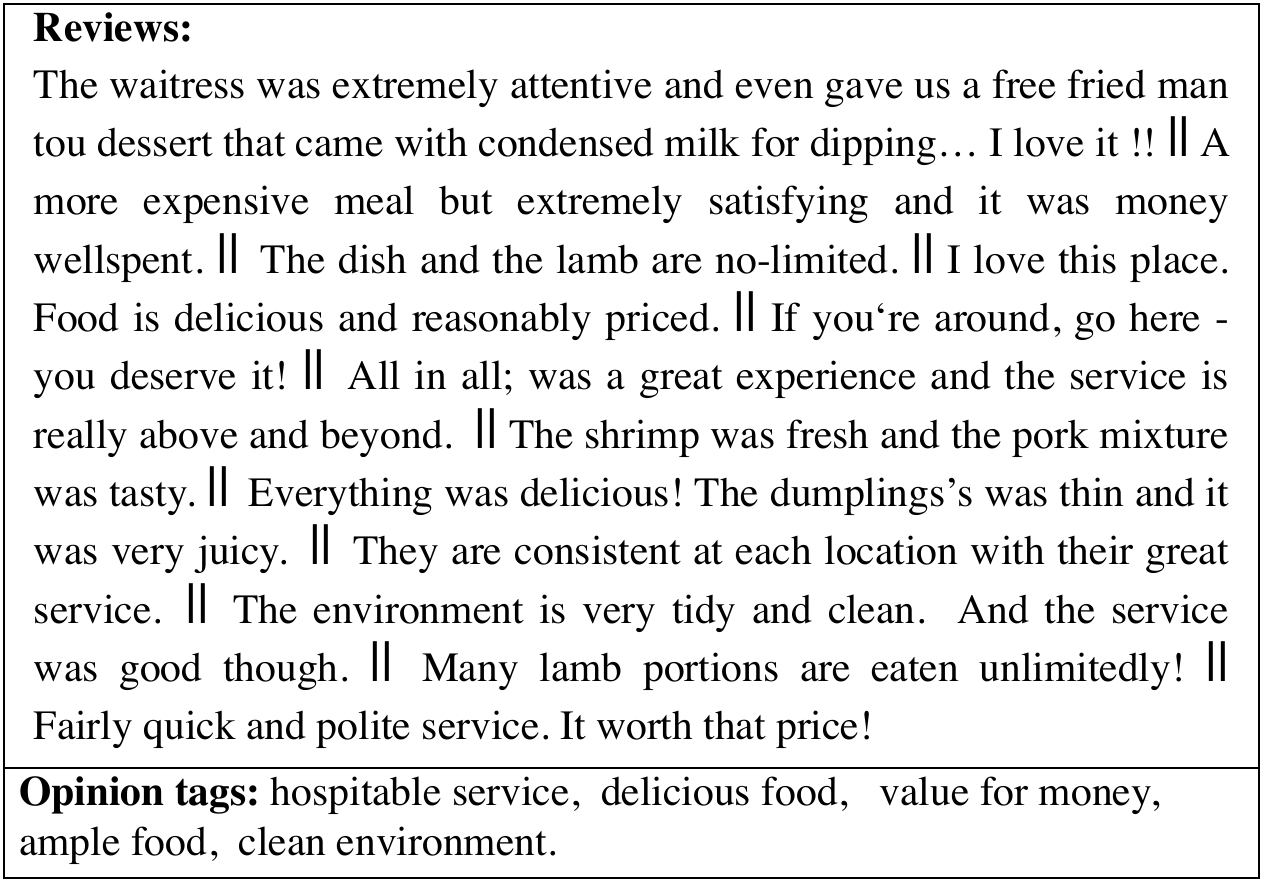}
	\caption{An example of a set of reviews and their corresponding opinion tags. Image source: \citep{li2021abstractive}.}
	\label{fig_ch2sec21_tagging}
\end{figure}

\emph{Opinion tags} refer to a ranked list of tags provided by the e-commerce platform that reflect the characteristics of reviews of an item; see Fig.~\ref{fig_ch2sec21_tagging}.
To assist consumers to quickly grasp a large number of reviews about an item, opinion tags are increasingly being applied by e-commerce platforms. 
Current mechanisms for generating opinion tags rely on either manual labelling or heuristic methods, which is time-consuming and ineffective. 
\citet{li2021abstractive} introduce the abstractive opinion tagging task, where systems have to automatically generate a ranked list of opinion tags that are based on, but need not occur in, a given set of user-generated reviews.

The abstractive opinion tagging task comes with three main challenges: 
\begin{enumerate*}[label=(\roman*)]
\item the noisy nature of reviews; 
\item the formal nature of opinion tags vs.\ the colloquial language usage in reviews; and 
\item the need to distinguish between different items with very similar aspects. 
\end{enumerate*}
To address these challenges, \citet{li2021abstractive} come up with an abstractive opinion tagging framework, named AOT-Net, that first predicts a salience score for each review, and given the salience scores, it groups all reviews into opinion clusters and ranks opinion clusters by cluster size. With the designed alignment feature and alignment loss, AOT-Net sequentially reads ranked opinion clusters and generates opinion tags with ranks.
To generate opinion tags in a personalized way, \citet{zhao2022personaot} select the information that users are interested in from reviews and then generated a ranked list of aspect and opinion tag pairs. The authors track user preferences not only using explicit feedback, i.e., reviews, but also using implicit feedback such as clicks and purchases in a heterogeneous graph neural network model.


\section{E-commerce users}
\label{ch2:sec2}

Over 55\% of online customers start to search on an e-commerce website as opposed to a generic web search engine~\citep{zhouwsdm2018}. 
Besides desktop clients, there are multiple e-commerce environments, e.g., mobile apps, smart watches, and interactive systems. 
These devices provide new means of interaction for users with e-commerce interfaces.
User behavior on e-commerce platforms can be divided into two types: implicit feedback and explicit feedback~\citep{brown2003buying,su2018detecting}. Implicit feedback is captured in transaction logs and includes clicks, purchases, browses, and engagements, etc.; explicit feedback of online shopping is captured in user comments, chat logs, and questions.
Following~\citet{lo2016understanding,zhouwsdm2018,gao2019product,chen2020jddc}, we list eight types of user behavior information from e-commerce platforms:

\begin{itemize}[leftmargin=*,nosep]

\item \textbf{Clicks.} As the entrance to an item page, a click on an item hints that the user is interested in the item. Click sources include the home page, shopping cart page, sale page, and the search result page, etc. 
\citet{zou2018drlunderreview} find that the more clicks, the bigger the interest from the user. 

\item \textbf{Purchases.} In e-commerce systems, purchases are very strong signals for recommendation. Most e-commerce platforms employ the \emph{Gross Merchandise Volume} (GMV) as a gold standard for measuring success. GMV indicates the total amount of purchases from merchandise sales as the target of optimization of e-commerce~\citep{anderson2002new,lee2001visualization}. 
Many recent studies use binary purchase information as the learning objective to characterize different levels of clicks~\citep{zhouwsdm2018}. 

\item \textbf{Browses.} On an item detail page, there are three browsable components: the main page (including basic information, title, price, pictures, etc.), the specification page (including more parameters and details), and comment page. The browsable components are helpful to understand users' interests, e.g., if a user browses the comments and specifications instead of only browsing the brief information, they have a higher probability of buying this item.

\item \textbf{Add-to-carts.} Adding to cart and ordering actions offer strong signals for e-commerce search and recommendation~\citep{su2018detecting}. Adding to cart usually reflects a strong sign of buying an item, whereas it may also reflect an interest shift phenomenon or high potential for re-purchase~\citep{zhouwsdm2018}. 

\item \textbf{Dwell time.} Dwell time is an effective signal to measure user engagement~\citep{yi2014beyond}. It denotes the length of time that a user spends on a web page before navigating to another page. Typically, the longer the dwell time, the more appealing the page. Dwell time is widely captured on e-commerce sites.

\item \textbf{Product-aware question answering.} As explained above e-commerce platforms allow consumers to ask product-aware questions to those whom bought the same product. Correspondingly, consumers can also answer these questions asked by other users on the platform. These questions and answers provide explicit feedback and opinions of the user~\citep{gao2019product}.

\item \textbf{Interactions with customer services.} Provided to customers before, during, and after a purchase, customer services give direct one-on-one interactions between a consumer and the e-commerce service provider via multiple channels, e.g., dialogues, emails, and messages.
Most user feedback from customer service is textual information. However, recently more and more multi-modal information, e.g., images, videos and audio messages is also included~\citep{zhao2021jddc}.

\item \textbf{Reviews and comments.}  As explained above, reviews and comments are prevalent in e-commerce platforms. Reviews and comments, written by consumers, explicitly reflect their opinions about specific products and services on the e-commerce platform.
\end{itemize}

\noindent Given these types of user behavior, recent research on analyzing user behavior on e-commerce platforms focuses on answering the following questions: 
\begin{itemize}[leftmargin=*,nosep]
\item How do people make their shopping decisions? What is the process from a user's click to their purchase in e-commerce?
\item What is the post-click behavior in e-commerce? What is the difference between macro-behavior and micro-behavior?
\end{itemize}

\noindent%
In this section, we analyze recent work on user behavior analysis in e-com\-mer\-ce:
\begin{enumerate*}[label=(\roman*)]
\item click behavior analysis (Section~\ref{02subsec221}); and
\item user engagement and post-click behavior (Section~\ref{subsec:pcbec}).
\end{enumerate*}

\subsection{From clicks to purchases}
\label{02subsec221}

As an e-commerce user interacts with an item, they express a certain degree of interest in the item.
When users browse an e-commerce platform, they may examine a specific item that is sufficiently relevant or intriguing. User clicks are an important signal for tracking a user's interest~\citep{chuklin-click-2015}. 
A user's online shopping behavior can be divided into two consecutive stages: item selection/clicks, and the decision to purchase the clicked item.
Users may have different intentions while shopping online, e.g., some wish to make a purchase as soon as possible while others are just looking around so as to get inspired. 
Therefore, \citet{wu2018turning} argue that clicks, as a kind of implicit feedback, should be integrated with other kinds of feedback to evaluate the ``relevance'' of items given a query on e-commerce portals.

During online shopping, users can add items to shopping carts and purchase them, but many platforms also facilitate additional types of activity. For instance, ``adding to favorites'' is a function to help users save some potentially interesting items for future purchase activities. 
To some extent, the degree of ``adding to favorites'' reflects the popularity of an item that can be exploited as a facet for item ranking for e-commerce search and recommendation~\citep{li2011towards}. 

\begin{figure}[!t]
	\centering
	\includegraphics[width=0.95\columnwidth]{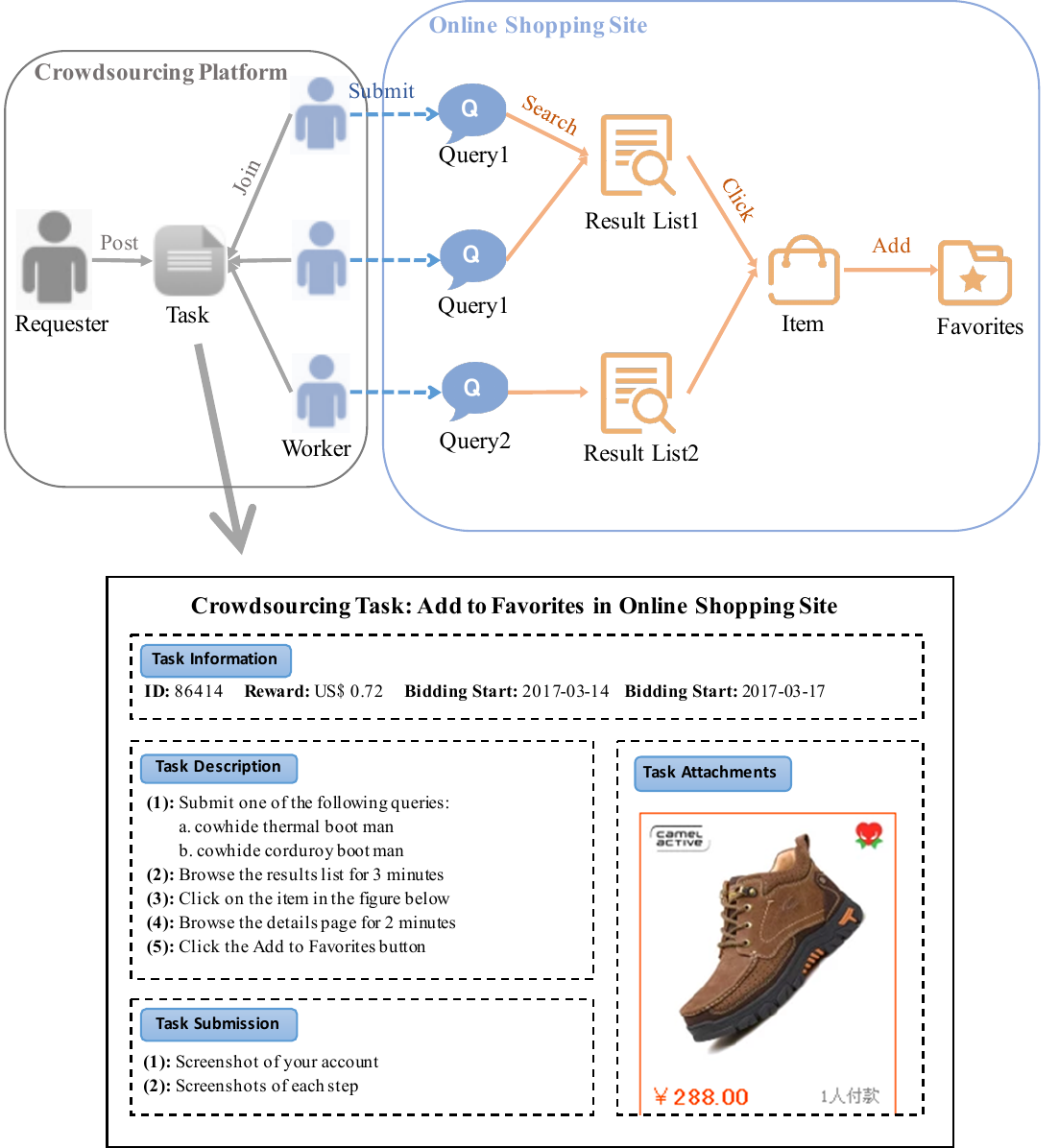}
	\caption{An example of crowdsourcing ``add to favorites'' task. Image source: \citep{su2018detecting}.}
	\label{fig_ch2sec21_a2fdetecttask}
\end{figure}

\begin{figure}[!t]
  \centering
  \subfigure[]{
    \label{fig_ch2sec21_a2fdetecta}
    \includegraphics[clip,trim=0mm 0mm 0mm 5mm,width=0.45\columnwidth]{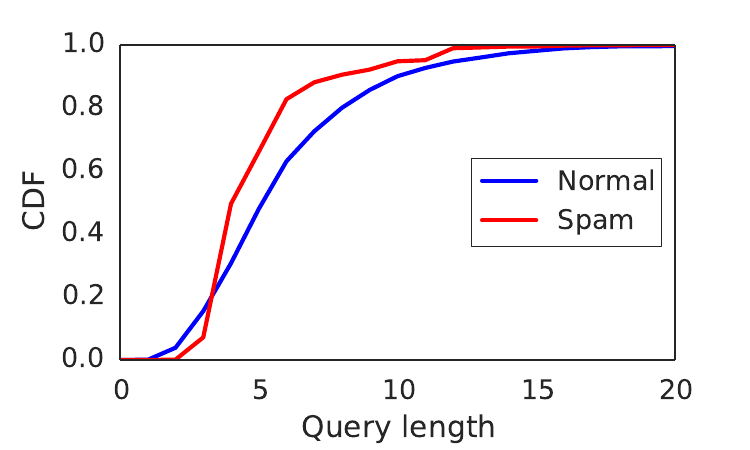}}
  \subfigure[]{
    \label{fig_ch2sec21_a2fdetectb}
    \includegraphics[clip,trim=0mm 0mm 0mm 5mm,width=0.45\columnwidth]{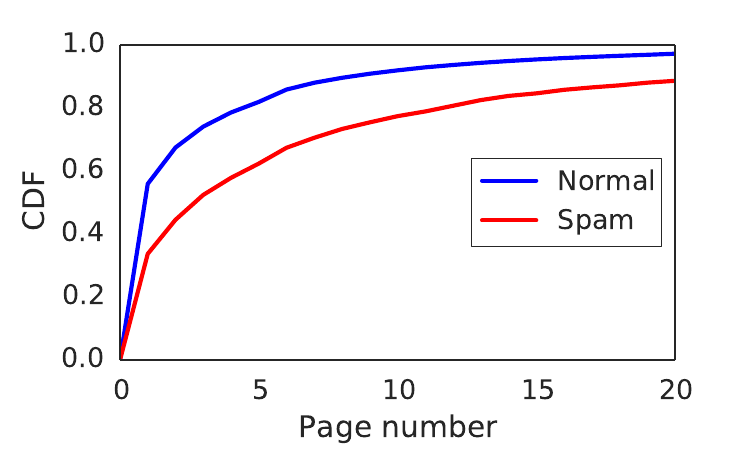}}
    \subfigure[]{
    \label{fig_ch2sec21_a2fdetectc}
    \includegraphics[clip,trim=0mm 0mm 0mm 5mm,width=0.45\columnwidth]{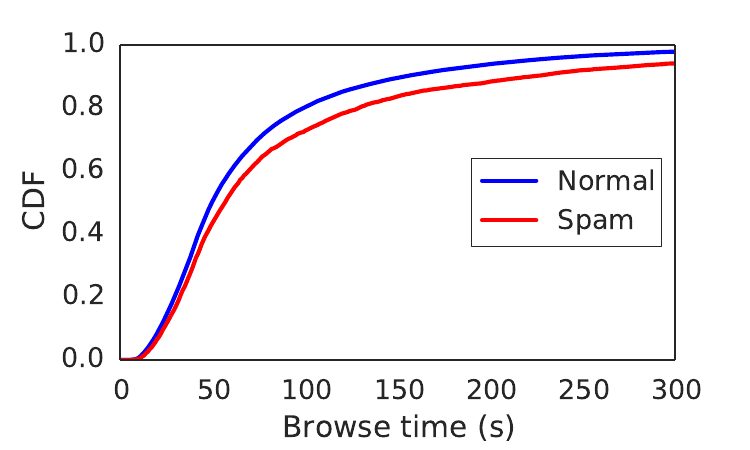}}
  \subfigure[]{
    \label{fig_ch2sec21_a2fdetectd}
    \includegraphics[clip,trim=0mm 0mm 0mm 5mm,width=0.45\columnwidth]{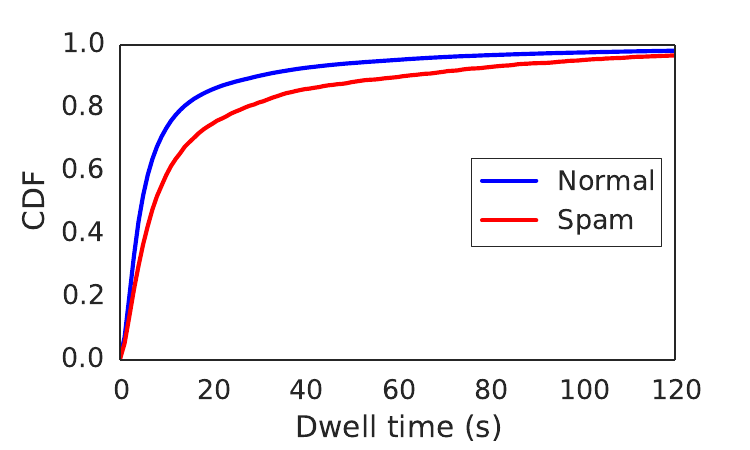}}
   \caption{Comparisons of behavior attribute distributions between normal and spamming ``add to favorites'' activities. Image source: \citep{su2018detecting}.}
  \label{fig_ch2sec21_a2fdetectacomp}
\end{figure}

To boost sales, some online retailers modify the ranking of their items' popularity with the usage of crowdsourcing platforms. For example, \citet{su2018detecting} investigate and detect such kind of activities in e-commerce, e.g., crowd workers need to follow some particularly designed guidelines to disguise themselves as normal users. An example of this crowdsouring ``add to favorites'' task is shown in Fig.~\ref{fig_ch2sec21_a2fdetecttask}. 
By simultaneously manipulating a number of crowdsourcing tasks and collecting user behavior, the authors compare behavioral attributes between normal activities and spamming activities. Fig.~\ref{fig_ch2sec21_a2fdetectacomp} shows these comparisons in terms of four behavioral attributes: query length, page number, browse time (time period between search and click), and dwell time (on detailed item pages). 

Different recommendation scenarios on an e-commerce platforms may yield different types of user click and purchase behavior. E.g., clicks on the follow-up recommendation results after adding an item to the shopping cart, and clicks on the recommended results listed on an item's detailed page~\citep{zhouwsdm2018}. 
The diversity in scenarios may make it harder to interpret clicks and their relation to purchase behavior.
In e-commerce search and recommendation, a purchase action is a natural ground-truth label for a click. 
If a user ends up purchasing an item after clicking on it, such a click indicates the user's strong interest in and satisfaction with the item. Accordingly, the conversion rate has been proposed as an important signal~\citep{zhouwsdm2018}:
\begin{equation}
\label{eq:conversion}
\begin{split}
\text{Conversion rate}=\frac{\text{Number of clicks that ended with an order}} { \text{Number of clicks}}.
\end{split}
\end{equation}
\noindent%
\emph{Purchase intent} represents a predictive measure of subsequent purchasing behavior.
Understanding purchase intent and how it is built up over time is important for personalized and contextualized e-commerce services. 
In recent years, many studies have explored the conversion from clicks to purchases~\citep{wen2019multi,guoziyi2018}. 
Moreover, to understand how user activities lead to purchase intent, both long and short-term purchase intent have been investigated~\citep{lo2016understanding,brown2003buying,kim2003combination,sismeiro2004modeling,swinyard2004activities,suh2004prediction,van2005predicting,young2004predicting}. 

Studies focusing on short-term purchase intent analysis have investigated user demographics~\citep{young2004predicting}, purchase patterns~\citep{kim2003combination}, item attributes~\citep{brown2003buying,van2005predicting}, and click streams~\citep{sismeiro2004modeling}. 
\citet{young2004predicting} find that the transaction, cost, and incentive programs are important predictors for determining the short-term intention to purchase clothing, jewelry, and accessories on e-commerce portals.
Furthermore, \citet{mcduff2015predicting} present a large-scale analysis of the connection between facial responses and purchases.

\citet{lo2016understanding} focus on long-term purchase intent analysis. 
The authors perform a large-scale long-term cross-platform study of user purchase intent and how it varies over time. 
They focus on four kinds of signals of user actions to detect purchase intent: closing-up on a piece of content, clicking through a link to an external website, searching for content, and saving content for later retrieval. 
The authors find that signals for purchase intent tend to slowly build up over time, and sharply increase about three to five days before a purchase. 
Moreover, users with a long-term purchase intent tend to save and click-through more content; these signals may be present for weeks before a purchase is made and they are amplified in the last three days before purchase.

Social interactions can also be used to improve understanding of consumer behavior~\citep{guo2011role,gunawan2015viral,hajli2017social,testa2018social}. 
Users may consult their social network when they need to purchase something they are unfamiliar with. 
Thus, although social relations only provide implicit signals, they have been found to be useful to understand purchase decisions~\citep{guo2011role}.
\citet{bhatt2010predicting} find that purchase intent from highly connected individuals is correlated with adoption by users in their social circle.
However, there is little evidence of social influence by these high degree individuals. The spread of purchase intent remains mostly local to first-adopters and their immediate friends. 
In a 2011 study of information passing in Taobao, \citet{guo2011role} verify that implicit information passing is present in the network, and that communication between buyers is a fundamental driver of purchasing activity.  
\citet{zhang2013predicting} present a system to understand the relation between users' Facebook profiles and purchase behaviors in eBay. 
Extensive analyses have been done on a benchmark dataset collected from Facebook and eBay; the authors find that there are significant correlations between social network information and online purchases.

\subsection{User engagement and post-clicks}
\label{subsec:pcbec}

User engagement is usually described as a combination of cognitive processes such as focused attention, affection, and
interest~\citep{mathur2016engagement}. 
In e-commerce, there is a long line of research that analyses user engagement~\citep[e.g., ][]{o2010development,vanderveld2016engagement,wu2017returning,zou2018drlunderreview}.
User engagement in e-commerce can be divided into two categories: short-term engagement and long-term engagement~\citep{zou2018drlunderreview}. 
Short-term engagement refers to the instant response (e.g., clicks and dwell time on an item page), which reflects the users' real-time preferences.
However, 
the systems may not only want to optimize for more clicks or purchases, but also to keep users in active interaction with the system (i.e., user stickiness), which is typically measured by \emph{delayed metrics}~\citep{lehmann2012models}.

Long-term user engagement is more complicated than short-term user engagement; it includes, e.g., dwell time on applications, depth of the page-viewing, and the internal time between two visits~\citep{wu2017returning}. 
Long-term user engagement reflects the user's desire to stay on the e-commerce portal longer and use the service repeatedly~\citep{zou2018drlunderreview}, i.e., the ``stickiness.''

After clicking an item via search results or recommendation results, the user enters the item page.
A user's post-click refers to the user's actions within the item page after the user clicks, including inner-item clicks (i.e., clicks within the item page), purchases, service contact, and thumbnail picture views~\citep{rosales2012post,yi2014beyond,mao2014estimating}.
Recent studies aim to characterize such post-click behavior on item pages as different post-click behavior has sharply different conversion rates~\citep{zhouwsdm2018,guoziyi2018,lalmas2018tutorial,wu2018turning}.

\begin{figure}[t]
	\centering
	\includegraphics[width = \columnwidth]{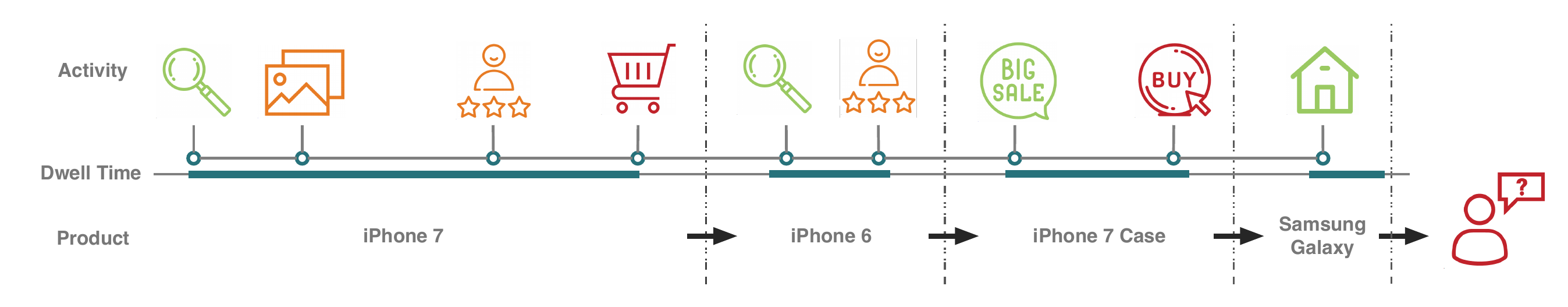}
	\caption{An illustrative example of post-click behavior from JD.com. Image source: \citep{zhouwsdm2018}.}
	\label{fig:micromintro}
\end{figure}

To illustrates post-click behavior on an e-commerce platform, Fig.~\ref{fig:micromintro} provides an example of observed data of a user during a short period. 
We see that the user first enters a product page for the iPhone 7 from a search result page. 
After reading the detailed description and comments, this user adds the item to their shopping cart. 
Then, the user shifts to a page for the iPhone 6 from the search result page and reads the comments. 
After that, they browse a page devoted to iPhone 7 cases from the sales page and order the case. 
Finally, they jump to a page about the Samsung Galaxy from the home page of the e-commerce site. 
During this period, two kinds of post-click behavior can be found: 
\begin{enumerate*}[label=(\roman*)]
\item from a coarse-grained perspective, the user interacted with the iPhone 7, the iPhone 6, iPhone 7 cases, and the Samsung Galaxy;  and
\item from a fine-grained perspective, each coarse-grained interaction includes a sequence of behavior that can indicate how the user locates the item page, whether the user clicks detailed information, whether a user adds-to-cart or orders an item, and how long the user dwells on an item~\citep{zhouwsdm2018}. 
\end{enumerate*}


\begin{figure}[t]
	\centering
	\includegraphics[width = 0.65\columnwidth]{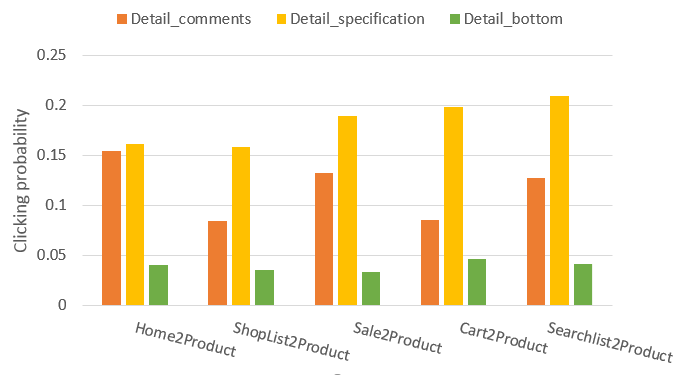}
	\caption{The relation between clicks and browsing modules. Image source: \citep{zhouwsdm2018}.}
	\label{fig:enter2detail}
\end{figure}

\begin{figure}[!t]
  \centering
  \subfigure[Dwell time vs.\ clicks]{
    \label{fig:box:a}
    \includegraphics[width = 0.4\columnwidth, height=1.7in]{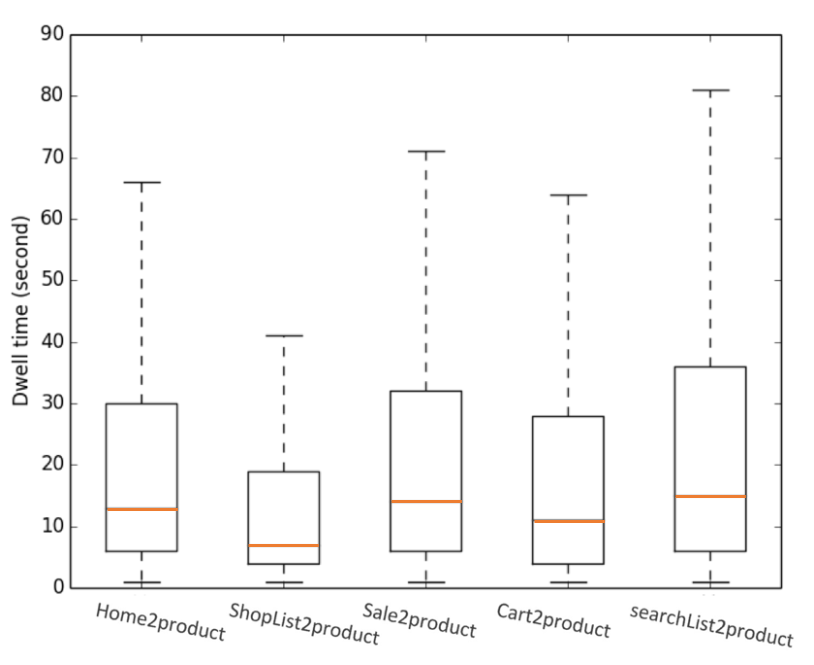}}
  \subfigure[Dwell time vs. browsing module]{
    \label{fig:box:b}
    \includegraphics[width = 0.4\columnwidth, height=1.7in]{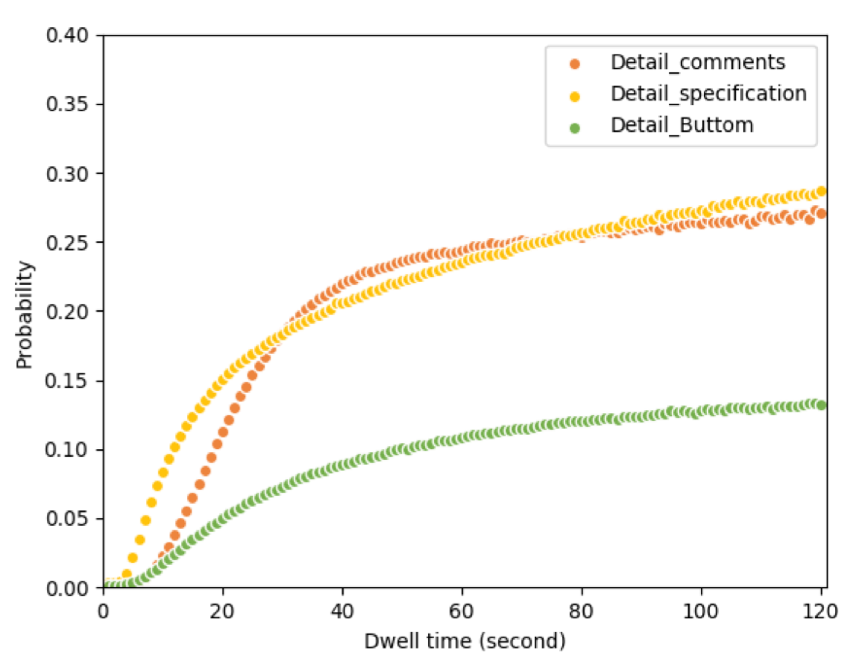}}
    \caption{Performance of dwell time, clicks, and browsable components. Image source: \citep{zhouwsdm2018}.}
  \label{fig:box}
\end{figure}

As mentioned in Section~\ref{ch2:sec2}, typically, there are three browsing modules on e-commerce sites: the main page (including basic information, title, price, and pictures), the specification page (including more parameters and details), and the comment page.
Fig.~\ref{fig:enter2detail} illustrates the relations between clicks and these browsable components~\citep{zhouwsdm2018}. 
We see that a user is more likely to buy an item if they produce more clicks on its different browsable components, i.e., a user may gather basic information from the main item page, review feedback from the comment page, and click images to check if the item satisfies their requirements.
\citet{guoziyi2018} show how dwell time is related to clicks and browsable components; see Fig.~\ref{fig:box}. 
The dwell time on an item is related to how a user locates the item. 
As shown in Fig.~\ref{fig:box:b}, the longer the dwell time, the more likely a user would visit detailed components, including reading comments and specifications.

\citet{zhouwsdm2018} investigate the relation between certain types of post-click behavior and the \ac{CVR}. They find that the post-click behavior ``Cart'' has the highest conversion rate, which means if a user adds an item to the cart, they are more likely to order it in the end. Similarly, if a user enters an item page from the list of items in the cart, they are also very likely to order it. 
When the dwell time is outside a certain range, the conversion rate drops. 
If the user stays much longer than they need to finish the page, they might have transferred their attention offline.
It is observed that users' interactions often exhibit a monotonic structure, i.e., the presence of a more explicit interaction (such as reviews) necessarily implies the presence of a more implicit signal (such as clicks)~\citep{wan2018item}.

\section{Discussion}
\label{ch2:discuss}

In this section, we have surveyed the infrastructure of e-commerce platforms, i.e., presentations and users. Specifically, we have introduced six information components that are widely applied on e-commerce platforms: search results, recommendation results, titles, product descriptions, question answering, and reviews.
We have highlighted studies about user behavior in e-commerce, including user clicks, purchases, engagement, and post-click behavior.

For e-commerce presentations, we have introduced basic concepts and identified key components of e-commerce interfaces. We have found that almost every e-commerce site provides six information components: search results, recommendation results, item titles, item features and descriptions, question answer pairs, and user reviews of the item.
Furthermore, we have summarized recent studies that focus on analyzing the effect of these information components. Empirical studies on these information components have revealed remarkably high correlations between user behavior and information displayed in e-commerce presentations.

For e-commerce users, we have observed complex user behavior from clicks to purchases.
According to empirical studies on e-commerce users, signals for purchase intent tend to slowly build up over time and sharply increase before a purchase. 
Studies also find that users are more likely to buy an item if they produce more clicks on its different browsable components.
If a user adds an item to a cart, they are more likely to purchase it in the end. Similarly, if a user enters an item detail page from the list of items in the cart, they are also very likely to purchase it.

To gain a deeper understanding of information discovery on e-commerce platforms, we list three research questions to guide the following three sections:
\begin{itemize}[leftmargin=*,nosep]
\item Can we model user behavior and profile users by using multiple types of user behavior, e.g., clicks, post-clicks, and purchases?
\item How can we understand frameworks and components of e-commerce search through interactions between users and search engines?
\item What are the principles and characteristics of e-commerce recommendations?
\end{itemize}
We will address these questions through discussions in Section~\ref{chapter:user},~\ref{chapter:search}, and~\ref{chp:rec}, respectively.
A considerable amount of relevant work about information components will be discussed in Section~\ref{chapter:qa} as they have a clear connection to question answering and dialogue generation in e-commerce.


\chapter{E-commerce user modeling}
\label{chapter:user}

In Section~\ref{chapter:basic}, we discussed work on e-commerce information infrastructures, focusing on e-commerce presentations, and on e-commerce users.
The unique characteristics of e-commerce users make modeling for e-commerce users essential when attempting to understand and support information discovery~\citep{lo2016understanding,Huangwww2018}. 
E-commerce user modeling can be separated into two types: \emph{user behavior modeling} and \emph{user profiling}. 
Given specific user behavior in various scenarios, e.g., click behavior, purchasing behavior, and post-click behavior, user behavior modeling focuses on learning a model of user behavior to predict the user's next preference.
In contrast, user profiling aims to predict a user's profile (e.g., age, gender, and occupation) given the user's behavior records.
In this section, we survey research on user behavior modeling and user profiling  in e-commerce.
First, we detail user behavior modeling approaches in Section~\ref{ch31:sec:ubm}. 
Next, we discuss studies on user profiling in e-commerce in Section~\ref{ch3:sec:upem}. 
Lastly, Section~\ref{ch3:secfuture} discusses emerging directions in e-commerce user modeling.

\section{User behavior modeling in e-commerce}
\label{ch31:sec:ubm}

In this section, we describe research on e-commerce user behavior modeling, including click behavior modeling (Section~\ref{ch31:subsec:cpr}), post-click behavior tracking (Section~\ref{ch21:subsec:pcbt}), and purchase intent modeling (Section~\ref{ch31:subsec:pm}).

\subsection{Click behavior modeling}
\label{ch31:subsec:cpr}

Click actions recorded in query logs have successfully been applied to extract important features in the context of ranking scenarios~\citep{agichtein2006improving}.
Regarding web search, click models have been proposed to help the search engine understand interactive user behavior~\citep{guo2009efficient,yilmaz2010expected,zhang2011user,chuklin-click-2015,borisov2016neural}.
Early research on the topic aimed to track a user's behavior by using probabilistic graphical models. 
More recently, neural networks have been applied to improve the performance of click models by representing user behavior to capture the user's information needs~\citep{borisov2016neural}.
Focusing on improving the effectiveness by exploiting information from user-system interactions, \citet{ferro2017including} explore embedding dynamic interactions into learning to rank frameworks.
Thereafter, curriculum learning and continuation methods have been successfully applied to exploit user interactions and facilitate rank learning~\citep{ferro2019boosting}.

Given the work mentioned above, click behavior modeling has received an increasing amount of attention in e-commerce scenarios~\citep[see, e.g.,][]{he2014practical,chapelle2015simple,chen2016deep,he2017neural,LiRCRLM17,wu2018turning,zhouwsdm2018,huang2019ecompred,gong2020edgerec,bian2021contra,wen2021hierarchy}.
Viewing click prediction as a binary classification problem, the researchers who conducted those early studies employed logistic regression to predict whether an item will be clicked~\citep{richardson2007predicting}, where handcrafted features are extracted from raw data to optimize a log-likelihood objective function for training.
Latent factor optimization approaches, e.g., factorization machines~\citep{FM}, have also been applied to use importance-aware and hierarchical structures purposed to manage dynamic user behavior~\citep{oentaryo2014predicting}. In Section~\ref{sec5:pm}, we detail studies about factorization machines in e-commerce recommendation. 

\begin{header}{CTR prediction metric}
The \emph{click-through rate} (CTR) is a widely applied evaluation metric for click prediction that reflects the probability of a click in a trial impression. 
Following~\citet{regelson2006predicting}, we established $p$ as the probability of a click, $P = \{p_1,p_2,\ldots,p_N\}$ as the set of product items, and $U = \{u_1,u_2,\ldots,u_M\}$ to represent the set of users.    
\end{header}
The maximum-likelihood estimate of $p$ refers to the number of observed successes divided by the number of trials, i.e., clicks/impressions.
Given a set of search or recommendation sessions $S$ and a query $q$, the probability of a product $\operatorname{CTR}(p{\mid}q)$ can be formulated as follows:
\begin{equation}
\label{eq:se3:qsctr}
\operatorname{CTR}(p{\mid}q) = \sum\limits_{{s^q} \in S} {{{{\Psi ^{{s^q}}}(p)} \over {{\mid}{s^q} \in S{\mid}}}},
\end{equation}
where $s^q$ denotes a session with $q$, and $\Psi ^{{s^q}}$ denotes an event of a click within $s^q$. 

\begin{header}{From shallow to deep models}
CTR has been widely applied as an evaluation metric for click modeling in e-commerce portals.
\citet{rendle2010pairwise} introduce a tensor-based method for CTR prediction; Bayesian approaches have also been used effectively for CTR prediction~\citep{graepel2010web}.
Starting in 2015, deep learning significantly improved CTR estimation by transferring traditional architectures and developing new ones. Deep neural networks effectively capture high-order feature interactions, resulting in better CTR prediction performance.
\citet{zhang2016deep} describe a deep neural network to learn patterns from categorical feature interactions. Similarly, \citet{chen2016deep,zhu2017optimized} employ neural network models with multiple fully-connected layers to predict user clicks. 
\citet{aryafar2017ensemble} investigate CTR prediction in promoted listings by using an ensemble learning approach to use different signals of listings.
Generally, these logistic regression models can effectively achieve memorization by applying cross-product transformations over sparse features. 
More recent work involves representing sparse features as dense vectors, which are concatenated to form an instance vector. This vector is then passed through a multi-layer perceptron, with a sigmoid output layer, to predict the click probability. These advancements have greatly enhanced model accuracy in CTR tasks~\citep{zhang2021deep}.
\end{header}

\begin{figure}[!t]
	\centering
	\includegraphics[width=0.9\columnwidth]{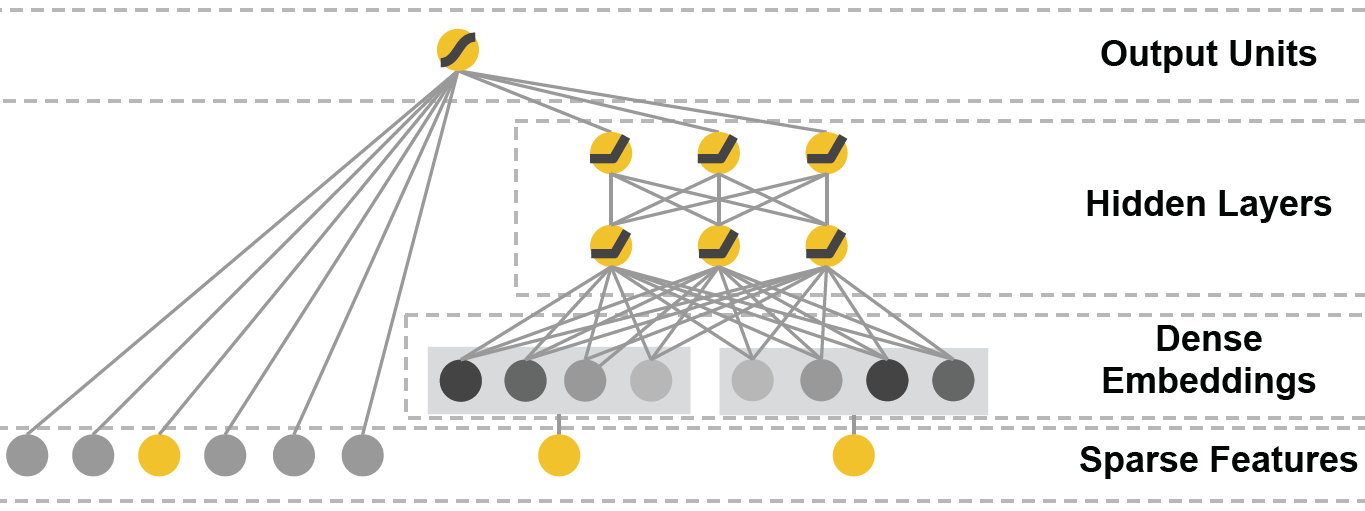}
	\caption{Wide \& Deep model architecture. Image source: \citep{cheng2016wide}.}
	\label{fig:wide_and_deep}
\end{figure}

\begin{header}{Wide \& Deep model}
Modeling the interactions between features, especially the interactions between low-order and high-order features, is essential for click prediction. The Wide \& Deep model~\citep{cheng2016wide} considers low- and high-order feature interactions simultaneously. 
Wide \& Deep pursues the balance between memorization and generalization. 
Owing to its simple structure, the ``strong'' features (i.e., feature combinations) of Wide \& Deep allow for the assignment of larger weights during training, thus endowing the model with stronger memory.
Besides the deep component based on an MLP, Wide \& Deep consists of another component, the wide component.    
\end{header}
It is a generalized linear model with an input feature set that includes raw features and a feature that has been transformed by the cross-product transformation and is defined as
\begin{gather}
	\phi_k(x) = \prod_{i=1}^d x_i^{c_{ki}},\quad c_{ki}\in \{0,1\},
\end{gather}
where $c_{ki}$ is a Boolean variable that is 1 if the $i$-th feature is part of the $k$-th transformation $\phi_k$, and 0 otherwise. Such a transformation allows the model to capture the interactions between the binary features, and adds nonlinearity to the wide component. 
The overall model architecture of Wide \& Deep is shown in Fig.~\ref{fig:wide_and_deep}.
Wide \& Deep has been shown to be effective in e-commerce recommendation scenarios; more details are provided in Section~\ref{sec5:nnm}.

The Wide\& Deep model is a representative of dual tower models for user behavior modeling. Similarly, DeepFM~\citep{guo2017deepfm}, DCN~\citep{wang2017deep}, xDeepFM~\citep{lian2018xdeepfm} and Autoint~\citep{song2020towards} have also been put forward for CTR prediction.
The deep neural network part in these dual tower models can always be regarded as a supplementary to
learn the residual signal of the feature interaction layer to approach the label, which yields stable training and the improved
performance.
In contrast, single-tower models like NFM~\citep{he2017neural} and the product-based neural network~\citep{Qu2018Product} have enhanced their modeling capacity due to their more sophisticated network structures, which allow them to capture complex feature interactions. However, they often struggle with issues such as getting stuck in poor local minima and exhibit a heavy reliance on careful parameter initialization. This sensitivity to initialization can affect their training stability and convergence, making optimization more challenging than for simpler models.

\begin{header}{Attention models for CTR}
Attention neural networks have been proposed to enhance the performance of CTR prediction. The deep interest network (DIN)~\citep{zhou2018deep} is the first model to introduce the attention network mechanism for user behavior modeling with CTR prediction. It assigns different weights to past behaviors based on their relevance to the target item. To capture dynamic interest evolution, the deep interest evolution network (DIEN)~\citep{zhou2019deep} has been proposed; it uses a two-layer GRU with an attentional update gate to model evolving user interests. Further advancements, like the behavior sequence transformer and the deep session interest network~\citep{feng2019deep}, use self-attention to model behavior dependencies and session-based representations, showing the importance of attention mechanisms in CTR prediction~\citep{xiao2020deep}.
More recent advances in user click models with attention  have focused on using deep neural networks to capture complex interactions given user profiles, item attributes, and contextual features~\citep{hou2023deep}. These models have shown great potential in improving the accuracy and scalability.
\end{header}

\begin{header}{Memory-based models}
With the accumulation of large amounts of user behavior data on large e-commerce platforms, effectively handling long behavior sequences is increasingly important. However, many models such as DIN~\citep{zhou2018deep} struggle with the time complexity when processing such sequences. To address this, \citet{ren2019lifelong} introduce the hierarchical periodic memory network; it uses a lifelong memory mechanism with multi-layer GRUs updating at different frequencies, capturing long-term and multi-scale temporal patterns. Similarly, the user interest center and the multi-channel user interest memory network are designed to handle long-term user interest modeling, providing a more systematic, industrial-level approach~\citep{pi2019practice}.
Multi-interest networks have also been studied to improve the robustness and consistency in user click modeling~\citep{cen2020controllable,chang2023twin}.
To mitigate noisy correlations and user intent vanishing during this procedure, attribute transition graphs and matching among various patterns need to be constructed. To this end, \citet{liu2023enhancing} characterize user intents with attribute patterns, where the frequent and compact attribute patterns serve as memory to augment session representations.
\end{header}

\begin{figure}[!t]
	\centering
	\includegraphics[width = 0.9\columnwidth]{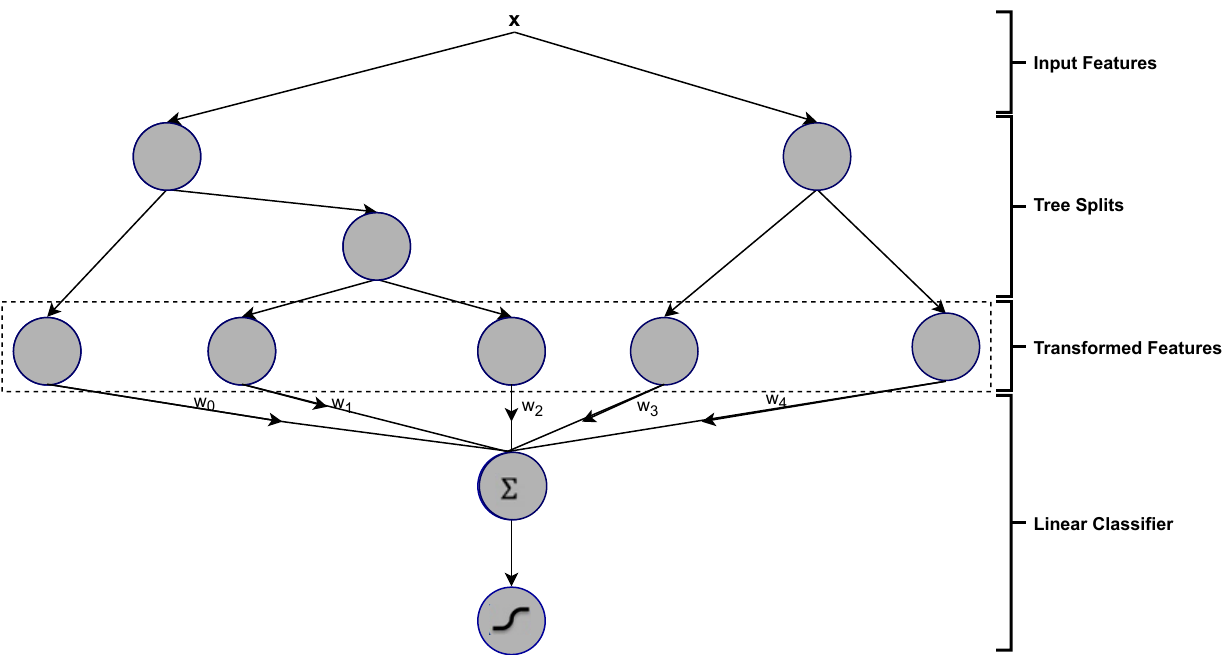}
	\caption{Overview of the hybrid click prediction model that uses GBDT and logistic regression. Image source: \citep{he2014practical}.}
	\label{ch31:fig:gbdt4ctr}
\end{figure}

\begin{header}{Hybrid models combining multiple factors}
To address the complexity of feature interactions, various hybrid models have been proposed. For example, the gradient boosting decision tree model~\citep[GBDT;][]{chen2016xgboost} has been applied successfully to predict user clicks~\citep{he2014practical}.
Fig.~\ref{ch31:fig:gbdt4ctr} illustrates the structure of a hybrid model with GBDT and logistic regression. The model concatenates the boosted decision trees, which transforms features and the sparse logistic regression classifier. 
\end{header}
The input is a structured embedding $x = (e_{i_{1}},e_{i_{2}},\ldots,e_{i_{n}})$ for each item $x$, where $e_{i}$ refers to the $i$-th unit vector, and $i_{n}$ is the index of the categorical features. The output of the model is a binary label $y\in\{+1,-1\}$, which indicates a click or no click. Given a labeled pair $(x,y)$, the authors denote the linear combination of active weights as $s(y,x,w)$, which can be calculated as follows:
\begin{equation}
s(y,x,w)=y\cdot w^{T}\cdot x=y\sum\limits_{j=1}^{n}{w_{j,i_{j}}},
\label{eq:se3:gbdt}
\end{equation}
where $w$ is the weight vector of the click score. Using stochastic gradient descent~\citep[SGD; ][]{saad1998online}, the authors inferred the likelihood function $p(y{\mid}x,w)$ by applying a sigmoid function over $s(y,x,w)$. 
Based on these transformed features, the authors applied logistic regression to predict a click or no click.
Boosted decision trees are able to aggressively reduce the number of active features with only moderate prediction accuracy degradation. The hybrid click model is widely applied in e-commerce recommendations for candidate ranking (see Section~\ref{sec5:pm} for more details.).

\begin{figure}[!t]
	\centering
	\includegraphics[width = 0.9\columnwidth]{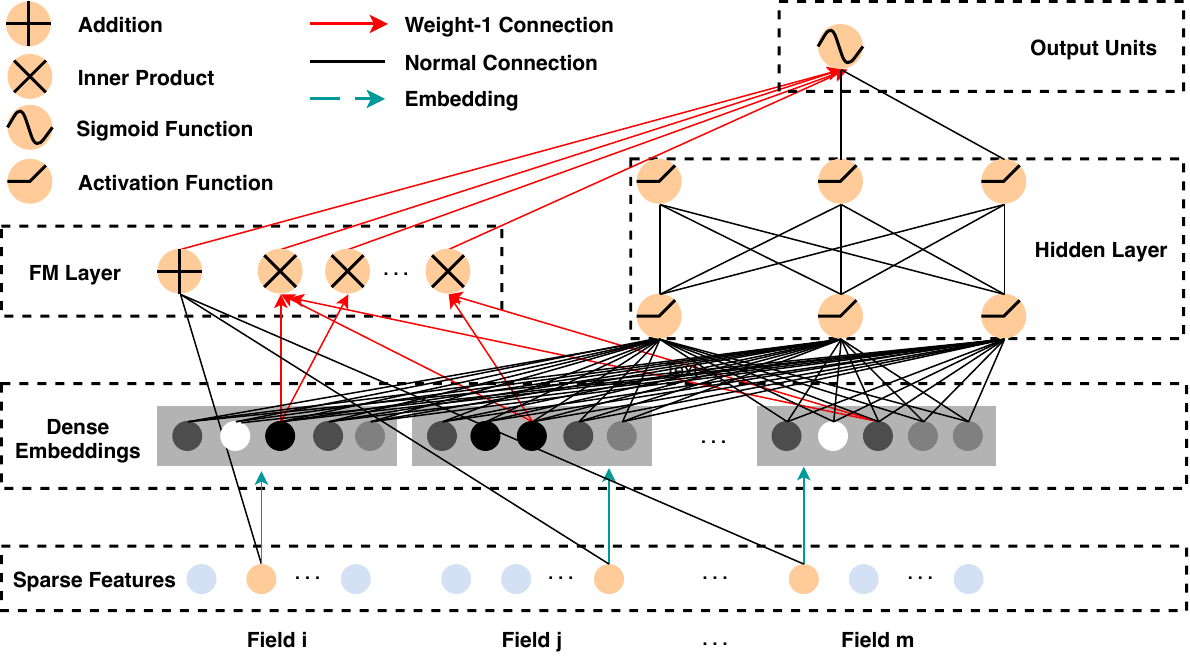}
	\caption{Architecture of the DeepFM model for CTR prediction. Image source: \citep{guo2017deepfm}.}
	\label{ch31:fig:deepfm}
\end{figure}

To explore the feature interactions hidden in data collections, \citet{guo2017deepfm} propose a neural network method, i.e., DeepFM, that combines the architectures of factorization machines and deep neural networks. 
As shown in Fig.~\ref{ch31:fig:deepfm}, DeepFM uses a wide and deep component to share the same raw input feature vector; this allows the model to learn low- and high-order feature interactions simultaneously. All parameters are jointly trained for the combined prediction model, as described by Eq.~\ref{eq:se3:deepfm}:
\begin{equation}
\label{eq:se3:deepfm}
\widehat y = \operatorname{sigmod} ({y_{FM}} + {y_{DNN}}),
\end{equation}
where $\widehat y \in (0,1)$ refers to the predicted CTR, $y_{FM}$ is the output of the FM component, and $y_{DNN}$ is the output of the deep component. 
The authors apply a feed-forward network in the deep component to learn higher-order feature interactions. 

More DeepFM model-based deep learning methods have been proposed to address the CTR prediction problem, including deep convolutional neural networks (CNNs)~\citep{chan2018convolutional} and deep interest neural networks~\citep{zhou2018deep,feng2019deep,li2019multi,zhou2019deep,chen2021deepctr,zhu2022tkdectr,guo2022icdectr,cheng2022dynamic}.
All of the above-mentioned deep neural networks have significantly contributed to the optimization of item ranking in e-commerce search and recommendation; this will be further discussed in Sections~\ref{ch4:sec:rs} and ~\ref{sec5:nnm}, respectively.

To ensure consistent evaluation and comparison of CTR prediction models, benchmark frameworks such as the open benchmarking for CTR~\citep{zhu2021open} and BARS-CTR~\citep{zhu2022bars} have been introduced. These frameworks provide a standardized way to evaluate model performance across different datasets, improving reproducibility and promoting further advancements in CTR prediction research.

\subsection{Post-click behavior tracking}
\label{ch21:subsec:pcbt}

As we have discussed in Section~\ref{subsec:pcbec}, post-click behavior plays an important role in modeling for e-commerce users in search (Section~\ref{chapter:search}) and recommendation scenarios (Section~\ref{chp:rec}). 
Multiple studies have focused on applying various types of interaction signals to model post-click behaviors in search and recommendation scenarios.
\citet{sculley2009predicting}~measure users’ post-click experience by evaluating the corresponding bounce rate. 
The model proposed by~\citet{zhong2010incorporating} uses both user clicks on the search page and post-clicks beyond the search page to provide an unbiased estimation of document relevance. 
\citet{lalmas2015promoting} investigate how viewport time can be used to measure user attention level as an engagement metric. 
\citet{o2016leveraging} use user interactions as signals within the clicked items to enhance the search results. 
\citet{wan2018item} determine the monotonic dependency between explicit user signals and more implicit signals to improve recommender systems. 
\citet{lu2018between} propose a preference prediction model to predict user actual preferences for the clicked items by taking into account multiple post-click interactions.

\begin{header}{Dwell time}
As we have discussed in Section~\ref{subsec:pcbec}, dwell time is the most common evaluation metric for the analysis of post-click user behavior~\citep{yin2013silence}. 
Accordingly,~\citet{yin2013silence} built a graphical model that focuses on using explicit user feedback and dwell time to predict user preferences in e-commerce recommendations. 
 \citet{yi2014beyond} show that integrating dwell time into the learning objective or learning weight results in better recommendation performance than pure predictions of the CTR. 
\citet{rosales2012post,chapelle2015simple} use dwell time as a proxy of post-click experience in online advertising to improve the ranking performance. 
\citet{bogina2017incorporating} explore the value of incorporating dwell time for session-based recommendations by boosting items above the preassigned dwell time threshold.
Modeling user behavior by taking into account dwell time has been shown to facilitate e-commerce recommendation performance.
In Section~\ref{sec5:sq}, we will discuss more studies that focused on modeling sequential user dynamics by using dwell time.    
\end{header}

\begin{figure}[!t]
	\centering
	\includegraphics[width = 0.95\columnwidth]{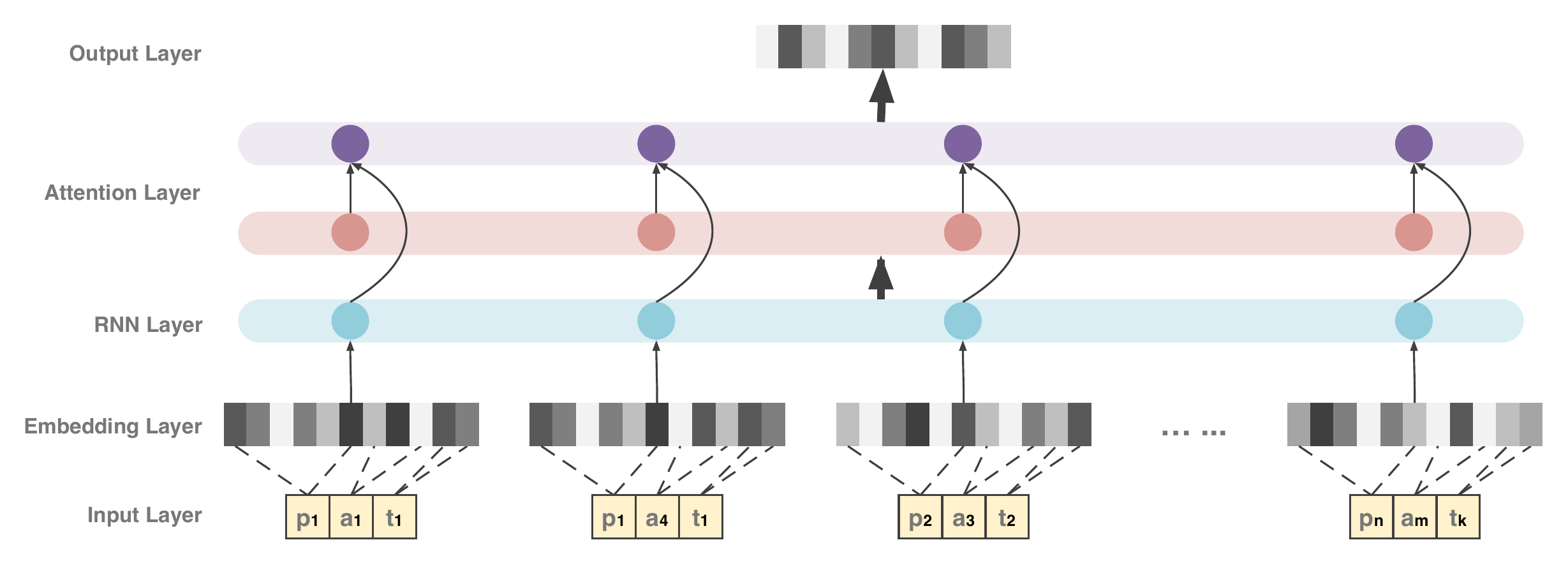}
	\caption{Micro-behavior modeling framework. Image source: \citep{zhouwsdm2018}.}
	\label{ch23:fig:microm}
\end{figure}

\begin{header}{User return modeling}
There is a limited amount of work on modeling user returns in e-commerce, especially when user returns depend heavily on the quality of the provided service~\citep{lo2016understanding,zhouwsdm2018}.
The model developed by \citet{zhouwsdm2018} provides rich user interfaces after a user clicks an item. For instance, it encourages users to visit different item modules or sub-pages, i.e., to read comments or click on pictures embedded within the item page; this generates a large amount of heterogeneous post-click behavior. 
Three problems pose a challenge for attempts to model micro-behavior on e-commerce platforms: 
\begin{enumerate*}[label=(\roman*)]
\item Sparseness and high dimensionality of the user representation; 
\item Sequential information of micro-behavior; and 
\item Diverse effects of micro-behavior.
\end{enumerate*}
To address these three challenges, \citet{zhouwsdm2018} propose the framework shown in Fig.~\ref{ch23:fig:microm}, which consists of five layers: an input layer, an embedding layer to solve the problems of sparseness and high dimensionality, an RNN layer to model sequential information, an attention layer to capture the diverse effects of micro-behavior, and an output layer.     
\end{header}
The input of the model comprises the data of a user, $u$, with a sequence of micro-behavior. 
Formally, the authors define it as the sequence $S_u =\{x_1, x_2, \ldots, x_n\}$, where each  $x_i$ is a tuple, i.e., 
\begin{equation}
x_t=(p_v,a_m,d_k),
\label{eq:input}
\end{equation}
where $p_v \in \mathbb{R}^V$ is a one-hot indicator vector where $p_v(i) = 1$ if $x_i$ is about the $i$-th product and other entities are zero. 
Similarly, $a_m \in \mathbb{R}^M$ and $d_k \in \mathbb{R}^K$ are indicator vectors for activities and dwell time, respectively. 
Each indicates a unique element in the product set $P$, activity set $A$, and dwell time set $D$, respectively. 
Here, the vocabulary sizes of $P$, $A$, and $D$ are $V$, $M$, and $K$, respectively, and there are $V \times M \times K$ tuples in total.
To address sparseness and high-dimensionality problems, the authors design an embedding layer to transform the input $x_t$ into a low-dimensional dense vector $e_t$:
\begin{equation}
e_t=\operatorname{concatenate}(W_{P}p_v,W_{A}a_m,W_{D}d_k),
\label{eq:312emb}
\end{equation}
\noindent where $W_P\in R^{d_P\times V}$, $W_A\in R^{d_A\times M}$, and $W_D\in R^{d_D\times K}$, where $d_P \ll N$, $d_A \ll M$, and $d_D \ll K$ are the number of latent dimensions for products, activities, and dwell time, respectively. 
The initial weights of $W_P$, $W_A$, and $W_D$ are trained by applying word2vec~\citep{mikolov2013efficient}. Additionally, the final embedding of $x_t$ is the concatenation of three embeddings. To capture the sequential information of micro-behavior, the authors build an RNN layer. The output of the embedding layer $e_t$ is the input of the RNN layer. The $t$-th hidden state unit output is calculated as
\begin{equation}
h_t=\sigma (W_{eh}e_t+W_{hh}h_{t-1}+b_t),
\label{eq:312emb2}
\end{equation}
where $\sigma (\cdot)$ is a non-linear activation function, e.g., ReLU, sigmoid, or tanh; $W_{eh}\in R^{d_h\times d_{e}}$, $W_{hh}\in R^{d_h\times d_h}$, and $b_i\in R^{d_h}$. 
To capture the effects of micro-behavior, the authors introduce an attention layer~\citep{attention1} that assigns proper weights to each hidden unit; this helps to obtain a more balanced output. 
The attention weight is mapped from the hidden layer vector to a real valued score by the function $\sigma (\cdot)$. To achieve sufficient expressive ability, the function $\sigma (\cdot)$ is typically implemented by a neural network layer. 
Then, the final output is an attention weighted pooling of the RNN layer. 
To exploit the different transition patterns between items and operations in micro-behavior modeling, \citet{meng2020sigirclick} incorporate item knowledge into a joint user modeling framework including a recurrent neural network and a graph neural network.
To incorporate the micro-behavior information in the iterative process of user behavior modeling, \citet{jianhao2022session} model a user session as a fine-grained sequence of micro-behaviors and proposed a self-attention mechanism to encode the dyadic relations of micro-behaviors. 

Experimental results have confirmed that post-click user modeling can provide deeper insights into user behavior, which is used to advance e-commerce search and recommender systems by successfully modeling the sequential dynamics in the candidate retrieval stage~\citep{zhouwsdm2018}.
However, post-click behaviors are often sparse in real-world scenarios, making it challenging to supplement large-scale implicit feedback. To address this, recent studies have integrated post-clicks with other user behaviors.  
\citet{wen2019leveraging} describe a generic probabilistic framework to fuse click and post-click feedback in recommender systems.
 \citet{wang2021clicks} reveal the importance of mitigating the clickbait issue
from click behaviors, and apply causal inference to establish a causal graph to reformulate the process.
In Section~\ref{sec5:sq}, we discuss further studies that have focused on integrating post-click tracking into e-commerce recommendation.    

\subsection{Purchase-intent modeling}
\label{ch31:subsec:pm}

\emph{Purchase-intent prediction} is another important task in e-commerce modeling~\citep{qiu2015predicting,kooti2016portrait,lo2016understanding,wan2017modeling}.
According to~\citet{bellman1999predictors}, the volume of online activities of a customer proves useful when predicting the occurrence of a future purchase.
Statistical models of customer purchase behavior have been studied for decades. 
Early research on purchase behavior modeling was based on statistical approaches, e.g., negative binomial distribution models~\citep{bearden1999handbook,kooti2016portrait}.
Features related to information gathering and the purchase potential (e.g., monetary resources and product values) also help to predict the purchase intention~\citep{bearden1999handbook,hansen2004predicting,pavlou2006understanding}.

\begin{header}{Feature-based methods}
User-aware features have been successfully applied to predict purchase intent in e-commerce scenarios.
\citet{qiu2015predicting} put forward a pipeline-based purchase-prediction approach that includes three main components. 
First, the authors use associations between products to predict the needs of customers; then, they combine collaborative filtering and a hierarchical Bayesian discrete choice model enable customer preference learning; lastly, they construct a support vector regression-based model to calculate the popularity of products.
After analyzing user behavior on~Pinterest, \citet{lo2016understanding} propose a predictor to detect a user's purchase intent. 
The authors apply five kinds of feature: demographics, activity, action-type, content, and temporal features.
\end{header}
\citet{wan2017modeling} also introduce a three-stage model to predict the purchase behavior on a real-world e-commerce portal. 
To identify who can be converted to regular loyal buyers and then targeted to reduce promotion cost,~\citet{liu2016repeat} describe a solution for repeat buyer prediction; they collect a large number of features to capture the preferences and behavior of users, characteristics of merchants, brands, categories, and items, and the interactions among them.
\citet{hendriksen-2020-analyzing} analyze the potential of long-term historical records (from logged-in users) to more accurately and reliably predict purchase intent.
Additionally, \citet{ariannezhad-2021-understanding} show that data that provides information on customer behavior in one channel (e.g., online) can help to predict purchase intent in other channels (e.g., offline).
Social media has become another important source of information to help explore consumer purchase intentions~\citep{mishne-2006-deriving,zhang2013predicting,ding2015mining}. 
\citet{zhang2013predicting} explore whether users' social media information is correlated with their e-commerce profiling categories. 
Accordingly, the authors use correlations to build machine learning algorithms to predict user purchase behavior.

\begin{figure}[!t]
	\centering
	\includegraphics[width = 0.65\columnwidth]{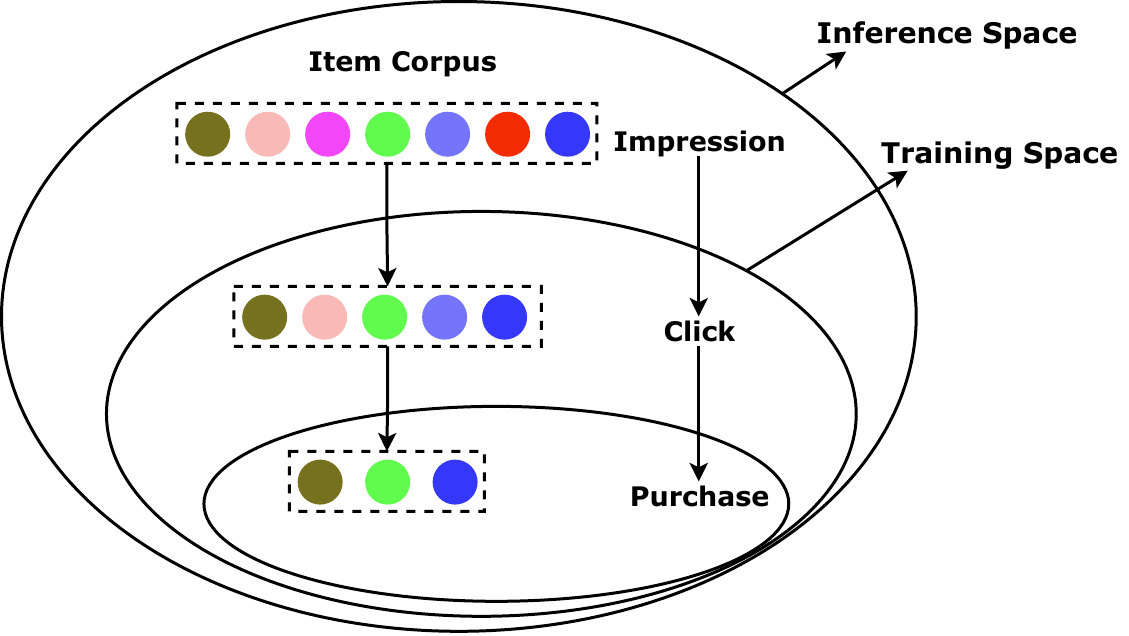}
	\caption{Illustration of the sample selection bias problem in conventional CVR prediction, where the training space only consists of clicked samples, whereas the inference space is the entire space of all items. Image source: \citep{wen2020entire}.}
	\label{ch32:fig:cvr2020entire}
\end{figure}

\begin{header}{CVR prediction methods}
\emph{Conversion rate} (CVR) prediction is another way to predict the user purchase intention on e-commerce platforms.
CVR calculates the proportion of users who will eventually convert after clicking~\citep{lee2012estimating,yang2016large,lu2017practical,wen2019multi,su2020attention,wen2020entire,yasui2020feedback,yang2021capturing,hou2021conversion,li2021conversionpred}.
Because conversions are extremely rare, CVR modeling is very challenging. 
CVR can be split into the following two categories: post-view conversion and post-click conversion, i.e., conversion after viewing an item without having clicked it, and conversion after having clicked the item, respectively. 
Most approaches focus on the task of post-click conversion.     
\end{header}
\citet{wen2019multi} propose a decision tree ensemble model, i.e., \emph{ldcTree}, that exploits deep cascade structures and applies cross-entropy based feature representations. 
Nonetheless, there are still three challenges in CVR estimation: \emph{data sparsity}, \emph{sample selection bias}, and \emph{delayed feedback}.
The data sparsity problem reflects the insufficiency of click samples in training data.
{Sample selection bias} refers to the systematic difference in the data distribution between the training space and inference space~\citep{wen2020entire}. 
Fig.~\ref{ch32:fig:cvr2020entire} illustrates the sample selection bias problem related to the development of an efficient industrial-level recommender system.
The delayed feedback problem indicates that the e-commerce platform can receive feedback with a delay after an item is impressed or clicked by a user~\citep{chapelle2014modeling,su2020attention}.

To address the data sparsity problem, \citet{ma2018entire} consider the entire space multi-task model, which aims to apply multi-task learning to accomplish two subtasks of predicting the post-view click-through rate and post-view conversion rate.
\citet{wen2020entire} observe that users always engaged in abundant purchase-related actions after clicking. 
Thus, they propose a deep neural network method within a multi-task learning framework to decompose post-click behavior to predict CVR. 
The authors distinguish between purchase-related actions and other actions, which, taken together, can be used to form a probabilistic sequential user behavior graph.

The above multi-task strategies are also helpful in alleviating the selection bias problem. 
Furthermore, \citet{zhang2020large} detail a doubly robust estimation method to debias CVR prediction.
\citet{yasui2020feedback} provide a dual learning method to simultaneously address the delayed feedback problem and the selection bias problem.
To reduce the variance of doubly robust loss to enhance model robustness, \citet{guo2021enhanced} enhance a more robust doubly robust approach for debiasing post-click conversion rate estimation.
But the authors do not directly control the bias and the variance in an effective way.
To address this problem, \citet{dai2022kddcvr} propose a generalized framework of doubly robust learning, which unifies the existing doubly robust methods. Based on this framework, two new doubly robust methods were proposed to control the bias and mean squared error.

\citet{chapelle2014modeling} describe a delayed feedback model to optimize CVR as a joint probability over the predicted CVR and the delayed time distribution.
\citet{yoshikawa2018nonparametric} extend this delayed feedback model to a non-parametric model. 
\citet{su2020attention} focus on post-click calibration in CVR modeling. The authors extract pre-trained embeddings from impressions/clicks to enhance the conversion models; they propose an inner/self-attention mechanism to capture the fine-grained personalized product purchase interests.
To estimate unbiased CVR in the online settings, \citet{yang2021capturing} introduce a elapsed-time sampling delayed feedback model to track relations between the observed conversion distribution and the true conversion distribution.
\citet{li2021conversionpred} use an idealized dataset for training a prophet model that can use the data properly, and then learn the actual model by imitating the prophet.
\citet{chen2022kddasymptotically} confirm the importance of dividing observed samples in a more granular manner, and hence propose an unbiased importance sampling method with two-step optimization to address the delayed feedback issue.

These purchase-intent modeling strategies have been applied in e-commerce searches and recommendations; we will discuss these strategies in more detail in Sections~\ref{ch4:sec33:l2r} and~\ref{ch5:crm}.

\begin{header}{CLTV prediction methods}
Lastly, \emph{customer lifetime value} (CLTV) prediction has also received attention in recent years. 
CLTV is an important task in e-commerce search and recommendation models.  
CLTV is defined as the sales, net of returns, of a customer over a 1-year period.  
The objective of CLTV prediction is to improve three key business metrics: 
\begin{enumerate*}[label=(\roman*)]
\item the average customer shopping frequency, 
\item the average order size, and
\item the customer churn rate. 
\end{enumerate*}
With CLTV prediction, e-commerce retailers can rapidly identify and nurture high-value customers~\citep{vanderveld2016engagement}.
Classic work on CLTV prediction applies handcrafted features and ensemble classifiers \citep[e.g., GBDT,][]{chen2016xgboost}.     
\end{header}
\vspace{-4mm}
\section{User profiling in e-commerce}
\label{ch3:sec:upem}

A \emph{user profile} refers to personal information about a specific user. 
Personalization plays an important role in web search and recommender systems. 
In e-commerce portals, user profiling is a critical module for e-commerce information discovery tasks, as it provides personalized content in search or recommendation results. 
User profiling can be defined as the process of exploring information about a user’s interest domain~\citep{dong2014inferring,kanoje2015user}. 
Information about a user can be used by e-commerce search and recommender systems to enhance the system effectiveness because it enables better user understanding (see Section~\ref{ch3:sec24:person} and~\ref{sec:re-ranking}). 
Given that it originated from work on the prediction of user purchase intention, research into user profiling in e-commerce has continuously garnered attention over the years~\citep[see, e.g., ][]{solomon1994buying,braynov2003personalization,hollerit2013towards,zhang2013predicting,gupta2014identifying,rahdari2017analysis,Huangwww2018}.
Early work on user profiling for e-commerce mainly focuses on information filtering, social media analysis, web searches, and fraud detection~\citep{solomon1994buying,fawcett1996combining,adomavicius1999user,kuflik2000generation,braynov2003personalization}. 
Most of these are rule-based strategies.
For example, \citet{fawcett1996combining} employ a rule-based user-profiling method to uncover indicators of fraudulent behavior. 
These indicators were used to create user profiles that were then applied as features of their proposed system, which combines evidence from multiple profilers to generate high-confidence alarms. 

\subsection{Types of user profiling}
User profiling can be classified as either \emph{profile extraction} and \emph{profile learning}~\citep{tang2010combination}. 
Profile extraction focuses on extracting information about a user, such as demographic data (e.g., age, gender, location) or basic behavior patterns. However, in e-commerce, profile extraction is often less critical because it provides only a fixed snapshot of the user, lacking the adaptability needed to capture evolving preferences and real-time behavioral changes.
In contrast, profile learning is more significant as an e-commerce modeling tool~\citep{kuflik2000generation,cufoglu2014user}. 
Profile learning methods can be grouped into three categories: 
\begin{enumerate*}[label=(\roman*)]
\item content-based methods, 
\item collaborative methods, and 
\item hybrid methods~\citep{cufoglu2014user}. 
\end{enumerate*}
Content-based methods infer user profiles based solely on the users' own previous behavior~\citep{kuflik2000generation}. 
Collaborative methods in user profiling focus on applying collaborative filtering approaches to infer profile information based on the behavior of users in a suitably defined neighborhood of similar users. Collaborative filtering methods can be classified as either \emph{memory-based methods} or \emph{model-based methods}~\citep{godoy2005user,cufoglu2014user}.  
Memory-based solutions estimate ratings for a user based on the entire collection of previous ratings of similar users~\citep{adomavicius2005toward,su2009survey}. 
In contrast to memory-based collaborative filtering methods for user profiling, model-based methods use the collection of ratings to learn a model to estimate user profiles~\citep{su2009survey}.
Hybrid methods have garnered attention because they combine content-based methods and collaborative methods~\citep{godoy2005user}.
Regarding research on personalized recommendations, to capture users' information and interest, \citet{liu2010iui} introduce a dynamic collaborative filtering method for news recommendation where the recommender constructs user profiles based on their past click behavior. 
The authors conduct a log analysis of the changes in user interest in news topics over time. 
By classifying users' news interests as either \emph{genuine interests} or the \emph{influence of local news trends}, the authors are able to 
\begin{enumerate*}[label=(\roman*)]
\item construct a Bayesian framework to model a user’s genuine interests based on their past click behavior, and 
\item predict current interests by jointly analyzing the genuine interest and the local news trends.
\end{enumerate*}

\subsection{User profiling with social media}
Social media is playing an important role in e-commerce user profiling.
On the one hand, social media provides a source for generating user profiles, especially when addressing the \emph{cold start} problem. 
\citet{mishne-deriving-2006} provided an early example of this idea, using a combination of text analysis and external knowledge sources to estimate the commercial tastes of bloggers from their posts.
{On the other hand, profiling information learned from social media data can be applied to explore the user's purchase intentions on e-commerce platforms.}
{Targeting users with no history on an e-commerce site, \citet{zhang2013predicting} focus on predicting the purchase behavior by proposing a 
feature-selection method to predict the product categories from which a user will buy.}
Representing user purchase intent as textual information that indicates a desire to purchase a product or service in the future, \citet{gupta2014identifying} propose a binary classification approach to identify the user purchase intention based on their social media posts.
\citet{ding2015mining} explore relationships between a user’s consumption habits and their social media data. They propose a consumption intention mining model (CIMM) based on CNNs. 
\citet{lo2016understanding} analyze user activities in social media to build a time-varying model to predict user purchase intent. The authors analyze Pinterest\footnote{\url{https://www.pinterest.com}} data to understand how the usage of an e-commerce platform relates to future user shopping behavior. They find that indicators of purchase intent tended to gradually build up over time and sharply increase 3 to 5 days before purchase. 
Multi-modal information has also been applied to infer user profiles. 
\citet{gelli2017personality} focus on automatically discovering actionable images for users according to their personality. 
By applying their model to a large-scale dataset, the authors find a significant correlation between personality
traits and affective visual concepts in the image content.

\subsection{Graph-based user profiling}
\begin{figure}[!t]
	\centering
	\includegraphics[width = 0.85\columnwidth]{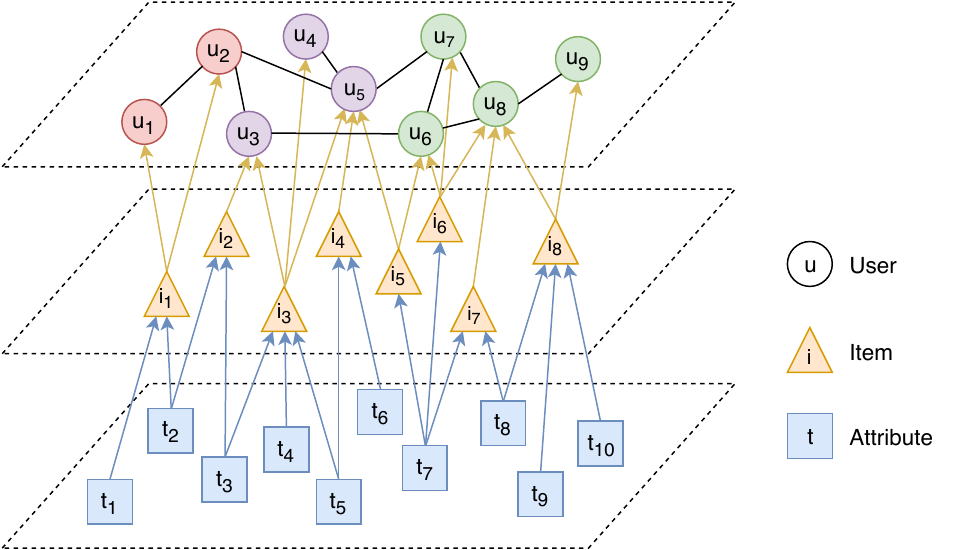}
	\caption{User profiling results in the form of a heterogeneous graph. Image source: \citep{chen2019semi}.}
	\label{ch33:fig:hgal}
\end{figure}

Most approaches to user profiling only use a single type of information.
In e-commerce modeling, heterogeneous graphs are also being used to work with user profiles. 
Fig.~\ref{ch33:fig:hgal} shows an example of a graph with heterogeneous information; particularly, three kinds of nodes were applied to represent three types of data, i.e., users, items, and attributes, respectively. 
\citet{chen2019semi} focus on applying rich interactions among data instances, i.e., co-click and co-purchase behavior on e-commerce platforms, to enhance user-profiling performance in e-commerce models.
Neighborhood features provide useful information that helps to infer user profiles, e.g., users that have similar co-purchase behavior on e-commerce platforms are likely to be of a similar age.
The authors propose a heterogeneous graph attention network to infer user profiles within a multi-type data environment; the network is able to model unsupervised information in a heterogeneous graph by encoding the graph structure and node features.
\citet{gu2020hierarchical} construct a hierarchical profiling framework to model users' real-time interests at different granularities.
A pyramid recurrent neural network model for hierarchical user profiling is constructed based on users’ micro-behavior; it is subsequently applied to model the types and dwell times of behavior to enable an effective formulation of users' real-time interests.
These graph-based user-profiling methods have been applied to enhance the re-ranking results in e-commerce recommendation systems. More details are discussed in Section~\ref{sec:re-ranking}.

\section{Emerging directions}
\label{ch3:secfuture}

\subsection{Graph learning for user behavior modeling}

Graph neural networks use neural networks to represent graph information~\citep{scarselli2009graph,bruna2013spectral,battaglia2016interaction}.
Graph convolutional networks extend CNNs to graph structured data; they have been shown to be effective on a range of graph classification tasks~\citep{bruna2013spectral,defferrard2016convolutional,zhao2021sigirwgcn,liu2022sigirgraph}, and when applied for semi-supervised classification~\citep{hamilton2017inductive,kipf2016semi} and link prediction~\citep{ma2018dynamic}. 
Most of these models have been designed for static graphs. However, \citet{ma2018dynamic} propose a dynamic graph neural network model that can model dynamic information.
\citet{xu2021transformer} enhance the capacity for learning complicated temporal dependencies in a graph, by proposing a transformer-style relational reasoning network with a dynamic memory updating mechanism. 
Graph neural networks that applied to recommender systems learn item embeddings within a large-scale item relationship graph~\citep{ying2018graph} that describes users, items, and pairwise relations in e-commerce scenarios~\citep{ma2018dynamic,gaowsdm2022rec}.
Graph learning can also be applied for post-click modeling and user purchase-intent modeling.
To predict fine-grained post-click CVR, \citet{bao2020gmcm} design a model to represent user micro-behavior as a purchase-related micro-behavior graph. 
The authors apply a multi-task learning framework to construct a graph-based micro-behavior conversion model that can capture the correlation between different types of micro-behavior. 
The proposed multi-task learning and inverse propensity weighting modules mitigate the data sparsity- and sample selection bias-related problems.
To prediction efficient and accurate CVR, \citet{wen2021hierarchy} propose a graph neural network to hierarchically model both micro- and macro-behaviors in a unified framework.
It seems likely that graph neural networks will continue to facilitate new ways of modeling, gaining insights into, and predicting, e-commerce user behavior in search (Section~\ref{ch4:sec24:lre}) and recommendation scenarios (Section~\ref{sec5:ebm}).

\subsection{Dynamic user behavior modeling and profiling}

Most studies on e-commerce user behavior modeling and user profiling are conducted under the assumption that a snapshot of user behavior is recorded on e-commerce portals. However, a user’s personal interests and behavior may change over time~\citep{koren2009collaborative,gao2013modeling,yin2014temporal,huang2015tencentrec,yin2015dynamic,jagerman2019people,wang2021clicks,huang2022clickrec}. Thus, modeling e-commerce users' temporal behavior is important for e-commerce search and recommendation system development.
Social networks and social media provide a rich source of information for temporal models of user behavior~\citep{mislove2010you,gao2013modeling,yin2014temporal,yin2015dynamic}. 
\citet{yin2014temporal} design a latent mixture model, which they named a temporal context-aware mixture model, to account for the intentions and preferences that drive user behavior.
It models the topics related to users' intrinsic interests, and the topics related to temporal context’ it then jointly analyzes the influences of the two factors to model user behavior in a unified way. 
To enable the dynamic learning of user profiles, \citet{cao2017you} developed a model that considers multiple information sources and their relations.
Similarly, \citet{liang2018dynamic-kdd} proposes a streaming profiling algorithm that initially applies a user-expertise tracking model to track the changes in the dynamic expertise of users; it then uses a keyword diversification algorithm to produce top-$k$ diversified keywords that allow the users’ dynamic expertise to be profiled at a specific timestamp.

Time and temporal phenomena are valuable sources of information that can facilitate the understanding and prediction of user behavior; this area of research is likely to continue to garner much attention in the near future.

\subsection{User modeling with insufficient data}

As we have discussed in this section, a wide range of click models and ranking methods can be applied to model user click behavior~\citep{chuklin-click-2015,borisov2016neural,ferro2017including,he2017neural,wu2018turning,ferro2019boosting,liu2022sigirgraph,vardasbi-2022-probabilistic}. 
However, the models developed to date tend to require fully labeled data to train the ranking models, although, in realistic e-commerce scenarios, not all of a user's behavior can be recorded. 
Similarly, regarding post-click modeling, purchase-intent prediction, and user-profiling tasks, the reality that there is a limited amount of behavioral data makes it difficult to work with existing click modeling and ranking solutions. 
For example, in the case of CVR tasks, the data sparsity problem arises when the number of training samples for the sequential behavior of the form ``click $\to$ purchase'' is insufficient to fit the large parameter space of the CVR task~\citep{wen2020entire,guo2021enhanced}.
How to enhance user modeling performance under the conditions of a limited amount of imperfectly labeled data remains an important open problem in e-commerce.

\chapter{E-commerce search}
\label{chapter:search}

E-commerce search, or simply ``product search,'' represents a special retrieval scenario where users submit queries to retrieve products using a search engine~\citep{ai2017learning}.
E-commerce search portals are gaining in popularity as many consumers choose e-commerce search on an e-commerce platform rather than generic web search~\citep{li2011towards}. 
Unlike in web search, in e-commerce search there can be millions of results to surface for a given search query~\citep{wu2018turning}.
We have discussed user behavior modeling and user profiling in Section~\ref{chapter:user} to understand how to explore and exploit information from user behavior. 
In this section, we focus on the other side of the coin, on search technologies that are based on users' interactive behavior. 
To learn about e-commerce search solutions, we discuss research on query understanding and ranking technologies for e-commerce search.
In Section~\ref{ch4:sec:ces} we summarize characteristics of e-commerce search.
In Section~\ref{ch4:sec33:eva} we recall key metrics used for evaluating e-commerce search.
In Section~\ref{ch4:sec:qes} we present studies on representing e-commerce search queries. 
Then, we detail e-commerce ranking approaches in Section~\ref{ch4:sec:rs}. 
Finally, we discuss emerging research directions in e-commerce search in Section~\ref{ch3:sec:fd}.

\section{Characteristics of e-commerce search}
\label{ch4:sec:ces}

Before detailing related work, we highlight characteristics of e-commerce search. We divide this section into two parts: an overview of e-commerce search, and challenges in e-commerce search.

\subsection{Overview of e-commerce search}
\label{ch3:sec11:overview}

Early e-commerce search approaches are based on traditional information retrieval theory and faceted search models~\citep{yee2003faceted,jansen2006effectiveness}. 
\citet{jansen2006effectiveness} explore the difference between ad-hoc search and e-commerce search. 
A grocery retrieval system has been developed by considering a discrepancy between consumers' shopping lists and retailers' stock information~\citep{nurmi2008product}.
Early e-commerce search systems rely on information that retailers make available: either semantic markup on unstructured HTML documents or a data feed provided in some predefined structured format.
Product resolution~\citep{balog2011investigation} focuses on recognizing webpages that represent the same product.
Based on the task of product resolution, \citet{duan2013probabilistic} propose a probabilistic mixture model for mining and analyzing product search logs. 
Similar setups can be also found in a product-aware keyword search system~\citep{duan2013supporting}.

Unlike traditional ad-hoc retrieval, e-commerce search relies on a decision mechanism about consumers' purchase behavior in e-commerce portals~\citep{li2011towards}.
There are two main stakeholders in e-commerce search, consumers and business owners, whose interests align but also conflict to a  certain extent~\citep{tsagkias-2020-challenges}.
In e-commerce search, customers do not just browse relevant items, but also try to locate an item that satisfies their specific purchase intent~\citep{li2011towards}.
While consumers aim to find the best quality at the lowest price, businesses want to maximize profit, which translates into higher prices for customers or lower costs for businesses.
E-commerce search typically requires more structured information (e.g., brands, categories, shops, etc.) than web search and more diversified personal definitions of ``relevance'' during search sessions.
On the one hand, as we have discussed in Section~\ref{ch2:sec2}, users come to e-commerce websites with a wide spectrum of intents. 
Hence, multiple user behavior discussed in Section~\ref{chapter:basic} and~\ref{chapter:user}, e.g., clicks, post-clicks, and purchases, etc., should be integrated to model the ``relevance'' in e-commerce search.
On the other hand, only a few products are actually purchased by the consumers and different individuals have different opinions even about the same product~\citep{ai2017learning}.
Thus, e-commerce search should consider users’ differences to satisfy the needs of all consumers.
In general, there are four unique characteristics in e-commerce search:

\begin{itemize}[leftmargin=*,nosep]
\item \textbf{Consumer query intent.} Similar to web search, queries in e-commerce search can be divided into three classes: navigational, informational and transactional~\citep{li2011towards}. However, e-commerce search queries take a different form. Specifically, navigational queries are product serial numbers and inquiries for customer support; informational queries include leaves in the product taxonomy and product attributes; and transactional queries are a mix of navigational and informational queries. Unlike traditional web search, there are three query intents for e-commerce search: \emph{target finding, decision making, and exploration}~\citep{su2018user}. Following the well-known web-search taxonomy due to~\citet{broder2002taxonomy}, \citet{su2018user} describe a hierarchical e-commerce search taxonomy to explore consumers' shopping intents, with \emph{shallow exploration}, \emph{targeted purchase}, \emph{major-item shopping}, \emph{minor-item shopping}, and \emph{hard-choice shopping}.
The authors find that consumers tend to conduct more focused searches in target finding sessions compared to those in the decision making and exploration sessions.
In target finding sessions, consumers tend to issue a few specific queries and browse only top ranked results; in decision making sessions, consumers tend to issue short queries, browse deep, and click more results; and in exploration sessions consumers issue many diverse queries but do not click often.
Given these search intents, customized search approaches for each type of search queries can be developed to improve the utility of e-commerce search.

\item \textbf{Heterogeneous consumer behavior.}  As we have described in Section~\ref{chapter:basic}, multiple types of user behavior can be observed in e-commerce platforms. During an online shopping journey, a consumer may have multiple targets at different stages. \citet{blake2016returns} observe that, during a journey, e-commerce search proceeds as a kind of ``funnel'' where, initially, search is along broad categories, and then it becomes more refined to obtain an item at the lowest cost given a consumer’s cost of search. Hence, e-commerce search approaches should be aware of the
stages in each consumer's journey. Meanwhile, the overall impact of heterogeneous consumer behavior also makes e-commerce search different from traditional web search. Users' micro behavior, post-click behavior, and engagement make the search intent dynamic and complicated during search sessions~\citep{zhouwsdm2018,wu2018turning}. 

\item \textbf{Online and offline ranking.} Traditional learning to rank methods sort documents according to their relevance to the query. 
E-commerce search has an intrinsic difference in the relevance in rankings: the notion of ``relevance'' is blurred~\citep{wu2018turning}. 
Users come to e-commerce platforms with a wide spectrum of intents. Some users wish to make a purchase as soon as possible while others are just wandering around the platform to get inspired. Hence, various kinds of signals, including clicks, favorites, adding carts, purchases, etc., should be integrated to model relevance in e-commerce search. 
E-commerce businesses having both online and physical presence bring a unique blend of infrastructure challenges~\citep{ariannezhad-2021-understanding}. 
Thus, users' shopping experiences lie in smooth transitions from offline to online and vice versa~\citep{tsagkias-2020-challenges}.  

\item \textbf{Business criteria and metrics.} Most e-commerce platforms apply \emph{Gross Merchandise Volume} (GMV) as the gold standard for measuring success, which indicates the total amount of sales during e-commerce activities. 
{Thus, one of the main targets of an e-commerce search algorithm should be to maximize the value of purchases per search session~\citep{wu2018turning}.}
Many e-commerce search engines apply a two-stage framework to resolve the whole process into two successive subtasks: a ranking problem and a classification problem~\citep{wu2018turning}.
This two-stage search process on e-commerce platforms makes the optimization more complicated in contrast to web search. In e-commerce, regulatory and business constraints decide which products can be shown to which consumers, whereas competing brands can have agreements with an online retailer to restrict showing their products with those of their competitors. Therefore, it is important to understand consumers' inventory gaps and provide alternatives in e-commerce search~\citep{tsagkias-2020-challenges}. 
\end{itemize}

\subsection{Challenges in e-commerce search}
\label{ch4:sec11:challenge}

Based on the above criteria, we see two main challenges in e-commerce search~\citep{rowley2000product,jansen2006effectiveness,li2011towards,duan2013probabilistic,ai2017learning,trotman2017architecture,wu2018turning}.
First, there exists a mismatch between users' queries and product representations where both use different terms to describe the same concepts~\citep{li2011towards}.
This mismatch problem is even more severe in personalized search when more personalized information needs to be considered during retrieval.
Second, the ranking problem in e-commerce search is challenging: multiple types of information sources make ranking products in e-commerce search more complicated than in web search.
Diverse relevance factors make it difficult to use traditional static-ranking evaluation metrics, e.g., NDCG and MAP, to measure the quality of rankings in e-commerce search.

Recent studies on e-commerce search that aim to tackle the above challenges,  focus on one of two aspects~\citep{trotman2017architecture,wu2018turning}: 
\begin{enumerate*}[label=(\roman*)]
\item matching optimization in e-commerce search, i.e., the vocabulary gap problem, representation-based matching, interaction-based matching, and matching in personalized search; and
\item ranking optimization in e-commerce search, i.e., learning to rank methods and evaluation metrics.
\end{enumerate*}
For real-world e-commerce search, a joint online and offline search framework with both semantic matching and ranking optimization modules is able to outperform traditional search systems at both semantic retrieval and personalized ranking scenarios~\citep{li2019semantic}.
In Section~\ref{ch4:sec:qes} and~\ref{ch4:sec:rs}, we summarize recent work on e-commerce search that is aimed at tackling the two research challenge listed above. Prior to that, we introduce the evaluation metrics used to assess e-commerce search in Section~\ref{ch4:sec33:eva}.

\section{Evaluation metrics}
\label{ch4:sec33:eva}

Evaluation in web search focuses on the relevance to a given query of documents. 
E-commerce search provides multiple signals to judge the saliency of items. 
Besides for relevance of an item to a given query, revenue-aware features are also considered in e-commerce search evaluation~\citep{wu2018turning}.

\subsection{Relevance-based metrics}
Relevance-aware evaluation metrics are in various information retrieval domains~\citep{schutze2008introduction}, including in e-commerce search.
It is common to see studies compute evaluation metrics based on the top 100 items retrieved by each e-commerce search model~\citep{ai2017learning,van2016learning,van2018mix}.
Mean average precision (MAP), hit ratio (HR), mean reciprocal rank (MRR), and normalized discounted cumulative gain (NDCG) are four widely-used relevance-aware metrics~\citep{van2018mix,wu2018turning}. 
These metrics are also widely applied in various information retrieval domains~\citep{schutze2008introduction}.
Average precision (AP) computes the average value of Precision over the interval from $0$ to $1$. Given $k$ candidate items,  $AP@k = \sum_{k = 1}^n {P@k \cdot rel(k)} $, where $P@k$ refers to Precision@k; $rel(k)$ indicates $1$ if the $k$-th item is relevant and $0$ otherwise $rel(k)=0$. 
Based on that, MAP calculates the mean of AP for all queries:
\begin{equation}
\label{eq:mapdef}
\mathit{MAP} = {1 \over Q}\sum\limits_{q = 1}^Q \mathit{AP}_q,
\end{equation}
where $Q$ denotes the number of queries. HR@$k$ refers to the fraction of queries for which the relevant item is included in the top-$k$ results, so we have:
\begin{equation}
\label{eq:hrdef}
\mathit{HR}@k = {{|Q_{rel}^k|} \over Q},
\end{equation}
where ${|Q_{rel}^k|}$ denotes the number of queries for which the relevant item is included in the top-$k$ results. MRR, also known as average reciprocal hit ratio~\citep{radev2002evaluating}, evaluates processes where a list of possible responses to a sample of queries ordered by probability of correctness:
\begin{equation}
\label{eq:mrrdef}
\mathit{MRR} = {1 \over Q}\sum_{i = 1}^Q {1/\mathit{rank}_i},
\end{equation}
where $rank_{i}$ refers to the rank position of the first relevant document for the $i$-th query.
DCG@$k$ evaluates the relevance of a document based on its position in the top-$k$ results:
\begin{equation}
\label{eq:dcgdef}
\mathit{DCG}_k = \sum\limits_{i = 1}^k {{{re{l_i}} \over {\log (i + 1)}}},
\end{equation}
where $rel_{i}$ indicates the relevance of the document at position $i$. Ideal DCG (IDCG) is the DCG score for the ideal ranking, which is ranking the items top down according their relevance up to position $k$. NDCG allows one to compare the performance across different queries, using normalization of DCG by IDCG: 
\begin{equation}
\label{eq:ndcgdef}
\mathit{NDCG}@k = {\mathit{DCG}_k \over \mathit{IDCG}_k},
\end{equation}

\subsection{Revenue-aware metrics}

GMV indicates the total income amount transacted from merchandise sales, whereas the overall revenue generated for the e-commerce site is proportional to the GMV~\citep{wu2018turning}.
Revenue-aware metrics are applied to e-commerce search evaluation to evaluate how the methods can improve the actual revenue of search sessions. 
To calculate the average revenue for every impression in each e-commerce search session, \citet{wu2018turning} introduce the average revenue metric, $Avg.Rev$ $(i,q)$,  for a query-item pair $(i,q)$ as follows:
\begin{equation}
\label{eq:avgrevdef}
Avg.Rev\left( i,q\right)  =\frac{price\left( i\right)  \times purchase\left( i,q\right)  }{\left| S_{q}|i\in S_{q}\right|  },
\end{equation}
{where $purchase\left( i,q\right)$ denotes the number of times that the item $i$ has been purchased in a search session for a query $q$; and $\left| S_{q}\right|$ is the set of search sessions for query $q$ where the item $i$ is impressed.}

\citet{wu2018turning} specify a metric named $Rev@k$ to evaluate the revenue in e-commerce rankers. $Rev@k$ calculates the average revenues that a prediction algorithm would generate for each session. Specifically, $Rev@k$ is calculated as follows:
\begin{equation}
\label{eq:revdef}
Rev@k(\rho ) = \sum_{s \in S} {\sum_{{r_s} \le k} {price({r_s}^{ - 1})} } \Phi ({r_s}^{ - 1}),
\end{equation}
where $\rho$ is the ranking order and $r_{s}  \le k$ denotes the top-$k$ ranked positions in the session $s$. $r_{s}^{-1}$ denotes the corresponding item at the position $r_{s}$, $price(i)$ indicates the price of item $i$, while $\Phi$ denotes a purchase event. Based on $Rev@k$, it is able to evaluate the revenue influence of the candidate rankers.

Empirical studies have been performed in benchmark e-commerce search datasets to find differences between relevance-based metrics and revenue-aware metrics~\citep{wu2018turning}. Relevance-based metrics primarily measure the success of retrieving relevant data from user logs, while revenue-aware metrics provide a clearer understanding of how ranking methods influence actual revenue in e-commerce scenarios. This difference shows that, although relevance is important, incorporating revenue-aware metrics offers more practical insights into the financial impact of search ranking methods within e-commerce settings.

The lack of annotated real-world benchmarks poses a challenge for conducting large-scale empirical studies in e-commerce scenarios. To address this issue, recent studies have started using both synthetic and semi-synthetic datasets in their experiments, allowing for more controlled analysis while approximating real-world conditions~\citep{xu2022product}.

\subsection{User engagement metrics}

{As we discussed in Section~\ref{subsec:pcbec}, user engagement is defined as the quality of user experience in interaction with a system, characterized by various attributes, e.g., positive affect, aesthetic and sensory appeal, attention, novelty, and perceived user control~\citep{mathur2016engagement}.
In recent years, engagement metrics have been applied to evaluate the quality of interaction between the user and e-commerce search engines~\citep{vanderveld2016engagement}.
User engagement metrics in e-commerce can be divided into two categories: short-term engagement metrics and long-term engagement metrics~\citep{zou2018drlunderreview}.} 

To evaluate the quality of short-term user-system interactions, short-term engagement metrics about instant clicks, purchases, and dwell time are used in e-commerce search evaluation. 
As we discussed in Section~\ref{ch31:sec:ubm} and~\ref{ch31:subsec:pm}, the \emph{click-through rate} (CTR) and \emph{conversion rate} (CVR) are the two most widely applied metrics to evaluate instant click and purchase prediction.
Meanwhile, dwell time and bounce rate are the two main metrics used in short-term post-click evaluations~\citep{lalmas2015promoting}.

As we discussed in Section~\ref{subsec:pcbec}, long-term user engagement reflects the user's desire to stay on the e-commerce portal longer and use the service repeatedly~\citep{zou2018drlunderreview}, i.e., the ``stickiness.''
Long-term user engagements measure versatile user behaviors based on a very large number of environmental interactions. 
Multiple long-term engagement metrics have been applied in previous studies. 
\citet{wu2017returning} employ cumulative clicks over time to estimate the long-term interactions between the user and the system. 
\citet{zou2018drlunderreview} apply three evaluation metrics to evaluate the long-term user interaction behaviors: 
\begin{enumerate*}[label=(\roman*)]
\item average clicks per session: the average cumulative number of clicks over a user visit; 
\item average depth per session: the average browsing depth that the users interact with the recommender agent; and 
\item average return time: the average revisiting days between.
\end{enumerate*}
In multilingual scenarios, a transformed query converts a secondary language query into a semantically equivalent query in the primary language, allowing it to fully use the search engines' abilities.
To evaluate transformed queries in multilingual e-commerce search, a set of behavior metrics based on user engagement specific to the existing search system was developed. 
Using these metrics, a query transformation system has been built and tested both offline and through online A/B tests on the Amazon platform, showing improvements in the multilingual search experience for customers~\citep{hu2020query}.

\section{Matching strategies in e-commerce search}
\label{ch4:sec:qes}

In this section, we showcase recent research on matching in e-commerce search, especially concerning e-commerce query understanding and processing.
We divide this section into five parts: 
\begin{enumerate*}[label=(\roman*)]
\item we recall the vocabulary gap problem in Section~\ref{ch3:sec24:vocabulary};  
\item we detail representation-based matching approaches in Section~\ref{ch4:sec24:rbm};
\item we detail interaction-based matching approaches in Section~\ref{ch4:sec24:lre}; 
\item we detail hybrid matching approaches in Section~\ref{ch4:sec434:hybrid}; and 
\item Section~\ref{ch3:sec24:person} describes studies on query processing in personalized e-commerce search.
\end{enumerate*}

\subsection{Vocabulary gap}
\label{ch3:sec24:vocabulary}

Users' shopping lists often differ from the product information maintained by retailers~\citep{nurmi2008product}.
\citet{duan2013probabilistic} find that while query languages such as SQL can be successfully applied to search in these product databases, their usage is difficult for non-experienced end users.
In direct search scenarios in e-commerce platforms, most consumers formulate queries using characteristics of the product they are interested in (e.g., terms that describe the product's categories, brands, and shops, etc.).
Hence, it is common to see a mismatch in e-commerce search between queries and product representations, where different tokens are used to describe the same product, i.e., the \emph{vocabulary gap problem}~\citep{van2016learning}.
To address this problem, early studies focus on rewriting verbose queries in e-commerce search~\citep{bendersky2009analysis,xue2010improving,singh2012rewriting}.
Selecting a subset of the original query (i.e., ``sub-query'') has been shown to be effective for improving these queries.
\citet{xue2010improving} formally model the distribution of sub-queries, where the sub-query selection procedure is modeled as a sequential labeling problem. A conditional random field model was applied to track the local and global dependencies between query words.
\citet{singh2012rewriting} present techniques to reduce long queries to effective shorter ones that lack superfluous terms. 
The authors describe a system that provides high quality product recommendations for null queries, where time-based relevance feedback is used to improve the fidelity of rewrites.
However, these approaches focus only on the query space and overlook critical information from the product space and the connection between the two spaces~\citep{van2016learning}.

To address the vocabulary gap problem, a series of studies have been proposed to match queries and product information in e-commerce search. 
Following~\citet{sarvi-2020-comparison}, matching solutions can be divided into \emph{representation-based}, \emph{interaction-based}, and \emph{hybrid} approaches. 
All of them have been shown to be effective for matching consumers' queries and product information~\citep{li2019semantic,yao2022reprbert}.

\subsection{Representation-based matching}
\label{ch4:sec24:rbm}

In early work, approaches to the task of \emph{entity finding} have been applied to address the vocabulary gap problem between queries and products, where products are viewed as retrievable entities~\citep{balog2010entity,gade2016overview,van2016learning}. 
There are two problems in entity finding for e-commerce search~\citep{de2007overview}.
First, entity finding retrieves entities of a particular type from multi-domain knowledge bases, whereas e-commerce search systems operate within a single but dynamic domain. Second, queries in e-commerce search contain a lot of free-form text~\citep{rowley2000product}, whereas in entity finding most queries are semi-structured with relational constraints~\citep{balog2010entity}.

To address the above two problems of matching optimization, representation learning methods have been applied to obtain better representations for each text associated with products by encoding both the query and the product title into single embedding vectors. Representation learning helps by flexibly adapting to the dynamic nature of e-commerce domains. 
It also effectively encodes the free-form text of queries and product descriptions into a common semantic space. 
Numerous neural text matching methods have been developed~\citep{onal-neural-2018,mitra2017introduction,lin-2021-pretrained}.
DSSM is one of the earliest deep learning-based models in text matching, in which each text is vectorized separately by a five-layer network~\citep{huang2013learning}; 
CDSSM replaces the full connection layer with a convolution layer and a pooling layer to generate text vectors~\citep{shen2014learning}. 
\citet{hu2014convolutional} proposed ARC-I, where convolution operations represent two concatenated texts for matching using a linear transformation.
CNTN also adopts convolution neural network to represent two texts, and it proposes the neural tensor network to model the similarities between two texts~\citep{qiu2015convolutional}. 
MVLSTM obtains representations for each text and adopts an interactive method to measure similarities between two texts using 3 similarity operations, i.e., cosine, bilinear, and tensor layer~\citep{wan2016deep}.
As there are important differences between web search and e-commerce search, learning query and product representations is not a solved problem. 
Hence, to discriminate products based on textual descriptions, the importance of learning semantic representations of products was soon realized~\citep{demartini2009vector,van2016unsupervised}. 

\begin{figure}[!t]
	\centering
	\includegraphics[width = 0.35\columnwidth]{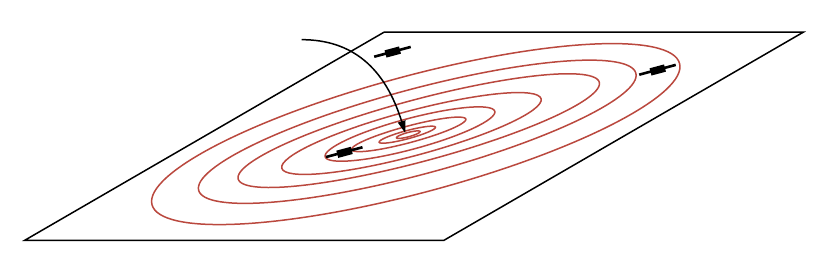}
	\caption{Illustrative example of how entities are ranked in vector space models w.r.t. a projected query. Image source: \citep{van2016learning}.}
	\label{ch41:fig:lse}
\end{figure}

\begin{figure}[!t]
	\centering
	\includegraphics[width = \columnwidth]{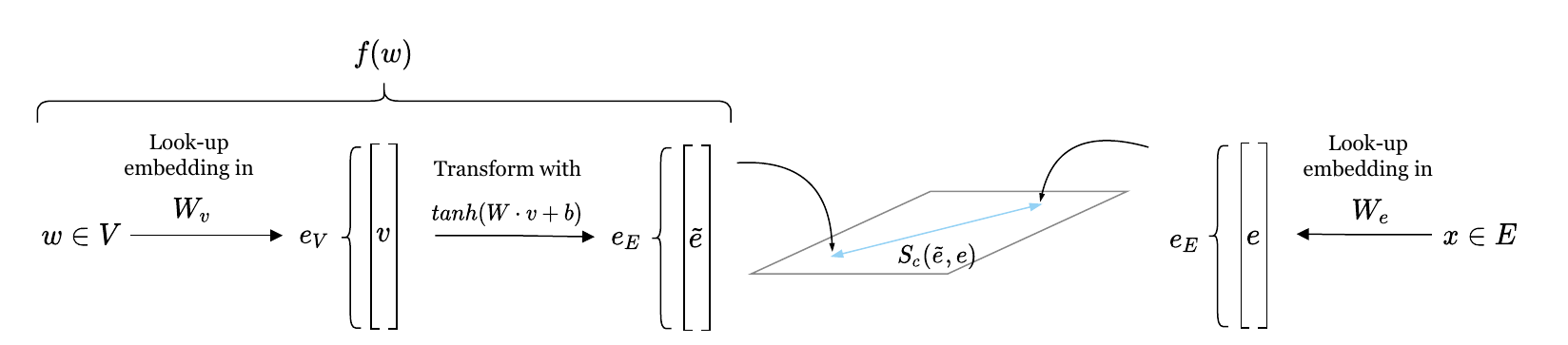}
	\caption{Schematic representation of the Latent Semantic Entities model for a single word $w$. Image source: \citep{van2016learning}.}
	\label{ch41:fig:lsem}
\end{figure}

\citet{van2016learning} propose an unsupervised distributed representation learning approach, namely latent semantic entities (LSE), to learn a unidirectional mapping between words and entities, as well as distributed representations of both words and entities.
Given a set of entities, the authors assume that each entity has a set of associated documents. LSE then learns a function that maps a sequence of words in the query from the vocabulary to an entity vector space. Thereafter, cosine similarity is applied to calculate a relevance score between candidate entities and the query. 
In Fig.~\ref{ch41:fig:lse}, we see how entities are then ranked according to the projected query. 
Specifically, the authors take the representation of a string of words to be the average of the representations of the words it contains. Then, a projection matrix is employed to map the average one-hot representations to a distributed representation. Similarly, distributed representations of entities are mapped to the same space. 
Fig.~\ref{ch41:fig:lsem} provides a schematic overview of the LSE model.
All the parameters are learned by using gradient descent.
Based on the LSE model, an increasing number of representation learning approaches have been proposed to address the vocabulary gap problem in e-commerce search~\citep{xu2018deep,van2018mix,zhang2019neural}.

In addition to learning semantic representations of products, e-commerce search, due to its specific task nature, also has multiple methods for optimizing matching results.  
\emph{Substitutable} products are those that are interchangeable in e-commerce platforms~\citep{Wang2018Improving}.
The substitutability relation among products can be determined in multiple ways.
\citet{van2018mix} find that product substitutability relations can facilitate the retrieval of relevant products impacted by the vocabulary gap problem, i.e., product substitutability can be integrated into product search either extrinsically or intrinsically.
The authors propose a two-stage framework to combine a textual matching method (the LSE model) with substitutability to infer representations of queries and products. Unlike previous work on entity representation learning, the authors integrate relations among entities within the latent semantic space inference. 

Users often browse multiple search results pages and make comparisons before purchase. 
Relevance feedback (RF) approaches have been proposed to extract the relevant topic~\citep{jin2013interactive}. However, the mismatch problem between queries and products still exists in the task of multi-page e-commerce search.
\citet{bi2019study} analyze different context dependency assumptions in multiple search result pages, and propose a context-aware embedding model to capture different types of dependency. 
The authors introduced three types of context dependency: 
\begin{enumerate*}[label=(\roman*)]
\item long-term context dependency, 
\item short-term context dependency, and 
\item long-short-term context dependency. 
\end{enumerate*}
Given these three types of context dependency, a context-aware embedding model is proposed.
By assuming that the users' preferences are associated with their implicit feedback, the embedding model, namely CEM, captures user preferences from their clicked items, which are implicit positive signals.

Another important aspect of e-commerce search concerns query reformulations by users. Based on query logs of eBay's search engine, \citet{hirsch2020query} offer a large-scale and in-depth study of users' query reformulations in e-commerce search. The authors analyze many aspects of search sessions composed of query reformulations, e.g., the number of reformulations and the distribution of their types, changes of search results pages as a result of the reformulations, clicks and purchases. An approach is proposed to predict if a query will be reformulated in an e-commerce search session.
The authors find that post-retrieval features and query performance predictors contribute the most to the prediction of reformulation.
By incorporating these features, the accuracy in predicting whether users will reformulate their queries can be significantly enhanced.
Based on an attention mechanism, the MMAN model is proposed to enhance query representations by extending category information; it includes three main modules: self-matching, char-level matching, and semantic-level matching~\citep{Yuan_2023}. Experiments show that these modules improve query representation, effectively handle long-tail queries, and achieve better semantic disambiguation.

Graph neural networks (GNNs)~\citep{scarselli2009graph,kipf2016semi} have been applied to e-commerce search to help infer query and product embeddings. 
GNNs derive node representations by aggregating features appearing in the neighborhood. 
\citet{niu2020dual} put forward a GNN-based method to learn representations of users and shops in e-commerce search. 
The authors propose a dual hierarchical graph attention network for e-commerce search. A heterogeneous graph is constructed to perform graph-based representation learning for both shops and queries, which includes both first-order and second-order proximities from various user interactions in e-commerce.
The proposed method can help to relieve the semantic gap between user queries and shop names by borrowing item neighbor title text. The proposed neighbor proximity loss provides strong additional guidance for learning graph topological structure.
A large-scale offline evaluation and online A/B tests demonstrate the significant superiority of this approach.
\citet{chang2021extreme} transfer the matching problem into an extreme multi-label classification problem~\citep{yu2022pecos}, aiming to tag input instances (i.e., queries) with the most relevant output labels of products. 
The authors suggest a tree-based sparse linear  model with n-gram TF-IDF features to augment the diversity of the matching results. 
For multi-lingual search scenarios in e-commerce, \citet{lu2021graph} detail a graph-based model with a graph convolution layer to fill the vocabulary gap.

Transformer-based pre-trained models like BERT~\citep{kenton2019bert} use stacked encoder layers that rely on a self-attention mechanism. 
BERT and BERT-like pre-trained models have been applied in product search \citep{peeters2020intermediate,liu2022towards,qiu2022pre,wang2023learning}. \citet{qiu2022pre} apply dual-tower pre-training strategies to optimize both user intent detection and embedding retrieval in e-commerce search. 
\citet{liu2022towards} examine the performance of multiple pre-training embedding methods and observed that query representation learning remains a bottleneck compared to product representation learning when using these semantic search training objectives.
Large-scale pre-trained language models, such as GPT-3~\citep{brown2020language}, also demonstrate promising performance across several benchmarks~\citep{kim2022ask}.

\subsection{Interaction-based matching}
\label{ch4:sec24:lre}
Encoding queries and products in representation-based matching methods are independent of each other. 
Mapping individual queries and products into
fixed-dimensional vectors may lose fine-grained matching information.
To tackle this challenge, interaction-based matching has been proposed.
This kind of method first matches different parts of the query with different parts of the document and then aggregates the partial evidence of relevance. 
In contrast to representation-based matching methods, interaction-based approaches usually build an interaction matrix between two documents and optimize it for matching~\citep{hu2014convolutional,pang2016text}. 
Interaction-based matching has first been applied in web-based retrieval.
\citet{hu2014convolutional} build an interaction matrix to conduct several convolution and pooling operations to extract matching features. 
\citet{guo2016deep} mention three factors in relevance matching -- exact matching signals, query term importance, and diverse matching requirements -- and design the architecture of their deep matching model. 
Similarly, MatchPyramid constructs an interaction matrix to capture matching patterns~\citep{pang2016text}. 
\citet{mitra2017learning} employ both distributed representations and local representations to obtain the final matching score.
Follow-up research has proposed a series of matching approaches specifically for e-commerce.
\citet{guo2019matchzoo} introduce a model based on MatchZoo. It is meant for short-text matching and replaces the matching histogram with a top-$k$ max pooling layer.
\citet{li2020deep} describe a product matching model, PMM, to make use of the information contained in titles and attributes of products. PMM consists of a product title matching module, and a product attributes matching module.
\citet{bi2019leverage} detail an end-to-end context-aware embedding model that can incorporate both long-term and short-term contexts to predict purchased items, unlike most approaches that focus on relevance feedback.

\subsection{Hybrid matching}
\label{ch4:sec434:hybrid}
Several hybrid matching models have been proposed to combine the strengths of representation- and interaction-based models.
\citet{mitra2017learning} propose a matching model, DUET, that integrates both local (interaction-based) and distributed (representation-based) features to calculate query-document relevance.
Pre-trained language models, such as BERT~\citep{kenton2019bert}, have further advanced hybrid approaches by capturing both contextualized token-level interactions and global semantic representations, achieving promising performance in tasks like search and recommendation~\citep{sun2019bert4rec,nogueira2019multi,lin-2021-pretrained}.
\citet{tracz2020bert} introduce a BERT-based model to use both types of matching in a similarity learning framework for product matching in e-commerce. 
\citet{yao2022reprbert} explore the deployment of BERT in online retrieval systems by distilling it into a representation-based architecture, while still maintaining the advantages of interaction-based processing for more precise matching.


\subsection{Matching in personalized search}
\label{ch3:sec24:person}

One of the primary characteristics of e-commerce search is its highly personal variance in queries.
First, multiple items could be topic-related with a consumer's query, but only a few are actually purchased, i.e., different individuals have different opinions even on the same
product. 
Hence e-commerce search without personalization is unlikely to satisfy consumers. 
Second, personalization has explicit benefits for e-commerce platforms by exhibiting the products that consumers would like to purchase. As we have seen in Section~\ref{ch4:sec:ces}, the definition of ``relevance'' for e-commerce search is not the same as for web-based search as most e-commerce platforms apply gross merchandise volume (GMV) as the gold standard for measuring success.
To the best of our knowledge, \citet{jannach2017investigating} have been the first to attempt to personalize product search by using personalized recommendation approaches.
However, the matching problem in personalized e-commerce search is more challenging as most platforms have approaches to matching products to queries that are far from perfect.
To address this problem, an increasing number of matching studies have been proposed.
Matching approaches in personalized product search can be classified into query-independent and query-dependent ones~\citep{liu2022category}.

\begin{figure}[!t]
	\centering
	\includegraphics[width = 0.35\columnwidth]{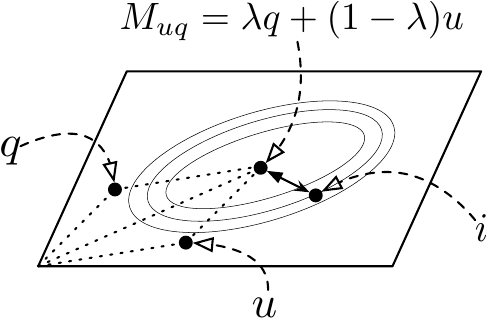}
	\caption{Personalized product search in a latent space with query $q$, user $u$, personalized search model $M_{uq}$ and item $i$. Image source: \citep{ai2017learning}.}
	\label{ch42:fig:ecvector}
\end{figure}

\begin{header}{Query-independent matching}
Query-independent matching methods embed users into a general profiling vector in the offline training stage~\citep{ai2017learning,ai2019explainable,liu2020structural}.
\citet{ai2017learning} design a deep neural network and jointly learn latent representations for queries, products, and users. A hierarchical embedding model is proposed for personalized e-commerce search. 
As illustrated in Fig.~\ref{ch42:fig:ecvector}, the authors project both queries and consumers into a single latent space and explicitly control their weights in a personalized product search model.
Following~\citet{van2016learning}, the authors design latent representations of queries and users to have good compositionality so that the personalized search model can be directly computed as a linear combination of query models and user models. 
Both queries and users are projected into a single latent space. Given a query $q$, the corresponding query intent is represented in $R^\alpha$; similarly, the user preference is represented in $R^\alpha$ given a user $u$.
As shown in Fig.~\ref{ch42:fig:ecvector}, the personalized product retrieval model is defined as ${M_{uq}} = \lambda q + (1 - \lambda ) u$, where $\lambda$ is a hyper-parameter that controls the weight of the query model $q$ and the user model $u$. To alleviate the mismatch problem during personalized product search, the authors put forward a hierarchical embedding approach to reflect the distributed representations of users, items, and queries. Their experiments are conducted with synthetic queries generated from product category information.    
\end{header} 

External structured knowledge graphs have been applied to enhance the personalized matching procedure.
Structured relationships among users, products, and queries have been jointly considered in graph neural network approaches. 
\citet{ai2019explainable} introduce a unified knowledge graph on multiple types of product data and conduct retrieval with it. 
A dynamic relation embedding module constructs a session-based knowledge graph and a soft matching algorithm extracts explainable paths with knowledge embeddings.
\citet{liu2020structural} exploit the structured representation learning scheme from user-query-product interactions with conjunctive graph patterns. Geometric operations, such as projections and intersections, are applied in the proposed graph neural networks. 
Derived from knowledge graph embeddings, \citet{liu2022category} use multiple vectors to encode the diverse preferences of users.
The authors used the category information to aggregate the multiple interests of users with category indications as references.
To exploit collaborative signals among products, users, and queries, \citet{cheng2022ihgnn} offer a hypergraph-based method from the ternary user-product-query interactions, by considering high-order features of neighbors.
Query-independent matching models can calculate user embeddings and store them in advance, which makes it convenient and efficient to apply in real-world search engines.
To tackle inconsistent user behavior in multi-stage e-commerce search systems, \citet{wang2023learning} employ external information to refine query-item matching. By mining various user interactions (ordered, clicked, unclicked items) within a post-fusion strategy, they generate more accurate semantic representations. This approach not only enhances retrieval efficiency, but also improves both offline recall and online conversion rates.

\begin{figure}[!t]
  \centering
  \subfigure[Purchase entropy]{
    \label{fig:ch42:zeroa}
    \includegraphics[width = 0.42\columnwidth, height=1.4in]{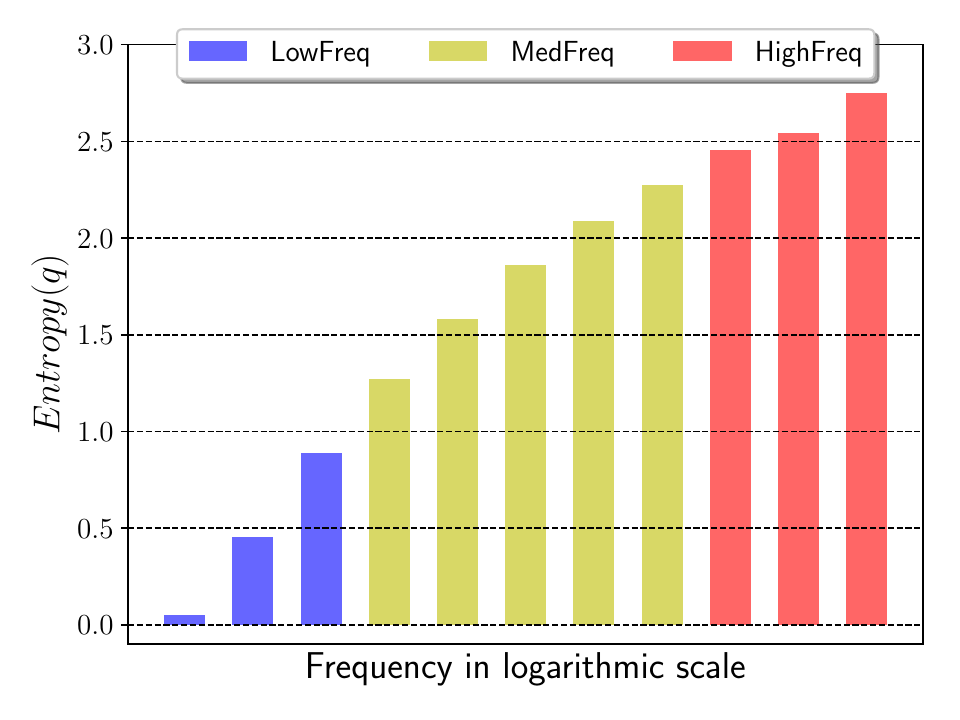}}
  \subfigure[Popularity model performance]{
    \label{fig:ch42:zerob}
    \includegraphics[width = 0.42\columnwidth, height=1.4in]{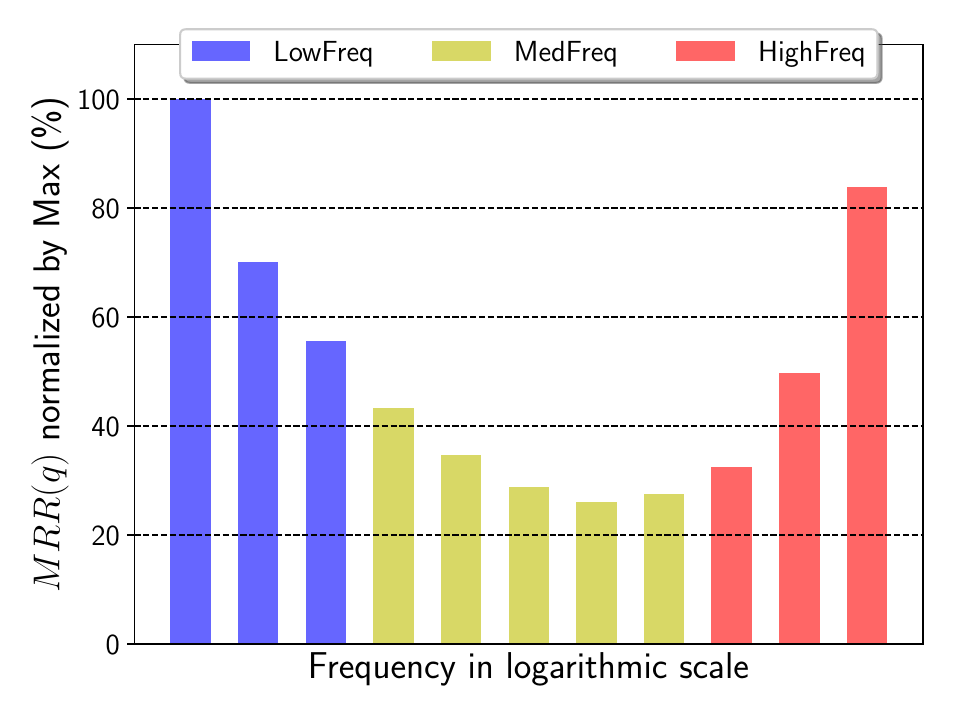}}
    \caption{The purchase entropy $\emph{Entropy}(q)$ and popularity model performance in $\emph{MRR}(q)$ for queries with different frequencies. Image source: \citep{ai2019zero}.}
  \label{fig:ch42:zeroattention}
\end{figure}

\begin{header}{Query-dependent matching}
To capture users' dynamic interests given a query, query-dependent matching approaches have been proposed.
To address the problem of when and how to conduct search personalization in product retrieval, \citet{ai2019zero} conduct an empirical analysis of the potential of personalization in product search with large-scale search logs sampled from a real-world e-commerce search engine. 
To analyze query specificity, the authors compute the purchase entropy of each query in the sampled e-commerce search logs as $\mathit{Entropy}(q) =  - \sum_{i \in {I_q}} {P(i|q)\log P(i|q)} $, where $I_{q}$ refers to the candidate item set for a query $q$.
In Fig.~\ref{fig:ch42:zeroa}, the authors provide the purchase entropy of queries on \emph{Beauty} products (e.g., facial cleanser) in the sampled search logs. They rank queries according to their frequencies on a logarithmic scale and split them into three groups: the queries with low frequency (LowFreq), with medium frequency (MedFreq), and with high frequency (HighFreq).
Queries with high frequencies have more potential for personalization, as more purchases on different items can be observed when the number of sessions increases.
The authors evaluate the differences between $P(i|q,u)$ and ${P(i|q)}$ by using the familiar $\mathit{MRR}$ metric, i.e., $\mathit{MRR}(q) = \sum_{u \in u} {\mathit{RR}(P(i|q),P(i|q,u)) \cdot P(u)}$, where ${\mathit{RR}(P(i|q),P(i|q,u))}$ reflects the the reciprocal rank of a ranked list produced by ranking with ${P(i|q)}$, using ${P(i|q,u)}$ as the ground truth. 
In Fig.~\ref{fig:ch42:zerob} the authors show the $\mathit{MRR}(q)$ of queries on Beauty products; the similarity between $P(i|q,u)$ and $P(i|q)$ is not monotonically correlated with query frequency. Accordingly, the potential of personalization varies significantly in different queries, which requires sophisticated models for personalization in product search.
\end{header}

To tackle this challenge, \citet{ai2019zero} introduce a zero-attention neural network model (ZAM) for personalized product search that conducts differentiated personalization for different query-user pairs.
This network builds on an embedding-based generative framework.
Query-dependent personalization is implemented by constructing user profiles with a zero attention strategy that enables it to automatically decide when and how to attend in different search scenarios.
Likewise, \citet{guo2018attentive} design a dual attention-based network to capture users' current search intentions and their long-term preferences.
\citet{bi2019study} study relevance feedback based on both long-term and short-term
context dependencies in multi-page product search with an end-to-end personalized product search model. 
\citet{xiao2019dynamic} propose a streaming Bayesian method to explicitly and collaboratively learn representations of different categories of entities in a joint metric space over time.

Sparseness is another critical challenge in tracking users' sequential behaviors in product search. 
Graph-based methods haven been proposed to tackle this challenge.
By using short-term user behavior, \citet{fan2022modeling} extend a structural relationship representation learning scheme~\citep{liu2020structural} to explore both local and global user behavior patterns.
\citet{zhu2022cross} integrate cross-domain transfer learning with a knowledge graph to  establish the underlying interest correlation. Their proposed method performs interest alignment across domains by explicitly modeling the long-term and short-term interactions between users and items, which capture the dynamics of product properties and user interests.

Pre-trained language models have also been applied to query-dependent matching in personalized product search. 
To conduct differential personalization in different contexts, \citet{bi2020transformer} propose a transformer-based embedding model (TEM); in TEM, personalization can vary from no to full effect. In contrast with ZAM, TEM  takes into consideration the interactions between purchased items so that it learns better dynamic representations of queries and items, which leads to better attention weights in personalized product search.
Based on TEM, a review-based transformer model, abbreviated as RTM, is designed to match user intents and items at the level of finer-grained information (e.g., their associated reviews)~\citep{bi2020learning}.
The authors conduct user-item matching at the review level so that the reason an item is ranked at the top can be explained; also, the importance of each user and item review during matching is dynamically adapted. RTM represents users and items based only on their reviews without the need for their identifiers, which improves the generalization ability during product search.
\citet{dai2023contrastive} describe a contrastive learning framework, CoPPS, to enhance user representations for personalized product search. By pre-training a sequence encoder with contrastive sampling and fine-tuning, CoPPS employs multiple data augmentation strategies to improve user modeling.

For real-world applications, \citet{zhang2020towards} jointly consider semantic matching and personalized matching at JD.com. 
The authors introduce a deep personalized and semantic retrieval model (DPSR) with a two-tower architecture: a multi-head design of a query tower and an attention-based loss function.
Similarly,~\citet{magnani2022semantic} describe a semantic retrieval model with a two-tower architecture in e-commerce search production at Walmart.com.
The authors select negative examples for training a large semantic retrieval model and use an approximate metric to evaluate the performance.

\section{Ranking strategies in e-commerce search}
\label{ch4:sec:rs}

Ranking optimization is another core task in e-commerce search. In this section, we introduce ranking strategies in e-commerce search. 

\begin{sidewaystable}[!htbp]
\centering
\caption{Comparison of ranking algorithms in terms of NDCG@10 for target variable ``Click Rate'', ``Cart Add Rate'', ``Order Rate'', and ``Revenue.''  Regu. is an abbreviation for Regulation, whereas lo. is an abbreviation for loss. Table source: \citep{karmaker2017application}}
\begin{tabular}{l cc cc cc cc}
\toprule
\multirow{2}{*}{\textbf{Algorithm}} 
& \multicolumn{2}{c}{\textbf{Click Rate}} 
& \multicolumn{2}{c}{\textbf{Cart Add Rate}}
& \multicolumn{2}{c}{\textbf{Order Rate}}
& \multicolumn{2}{c}{\textbf{Revenue}}
\\
\cmidrule(lr){2-3} \cmidrule(lr){4-5}  \cmidrule(lr){6-7}  \cmidrule(lr){8-9}
& \textbf{Train}     & \textbf{Test}   & \textbf{Train}     & \textbf{Test} & \textbf{Train}     & \textbf{Test} & \textbf{Train}     & \textbf{Test}
\\ 
\midrule
RankNet~\citep{burges2005learning} & 0.6857 & 0.6855 & 0.4399 & 0.4402 & 0.7158 & 0.7142 & 0.7577 & 0.7578
\\
RankBoost~\citep{freund2003efficient} & 0.5899 & 0.5904 & 0.4073 & 0.4043 & 0.5007 & 0.4994 & 0.5663 & 0.5639
\\
AdaRank~\citep{xu2007adarank} & 0.6877 & 0.6857 & 0.4464 & 0.4401 & 0.7334 & 0.7349 & 0.757 & 0.7566
\\
Random Forest~\citep{breiman2001random} & 0.6378 & 0.6125 & 0.4588 & 0.4296 & 0.5707 & 0.5288 & 0.6463 & 0.5959
\\
LambdaMART~\citep{burges2010ranknet} & 0.8426 & \textbf{0.8291} & 0.7664 & \textbf{0.7324} & 0.7728 & \textbf{0.7687} & 0.8183 & \textbf{0.7998}
\\
\small{Logistic Regression} \tiny{(L1 regu.)}~\normalsize{\citep{fan2008liblinear}} & 0.6284 & 0.6272 & 0.4274 & 0.4252 & 0.6677 & 0.6632 & 0.6873 & 0.6822
\\
\small{Logistic Regression} \tiny{(L2 regu.)}~\normalsize{\citep{fan2008liblinear}} & 0.5889 & 0.5866 & 0.4066 & 0.4025 & 0.5045 & 0.4983 & 0.5751 & 0.5675
\\
\small{SVM Classifier} \tiny{(L1 regu.+L2 lo.)}~\scriptsize{\citep{fan2008liblinear}} & 0.6366 & 0.6317 & 0.4348 & 0.4331 & 0.6870 & 0.6794 & 0.7105 & 0.7059
\\
\small{SVM Classifier} \tiny{(L2 regu.+L1 lo.)}~\scriptsize{\citep{fan2008liblinear}} & 0.4596 & 0.4594 & 0.3274 & 0.3219 & 0.4281 & 0.4289 & 0.4503 & 0.4462
\\
\small{SVM Regressor} \tiny{(L2 regu.+L2 lo.)}~\scriptsize{\citep{smola1997support}} & 0.2358 & 0.2341 & 0.1909 & 0.1914 & 0.2100 & 0.2087 & 0.2030 & 0.2027
\\
\small{SVM Regressor} \tiny{(L2 regu.+L1 lo.)}~\scriptsize{\citep{smola1997support}} & 0.2876 & 0.2865 & 0.2110 & 0.2096 & 0.2078 & 0.2038 & 0.2093 & 0.2121
\\
\bottomrule
\end{tabular}
\label{tabSeaLtr}
\end{sidewaystable}

\label{ch4:sec33:l2r}

As we have discussed in Section~\ref{ch3:sec11:overview}, learning to rank approaches have been applied to e-commerce search. 
Unlike web search scenarios based on relevance judgments, e-commerce search has multiple implicit and explicit signals that need to be integrated during the ranking process. 
In e-commerce search, learning to rank methods need to tackle a series of challenges~\citep{karmaker2017application}.
Learning to rank methods assume that the scoring of items to be ranked is a parameterized function of multiple features computed based on the given query and the items, whereas parameters are used to control the weights of features. 
A large number of product and query features, such as brands, rating, categories, etc., are important to obtain useful representations in e-commerce learning to rank models. 
Also, product popularity related features have been shown to be effective in optimizing ranking results of e-commerce search.
\citet{karmaker2017application} study the performance of multiple learning to rank strategies in e-commerce search, and find that LambdaMART~\citep{burges2010ranknet} is able to learn a balance between these two kinds of features.
Detailed comparisons are listed in Table~\ref{tabSeaLtr}.
In addition, user engagement behavior, such as clicks and orders, also plays an important role in optimizing ranking results~\citep{karmaker2017application,wu2018turning}. 
In contrast to web search which only has clicks to judge the relevance, e-commerce search contains four prominent relevance feedback behavior: clicks, cart-adds, orders, and revenue.
Thus, various training objectives can be considered in optimizing e-commerce ranking results. 

Another main challenge for e-commerce learning to rank methods is the lack of labeled information. 
In web search, high-quality labeled information obtained by eliciting relevance ratings from human experts or crowdsourcing, makes learning to rank methods effective. 
However, in the context of e-commerce search it is infeasible to deermine a standard method to obtain ground-truth information.
Intuitively, crowd sourcing annotations seem to provide labels for e-commerce learning to rank methods.
\citet{karmaker2017application} and \citet{alonso2009relevance} study the reliability of the relevance judgements provided by crowd workers for e-commerce queries.
The authors find that crowdsourcing fails to provide reliable relevance judgements for e-commerce queries.
Thus implicit feedback from e-commerce users is an important source of information that helps reveal the saliency of products for a given query.

\citet{wu2018turning} divide the ranking optimization problem into two successive stages: click optimization and purchase optimization. 
We have discussed previous studies on click modeling and purchase modeling in Section~\ref{ch31:subsec:cpr} and~\ref{ch31:subsec:pm}, respectively.
Inspired by these approaches,~\citet{wu2018turning} apply a list-wise learning to rank model~\citep{cao2007learning} to optimize clicks by jointly considering item positions and query-level structures; a binary classification approach is applied to predict the purchase behavior. 

Online learning to rank approaches have been applied to optimize ranking results in e-commerce search~\citep{hu2018reinforcement}.  
Unlike traditional static learning to rank models, online learning to rank methods optimize the production ranker interactively by exploiting users' implicit feedback~\citep{zoghi-copeland-2015,oosterhuis-balancing-2017,schuth-multileave-2016}.
Online learning to rank approaches can be divided into two groups: the first is to learn the best ranking function from a function space~\citep{hofmann2016online,hofmann2013balancing,yue2009interactively}; the second group directly learns the best list under some model of user interactions~\citep{radlinski2008learning,schuth-multileave-2016,oosterhuis-balancing-2017}.
As part of the first group of online learning to rank methods, \citet{hu2018reinforcement} propose a reinforcement learning method for ranking optimization in e-commerce searching scenarios. The authors formulate the multi-step ranking procedure in e-commerce search as a search session Markov decision process (SSMDP).
An algorithm, named deterministic policy gradient with full backup estimation (DPG-FBE), is then proposed for the problem of high reward variance and unbalanced reward distribution of SSMDP.
To integrate search results from heterogeneous sources,~\citet{takanobu2019aggregating} introduce a search result aggregation method that formulates a semi-Markov decision process, where a low-level policy is applied to represent items and a high-level policy is used to select rankers. 
Based on a large-scale real-world dataset, \citet{pkddAnwaarRK20} employ a counterfactual risk minimization (CRM) approach to directly optimize the ranking list from the log data.

In recent years, deep neural-based retrieval models~\citep{mitra2017introduction,kenter-neural-2017,onal-neural-2018} have been used in the ranking step of e-commerce search. 
\citet{magnani2019neural} enhance the deep neural network model using different types of text representation and loss function at Warmart.com.
\citet{zhang2019neural} detail an e-commerce ranking model with interaction features between the query and a graph of products maintaining product interactions.
Pre-trained foundation models have been shown to be effective in real-world web search scenarios~\citep{zou2021pre,lin-2021-pretrained,chu2022h}.
In e-commerce search, \citet{wu2022multi} apply BERT for product ranking within a multi-task learning framework.
The authors use the probability transfer method in the framework to model multiple sequential engagement behaviors.
By integrating semantic matching features output by the domain-specific BERT, the authors confirm the effectiveness of their proposed approach on real-world e-commerce search data.

\section{Emerging directions}
\label{ch3:sec:fd}

\subsection{Multi-modal e-commerce search}

Along with the rise of deep neural networks, multi-modal search has increasingly received attention~\citep{zhang2013attribute,mao2014deep,yang2017visual,liu2018attentive,balaneshin2018deep,guo2018multi,zhang2018visual,qu2021dynamic,wei2021universal,tan2022bit}.
The key of multi-modal search is to find an effective mapping mechanism to project data from different modalities into a common latent space.
Multi-modal search approaches can be classified into hashing-aware models and semantic-aware models. 
The former type of methods map various modalities in the original space to a Hamming space using hash functions~\citep{cao2017transitive,luo2018scalable,zhu2020deep,tan2022bit}, whereas the latter ones project the multi-modal data into a low-dimensional space by learning a mapping function~\citep{wang2016effective,laenen2018web,qu2021dynamic,wang2022siamese}.

As we discussed in previous sections, most approaches to e-commerce search focus on the textual matching problem. 
With the rise in online photos and openly available image datasets this is changing. 
\citet{yang2017visual} propose a novel end-to-end approach for scalable visual search infrastructure at Ebay. 
Similar platforms can be found at Alibaba~\citep{zhang2018visual} and JD.com~\citep{li2018design,wang2020metasearch}.
Early studies into multi-modal e-commerce search have mostly focused on the fashion category~\citep{yang2017visual,zhang2018visual,li2018design}. 
However,~\citet{wang2020metasearch,sigir21DaganGN21,liu2022pretraining} confirmed that multi-modal search is widespread across many e-commerce categories, especially categories that involve aspects that are harder to express verbally, but can naturally be captured visually, such as style, type, and pattern.

In multi-modal search, multiple modalities, such as text and images, can be found in both queries and products. Most multi-modal search solutions optimize a function to project multi-modal data into a low-dimensional space after unified representation learning.
\citet{laenen2018web} present a multi-modal search paradigm for e-commerce search that results in an improved shopping experience. The authors reason with both images and languages through a common embedding space.
\citet{guo2018multi} formulate the e-commerce personalized search problem based on the relevance between images and text with respect to the query and the user preferences from both textual and visual modalities.
A transition-based product search method has been proposed, where the multi-modal feature space is initialized based on the textual and visual features of products.
An interpretable multi-modal e-commerce retrieval framework has been proposed for fashion products~\citep{liao2018interpretable}; 
the authors bridge the gap between deep features and meaningful fashion concepts. They propose a hierarchical similarity function to accurately characterize the semantic affinities among fashion items. 

\citet{sigir21DaganGN21} highlight various differences between visual and textual search, which can be summarized in the following  challenges in multi-modal e-commerce search:
\begin{itemize}[leftmargin=*,nosep]
\item \textbf{Heterogeneous resources.} User queries submitted on e-commerce platforms can be real-world images shot while the user is engaging with the platform. 
\item \textbf{Large-scale sparse data.} Large multimedia corpora make scalability and efficiency key requirements for e-commerce search. 
Visual queries are more specific than text queries, which results in a smaller number of retrieved results and sparse coverage of categories.
\item \textbf{Limited user engagement.} In multi-modal e-commerce search, user engagement behavior, e.g., clicks, dwell time, purchases, etc., is substantially sparser than in textual search. 
\item \textbf{Ambiguous user intents.} Two different main use cases exist in multi-modal e-commerce search: target finding and decision making~\citep{sigir21DaganGN21}. The two use cases reflect the navigational intent and informational intent, respectively.
\end{itemize}

\noindent%
Additionally, annotating new product images and retraining a new feature representation model on large-scale data is expensive. 
To address this problem, few-shot multi-modal product search has been proposed to update the feature representation and index model with few-shot data and employ a fast learning strategy for new categories. 
\citet{wang2020metasearch} describe a framework for few-shot incremental search via meta-learning, with a multi-pooling feature extractor to extract discriminative multi-modal features.
Based on pre-trained language models, \citet{liu2022pretraining} detail an effective contrastive learning framework to learn representations of multi-modal search sessions based on multi-view heterogeneous graph networks.
\citet{liu2023multimodal} employ BERT with a self-distillation framework for product understanding by integrating visual and textual information.

Cross-modal search is receiving more and more attention~\citep{qu2021dynamic,wang2022siamese,tan2022bit}. 
Unlike multi-modal search, cross-modal search aims to solve the discrepancy problem between different modalities in search sessions~\citep{wang2022siamese}.
Although much progress has been made in bridging multiple modalities, it still remains challenging because of the difficult intra-modal reasoning and cross-modal alignment~\citep{qu2021dynamic}. 
As complicated relations exist among products in e-commerce, it is more difficult to explore these multi-hop interactions in cross-modal product search scenarios.
Moreover, the extremely high computational costs brought by multi-modal input limits its usefulness in large-scale cross-modal retrieval in e-commerce applications.

\subsection{Conversational e-commerce search}

Originating from early studies on interactive information retrieval~\citep{croft1987i3r,belkin1995cases}, conversational search refers to the process of interacting with a dialogue system to search for information.
Conversational search allows users to express their information needs by directly conducting conversations with search engines.
Unlike traditional query-based search engines, conversational search systems capture users' intent by taking advantage of the flexibility of mixed-initiative interactions and by providing useful information more directly using human-like responses~\citep{radlinski-2017-theoretical,Vtyurina:2017:ECS:3027063.3053175,ren2021wizard,vakulenko2021large}. 
User studies have been conducted to study whether conversational search is needed and what it should look like.
\citet{Trippas:2018:IDS:3176349.3176387} conduct a laboratory-based observational study, where pairs of people perform search tasks communicating verbally.
The authors find that conversation search paradigms are more complex and interactive than traditional search scenarios. Moreover, it is difficult to simulate human-human interactions in a conversational search session.
To address this problem, \citet{ren2021wizard} collect a dataset of human-human dialogues about conversational search in a wizard-of-oz fashion, namely wizard of search engine (WISE), where two workers play the role of seeker and intermediary, respectively. 

In a conversational search session, the user first initializes the conversation with a request, then the conversational agent iteratively asks the user about their preferences and estimates user interest based on their feedback.
Finally, the agent retrieves the information and generates the response. 
Many studies focus on the second stage, including generating clarifying queries~\citep{Hamedwww2020,liu2021learning,ghanem2022question} and understanding users' intent~\citep{wu-yan-2018-deep,gao-2021-advances,TRIPPAS2020102162}.
\citet{radlinski-2017-theoretical} consider the question of what properties would be desirable for a  system so that the system enables users to answer a variety of information needs in a natural and efficient manner.
\citet{azzopardi2018conceptualizing} outline the actions and intents of users and systems and explain how these actions enable users to explore the search space and resolve their information needs.
\citet{ghanem2022question} propose a method to generate user queries for story-based reading comprehension skills.
The response generation problem in conversational search is also being addressed.
\citet{ren2021wizard} describe a modular end-to-end neural architecture to transfer the output from intent understanding to improve response generation.
\citet{ye2022structured} design an interconnected network to co-generate structured and natural responses that allow for bidirectional semantic associations to generate responses.

In e-commerce, conversational search systems can help consumers access products through instant conversational interactions on mobile phones or other smart devices. Also, conversational e-commerce search alleviates the burden of reformulating queries and browsing through dozens of products in e-commerce search~\citep{zou2022learning}. Moreover, conversational e-commerce search provides a natural way to collect explicit feedback from users to understand their preferences~\citep{zamani2022conversational}.
As an early attempt, \citet{zhang2018towards} detail a paradigm for conversational product search to ask users about their preferred values of an aspect and adopt a memory network to retrieve search results. 
Based on that, \citet{bi2019conversational} focus on product-seeking conversations and proposed an aspect-value likelihood model for negative feedback, with a multivariate Bernoulli distribution to generate explainable e-commerce aspects.
To improve the quality of representation in conversational product search, \citet{zou2022learning} integrate the representation learning of user, query, item, and conversation into a unified generative framework.

There are several open questions and research directions for future work in conversational e-commerce search.
First, obtaining data in real-world scenarios is  a challenging problem. Several studies, including, e.g., ConvPS~\citep{zou2022learning}, have applied simulated data in their experiments. It would be useful to extend observational experiments to a wizard-of-oz setting by establishing a real-world dataset in the future. We also observe that the ongoing trend in simulating users in task-oriented dialogues could provide more useful insights~\citep{sunzhang2021simulator}.
Second, modeling user-system interactions is a crucial aspect of conversational e-commerce search.
\citet{vakulenko2021large} reveal the complicated situation for large-scale dialogue analysis specifically focusing on the patterns of mixed-initiative. 
In contrast with traditional web search scenarios, e-commerce search needs to analyze more complicated user engagement behavior (see Section~\ref{ch4:sec11:challenge} and~\ref{ch2:sec2}). 
Hence, in future work, both short-term and long-term user engagement interactions may need specialized attention in conversational e-commerce search.
Third, how to evaluate the performance of conversational e-commerce search is still an open question.
We have demonstrated a wide range of evaluation metrics applied in e-commerce search, e.g., engagement-based metrics and revenue-based metrics (see Section~\ref{ch4:sec33:eva}). 
Therefore, it is even more challenging to measure the success of conversational e-commerce search in terms of all these various kinds of metrics.
Measuring interactivity is critical in conversational information seeking~\citep{zamani2022conversational}.
Keeping track of how well a user understands the system and vice versa can be another important research direction.
Fourth, personalization conversational e-commerce search needs more attention in future work. 


\subsection{Generative retrieval in e-commerce}
Generative retrieval has emerged as a novel retrieval paradigm in recent years. During the training phase, documents and their corresponding document identifiers (docids) are encoded into the model's parameters. Given a query, the model can directly generate the relevant DocIDs~\citep{tay2022transformer}.
This approach uses the model's ability to memorize and use document representations, enhancing retrieval efficiency and accuracy by bypassing traditional retrieval pipelines.

For indexing strategies, the approach typically involves using a sequence-to-sequence (seq2seq) model to learn the mapping from queries to docids. 
In addition, a range of tasks have been introduced to enhance the indexing performance, including learning a mapping from documents to docids~\citep{tay2022transformer}, constructing pseudo-queries for documents and mapping them to docids~\citep{wang2022neural}, as well as training the model to rank different documents corresponding to the same query~\citep{li2024learning}.

Document representations are another key component of the generative retrieval model. Since docids are the direct output of model, the quality of these representations directly determines the prediction accuracy of generative models.
Existing docids can generally be categorized into two types: numeric-based and text-based.
\citet{tay2022transformer} introduce three types of numeric docids: 
\begin{enumerate*}[label=(\roman*)]
\item unstructured atomic docids, where each document is assigned a random and unique integer identifier, without any structural or semantic information; 
\item naively structured string docids, where each document is assigned a random and unique numeric string; and 
\item semantically structured docids, which use a hierarchical $k$-means method, allowing relevant documents to share the same prefix, thereby introducing semantic structure.
\end{enumerate*}
\citet{wang2022neural} propose that the meaning of a docid depends on both the position and the prefix context.
Therefore, they propose a prefix-aware weight-adaptive decoder to adapt to different docids.
\citet{sun2024learning} detail learnable document representations by using a discrete autoencoder to encode documents into short, discrete docids. It optimizes these docids by converting documents into docids through an encoder, then training the model to minimize the reconstruction loss of converting docids back to the original documents.
Text-based docids are another popular approach, as they can inherit the language model’s text generation ability and do not require learning a new docid vocabulary.
The title can be regarded as an intuitive abstract text that represents the content of a document. 
\citet{de2020autoregressive} describe how to use the document title as a docid and achieve good retrieval performance. 
However, an arbitrary document may not have an informative and structured title like those in Wikipedia; \citet{zhou2022ultron} combine keywords from both the URL and the title to form a docid. 
Additionally, some work has attempted to incorporate more content into text-based docids, including titles, URLs, document substrings, pseudo-queries, and more~\citep{li2023multiview}.
The substrings in the document can also represent the information stored in the document.
\citet{bevilacqua2022autoregressive} use arbitrary n-grams in documents as docids, and retrieve documents using a pre-built FM-indexer.

\chapter{E-commerce recommendation}
\label{chp:rec}

Recommendation methods refer to information filtering techniques, which can be traced back to the 1980s~\citep{malone-1987-intelligent}, i.e., even before the web had been developed. 
Significant research efforts have been devoted to recommendation in various domains, such as email, movies, books, and music. 
E-commerce has its unique properties; recommendation methods developed for other domains may not be suitable for e-commerce scenarios. 

In this section, we focus on recommendation techniques for e-commerce.
First, we summarize the key characteristics of e-commerce recommendation, for which a two-stage framework is developed.
This forms the mainstream solution for e-commerce recommender systems.
We then review models developed for the two stages. 
We discuss evaluation methodologies for e-commerce recommendation and conclude with a description of future research directions to help build better e-commerce recommendation solutions.

\section{Characteristics of e-commerce recommendation}
\label{ch5:cha}

Recommendation techniques have been applied extensively in e-commerce, in many different scenarios.
They play a crucial role in e-commerce by facilitating users to find desired products and to help boost revenue. 
E-commerce has three key characteristics that affect recommendation techniques:
\begin{itemize}[leftmargin=*,nosep]
\item \textbf{Large volume of products.} The first key characteristic of e-commerce is the large volume of products, which brings challenges to algorithm scalability, as recommendation algorithms need to quickly scan products and select the ones that are of interest to a user. 
It is common that an e-commerce platform contains millions of products~\citep{kersbergen-2021-learnings}. This sheer number of items inevitably poses a computational challenge to recommender systems. The training of models needs to be efficient to be able to quickly (e.g., hourly or at least daily) refresh the model given new user behavior data and latency needs to be sufficiently low in order to be able expose the large catalog to a large number of users.

\item \textbf{Sparsity.} The second characteristic is the sparsity of behavior, since a user can only consume (e.g., purchase or click) a few products. 
The main aim of e-commerce recommendation is to satisfy the information needs of users in viewing or purchasing products and services. Behavioral data (e.g., browsing, purchases, and clicks) is an important and useful data source for learning the personalized tastes of users. Given the huge volume of products, it is impossible for a user to interact with most items. 
The products with which a user actually interacts are typically just a small fraction of the entire catalog~\citep{li-2023-next}. This results in a significant sparsity issue that forces recommendation methods to learn from user behavior on a limited set of products and to generalize their predictions to all other products.

\item \textbf{Data richness.} The third characteristic is the richness of product and user data (see Section~\ref{chapter:basic}). 
In addition to user behavior data that directly reflects user preferences, rich information about products (e.g., product name, description, categories, images, etc.) and users (e.g., age, gender, occupation, income level, etc.) is available in e-commerce scenarios. 
Moreover, there is much contextual information associated with user behavior, such as time, location, last purchase, and submitted query in the session etc.~\citep{chen-top-n-2017}. This auxiliary information is useful for inferring why a user chooses an item, and is particularly beneficial for cold-start scenarios, where a user (or an item) has very few interactions.

\end{itemize}

\noindent%
As we will see below, many developments in e-commerce recommendation have addressed the three characteristics listed above, resulting in a large number of technical achievements that can be directly put to practical use. 

\begin{figure}[!t]
	\centering
	\includegraphics[width=0.8\columnwidth]{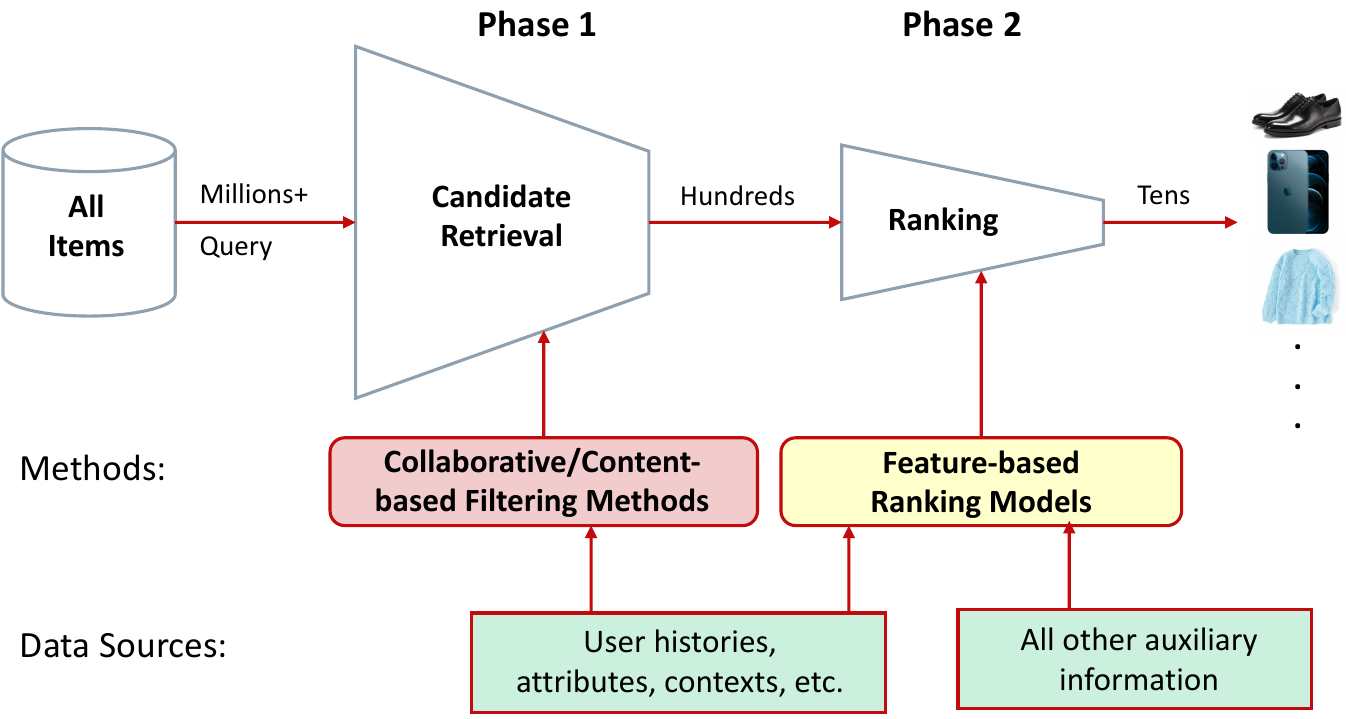}
	\caption{Illustration of a two-stage recommendation solution with candidate retrieval and candidate ranking. Image source: \citep{ren2018information}.}
	\label{fig:recommendation_overview}
\end{figure}

We present a two-stage solution that has been commonly used in industrial recommender systems~\citep{cheng2016wide,Wang2018Improving, Covington:2016}.
Fig.~\ref{fig:recommendation_overview} illustrates a two-stage recommendation framework with two main phases: \emph{candidate retrieval} and \emph{candidate ranking}.
Given an information need (either expressed explicitly, e.g., by a user expressing interest in a product category, or implicitly, e.g., through the event of a user visiting a recommender system), the first phase of candidate retrieval goes through the entire product catalog, which may contain millions of products, and selects a small set of products that best match the information need. 
Then the selected candidates (usually in the order of hundreds) are passed to the second phase of candidate ranking, which ranks the candidates to produce the final top-$k$ products to the user. 
Such a two-stage architecture can strike a balance between efficiency and effectiveness -- it not only supports fast retrieval from a large-scale catalog by using an efficient and light-weight candidate retrieval model, but also maintains good recommendation performance as the final ranking is determined using a potentially fine-grained ranking model.
We survey studies into each stage in Section~\ref{ch5:crm} and~\ref{ch5:ranking}, respectively. 

It is worth mentioning that the pipeline sketched in Fig.~\ref{fig:recommendation_overview} may involve a third stage, re-ranking, to refine the list produced by the ranking model to meet additional criteria or constraints, such as diversity, novelty, or fairness~\citep{abdollahpouri2019popularity,Wilhelm2018Practical,Gogna2017Balancing,Pei2019Personalized,ai2018learning}.
We briefly introduce recent progress in re-ranking models in Section~\ref{sec:re-ranking}.

\section{Candidate retrieval models}
\label{ch5:crm}

In this section, we survey candidate retrieval models ranging from traditional heuristic methods (Section~\ref{ch5:crm:hbm}) to recent advanced embedding-based methods (Section~\ref{sec5:ebm}).

\subsection{Heuristic methods}
\label{ch5:crm:hbm}

Heuristic-based methods are commonly used for candidate retrieval because they are simple and easy to implement. These methods are based on a manually heuristic rather than on optimization with an objective function. 
Early studies have presented various retrieval strategies to search candidate products. 
For example, selecting high sale or promotion items has been widely adopted in practical recommender systems, whereas several researchers use principles from economics to perform item selection for candidate retrieval~\citep{zhao2017multi,zhang2016economic}. 

There are many heuristic methods that aim at mining item or user relations for candidate retrieval. We detail approaches to these methods from three perspectives:
\begin{enumerate*}[label=(\roman*)]
\item neighborhood-based methods,
\item graph-based method
and
\item methods based on complementary and substitutable items.
\end{enumerate*}

\OurParagraph{Neighborhood-based methods.}
Neighborhood-based methods first compute similarity between items or users, and then make recommendations by aggregating information from similar users or items. Neighborhood-based methods can be classified into \emph{user-oriented} and \emph{item-oriented}. 
The paradigm of user-oriented methods~\citep{grouplens1994,grouplens1997} can be summarized as follows~\citep{adomavicius2005toward}:
 \begin{equation}
    \label{eq:userCF}
    \begin{split}
        \hat{r}_{u,i} = Agg( \{ r_{u' , i} \}_{u' \in S_{u}}  ).
    \end{split}
\end{equation}
where $\hat{r}_{u,i}$ is the rating of user $u$ for the target unrated item $i$ that we seek to estimate, $S_u$ is the set of similar users of $u$, and $r_{ u' , i}$ is the rating of similar user $u'$ on the target item $i$. $Agg(\cdot)$ is a function that aggregates information from similar users. The process consists of two steps: 
\begin{enumerate*}[label=(\roman*)]
\item finding similar users, and 
\item aggregating the information of similar users. 
\end{enumerate*}
Recent work has explored various approaches for similarity computation, including but not limited to Pearson correlation coefficients, information entropy, and mean squared difference~\citep{shardanand1995social} between the ratings given by two users~\citep{grouplens1994,shardanand1995social}. Spearman rank correlation has been used to measure item rank similarity rather than value similarity \citep{herlocker1999algorithmic}. 
For the aggregation function $Agg(\cdot)$, a linear weighted combination is a commonly used strategy~\citep{grouplens1994,shardanand1995social,herlocker1999algorithmic}. The similarity score is usually used for setting combination weights, as similar users naturally deserve a larger contribution to a prediction.

Inspired by the success of user-oriented methods, item-oriented methods have also been explored~\citep{sarwar2001itemCF,karypis2001itemCF}. Item-oriented methods have a quite similar process as user-oriented methods except that they aim at mining item similarity rather than user similarity. Recent studies empirically show that item-oriented methods achieve better performance than user-oriented methods~\citep{mclaughlin2004useranditemcf,sarwar2001itemCF,karypis2001itemCF}.

In summary, neighborhood-based methods have shown several advantages: they are simple, efficient, highly explainable, and easily deployed. 
However, their disadvantages are also obvious: they have a heavy reliance on human expertise, lack of flexibility, and they suffer from data sparsity. 
Although they may not perform as well as recent, more advanced methods, they usually serve as a benchmark for candidates generation in real-world recommender systems.
\begin{figure}[!t]
	\centering
	\includegraphics[width=0.6\columnwidth]{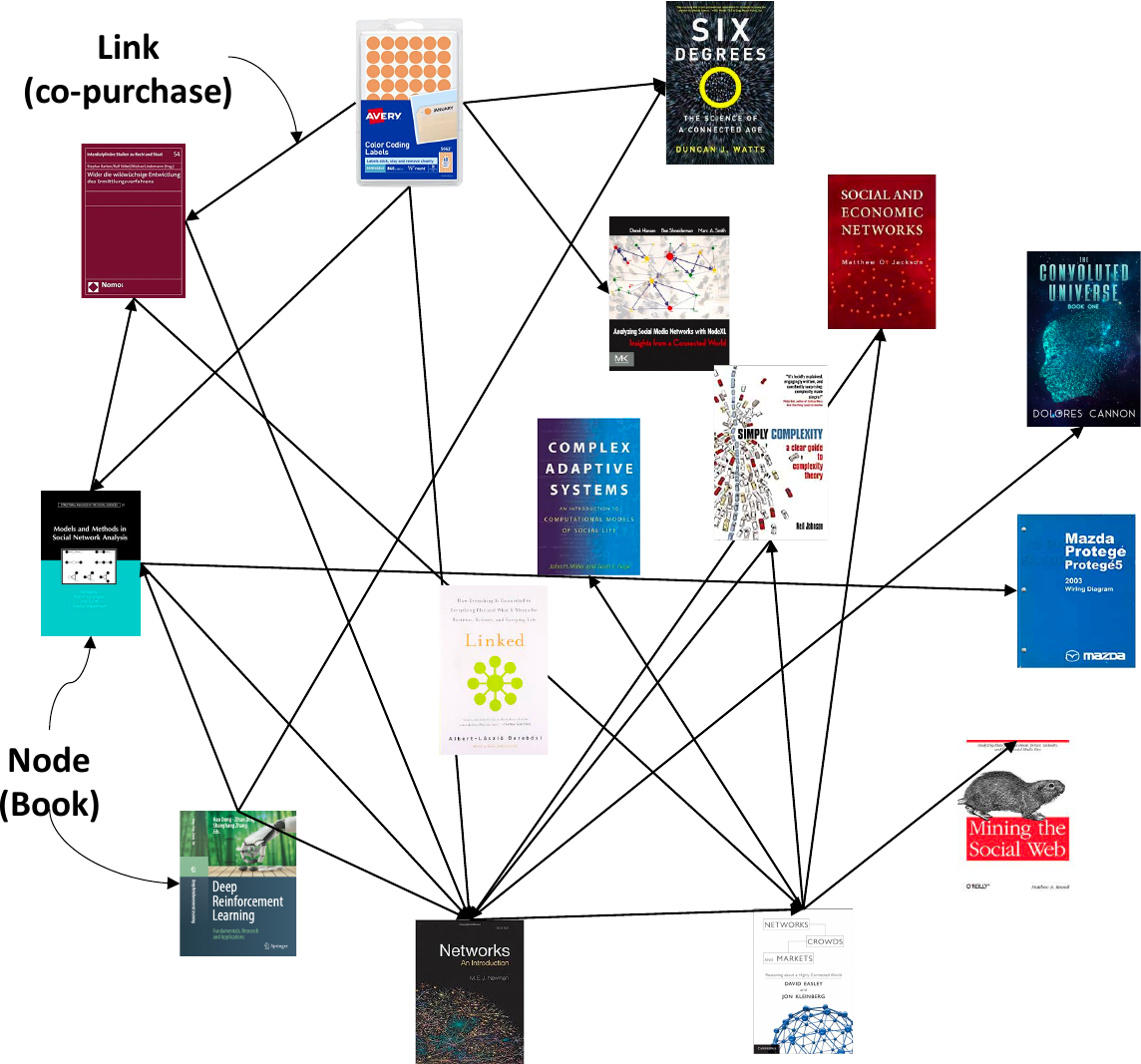}
	\caption{An item graph. Each node represents an item and each edge indicates the existence of  a co-occurrence relation between the two items. Image source: \citep{leem2014impact}.}
	\label{fig:ch5graph}
\end{figure}

\OurParagraph{Graph-based methods.}
Another type of heuristic strategy is to explore similar items through an item graph. 
As shown in Fig.~\ref{fig:ch5graph}, the graph can be constructed from item-item co-occurrence or user-item interaction information, where each node represents a user or an item while each edge indicates a certain relation between these objects. 
Based on the graph, it is easier and more effective to evaluate the similarity between items via the closeness of two items in the graph. 
This has inspired a number of recent publications on graph-based retrieval. 
For example, \citet{leem2014impact} conduct information propagation on the graph to calculate item similarity, whereas \citet{eksombatchai2018pixie} directly perform random walks from seed nodes to select relevant items.

\OurParagraph{Complementary and substitutable items.}
Beyond item similarity, more complicated item relations concerning complementarity of items and substitutability of one item for another have also been considered~\citep{Wang2018Improving,zhang2018quality,chen2020try,guoziyi2018}. 
As explained in Section~\ref{subsubsection:3122}, complements indicate items that might be purchased together, while substitutes indicate items that are interchangeable. 
Mining complements and substitutes is beneficial to satisfy the true need of users, and further increases the click-through rate and user stickiness~\citep{Wang2018Improving,guoziyi2018}. However, it is challenging as we often lack ground-truth labels of such relations. To deal with this problem, \citet{Wang2018Improving} use co-view and co-purchase statistics, as weak relation signals to supervise the learning of the complements and substitutes. They further integrate the learned relations into a vanilla recommendation model and observe improvements in recommendation performance.
\citet{guoziyi2018} introduce a graph convolutional neural network that decouples item semantics for inferring complementary and substitutable items. 
The decoupled graph neural network contains a two-step knowledge integration scheme.
\citet{chen2020try} design attribute-aware collaborative filtering to perform substitute recommendation by addressing issues from both personalization and interpretability perspectives.

\begin{figure}[!t]
	\centering
	\includegraphics[width=0.6\columnwidth]{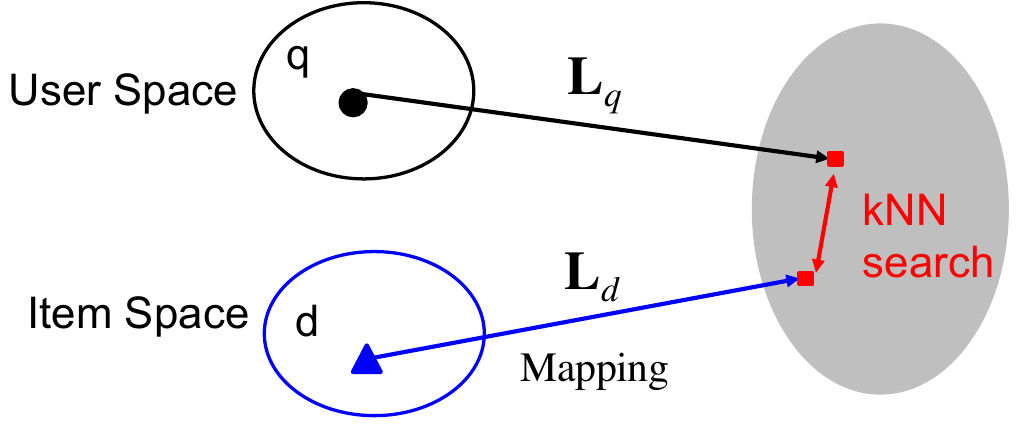}
	\caption{Illustration of the workflow of embedding-based methods: They first map users and items into a common embedding space; then they use the KNN algorithm to search the candidate items having the smallest embedding distance with the target user.
	Image source: \citep{andoni2006near}.}
	\label{fig:ch5gemb}
\end{figure}

\subsection{Embedding-based methods}
\label{sec5:ebm}
Embedding-based methods are another type of candidate retrieval methods.
As shown in Fig.~\ref{fig:ch5gemb}, this kind of method first maps users and items into a common embedding space, and then uses approximate the K-Nearest Neighbor (KNN) algorithm~\citep{andoni2006near} to search for candidate items with the smallest embedding distance to the target user. 
The key of these methods lies in learning high-quality embeddings for each user and item. In what follows, we introduce techniques for learning embeddings.

\begin{header}{Matrix factorization}
Matrix factorization (MF) is a classic embedding-based method. 
The basic assumption behind MF is that the user-item interaction matrix has a low-rank structure. 
MF delineates each user and item as an embedding vector and then predicts the preference between each user-item pair as the inner product of their embedding vectors. 
Let $p_u \in \mathcal{R}^k$, $q_i \in \mathcal{R}^k$ denote the embedding vector of user $u$ and item $i$, respectively. 
MF makes a prediction for the user-item pair $(u,i)$ as follows:    
\end{header}
 \begin{equation}\label{eq:sec5-MF-pre}
    \hat{r}_{ui} = p_{u}^\top q_i.
\end{equation}
MF can be optimized by minimizing the deviation between the predictions and the user-item interactions. 
Formally, we have the following objective function:
\begin{equation}
{\min _{P,Q}}\sum\limits_{(u,i)} l ({r_{ui}},{\hat r_{ui}}) + \lambda {L_{reg}}(P,Q),
\end{equation}
where $r_{u,i}$ denotes a user-item interaction. 
This can be explicit feedback (e.g., user ratings), which directly reflects the user preference, or implicit feedback (e.g., purchases and clicks), which indicates whether the user interacts with the item. 
$l(\cdot,\cdot)$ denotes the selected error function between the prediction and the ground truth label. 
It can be selected from mean squared error (MSE) loss~\citep{Koren2009Matrix}, binary cross-entropy loss~\citep{He2017NCF}, hinge loss~\citep{wu2016collaborative}, and Poisson likelihood~\citep{PoissionMF}. $L_{reg}(\cdot,\cdot)$ denotes a regularizer for the embeddings to avoid over-fitting. 
Here we collect $p_u$ (and $q_i$) for each user (and item) as a matrix $P$ (and $Q$).

Compared with heuristic-based methods, MF is a more generic method that can adaptively learn user preferences from their history behaviors and require no manually crafted heuristic design. 
The use of MF has brought a revolution, pushing research attention from previous heuristic-based methods towards embedding-based methods. 

\begin{header}{Information-enhanced embedding models}
Despite the prevalence of MF, it is still insufficient to yield accurate embeddings. The reason is that MF directly projects user/item IDs to an embedding space, making MF reliant on the behavioral signal from the objective function. 
Hence, MF models do not perform well for inactive users or items with very few interactions.
To deal with this problem, several methods have been proposed to enrich the representation with supplementary information. 
We can divide them into three groups:    
\end{header}
\begin{enumerate*}[label=(\roman*)]
\item neighborhood-enriched embedding methods
\item feature-enriched embedding methods
and 
\item graph-enriched embedding methods.
\end{enumerate*}
We discuss each of these groups in detail as follows:

\OurParagraph{(i) Neighborhood-enriched embedding.} 
Beyond the user ID, \citet{bell2007lessons,koren2008svd++,kabbur2013fism} enrich a user's representation with their rating history. 
This kind of method generates user-item preference scores as follows:
\begin{equation}
    \label{eq:sec5-svd++}
    \hat{r}_{ui} = q_{i}^\top \left(p_u + |\mathcal{N}(u)|^{-\alpha} \sum_{j\in \mathcal{N}(u)} y_j\right),
\end{equation}
where $\mathcal{N}(u)$ denotes a set of items with which the user $u$ has interacted. 
Each item $j\in \mathcal{N}(u)$ is mapped into a common embedding space to get a vector representation $q_j$. 
Neighborhood-enriched embedding methods can be considered as a combination of MF-based methods and neighborhood embedding methods.
MF-based and neighborhood methods make predictions from different perspectives, which results in different strengths and weaknesses.
Neighborhood methods struggle to detect all associations captured by  interactions, whereas MF-based methods are poor at detecting associations among sparse neighborhoods.
Hence, endowing MF with neighborhood methods fosters its merits and circumvents its weaknesses.

\begin{figure}
    \centering
    \includegraphics[width=0.6\columnwidth]{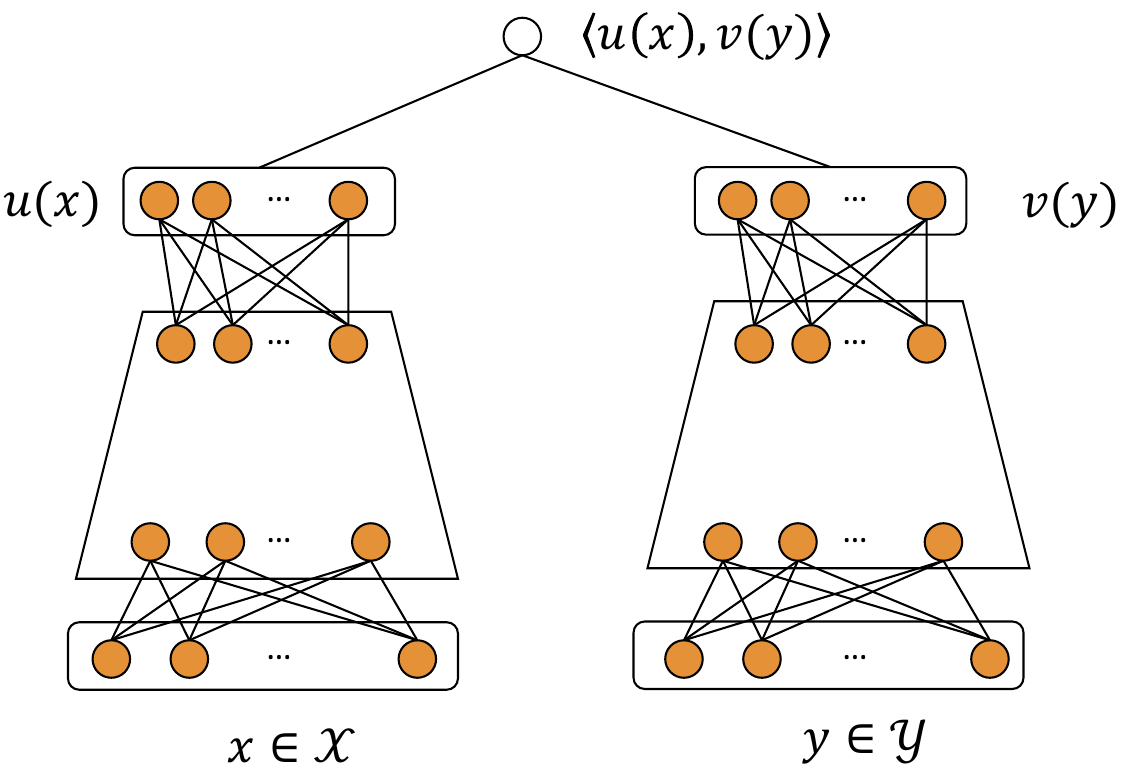}
    \caption{Illustration of the two-tower architecture for learning users' and items' representations. Image source: \citep{xu2016tag-dssm}.}
    \label{fig:sec5-dssm}
\end{figure}

\OurParagraph{(ii) Feature-enriched embedding.} 
Another way to improve MF is to use rich user and item features, e.g., a user's age, gender, education, revenue, product tags, category, and price. 
These features are valuable in enriching the representations of users and items, which  further boosts the recommendation performance. 
Fig.~\ref{fig:sec5-dssm} summarizes the architecture of feature-enriched candidate retrieval methods as a two-tower structure~\citep{yi2019youtube-2tower,fan2019mobius,xu2016tag-dssm}. 
The left tower is the user tower that translates a user's features into the user's embedding representation; the right tower is the item tower generating the item representation. 
Let $x_u$ denote the features of user $u$, while $x_i$ denotes the features of item $i$. The function of the two towers can be depicted as follows:
\begin{equation}
p_u = f_U(x_u),\, q_i = f_I(x_i),
\end{equation}
where $f_U(\cdot)$ and $f_I(\cdot)$ denote the translation function with regard to the implementation of the two towers. 
Linear models and neural networks (e.g., MLP~\citep{huang2013learning}, LSTM~\citep{song2016TDDSSM}, CNN~\citep{shen2014learning} and auto-encoders \citep{wu2016collaborative,liang2018variational}) can be used as the translation function. 
Given user and item embeddings, the two-tower model makes a prediction for each user-item pair. Cosine similarity and inner product are usually adopted to measure the embedding distance, so we have:
\begin{equation}
   \label{eq:sec5-dssm-pre}
    \begin{split}
      \text{Cosine: } & \hat{r}_{ui} = \frac{q_{i}^\top p_u }{\|p_u\|\cdot\|q_i\|}, \\
      \text{Inner product: } & \hat{r}_{ui} = q_{i}^\top p_u.
    \end{split}
\end{equation}
 
\OurParagraph{(iii) Graph-enriched embedding.} 
A drawback of the aforementioned methods is that they regard each user or item as an ``island'' and fail to explicitly encode their relations into representations. 
These relations, e.g., item-item occurrences or user-item interactions, are important in revealing correlations between users and items, which provide valuable signals for recommendation. 
To address this problem, several studies have established specific graphs between users and items, and then conduct graph representation learning to generate user and item embeddings. 
One representative method is EGES~\citep{wang2018billion}. 
EGES constructs an item-item graph based on user behavior sequences, where two items are connected if they consecutively occur in one sequence. EGES then applies DeepWalk~\citep{perozzi2014deepwalk} on the item graph to generate item representations. 
Similarly, other related methods perform graph convolutional networks to enrich the representation learning of both users and items~\citep{wang2019NGCF,ying2018graph,LightGCN}.
A knowledge graph with informative relations between items has been exploited to learn better embeddings~\citep{wang2019kgat,wang2020ckan}.

LightGCN~\citep{LightGCN} is a light graph neural network (GNN) model for recommendation, in which only the item and user embedding need to be learned, whereas non-linear operations are not considered. 
The basic concept of LightGCN is to learn the user or item representation by aggregating the information from multi-order neighbors in the user-item interaction graph. 
Assuming that the user and item embedding are $\mathbf{e}_{u}^{0}$ and $\mathbf{e}_i^{0}$ for user $u$ and item $i$ respectively, LightGCN takes the following  aggregation to get the representations of different layers:
\begin{equation}\label{lightgcn}
    \mathbf{e}^{k+1}_u = \sum_{i \in \mathcal{N}_u} \mathbf{e}_{i}^{k}, \,
    \mathbf{e}^{k+1}_i = \sum_{u \in \mathcal{N}_i} \mathbf{e}_{u}^{k},
\end{equation}
where $k$ represents the $k$-th layer, $\mathcal{N}_u$ represents the neighbors of user $u$, and $\mathcal{N}_i$ represents the neighbors of item $i$. 
Then, the final representation of users and items are computed as follows:
\begin{equation}
    \mathbf{e}_u = \sum_{k=0}^{K} \alpha_k \mathbf{e}_{u}^{k}, \, \mathbf{e}_i = \sum_{k=0}^{K} \alpha_k \mathbf{e}_{i}^{k},
\end{equation}
where $a_k$ is a hyper-parameter to control the contributions of different layers. 
Eventually, it takes the dot product of the two representations as the prediction score. 

The quality of data affects the upper bound of the performance of the learning-based models. However, user behavior data in e-commerce recommendation is usually very sparse. 
To address this problem, data augmentation is a common strategy to increase the diversity of data. 
SGL~\citep{wu202SGL} designs three types of methods to augment the training data for graph-based methods. The augmented data is used for an additional unsupervised task which maximizes the agreement of positive pairs.

\subsection{Session-based recommendation}
\label{sec5:sq}

Modeling sequential dynamics is important for candidate retrieval in e-commer\-ce recommendation. Sequential (or session-based) recommendation takes behavior sequences as the input, and then predicts the user's next click or purchase behavior. 
Sequential methods can be grouped into two classes: Markov chain-based and neural network-based sequential recommendation methods.
 
Early studies on sequential recommendation methods often use Markov chains by assuming that the user's next action only depends on the previous one. 
As a representative method, the factorizing personalized Markov chains model (FPMC) extends MF by modeling the effects of sequential-consecutive actions~\citep{FPMC}. 
Since the previous action is a critical factor affecting the user's next decision, FPMC achieves a significant gain over MF-based models. 
Following FPMC, \citet{he2016fusing} use a higher-order Markov chain to capture the sequential dependence among non-consecutive behavior. 
It is hard to use Markov chain-based methods for capturing complicated and long-term dependencies in sequential data, and this limits their performance in e-commerce recommendation.

More recently, deep neural networks have been used in sequential recommendation due to their powerful expressive ability on capturing behavior dependences (discussed in Section~\ref{chapter:user}). 
Generally speaking, neural networks for sequential recommendation can be divided into three types: 
\begin{enumerate*}[label=(\roman*)]
\item RNN-based methods that model sequential dependence with RNN (or improved versions such as LSTM and GRU) to capture both long-term and short-term dependencies~\citep{quadrana2017personalizing,jannach2017recurrent,hidasi2016parallel}; 
\item CNN-based methods that concatenate the embedding of the previous item in a sequence as to a matrix~\citep{tang2018personalized}; and
\item attention-based methods that introduce attention mechanisms into sequential recommendation by considering various types of user behavior with different influences~\citep{kang2018self,sun2019bert4rec}. 
\end{enumerate*}

\subsection{Next-basket recommendation}
The \ac{NBR} task uses information from previous sessions. 
It is defined as recommending a group of items to a user based on their shopping history, where the history is a time-ordered  sequence of baskets that they have purchased in the past~\citep{li-2023-next}. 
Each basket is a set of items with no particular order. 
This formulation fits the grocery shopping setting well, where a user's purchase history occurs naturally in the form of such baskets. 
Variations of the \ac{NBR} task where the order of items in a basket may be relevant, can, e.g., be found in music (playlist recommendation), in travel (recommending holiday packages), and in research and education (recommending reading lists).
Two main characteristics of the grocery shopping scenario make the \ac{NBR} task in this domain distinct from other retail domains: 
\begin{enumerate*}[label=(\roman*)]
\item users shop for grocery items repeatedly and on a regular basis, and 
\item grocery items have a short life time and are repurchased frequently by the same user~\citep{liu-2019-characterizing}. 
\end{enumerate*}

In the grocery shopping domain, it has been found to be useful to distinguish between \emph{repeat items}, i.e., items that a user has consumed before, and \emph{explore items}, i.e., items that a user has not consumed before~\citep{ariannezhad-2022-recanet,li-2023-next}.
In particular, for repeat items, the set of candidate items that needs to be considered for an individual user is usually in the low hundreds~\citep{ariannezhad-2021-understanding} as opposed to the full item catalog that needs to be considered for explore items, whose size may exceed 50,000 items in grocery shopping~\citep{ariannezhad-2020-demand,sprangers-2023-parameter}.
This fact makes the task of retrieving repeat items to be included in the next basket to be recommended to a user considerably easier than the task of retrieving explore items.
Frequency-based, nearest neighbor-based, and deep learning-based methods have all been used for the \ac{NBR} task, and for the candidate retrieval phase in particular~\citep{li-2023-next}.
As a rule of thumb, being biased towards the easier repetition task is an important strategy that helps to boost the overall NBR performance. 
Deep learning-based methods do not effectively exploit the repetition behavior. 
Indeed, they achieve a relatively good exploration performance, but they are not able to outperform simple frequency-based 
baselines in several cases. 
Some recent state-of-the-art \ac{NBR} methods are skewed towards the repetition task and outperform frequency-based baselines. However, the improvements they achieve are limited, especially considering the complexity and computational costs, e.g., for the training process~\citep{yu-2020-predicting} and for hyper-parameter search~\citep{DBLP:conf/um/FaggioliPA20,hu-2020-modeling}.

A number of variations of the \ac{NBR} task have recently been considered, each with their own challenges for retrieving candidate items.
The \emph{within-basket recommendation} task uses information from previous sessions as well as information from an incomplete basket to which additional items could potentially be added~\citep{ariannezhad-2023-personalized}. 
Another variation concerns an \emph{item-centered} scenario (as opposed to the familiar user-centered scenario), where the input is an item and the task is to identify users who might be interested in consuming the item~\citep{li-2023-who}. 

\section{Candidate ranking models}
\label{ch5:ranking}

Given generic feature vectors as input, work on candidate ranking strategies has mainly focused on modeling interactions between features. 
According to the types of interaction function they adopt, existing methods can be divided into linear, polynomial, and neural network models.

\subsection{Linear models}

Early studies on candidate ranking usually apply linear models, such as \emph{logistic regression} (LR)~\citep{kleinbaum2002logistic,hosmer2013applied} and naive Bayesian methods~\citep{hastie2009elements}. 
In contrast to other complicated models, linear models are straightforward, efficient, and explainable. Although linear models may not perform as well as deep neural networks, they indeed lay the foundation for recent advances in e-commerce recommendation~\citep{peng2002introduction,kiseleva2016beyond,bernardi2015continuous}. 
In e-commerce recommendation, logistic regression (LR) is one of the most popular methods that formulate the task as a classification task to rank through predicting the probability of an item to be interacted.
LR first collects features of users (e.g., age and gender) and items (e.g., price and categories) into a number of feature vectors, and then applies a linear combination function to map the feature vector into the final predicted score.
Similarly, \citet{kiseleva2016beyond} employ a naive Bayesian ranking strategy in e-commerce recommendation by considering contextual user profiling.

\subsection{Polynomial models}
\label{sec5:pm}

The performance of linear models is limited because of high space complexity and the inability of high-level feature modeling.
To address these two problems, factorization machines (FM) have been proposed~\citep{FM}. Factorization machines factorize parameters $w_{i,j}$ into an inner product of two latent vectors, i.e., $w_{i,j} \equiv \left<v_i,v_j\right>$, where $v_i$ denotes a latent vector of the $i$-th feature. 
With different types of knowledge, the feature interactions across multiple fields should have different weights in recommendation. 
To tackle this challenge, field-aware factorization machines (FFM)~\citep{FFM} have been proposed to capture field-aware weights and distribute a single latent vector to multiple fields. 

A drawback of FM and FFM is that they only capture second-order feature interactions but neglect higher order interactions, which are widely observed in e-commerce scenarios. 
As described in Section~\ref{ch31:subsec:cpr},~\citet{he2014practical} proposed a hybrid model by combining GBDT and logistic regression for click-behavior modeling.
The hybrid model is able to use boosted decision trees (i.e., GBDT) to conduct feature interactions into logistic regression for e-commerce recommendation. 
In this hybrid model, GBDT adaptively conducts feature selection and higher-order feature interactions.

\subsection{Neural network models}
\label{sec5:nnm}

Deep neural networks have been used in e-commerce recommendation because of their powerful expression ability of capturing complicated feature interactions. 
Up to now, research on neural network models can be categorized into the following three research directions:
\begin{enumerate*}[label=(\roman*)]
	\item The first direction aims at developing feature interaction modules based on neural networks, e.g., adding more neural network layers or combining the superiority of different neural networks.
	\item The second direction aims at enhancing the expression by using deep neural networks using FMs.
	\item The third direction aims at using the attention mechanism in capturing diverge and dynamic contributions of feature interactions.
\end{enumerate*}
Below, we will detail the recent advances along these research directions.

\begin{header}{Neural feature interactions}
\citet{DeepCross} propose the \textit{deep crossing} model, which can be considered as the first end-to-end deep learning framework for recommender systems. 
Deep crossing enjoys the merits of deep learning in coping with various features and capturing complex feature interactions. It consists of the four components: an embedding layer, a stacking layer, multiple residual units layer, and a scoring layer.  
The goal of the embedding layer is to transform per individual sparse features into dense vectors in latent space via neural networks. The stacking layer concatenates different embedding features from the embedding layer and generates a new vector for all features.
The scoring layer servers as an output layer with logistic regression to generate the final predicted score.
\end{header}

Collaborative filtering can be reconsidered from the perspective of deep learning.
Traditional collaborative filtering methods employ the inner product of a user's latent vector and an item's latent vector for rating prediction.
Neural collaborative filtering (NCF)~\citep{He2017NCF} has been proposed to replace the inner product operation with a neural network.

By learning frequent co-occurrences of features, deep neural networks may have poor memorization, i.e., these models easily over-generalize and recommend less relevant items when user-item interactions are sparse. 
To address this problem, \citet{cheng2016wide} introduce the Wide\&Deep model for recommendation.
The detailed model architecture of Wide\&Deep has already been presented in Section~\ref{ch31:subsec:cpr}.
Wide\&Deep maintains a balance between memorization and generalization: its wide component can effectively memorize sparse feature interactions, while the deep neural networks can generalize the previously unseen feature interactions through low-dimensional embeddings.

Follow-up studies improve either the wide component or the deep component in the Wide\&Deep model. 
The Deep\&Cross Network (DCN)~\citep{wang2017deep} replaces the wide component with a well-designed cross network.
DCN applies an explicit feature crossing mechanism with multiple cross layers. 
Later studies seek to apply automating machine learning (AutoML) to model the selection of feature interactions~\citep{Su2021Detecting,Ze2021A,ZhaO2021AutoLoss}.
SIF~\citep{Yao2020Efficient} uses one-shot architecture search methods to search proper interaction functions (e.g., inner product, plus/minor, max/min pooling, outer product, and concatenation) for collaborative filtering models.
AutoFIS~\citep{Liu2020AutoFIS} continuously searches effective feature interactions by incorporating architecture parameters to identify important feature interactions.
AutoGroup~\citep{Liu2020AutoGroup} groups useful features into sets using AutoML and then generates interactions from each set.

\begin{figure}
	\centering
	\includegraphics[width=0.8\columnwidth]{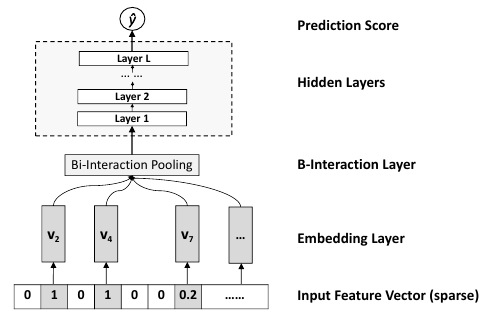}
	\caption{Overview of neural factorization machines. Image source: \citep{he2017neural}.}
	\label{fig:NFM}
\end{figure}

\begin{header}{Endowing factorization machines with neural networks}
Many models have been proposed to integrate \acp{FM} with deep neural networks to make full use of their advantages in feature combination.
Neural factorization machines~\citep[NFMs;][]{he2017neural} enhance \acp{FM} by modeling higher-order and non-linear feature interactions.
NFM introduces a bi-interaction pooling operation in neural network modeling, and presents a new neural network perspective for \acp{FM}. 
Given the embedding set of all features $\mathcal{V}_x$, the bi-interaction layer in NFM converts $\mathcal{V}_x$ to one vector:
\end{header}
\begin{gather}
\label{eq:NFM_BI}
	f_{BI}(\mathcal{V}_x) = \sum_{i=1}^{n-1} \sum_{j=i+1}^n (x_i v_i) \odot (x_j v_j),
\end{gather}
where $\odot$ denotes an element-wise product of two vectors. 
The output of bi-interaction pooling is a $d$-dimensional vector that encodes the second-order interactions between features in the embedding space. 
By stacking non-linear layers above the bi-interaction layer, NFM can effectively model higher-order and non-linear feature interactions. In contrast to traditional deep learning methods that simply concatenate or average embedding vectors in the low level, the use of bi-interaction pooling encodes more informative feature interactions. Fig.~\ref{fig:NFM} illustrates the architecture of NFM.

As demonstrated in Section~\ref{eq:se3:qsctr}, DeepFM~\citep{guo2017deepfm} aims to learn both low and high-order feature interactions, and consists of two components: the FM component and the deep component.
Compared with Wide\&Deep, DeepFM replaces its wide component with FM to remedy its shortcoming in automatic feature combination. Another difference is that DeepFM shares the feature embedding between the FM and deep component. 
Besides NFM and DeepFM, many other neural networks have been proposed based on \acp{FM}: FNN~\citep{zhang2016deep} directly stacks \acp{FM} with neural networks; PNN~\citep{qu2016product} models both bit-wise interactions and vector-wise feature interactions; and xDeepFM~\citep{lian2018xdeepfm} extends deepFM with explicit high-order feature interactions. 

\begin{header}{Attention mechanisms}
Attention mechanisms have been applied in recommender systems and achieved great success~\citep{AFM,LiRCRLM17,zhang2019deep}.
AFM~\citep{AFM} is an early attempt to introduce an attention mechanism to recommendation. It can be regarded as an extension of NFM. 
The sum pooling operation in NFM treats all pairwise feature interactions equally, which may produce suboptimal results.
To address this problem, AFM uses the attention mechanism on feature interactions by performing a weighted sum on the interacted vectors.
The output of the attention-based pooling layer is projected into the prediction score. 
In session-based or sequential recommendation scenarios, 
\citet{LiRCRLM17} propose an encoder-decoder model, neural attentive recommendation machine (NARM), to emphasize a user's main purpose in the current session.
The authors adopt a hybrid encoder structure with a global component for modeling long-term purposes and a local component for modeling short-term purposes.
Based on the combined session representation, a bi-linear matching scheme is then applied to compute the recommendation scores for each candidate item.
Following NARM, \citet{ren2018repeatnet} put forward the RepeatNet model to deal with the phenomenon of repeat consumption behavior.
The authors incorporate a repeat-explore mechanism into neural networks, which can select items from a user's history and suggests them at the appropriate moment.
In standard embedding paradigms, user features are compressed into fixed-length representation vectors. However, fixed-length vectors limit capturing the diverse interests of a user from historical behavior. 
\citep{zhou2018deep} introduce the deep interest network (DIN) to tackle this challenge by designing a local attention unit. The local attention unit in DIN adaptively learns the representation of user interests from historical behavior by taking account of the relevance of historical behavior.
\citet{Yuan2021Looking} unify CTR prediction models using a discrete choice model based on the self-attention mechanism. The authors regard feature interaction as the individual's comprehensive measurement of the influence of different factors on the decision-making process.
\end{header}

\subsection{Retraining strategies}
\label{sec:re-training}

E-commerce recommender systems rely on knowledge gleaned from historical interactions. As the number of collected interactions grows, recommendation models must be regularly retrained to reflect users' dynamic preferences. 
There are two intuitive heuristic retraining methods:    
\begin{itemize}[leftmargin=*,nosep]
    \item \textbf{Full retraining} simply merges the old data and new data to perform a full model training. The method is designed to capture both short-term and long-term features of the recommender system based on all the data it has accumulated. 
    \item \textbf{Fine-tune retraining} refers to using the parameters of the old model that were optimized by the old data to initialize the new model and train it with the new data. It reduces the time and storage overhead of retraining, making life-long updating feasible. 
\end{itemize}
Numerous deep learning-based retraining methods have been proposed. 
\citet{SML} propose a sequential meta-learning model (SML), including a meta-learning retraining framework for vanilla matrix factorization models. SML model captures the trend of changes between two distinct retraining phases at adjacent times.
Recently, graph convolutional neural network (GCN) has become the cutting-edge technique for recommendation~\citep{LightGCN,ying2018graph}. 
However, GCN-based recommender models encounter challenges regarding model retraining as GCN-based models take more time to converge.
CIGC~\citep{CIGC} has been proposed with two novel operators: incremental graph convolution and colliding effect distillation. 
The incremental graph convolution estimates the output of fully retraining the graph convolution using only new data; whereas the colliding effect distillation uses causal inference to retrain the representations of users (or items) that have no new data.

\section{Re-ranking strategies}
\label{sec:re-ranking}

The two-stage recommendation framework faces several problems in e-commerce recommendation:
\begin{enumerate*}[label=(\roman*)]
	\item The point-wise objective functions (e.g., log-loss) at the ranking stage often become sub-optimal because of the neglect of mutual influences between items.
	\item Users with different preferences may exhibit different behavior patterns. 
	\item Recommendation result diversification has not yet been well addressed during the ranking stage. 
	\end{enumerate*}
Therefore, before exposing recommended items to users, most e-commerce recommender systems refine the results through an additional re-ranking stage. 
Generally, the goal of re-ranking is to enhance the recommendation results through 
additional criteria or constraints~\citep{Chen2017Improving,zehlike2017fa,abdollahpouri2019popularity}.
Re-ranking methods can be categorized into heuristic strategies and list-wise objective functions.
The former type of method is based on determinantal point processes~\citep{Wilhelm2018Practical,Chen2017Improving} and maximal marginal relevance~\citep{Carbonell1998The}. These methods have been widely applied in e-commerce to avoid the items with the same category being presented consecutively for greater diversity. While for list-wise objective functions, DLCM~\citep{ai2018learning} and PRM~\citep{Pei2019Personalized} have been proposed with specific list-wise optimization objectives.

\section{Emerging directions}

We discuss five emerging research directions in e-commerce recommendation: 
\begin{enumerate*}[label=(\roman*)]
\item structured recommendations,
\item conversational recommendation, 
\item reasoning recommendations and explanations, 
\item biases and debiasing in recommendation, and 
\item unifying recommendation and search.
\end{enumerate*}

\subsection{Structured recommendations}

The task of structured recommendations is to predict the next  structured item sets instead of the next item. we discuss three categories of structured recommendations: slate recommendation, playlist recommendation, and next-basket recommendation.

In applications like music or bundle recommendations, the objective is to provide users with a ``slate'' -- a combination of items -- to maximize their engagement with the recommended content. This task raises critical questions, including the consideration of metrics such as diversity and the computational challenges posed by the combinatorial nature of slates. Reinforcement learning (RL) is extensively applied in slate recommendation \citep{Ie2019ReinforcementLF, Sunehag2015DeepRL, deffayet-2023-generative, DBLP:conf/kdd/TomasiCKCRD23}. However, due to the combinatorial complexity of actions, RL typically necessitates simplifying assumptions, such as the user selecting the optimal item \citep{Ie2019ReinforcementLF}. An alternative approach involves integrating a separate user preference model to optimize slate assembly and subsequently training the RL model \citep{DBLP:conf/kdd/TomasiCKCRD23}. \citet{DBLP:conf/nips/SwaminathanKADL17} investigate off-policy evaluation and optimization via inverse propensity scores for slate interactions. \citet{DBLP:conf/www/MehrotraLKLH19} constructe a hierarchical model to assess user satisfaction in slate recommendation systems.

Music playlist recommendation can be considered as a special case of music recommendation, focusing on delivering a curated list of songs to users. The order and characteristics of music tracks significantly influence the playlist's overall quality. An earlier study employs time-series-based machine learning to address the challenge of recommending music playlists \citep{Choi2016TowardsPG,Irene2019AutomaticPG,Monti2018AnEA,Vall2018TheIO,Kim2018TowardsSM,DBLP:conf/dasfaa/YangZXYZZ19,}. 
\citet{Choi2016TowardsPG} use a recurrent neural network (RNN) for music playlist generation, emphasizing track transition qualities. 
\citet{Monti2018AnEA} implement an ensemble of RNNs, using pre-trained embeddings for album and title representation. 
\citet{Irene2019AutomaticPG} predict user preferences by analyzing manually created playlists, employing both RNN and convolutional neural network (CNN) models.
Later studies have employed reinforcement learning (RL)-based methods to capture users' long-term preferences \citep{Hu2017PlaylistRB,DBLP:conf/ismir/ShihC18,DBLP:conf/gcce/SakuraiTOH20,DBLP:conf/lifetech/SakuraiTOH21,DBLP:journals/sensors/SakuraiTOH22,DBLP:conf/kdd/TomasiCKCRD23,DBLP:conf/atal/LiebmanSS15}. 
\citet{DBLP:conf/atal/LiebmanSS15} use a novel reinforcement-learning-based music recommendation system that generates playlists by considering both song preferences and transitions. \citet{Hu2017PlaylistRB} enhance playlist generation performance by integrating user feedback into the recommendation reward function. 
\citet{DBLP:conf/ismir/ShihC18} incorporate novelty and popularity indices into the reward function, resulting in playlists with a mix of new and well-known tracks. 
\citet{DBLP:journals/sensors/SakuraiTOH22} use informative knowledge graphs to enhance reinforcement learning optimization, and allowing users to customize flexible reward functions to discover new music genres. 
\citet{DBLP:conf/kdd/TomasiCKCRD23} present a reinforcement learning framework optimizing directly for user satisfaction via the use of a simulated playlist-generation environment.

Next-basket recommendation (NBR) is a task focused on predicting a user's next shopping basket based on their past purchase history, aiming to enhance user experience and satisfaction.
There are mainly three families of NBR methods. First are conventional NBR methods, such as those employing pattern mining \citep{DBLP:conf/icdm/GuidottiRPGP17}, KNN models \citep{hu-2020-modeling,DBLP:conf/um/FaggioliPA20}, and Markov chain
models \citep{FPMC}. \citet{hu-2020-modeling} and \citet{DBLP:conf/um/FaggioliPA20} model temporal patterns across frequency data and integrate this with neighbor information or user-wise collaborative filtering. \citet{FPMC} use matrix factorization and Markov chains to model users’ general interest and basket transition relations. 
Second are latent representation methods, which use representation learning techniques to capture implicit patterns in data. For instance, \citet{DBLP:conf/sigir/WangGLXWC15} apply aggregation operations to learn a hierarchical representation of user's last basket to predict the next basket. Third are deep learning-based method. Recurrent neural networks (RNNs) have been extensively applied in next-basket recommendation, demonstrating their efficacy in learning long-term trends by modeling the whole basket sequence. For instance, \citet{DBLP:conf/sigir/YuLWWT16} use max/avg pooling to encode baskets and \citet{DBLP:conf/kdd/HuH19} adapt an attention mechanism and integrate frequency information to improve performance. 
Some methods \citep{DBLP:conf/ijcai/LeLF19,DBLP:conf/aaai/WangH0SOC20} use item relations to enhance representation. 
\citet{yu-2020-predicting} employ graph neural networks (GNNs) to model item-item relations between baskets and a self-attention mechanism to discern temporal dynamics. 
Some methods \citep{DBLP:conf/sigir/BaiNZZDW18,DBLP:journals/corr/abs-2109-11654,DBLP:conf/dasfaa/LengYXX20,DBLP:conf/sigir/SunBDL0L20,DBLP:journals/jcst/WangZNG19} use auxiliary information, including product categories, amounts, prices, and explicit timestamps.

\subsection{Conversational recommendation}

The task of a \ac{CRS} is to provide recommendations to users through conversational interactions. 
\acp{CRS} are increasingly attracting attention~\citep[see, e.g.,][]{zhao2013interactive,ConversationRS-KDD16,yu2019visual,zou2020neural,mangili2020bayesian,sun2018conversational,lei20estimation,lei2020interactive,zhou2020improving,liu2020towards,zhou2020topicguided,zhang2020conversational,li2020seamlessly}. 
According to~\citet{gao-2021-advances}, the task of \ac{CRS} is formally defined as follows:
\begin{quote}
\emph{A recommendation system that can elicit the dynamic preferences of users and take actions based on their current needs through real-time multi-turn interactions.}
\end{quote}
Building on advances in interactive recommendation~\citep{christakopoulou2018q,wang2017factorization,liu2020diversified}, 
early studies on \acp{CRS} formulate the task as a specific application of task-oriented multi-turn dialogue systems (TDS)~\citep{DBLP:conf/ccks/LeLWWLJ18,dhingra-EtAl:2017:Long1,wen2017network,zhang2019memory}.
Studies into \acp{CRS} follow one of two main types of strategy: attribute-aware and topic-guided.

Attribute-aware \acp{CRS} aim to answer three main research questions: ``whe\-ther to ask or recommend,'' ``which attributes to ask'' or ``which items to recommend.'' 
Early work on attribute-aware \acp{CRS} obtains user preferences based on asking about items directly~\citep{zhao2013interactive,wang2018online,ConversationRS-KDD16,zou2020neural,DBLP:conf/aaai/VendrovLHB20}, or asking attributes through a heuristic method~\citep{christakopoulou2018q,zhang2018towards,DBLP:conf/sigir/LuoYWS20}. 
There are two main kinds of attribute-aware \acp{CRS} solutions. One kind asks a fixed number of questions and makes a recommendation at the last turn~\citep{lei20estimation,lei2020interactive}; whereas the other predicts a specific turn to recommend items. 
Reinforcement learning strategies have successfully been applied to attribute-aware \acp{CRS}.
\citet{liu2020towards} and \citet{li2020seamlessly} focus on cold-start users in conversational recommendation and extend bandit-based algorithms to balance the trade-off between exploration and exploitation. 
\citet{zou2022improving} propose TSCR, a transformer-based sequential conversational recommendation method that captures the sequential dependencies in dialogues to enhance recommendation accuracy.
\citet{deng2023unified} propose a novel unified multi-goal conversational recommender system, named UniMIND, which unifies these goals into a single sequence-to-sequence (Seq2Seq) paradigm and employs prompt-based learning strategies to facilitate multi-task learning.

Topic-guided \acp{CRS} interact with users through natural language conversations with fluent responses and precise recommendations~\citep{DBLP:conf/nips/LiKSMCP18,zhou2020topicguided,DBLP:conf/emnlp/ChenLZDCYT19,liu2020towards,zhou2020improving,DBLP:conf/emnlp/MaTH21,DBLP:conf/wsdm/ZhouZZWJ022}. 
Unlike attribute-aware \acp{CRS}, topic-guided \acp{CRS} focus on making recommendations using free text, which creates considerable flexibility to influence how a dialogue continues.
External knowledge has been applied in topic-guided \acp{CRS}~\citep{DBLP:conf/emnlp/MaTH21,zhou2020improving,DBLP:conf/emnlp/ChenLZDCYT19}.
\citet{DBLP:conf/emnlp/ChenLZDCYT19} integrate a recommendation system and a dialogue system via an end-to-end framework to bridge the gap between the two systems.
\citet{DBLP:conf/nips/LiKSMCP18} use an auto-encoder for recommendation and a hierarchical RNN for response generation.
\citet{zhou2020topicguided} propose a topic-guided \acp{CRS} method that incorporates topic threads to enforce transitions actively toward a final recommendation.
More recently, external knowledge graphs have been shown to be effective in improving the performance of topic-guided conversational recommendation systems.
\citet{DBLP:conf/emnlp/ChenLZDCYT19} apply knowledge graphs to enhance the semantics of contextual items for recommendation.
\citet{zhou2020improving} incorporate both word-oriented and entity-oriented knowledge graphs.
\citet{DBLP:conf/emnlp/MaTH21} perform tree-structured reasoning on a knowledge graph for recommendation.
\citet{zhang2022analyzing} focus on user reformulation behaviors to improve the robustness of conversational agents.
\citet{ren2022variational} explore user preferences in conversational recommendation and propose a variational reasoning mechanism to jointly track both short-term and long-term user behaviors.
\citet{zhang2023variational} present the first attempt to explicitly address the problem of dynamic reasoning over incomplete knowledge graphs. However, no study is capable of fusing recommendation and response generation in an end-to-end manner, which limits the potential for mutual reinforcement between these two tasks. Additionally, a lack of interpretability in current conversational recommendation system (CRS) models further hinders their ability to fully align with user needs. 
Models are typically trained on conversational recommendation datasets, but the assumption that the standard items and responses in these benchmark datasets are optimal leads to a tendency for CRSs to replicate the logic of the recommenders found in the data, rather than truly addressing the evolving needs of the users. This misalignment remains a significant challenge in advancing more user-centric and adaptable conversational recommendation systems.

Although \acp{CRS} have many merits, their evaluation is still a thorny issue.
Recent studies have evaluated \acp{CRS} either through offline evaluation or human evaluation~\citep{Lamel2000TheLA,li2015toward}.
Offline evaluation evaluates a dialogue system based on test sets, whereas human evaluation reflects the overall performance of the agent through in-field experiments~\citep{Black2011SpokenDC,gilotte2018offline} or crowd-sourcing~\citep{zhou2020topicguided,DBLP:conf/nips/LiKSMCP18}.
However, offline evaluation is often limited to single turn assessments, while human evaluation is intrusive, time-intensive, and is not scalable~\citep{zhao2019toward,siro-2022-understanding}.
As an alternative, user simulators that mimic user behavior are able to provide broad insights to generate human-like conversations for assessing \acp{CRS}~\citep{afzali2023usersimcrs}.

\subsection{Explainable e-commerce recommendation}

Although recommendation models can generate relevant items for users in many e-commerce applications, it is often ambiguous to understand why an item is recommended to a user.
Hence it is necessary to develop explainable recommendation strategies to generate not only high-quality recommendations but also intuitive explanations. 
Recent years have witnessed a growth in the number of publications on explainable recommendation. \citet{zhang2014explicit} generated textual sentences as recommendation explanation to help users understand each recommendation result. 
\citet{chen2018visually} propose visually explainable recommendations where particular regions of a recommended image are highlighted as the visual explanations for users. \citet{sharma2013social,quijano2017make} generate a list of social friends who also like the recommended product as social explanations for target user, whereas \citet{gao2019bloma} generate the recommendation described by a set of topics. 

Several researchers have started to generate explanations for deep recommendation models. For example, several studies use knowledge graphs for interpretation. They construct multi-hop paths from users to items along the knowledge graph, which indicates a specific explainable user-item relation \citep{hu2018leveraging,wang2019explainable,xian2019reinforcement}. Besides, \citet{chen2021neural} propose a neural collaborative reasoning system integrating the power of representation learning and logical reasoning. However, research on explainable deep recommendation models is relatively new and deserves to be further explored in e-commerce.

\subsection{Biases and debiasing in recommendations}

Many recommendation solutions about fitting user behavior may deteriorate owing to biases in behavior inherent in e-commerce recommendation ~\citep{LightGCN,sun2019bert4rec}. 
In e-commerce scenarios, user behavior is observational rather than experimental, which is often affected by many factors, e.g., self-selection of the user (selection bias)~\citep{marlin2007collaborative}, systematic exposure mechanisms (exposure bias)~\citep{ovaisi2020correcting}, public opinions (conformity bias)~\citep{krishnan2014methodology,liang2016modeling} and the display position (position bias)~\citep{joachims2007evaluating}. 
These biases make the data deviate from reflecting true preferences of users in recommender systems.
Efforts to debias recommendation can be divided into three major categories: 
\begin{enumerate*}[label=(\roman*)]
\item data imputation, which assigns pseudo-labels to missing data to reduce variance~\citep{steck2013evaluation}, 
\item inverse propensity scoring (IPS), which reweighs the collected data for an expectation-unbiased learning~\citep{sun2019debiasing,wang2016learning}, and \item generative modeling, which assumes the generation process of data and reduces biases~\citep{liang2016modeling}.
\end{enumerate*}
Most approaches lack the universal capacity to account for mixed or even unknown biases. 
To bridge the gap, \citet{chen2021autodebias} propose a universal debiasing framework that not only account for multiple biases and their combinations, but also
frees human efforts to identify biases.  
\citet{huang2022clickrec} introduce DANCER, a debiasing method that accounts for dynamic selection bias and user preferences, demonstrating its improved rating prediction performance over static bias methods.
\citet{10.1145/3583780.3615011} explore the use of uncertainty estimates in ranking scores to reduce societal biases in retrieved documents while minimizing utility loss. 
They propose an uncertainty-aware, post hoc bias mitigation method that outperforms baselines in terms of utility-fairness trade-offs, controllability, and computational costs, without requiring additional training.
Although recent years have seen a surge in research efforts devoted to recommendation biases, biases are still an important problem in e-commerce recommender systems. Sophisticated meta models to capture complex patterns and exploration of dynamic biases in recommendation should provide helpful insights.

\subsection{Unifying recommendation and search}
\label{sc5:urec}

Search and recommendation in e-commerce have similar characteristics, except for the different representation of ``contexts'' -- search aims at retrieving relevant items for matching a query while recommendation aims at finding items for matching a user's preferences. However, researchers usually conduct separate studies on them and use different techniques and training data for the two tasks. Thus, building a unified model for search and recommendation has the potential to improve both tasks as more comprehensive user behavior data can be used. One practical way to unify the two tasks is as part of the aforementioned conversational recommendation scenario, and the other is personalized search, which we detail next.

Early search engines, like Google and AltaVista, retrieved personalized results based on keywords. 
Personalized search has become far more complex with the goal to ``understand exactly what you mean and give you exactly what you want.'' Concretely, a personalized search engine not only focuses on retrieving items that satisfy the user’s current information needs, which is usually related to the query topic, but also considers user personality and aims at retrieving items that meet user preference. 
To achieve both goals, it is critical to model interactions between users, items and queries. 
\citet{ai2017learning} use a hierarchical embedding model to linearly combine the item-query matching scores with item-user preference scores; 
\citet{guo2019attentive} explore long and short term user preference learning model for personalized search; 
\citet{10.1145/3459637.3482489} integrate user behavior in search and recommendation into a heterogeneous behavior sequence and use a joint model to handle both tasks based on this unified sequence;
\citet{si2023search} use users' search interests for recommendations; they separately learns similar and dissimilar representations from search and recommendation behaviors using transformer encoders.
\citet{liu2020structural} construct a specific user-item-query graph and conducts node representation learning on the graph. 
\citet{10.1145/3488560.3498414} propose a method that jointly predict user clicks for both search and recommendation scenarios by constructing a unified graph to share user and item representations uniformly. Such graph embedding techniques open the potential to integrate both node information and topological structure information, which can capture high-order user-item-query interactions.

\subsection{Large language models in recommendation}
\label{sc6:llmrec}
Large language models (LLMs) have exhibited strong capabilities in understanding and processing text.
Their application to recommendation systems is actively being explored.
The main benefit of using LLMs in recommendation systems is their ability to produce high-quality representations of text features and make use of the wide range of knowledge they hold~\citep{liu2023pre}. LLM-based models can capture context more accurately, allowing them to better understand user questions, product descriptions, and other textual information.
Studies that apply LLMs to recommendation systems can be divided into two categories: discriminative strategies and generative strategies.

For studies into discriminative strategies, to improve the quality of vector representations for queries and products, and fully use the external knowledge stored in LLMs, a common approach is to fine-tune the original models, adapting them to recommendation tasks in order to obtain high-quality representations. 
\citet{qiu2021u} propose a novel U-BERT approach that utilizes a pre-training and fine-tuning framework to learn user representations. By using content-rich domains, U-BERT compensates for users' features in domains where behavior data is insufficient, improving recommendation performance.
Similarly, \citet{wu2021userbert} use unlabeled user behavior data and incorporate two self-supervised tasks: masked behavior prediction and behavior sequence matching for user model training.

Compared to discriminative models, generative models have better natural language generation capabilities. Therefore, most generative models typically translate recommendation tasks into natural language tasks, allowing the model to directly output recommendation results through fine-tuning or in-context learning.
\citet{sun2023chatgpt} introduce a sliding window prompt strategy for ranking candidates. This strategy ranks items within a window at each step, sliding the window from back to front multiple times to generate the final ranking results. This approach helps improve ranking performance by iteratively refining the candidate list.
\citet{kang2023llms} investigate the ability of LLMs to predict user ratings based on past behavior, comparing them with traditional collaborative filtering methods.

The application of LLMs to e-commerce recommender systems is still in its early days. Many challenges remain, in terms of evaluation, effectiveness, efficiency, and transparency.


\chapter{E-commerce QA and conversations}
\label{chapter:qa}

Section~\ref{chapter:basic} provides insights on how natural language processing technologies have been widely applied in e-commerce platform interfaces to help consumers better communicate with those platforms.
This section zooms in on question answering (QA) services and dialogue systems on e-commerce platforms. 
We divide this section into three parts: e-commerce question answering, e-commerce dialogue systems, and emerging directions. 
We first detail characteristics and approaches to e-commerce question answering (Section~\ref{ch6:sec1}).
Then, we demonstrate recent studies on dialogue systems applicable in e-commerce customer services (Section~\ref{ch6:sec2}).
Lastly, we describe emerging directions in e-commerce question answering and dialogue systems (Section~\ref{ch6:sec4}).

\section{Question answering in e-commerce}
\label{ch6:sec1}

In this section, we describe related work on e-commerce question answering. We divide this section into three parts: we first introduce studies on question answering in Section~\ref{ch6:sec1.1}, then in Section~\ref{ch6:sec1.2} we formulate characteristics of product-aware question answering; finally, we detail approaches to e-commerce question answering in Section~\ref{ch6:sec1.3} and~\ref{ch6:sec1.4}. 

\subsection{Introduction to question answering}
\label{ch6:sec1.1}

QA systems~\citep{simmons1965answering} are meant to facilitate users' access to information. 
For many web-based applications QA services provide a proper answer to a given question from the user~\citep{heilman2010tree,li2002learning}. 
Question answering research has received much attention in the past decades, including approaches to question classification, answer selection, answer generation, and answer summarization~\citep{li2002learning,heilman2010tree,liu2016retrieving,geigle2016scaling,song2017summarizing}.
QA systems have various classifications. 
QA systems can be divided into open-domain and domain-specific QA systems~\citep{chen2020open}.
Open-domain QA focuses on answering questions relying on knowledge and ontologies~\citep{ferrucci2010building}, whereas domain-specific QA focuses on providing proper answers in a specific scenario, e.g., customer service, hotel booking, etc.  
QA systems can also be divided into retrieval-based and generation-based QA systems according to how they generate answers~\citep{yang2015wikiqa}. 
The former searches and extracts potential answers via search engines, whereas the latter applies generation-based methods to give proper answers to the questions.
And finally, according to the answers, QA systems can be divided into factoid QA and non-factoid QA~\citep{song2017summarizing}.
Factoid QA systems return a concise answer to the given question, whereas 
non-factoid QA systems provide more subjective answers to the given questions.

Early work on QA distinguishes between four categories of QA system: list-structured database systems, graphical database systems, text-based systems, and logical inference systems~\citep{simmons1965answering}. 
All these systems have a limited scope with their rule-based strategies. 
Search engines remain integral components of QA systems.
With the development of information retrieval, research ``re-discovered'' QA systems in the late 1990s~\citep{jurafsky2000speech}.
TREC has launched dedicated QA tracks in 1999, with the purpose of advancing research into QA systems~\citep{srihari1999information,voorhees1999trec}.
A typical retrieval-based TREC QA system has three main components: question processing, passage retrieval, and answer processing~\citep{jurafsky2000speech}. 
For each step, sub-tasks must be considered, e.g., query formulation or answer type detection. Based on this framework, several approaches have been proposed to address research tasks in each component~\citep{brill2001data,li2002learning}.
Early studies on QA focus on factoid QA systems that generate concise answers~\citep{srihari1999information,jurafsky2000speech,brill2001data}.
Retrieval-based methods effectively answer these concise and simple questions~\citep{jurafsky2000speech,ahn2004using}.
However, complicated questions are found difficult to be addressed by pure retrieval-based methods~\citep{lin2006role}.
Therefore, integrating natural language understanding and knowledge-based reasoning techniques is essential for retrieval-based QA strategies in answering complicated questions.
%

In TREC-QA 2004, questions are grouped into topics, which motivates research on fact identification from reference knowledge resources, e.g., Wikipedia~\citep{ahn2004using}.
Wikipedia can be considered a generic collection of articles with real-world facts for open-domain QA systems. 
With the development of knowledge bases, innovations have occurred in the context of QA from knowledge bases with the creation of resources like web questions and short questions~\citep{berant2013semantic,bordes2015large}.
However, inherent limitations such as incompleteness and fixed schemas have persisted in traditional knowledge-based QA systems. 
Thus, in the 2000s QA work increasingly on systems that are able to generate answers from raw text explored, especially using Wikipedia~\citep{ahn2004using,buscaldi2006mining,ferrucci2010building,ryu2014open}.
As far as we know, \citet{ahn2004using} are the first to combine Wikipedia as a text resource with other resources in QA.
Similarly,~\citet{ryu2014open} perform QA using a Wikipedia-based knowledge model by combining articles with other answer-matching components. 
\citet{ferrucci2010building,baudivs2015yodaqa} integrate web-based and Wikipedia-based articles as knowledge resources into  highly developed full-pipeline QA platforms.
%

In more recent years, QA models increasingly apply deep neural networks to understand questions and generate answers.
\citet{yin2015neural} present an end-to-end neural network model, neural generative question answering (GENQA), that can generate answers to simple factoid questions.
Subsequently, a bi-directional attention flow mechanism has been proposed to obtain query-aware passage representations~\citep{seo2016bidirectional}.
\citet{chen2017reading} develop a system for question answering from Wikipedia, DrQA, that is composed of a two-stage retrieval-reader QA framework.
DrQA includes a document retriever module based on bigram hashing and TF-IDF matching. 
It also contains a document reader module where a multi-layer recurrent neural network is trained to detect answer spans in those few returned documents.

Following~\citet{chen2017reading}, most open-domain QA systems apply a two-stage retrieval-reader framework in their QA mechanisms~\citep{wang2018r,sun2018open,lin2018denoising,pang2019has,lee2019latent,guu2020retrieval,karpukhin2020dense,izacard2021leveraging,mao2020generation,sachan2021end,NEURIPS2021_da3fde15,yu-etal-2022-kg,DBLP:conf/emnlp/KediaZL22,DBLP:conf/emnlp/Ju00Z022,DBLP:conf/acl/WangY023}.
These studies employ a determinate retrieval function and treat each passage independently in the retrieval stage.
\citet{wang2018r} propose a reinforcement learning-based ranking strategy in the retrieval stage.
\citet{sun2018open} offer a graph convolution-based neural network by operating over heterogeneous graphs of knowledge base facts and text sentences. 
In contrast with previous kowledge-based open-domain QA systems, the authors propose heterogeneous update rules that handle knowledge base nodes differently from the text nodes.
\citet{lin2018denoising} design a coarse-to-fine denoising model to extract correct answers from multiple paragraphs in the noisy data. 
Their model employs a paragraph selector to filter out those noisy paragraphs and keep informative paragraphs.
Similarly, \citet{pang2019has} describe a three-level probabilistic formulation model for open-domain QA.
Word-level matching strategies are usually applied in the retrieval stage to match keywords represented in high-dimensional and sparse vectors. 
Dense passage retrieval has successfully been applied to open-domain QA to improve the matching performance as it is complementary to sparse representations in the retrieval stage.  
\citet{karpukhin2020dense} train a dense embedding model using only pairs of questions and passages.
\citet{izacard2021leveraging} detail an effective two-step dense passage retrieval method; the authors retrieve supporting passages using either sparse or dense embeddings and then employ a sequence-to-sequence model to generate the answer.
\citet{zhu2021adaptive} use a partially observed Markov decision process (POMDP) to re-formulate the QA problem using a reinforcement learning method to optimize the interactions between different components. 
\citet{yu-etal-2022-kg} use a knowledge graph to establish relational dependencies among retrieved passages and employ a graph neural network to re-rank retrieved passages for each query.
More recent work has proposed to improve reader performance and thereby improve QA performance. \citet{DBLP:conf/emnlp/KediaZL22} introduce a method for fusing information across multiple passages within a transformer encoder using global representation tokens. \citet{DBLP:conf/emnlp/Ju00Z022} design a knowledge graph enhanced passage reader that fuses graph and contextual representations into the hidden states of the reader model. \citet{DBLP:conf/acl/WangY023} enhance the fusion-in-decoder (FiD) framework by incorporating a process to distinguish between relevant and spurious passages, thereby improving the model's reasoning and performance in open-domain QA.
 
A model that matches the question with a passage using gated attention-based recurrent networks has been shown to be effective on QA benchmark datasets~\citep{wang2017gated}.
QANet combines local convolution with global self-attention for reading comprehension, which improved the reading comprehension performances~\citep{yu2018qanet}.
More recent studies have shown that pre-trained language models effectively understand questions and answers in QA systems~\citep{guu2020retrieval,mao2020generation,sachan2021end,NEURIPS2021_da3fde15}.
\citet{mao2020generation} augment a query in open-domain QA using text generation of a pre-trained language model.
\citet{sachan2021end} propose a QA method with an unsupervised pre-training of the retriever with a supervised fine-tune procedure. 

Many benchmark QA datasets have been proposed.
Several QA benchmark datasets, such as SQuAD~\citep{rajpurkar2016squad}, TriviaQA~\citep{joshi2017triviaqa}, and SearchQA~\citep{dunn2017searchqa}, only evaluate the reasoning ability within a single paragraph, whereas the other relevant documents or paragraphs are neglected.
These datasets employ knowledge bases for multi-hop reasoning, and are therefore constrained by the schema of knowledge bases.
\citet{yang2018hotpotqa} introduce an open-domain QA benchmark dataset, HotpotQA, which requires reasoning over multiple documents without constraining itself to a knowledge base.
To understand how the questions and answers are distributed in open-domain QA, \citet{lewis2020question} perform a large-scale analysis on open-domain QA benchmark datasets, and provide annotated subsets of test sets indicating whether test-time questions are duplicates of training time questions.

\subsection{Characteristics of e-commerce question answering}
\label{ch6:sec1.2}
E-commerce QA services focus on answering product-aware questions asked by e-commerce users.
Early studies on e-commerce QA focus on providing answers automatically from reviews by heuristic methods~\citep{li2009answering,moghaddam2011aqa,yu2012answering}.
With the development of both QA techniques and e-commerce services, e-commerce QA has received increasing attention in recent years~\citep{mcauley2016addressing,yu2017modelling,yu2018aware,fan2019reading,zhang2020answer,gao2019product,gaoshen2021tois,feng2021multi,deng2022toward}.

Distinct characteristics of e-commerce QA, as opposed to open-domain QA, are: 
\begin{enumerate*}[label=(\roman*)]
\item Domain-specific aspects are the first e-commerce QA characteristic.
E-commerce QA systems rely on exploiting domain-specific information from product descriptions.  
Different products make different product-related aspects relevant or popular~\citep{mcauley2016addressing}. 
These product-aware aspects can help distinguish products and answer questions.
\item There is a large number of consumer reviews,  which can be used as a data source to help people form opinions and decisions~\citep{liu2016retrieving,mcauley2016addressing}.
With the growth of those opinionated reviews, e-commerce users rely on advice from reviews before making purchase decisions.
Reviews have been used as supporting data and candidate answers to supervise QA prediction models~\citep{yu2018aware}.
\item There is a variety of answer sources.
Most e-commerce QA services focus on extracting answers from reviews~\citep{mcauley2016addressing}, and many e-commerce sites provide question answer pairs as knowledge bases for QA.
\end{enumerate*}

Text generation approaches have been studied to generate answers to given questions and reviews.
Question reranking and answer reranking also have been studied~\citep{yu2017modelling,zhang2020answer}. 
E-commerce QA research can be divided into two directions: extractive product-aware QA and generative product-aware QA. The former focuses on extracting sentences or passages from reviews to answer questions, whereas the latter applies textual generation approaches to generate answers.  
We detail each type of e-commerce QA study in Section~\ref{ch6:sec1.3} and~\ref{ch6:sec1.4}, respectively.

\subsection{Extractive product-aware QA}
\label{ch6:sec1.3}

Most e-commerce QA systems extract relevant sentences or fragments from the input text to answer the question given by the consumer. 
Early studies automatically extracted answers from reviews by heuristic unsupervised methods~\citep{li2009answering,moghaddam2011aqa}.
Follow-up work mainly focuses on the matching between questions and reviews or candidate answers~\citep{mcauley2016addressing,yu2018aware}. 
\citet{yu2012answering} proposed a framework for opinionated QA, which organizes reviews into a hierarchical structure and retrieves review sentences as the answer. 
The authors then use such a hierarchical structure to help retrieve questions and relevant review fragments.
A joint optimization approach is proposed by simultaneously considering review salience, coherence, and diversity to rank fragments.
\citet{liu2016retrieving} find a concise set of questions addressed by a given review and cover its main points to help the user quickly comprehend the reviews.
The authors propose a two-stage framework, where a probabilistic retrieval model is used to retrieve candidate questions and a matching procedure between answers and questions is used to bridge the vocabulary gap between reviews and questions.

Some products, such as clothes and paintings, may not have proper names.
Different strategies have been considered to replace the external knowledge of e-commerce to address this problem.
\citet{mcauley2016addressing} propose an answer prediction model by incorporating an aspect analytic model to learn latent aspect-specific review representation for predicting the answer. 
\citet{wan2016modeling} address ambiguity, subjectivity, and diversity problems in consumer reviews. 
By using multiple answers in a supervised framework, the authors provide more accurate answers to objective and subjective questions. 
The authors also release a large-scale e-commerce QA dataset consisting of 135 thousand products from Amazon, 808 thousand questions, 3 million answers, and 11 million reviews.
\citet{carmel2018product} focus on subjective questions from Amazon customers, which can relate to various intent types such as product usage, recommendations, and opinions.
The authors apply automatic QA methods, enhanced with community QA approaches to retrieve the most relevant answer found in reviews and QAs to address this problem.
\citet{yu2018aware} propose an answer prediction model by incorporating an aspect analytic model to learn latent aspect-specific review representation for predicting the answer.
The authors establish the advantage of generating aspect-specific representations for new questions, which they use to develop a predictive answer model to capture intricate relationships among question texts and review texts.
The proposed model uses reviews as a knowledge source to predict the answer by classifying answers into two types, binary (i.e. ``yes'' or ``no'') and open-ended responses.
As the amount of labeled data is limited in customer reviews,~\citet{das2019learning} propose an adversarial review-based approach to answer subjective and specific product-aware questions in a weakly supervised setting. 
Reading comprehension has been found to be useful to help extract relevant answers from e-commerce reviews~\citep{fan2019reading,xu2019review,zhang2020answering,chen2019answer}.
Using the raw text of product-aware questions and customer reviews, \citet{fan2019reading} introduce an end-to-end neural network model to synthesize multiple review representations.
\citet{chen2019answer} design a multi-task attentive model, namely QAR-Net, to identify plausible answers from product reviews for user questions. 
QAR-Net can use generated question answer pairs to help question-review matching. 

Pre-trained language models help understand the content of questions and reviews.
\citet{xu2019review} apply a BERT-based fine-tuning approach to extract answers from reviews.
\citet{mittal2021distantly} use a pre-trained language model to learn a relevance function by jointly learning unified syntactic and semantic representations of questions and reviews.
A QA dataset for review comprehension with subjectivity labels for questions and answers has also been exploited~\citep{bjerva2020subjqa}. 
Besides user reviews, another type of information, namely product details provided by the manufacturer, has been considered an auxiliary information source for addressing product-related questions~\citep{zhang2020answering}.
In order to alleviate the unavailability of labeled data, \citet{DBLP:conf/emnlp/JainRA23} introduce a distant supervision based model to prepare training data without manual effort.

\begin{figure}[!t]
	\centering
	\includegraphics[width = 0.9\columnwidth]{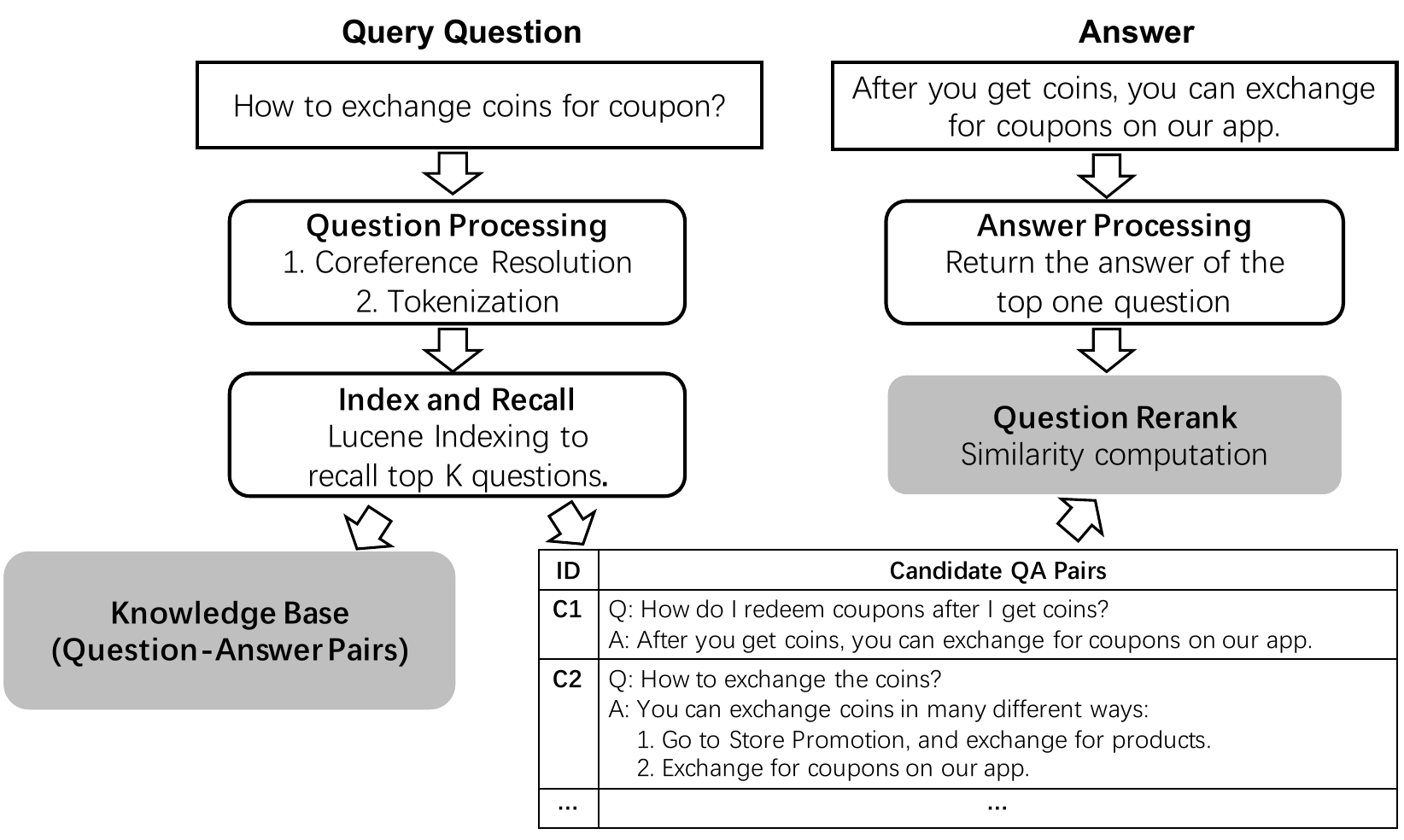}
	\caption{An overview of the retrieval-based QA system in Alibaba. Image source: \citep{yu2017modelling}.}
	\label{ch62:fig:frameworkalime}
\end{figure}

Review-based QA approaches extract answers from customer reviews, which can partially address users' questions.
However, there are many products with few or no reviews available.
By collecting question answer pairs from real users, many e-commerce platforms develop retrieval-based QA systems for automatically answering frequently asked questions (FAQs) in the e-commerce industry~\citep{yu2017modelling,song2020tcnn,song2021online,zhang2020answer}.
In Fig.~\ref{ch62:fig:frameworkalime}, we see the retrieval-based QA framework applied by Alibaba~\citep{yu2017modelling}.
Given a collection of question answer pairs (i.e., the knowledge base in Fig.~\ref{ch62:fig:frameworkalime}), a key component is the question rerank module, which reranks candidate questions in a question answering knowledge base to find the best match given a question from a user.
Based on such a framework, \citet{yu2017modelling} formulate e-commerce QA as a paraphrase identification problem, where the target is to identify semantic relations of the given sentence pairs. 
The authors describe a transfer learning QA strategy to adapt the shared knowledge learned from a resource-rich source domain to a resource-poor target domain.
Amazon has presented a large review-based QA dataset, namely AmazonQA, based on their real-world community QA platform~\citep{gupta2019amazonqa}. 
AmaonQA uses consumer reviews as the data resource and extracts snippets to answer questions. 
\citet{song2020tcnn} improve the matching performance in retrieval-based e-commerce QA by introducing a multi-layer triple convolutional neural network model. 
Also, a sub-graph searching mechanism is shown to improve the efficiency of retrieval-based e-commerce QA~\citep{song2021online}.
\citet{zhang2020answer} focus on answer selection in retrieval-based e-commerce QA. 
Using graph neural networks, the authors jointly model multiple semantic relations, including semantic relevance between the question and answers, textual similarity among answers, and textual entailment between answers and reviews.
\citet{rozen2021answering} detail an answer prediction approach that uses similar questions about other products.
The authors calculate contextual product similarity to determine whether two products are similar in the context of a specific question.
Two large-scale datasets, including a question-to-question similarity dataset from Amazon and a corpus of question answer pairs from Amazon, have been released with the publication.

\subsection{Generative product-aware QA}
 \label{ch6:sec1.4}

Many e-commerce portals have provided question answering services that assist users in posing product-aware questions to other consumers who have purchased the same product before.
Users must read the product's reviews to find the answer themselves. 
Given product attributes and reviews, following a cascading procedure, an answer is manually generated:
\begin{enumerate*}[label=(\roman*)]
\item a user skims reviews and finds relevant sentences; 
\item they extract functional semantic units; and 
\item and the user jointly combines these semantic units with attributes and writes an appropriate answer.
\end{enumerate*}
With a rapidly increasing number of reviews this process needs support~\citep{gao2019product}.
Several strategies have been proposed to automatically generate answers using the product's reviews to alleviate the burdens of customers~\citep{gao2019product,chen2019driven,deng2020opinion,lu2020chime,feng2021multi,deng2022toward}.
The task on which these approaches focus is \emph{generative product-aware QA} given reviews and product attributes.

In first attempts, \citet{gao2019product,chen2019driven} propose the task of \textit{product-aware answer generation}, where a product-related question answering model is applied to incorporate customer reviews with product attributes.
The authors formulate the research problem in generative e-commerce QA: for a product, there is a question $X^q = \{x^q_1, x^q_2, \dots, x^q_{T_q}\}$, $T_r$ reviews $X^r = \{x^r_1, x^r_2, \dots, x^r_{T_r}\}$ and $T_a$ key-value pairs of attributes $A = \{(a^k_1, a^v_1), (a^k_2, a^v_2), \dots, (a^k_{T_a}, a^v_{T_a})\}$, where $a^k_i$ is the name of $i$-th attribute and $a^v_i$ is the attribute content.
Each attribute, including key $a^k_i$ and value $a^v_i$, is represented as a single word in the generation task.
Given a question $X^q$, an answer generator reads the reviews $X^r$ and attributes $A$, then generates an answer $\hat{Y} = \{\hat{y}_1, \hat{y}_2, \dots, \hat{y}_{T_y}\}$.
The goal is to generate an answer $\hat{Y}$ that is grammatically correct and consistent with product attributes and opinions in the reviews.

\begin{figure}[!t]
	\centering
	\includegraphics[width = 0.9\columnwidth]{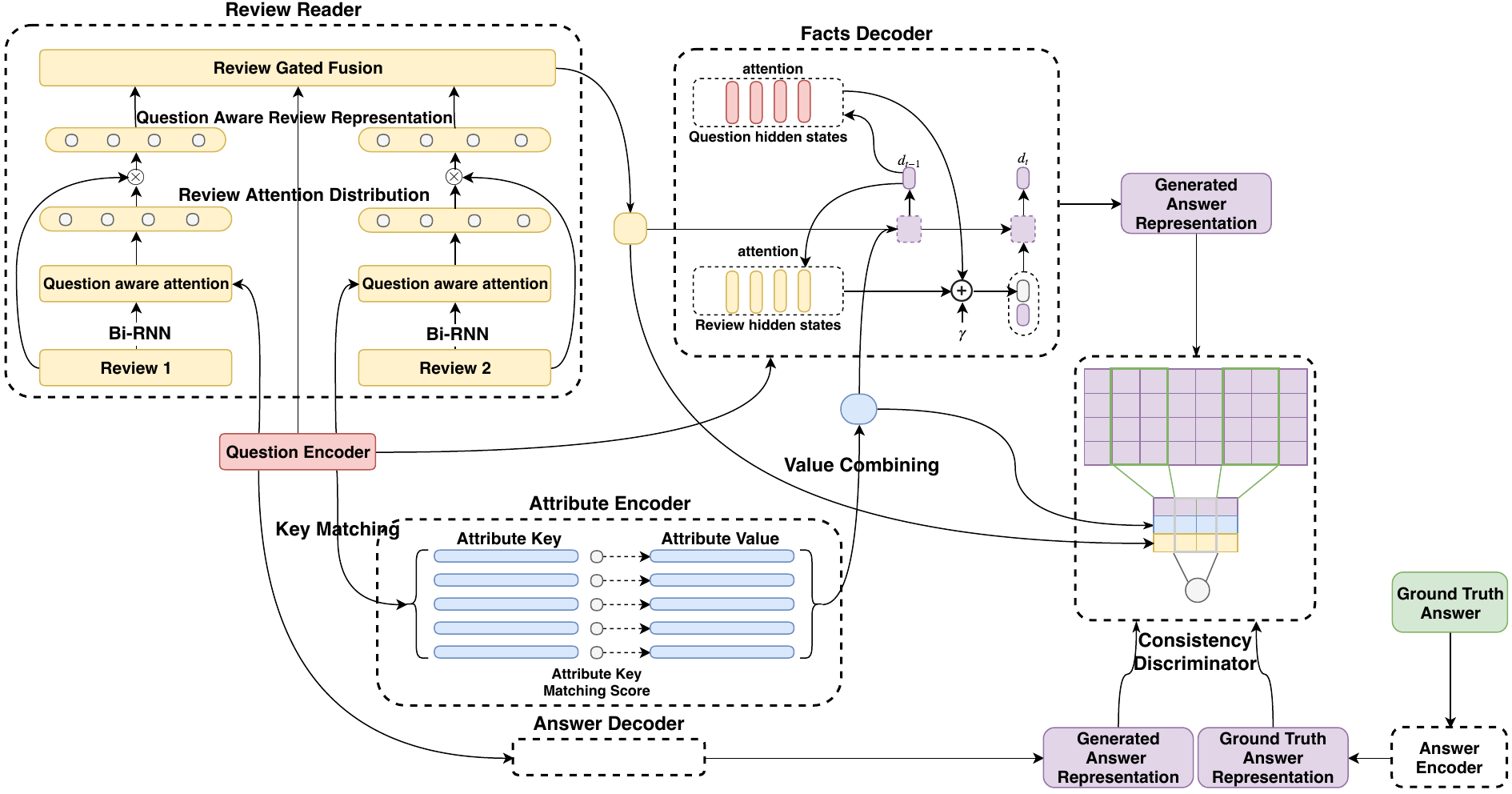}
	\caption{Overview of the product-aware answer generator model. Image source: \citep{gao2019product}.}
	\label{ch63:fig:framegao}
\end{figure}

Fig.~\ref{ch63:fig:framegao} provides an overview of the product-aware answer generator, the PAAG model, proposed by \citet{gao2019product}. PAAG has four parts: 
\begin{enumerate*}[label=(\roman*)]
\item a \textit{review reader} reads the review to extract relevant semantic parts;
\item an \textit{attribute encoder} encodes the attribute key-value pairs using a key-value memory network;
\item a \textit{facts decoder} generates the final answer according to the facts learned by the two modules introduced before; and 
\item a \textit{consistency discriminator} distinguishes whether the generated answer matches the extracted facts, and we also use the result of the discriminator as another training signal. 
\end{enumerate*}
A generative e-commerce QA dataset extracted from JD.com is released with the publication.
Similarly, \citet{chen2019driven} formulate a noise-tolerant solution based on convolutional neural networks to generate natural answers.
\citet{deng2020opinion} exploited opinion information reflected in the reviews. The authors generated opinion-aware natural answers using multi-task learning to integrate opinion detection and answer generation simultaneously.

\begin{figure}[t]
\centering
\includegraphics[width=0.85\textwidth]{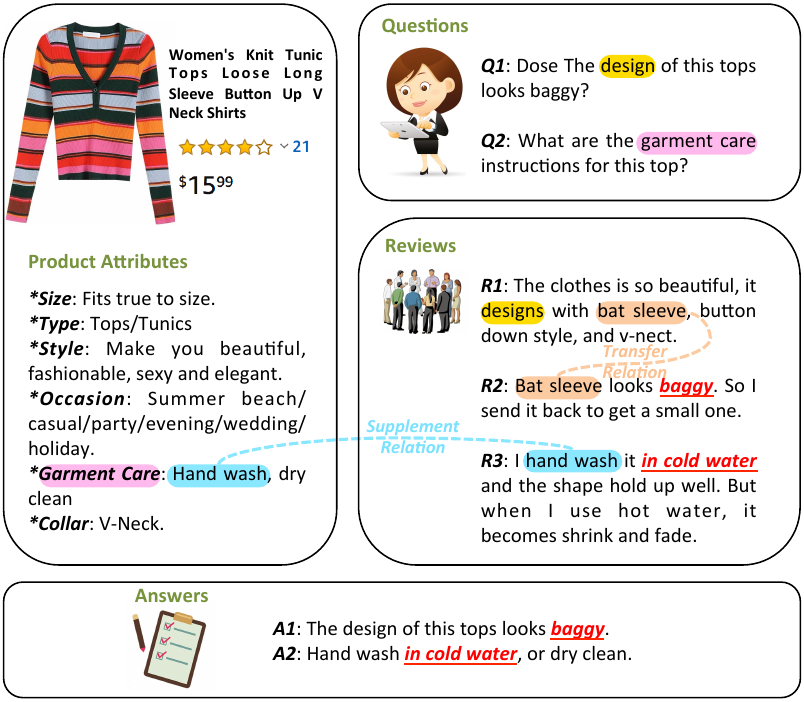}
\caption{Examples of the multi-type text relation for product-aware question answering. Image source: \citep{feng2021multi}.} 
\label{fig:64fengexample}
\end{figure}

It is necessary to consider the text information from different reviews and attributes to answer specific questions in the wild. 
In Fig.~\ref{fig:64fengexample}, \citet{feng2021multi} provide examples to demonstrate the multi-type text relation for product-aware question answering.
As an example, \textit{Q1} asks \textit{``Does the design of this top look baggy?''} \textit{R1} and \textit{R2} do not answer this question directly. But they provide a common entity \textit{``bat-like sleeve.''} If we transfer the information provided by \textit{R1} and \textit{R2} to answer \textit{Q1} indirectly, it is easy to generate the answer that \textit{``The design of this tops looks baggy.''} 
By integrating, understanding, and reasoning over the information of reviews and product attributes we may generate more accurate and pleasing answers to complex questions.
A major limitation of most generative QA approaches is that they analyze each review and the corresponding attribute of the product individually, i.e., they neglect the relationship between different reviews/attributes of the product. 
\citet{feng2021multi} propose a review-attribute heterogeneous graph neural network, RAHGNN, for product-aware answer generation to sufficiently understand and reason about the related information and its inner logic in multiple types of text.
Most generative product-aware QA methods neglect personalization as it is insufficient to provide the same ``completely summarized'' answer to all customers.
As an exception, \citet{deng2022toward} describe a personalized answer generation method, PAGE, to model multi-perspective user preferences in personalized product question answering.

\section{Dialogue systems in e-commerce}
\label{ch6:sec2}

Dialogue systems have increasingly attracted attention in e-commerce.
This section introduces studies on dialogue systems that can be applied to e-commerce platforms.
Following previous work investigating this problem~\citep{Chen2017A,ziwwsdm1}, we divide this section into three parts. 
We introduce recent studies on dialogue systems in Section~\ref{ch6:sec3:intro}, then detail task-oriented dialogue systems in e-commerce in Section~\ref{ch6:sec3:tds}, and discuss knowledge-grounded conversational agents in Section~\ref{ch6:sec3:kgc}.

\subsection{Introduction to dialogue systems}
\label{ch6:sec3:intro}

Dialogue systems are being considered in numerous applications, from e-commerce technical support to personal assistant tools~\citep{song2017summarizing,Chen2017A,ziwwsdm1,sun2016conversational,zhang2018towards,lei2018sequicity,liu2018knowledge,meng2020dukenet,sun2021conversations,shen2021vida,zhao2021jddc,liu2021conversational,ren2022variational,li2022knowledge,yu2022xdai}. 
The goal of creating an automatic human-computer conversational system as an assistant or chat companion is no longer an illusion now that two important factors have been seen progress.
First, many conversation logs are now accessible, making it possible for machines to learn how to respond to input utterances. 
Second, deep generative neural network models, such as sequence-to-sequence and generative adversarial networks, are now able to capture complex patterns in large volumes of data~\citep{Chen2017A}.
Based on these two factors, studies on dialogue systems focuses on methods to provide a natural and coherent response given an utterance from a user~\citep{Young2013POMDP, Ritter2011Data, Banchs2013IRIS, Ameixa2014Luke}.

Dialogue systems can be divided into chitchat systems, task-oriented dialogue systems, and knowledge-grounded conversations~\citep{Chen2017A}.
Chitchat agents are applied widely in open-domain dialogue systems, where dialogue systems interact with humans to provide reasonable and natural responses for open-domain dialogues~\citep{ziwwsdm1,yan2017building}.
Chitchat messages usually represent user experiences and preferences, playing an essential role in many real-world applications. 
\citet{yan2017building} reveal that most utterances in the online shopping scenario are chitchat messages.

\begin{figure}[t]
\centering
\includegraphics[width=0.8\textwidth]{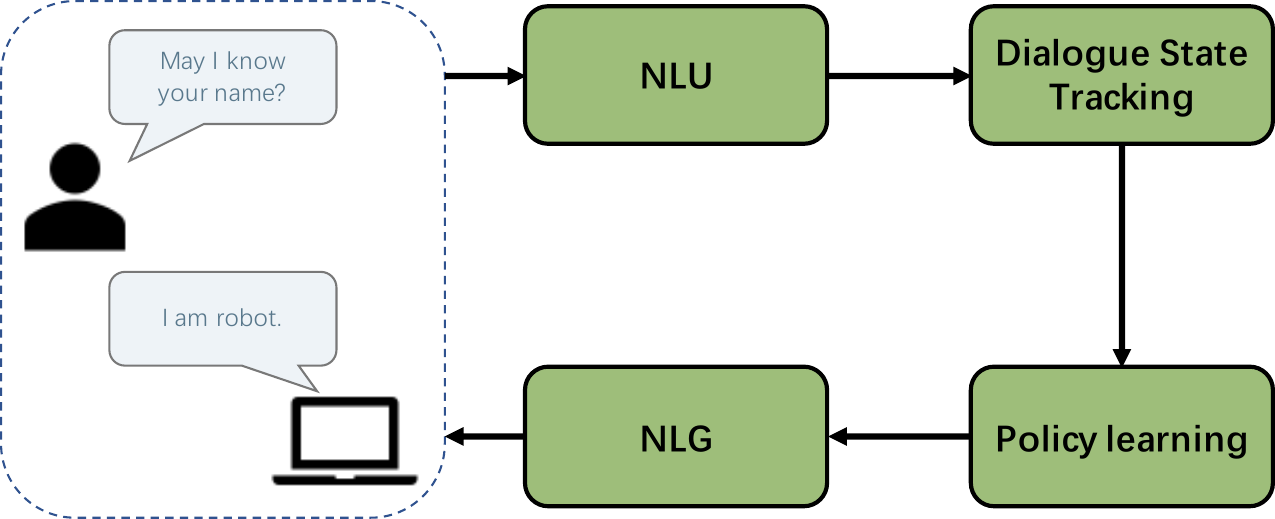}
\caption{Traditional pipeline for task-oriented dialogue systems. Image source:~\citep{Chen2017A}.} 
\label{fig:64framework5}
\end{figure}

Task-oriented dialogue systems aim to complete a specific task, e.g., restaurant reservation,  along with a response generation process.
Fig.~\ref{fig:64framework5} shows the four individual modules on which traditional task-oriented dialogue systems are based: natural language understanding, dialogue state tracking, policy learning, and natural language generation~\citep{wen2017latent,mrkvsic2015multi}.
Given an utterance from a user, the system generates a proper response to address the user's intention. In recent years, end-to-end task-oriented dialogue generation methods have been proposed to address the overall purpose more efficiently~\citep{wen2017network,lei2018sequicity,wu2019transferable}.

Knowledge-grounded conversations focus on generating a response with the correct knowledge to address the user's utterance~\citep{meng2020dukenet}.
Work on knowledge-grounded conversations can be categorized into two groups. 
Methods in the first group use \textit{structured knowledge}~(given knowledge graphs) ~\citep{zhou2018commonsense,liu2018knowledge,tuan2019dykgchat, wu2019proactive, moon2019opendialkg,wu2020diverse, zhou-etal-2020-kdconv, wang2020improving, wu2020improving, xu2020knowledge, xu2020conversational, jung2020attnio,xu-etal-2021-discovering}.
Those in the second group use \textit{unstructured knowledge}, such as \textit{document-based unstructured knowledge}~(given a whole document, e.g., a Wikipedia article)~\citep{chuanmeng2020refnet, ma2020survey, ma2020compare, tian-etal-2020-response, ren2019thinking,gopalakrishnan2019topical, zekangli2018incremental, qin-etal-2019-conversing, moghe2018towards, zhou2018dataset} or \textit{piece-based unstructured knowledge}~(given some separate pieces of knowledge, e.g., Foursquare tips)~\citep{ghazvininejad2018knowledge, dinan2018wizard,meng2020dukenet, kim2020sequential, Lian2019Learning,zheng2019enhancing,zheng2020difference, chen2020bridging, zheng2020approximation, zhao2020knowledge,yu2022xdai}.

With respect to generating responses, dialogue systems can be divided into retrieval-based and generation-based dialogue systems. 
The former retrieves several response candidates from a prebuilt index and then selects an appropriate one as a response. In contrast, the latter directly synthesizes a reply via natural language generation techniques~\citep{hvred,tao2021survey}.

Retrieval-based dialogue systems retrieve several response candidates from a prebuilt index and then select an appropriate one as a response.
Social networks have accumulated a significant amount of conversational data among humans on the web, motivating researchers to investigate data-driven approaches to re-use human conversations and select a response for new input from candidates~\citep{tao2021survey,xu2021response}.
Retrieval-based dialogue generation methods outperform their generation-based counterparts in response fluency and informativeness. 
They power a series of real-world applications, e.g., XiaoIce from Microsoft~\citep{zhou2020design}.
Learning to rank and matching approaches have been widely applied in the retrieval process~\citep{yan2016learning,wu2017sequential,zhou2018multi,yang2018response,yuan2019multi,su2020dialogue,tao2021survey,lin2022task}.
A core task in retrieval-based dialogue systems is response selection.
Studies into retrieval-based response selection can be divided into three types: representation-based, interaction-based, and pre-trained language model-based methods. 
Representation-based methods are composed of a representation-matching paradigm and consist of a representation layer and a matching layer~\citep{yan2016learning,wu2018response,wang2017deep,zhou2018dataset,zhou2018multi,yan2018response,xu2021topic}.
Interaction-based methods use context-response interactions to match potential responses~\citep{tao2021survey}.
These methods follow a representation-matching-aggregation paradigm, formulating an interaction function to calculate the interaction between the two representation matrices of input utterances.
The interaction function has two main types of definition: similarity-based and attention-based  methods~\citep{tao2021survey}.
Similarity-based methods calculate the similarity of each word pair between the context message and the response candidate~\citep{wu2017sequential,zhang2018modeling,zhou2018multi}.
Attention-based methods, however, use an attention mechanism to match the context message and the candidate's response~\citep{chen2019sequential,humeau2019poly,yuan2019multi}.
In recent years, pre-trained language models have been applied in retrieval-based dialogue systems due to their strong ability for language representation and understanding.
These approaches employ an attention-based strategy to unify the representation, interaction, and aggregation operations by feeding the concatenation of context utterances and the candidate responses into a pre-trained multi-layer self-attention network~\citep{whang2020effective,gu2020speaker,xu2021learning,han2021fine,tao2021pre,feng2022reciprocal,li2022unsupervised}. 

Generation-based dialogue systems generate natural-sounding replies automatically to exchange information (e.g., knowledge, sentiments, etc.) and complete a variety of specific tasks in a conversational interaction process~\citep{Young2013POMDP,shawar2007}. 
End-to-end textual generation models~\citep{shang,vinyals2015neural,sordoni2015,li2016a,li2016b,serban2016building} have proved capable in multiple dialogue systems applications with promising performance.
Most end-to-end neural generation models apply an encoder-decoder architecture based on a recurrent neural network, which directly maps an input context to the output response.
Several approaches have been proposed to softly model language patterns, such as word alignment and repeating into sequence-to-sequence structure~\citep{bahdanau2015neural,gu2016incorporating,serban2017multiresolution,cao2017}.
\citet{gu2016incorporating} propose a copy mechanism to consider additional copying probabilities for contextual words in forum conversations. 
\citet{serban2017multiresolution} decode coarse tokens before generating the complete response.
Variational neural networks perform efficient inference and learning in models with directed probabilities on a large-scale dataset~\citep{Kingma2014Auto, Kingma2014Adam}. 
\citet{cao2017} tackle the boring output issue of deterministic dialogue models by introducing a latent variable model for a one-shot dialogue response.
\citet{hvred} propose HVRED to use the latent variable at the sub-sequence level in a hierarchical setting, whereas \citet{ziwwsdm1} add a hierarchical structure and a variational memory module into a neural encoder-decoder network.
\begin{figure}[t]
\centering
\includegraphics[width=0.9\textwidth]{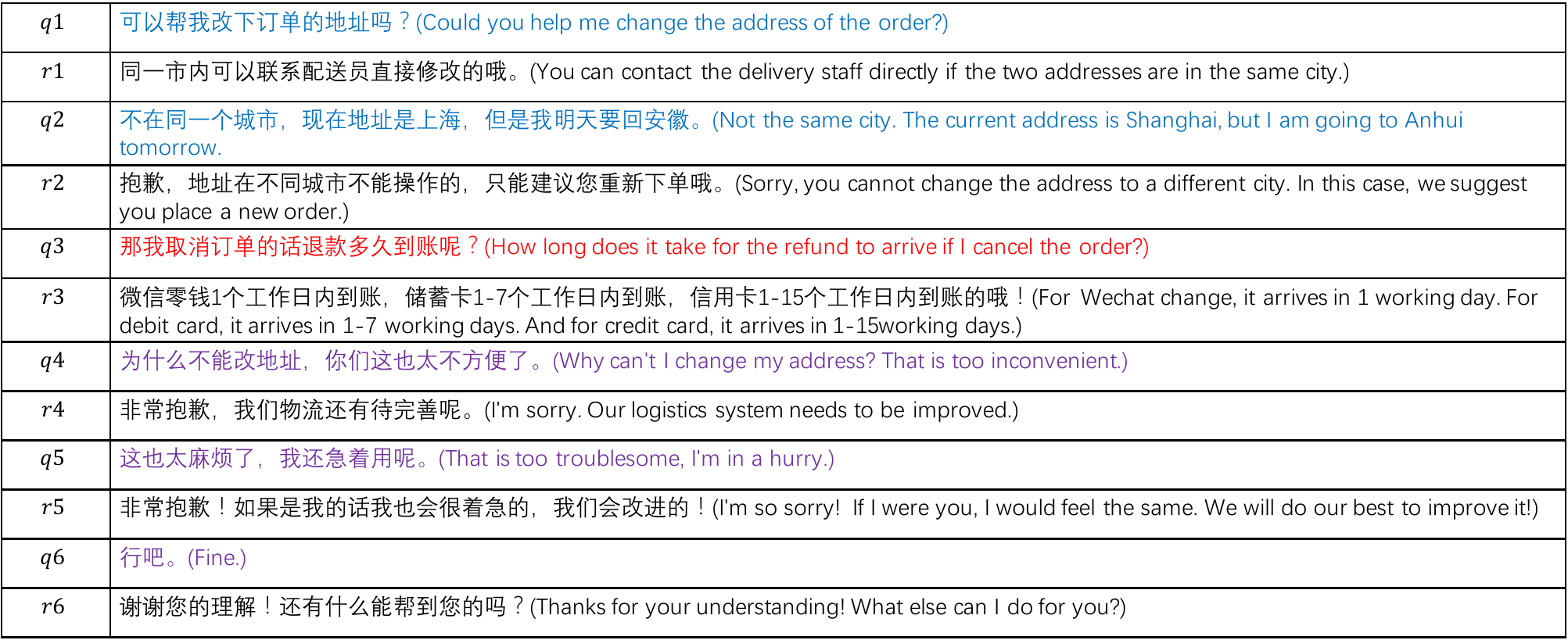}
\caption{An example e-commerce dialogue in the JDDC 1.0 dataset. Image source: \citep{chen2020jddc}.} 
\label{fig:64table}
\end{figure}
In e-commerce platforms, after-sale customer service is the main application scenario for dialogue systems. 
E-commerce dialogues need to address three targets: 
\begin{enumerate*}[label=(\roman*)]
\item task completion, such as changing the order address, providing the receipt, and returning the order; 
\item knowledge-based response selection and generation, such as checking the status of the delivery, answering the request about the refund period; and 
\item empathetic response generation, such as satisfying the consumers' request and replying to consumers' complaints.
\end{enumerate*}
The JDDC datasets have been collected from JD.com, one of the largest e-commerce platforms in China.
Fig.~\ref{fig:64table} shows a typical example dialogue from JDDC 1.0 dataset.
The blue text shows the target completion task, the red text indicates the knowledge-based response generation task, and the purple text shows the empathetic response generation task.
Another characteristic of e-commerce dialogue systems is the phenomenon of multiple modalities.
Text and images are often used in customer service dialogues~\citep{zhao2021jddc,yuan2022mcic}.
\begin{figure}[t]
\centering
\includegraphics[width=0.9\textwidth]{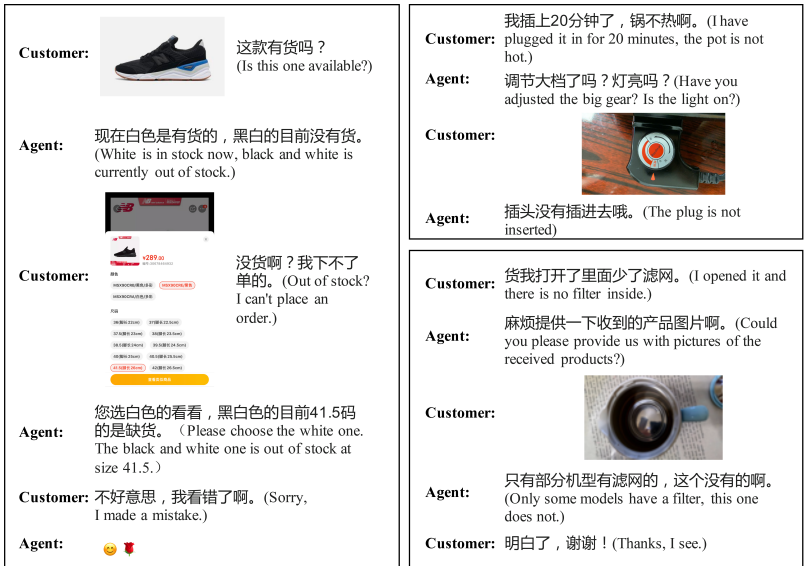}
\caption{Three segments of dialogue sampled from our JDDC 2.0 corpus. Image source: \citep{zhao2021jddc}.} 
\label{fig:64table2}
\end{figure}

Fig.~\ref{fig:64table2} shows another example from the JDDC 2.0 dataset to demonstrate the multi-modality characteristic of e-commerce.
In this figure, three dialogue segments show images that are used to distinguish different product models for the same brand or used for identifying the location and cause of product failures.
According to the above three targets in e-commerce dialogues, we detail e-commerce dialogue systems from three angles: task-oriented dialogue systems, knowledge-grounded dialogue systems, and empathetic dialogue systems in Sections~\ref{ch6:sec3:tds},~\ref{ch6:sec3:kgc}, and~\ref{ch6:sec3:emp}, respectively.

\subsection{Task-oriented dialogue systems}
\label{ch6:sec3:tds}

Methods underlying task-oriented dialogue systems can be divided into pipeline methods and end-to-end methods.
As shown in Fig.~\ref{fig:64framework5}, the pipeline of task-oriented dialogue systems can be divided into natural language understanding (NLU), dialogue state tracking (DST), and dialogue policy learning (DPL), and natural language generation (NLG)~\citep{Chen2017A}.
For each stage, a number of pipeline-based approaches have been proposed, even though a lot of domain-specific handcrafting in traditional task-oriented dialogue systems is required, which, moreover, is difficult to adapt to new domains~\citep{bordes2016learning}.
Recently, end-to-end neural network solutions have been widely applied to the task~\citep{bordes2016learning,zhao2016towards,wen2017network,jin2018explicit,gao2018neural,madotto2018mem2seq}.

In NLU, the dialogue system maps the input utterance into semantic slots. These semantic slots are pre-defined in different application scenarios.
NLU includes two challenging problems: intent detection and slot filling.
Intent detection methods for language understanding are performed to detect the user's intent~\citep{deng2012use,tur2012towards}.
Deep neural networks have also been applied to detect the user's intent.
\citet{huang2013learning} use convolutional neural networks (CNN) to detect the user's intent; see also~\citep{shen2014learning}.
Slot filling is usually set as a sequence labeling problem, where words are assigned with semantic labels~\citep{Chen2017A}.
Deep belief networks have been successfully applied to address the filling problem~\citep{deng2012use,deoras2013deep}.
Subsequently, recurrent neural networks have been shown to be effective in slot filling~\citep{mesnil2013investigation,sarikaya2011deep,yao2013recurrent,yao2014spoken}.
Unlike other NLU approaches,~\citet{liang2020moss} jointly formulate intent detection and slot filling as a sequence generation problem.
\citet{rastogi2020towards} provide a schema-guided paradigm for NLU.
Recently, pre-trained language models have been applied to enhance NLU in task-oriented dialogue systems.
\citet{wu2020tod} propose a self-supervised language model trained on multiple task-oriented dialogue system benchmarks. \citet{zhang2021effectiveness} design a pre-trained model for few-shot NLU by fine-tuning BERT on a small set of labeled utterances.
\citet{10.1145/3477495.3532069} explore tree-induced semi-supervised contrastive pre-training for NLU in task-oriented dialogue systems. 
The authors improve the NLU performance by injecting structural-semantic information to enhance the representation of dialogues.
To explore more knowledge from long sequences in dialogue context, \citet{zhong2022dialoglm} formulate a window-based pre-trained model for NLU based on the sequence-to-sequence model architecture.

In a conversation, a \emph{dialogue state} refers to a full and temporal representation of each participant's intention~\citep{goddeau1996form}. 
In task-oriented dialogue generation, dynamically tracking dialogue states is the key to generating coherent and context-sensitive responses.
In DST, we use a dialogue state $H_{t}$ to denote the representation of the dialogue till time $t$. 
Traditional approaches to DST focus on searching hand-crafted rules to select the most likely results~\citep{wang2013simple,young2010hidden,williams2012belief}, where the dialogue state tracking is transferred to a slot filling problem.
This type of slot filling problem has also been addressed by approaches using conditional random fields~\citep{lee2013structured,lee2013recipe,ren2013dialog} and maximum entropy~\citep{williams2013multi}.
However, relying on the most likely results from an NLU module~\citep{Perez17}, these rule-based systems hardly model uncertainty, which is prone to frequent errors~\citep{williams2014web, Perez17}.

Unlike rule-based state tracking methods, \citet{young2010hidden} propose a distributional dialogue state for statistical dialog systems and maintain a distribution over multiple hypotheses facing noisy conditions and ambiguity.
Neural networks have been successfully applied to dialogue state tracking~\citep{henderson2013deep,mrkvsic2015multi,mrkvsic2017neural}.
In task-oriented dialogue systems, end-to-end neural networks are employed for tracking dialogue states via interacting with an external knowledge base~\cite{wen2017network,eric2017key,bordes2016learning,williamszweig2017Long}.
\citet{wen2017network} divide the training procedure into two phases: dialogue state tracker training and complete model training. 
\citet{mrkvsic2017neural} propose a dialogue state tracker based on word embedding similarities.
\citet{eric2017key} implicitly model a dialogue state through an attention-based retrieval mechanism to reason over a key-value representation of the underlying knowledge base. 
\citet{bordes2016learning} track the dialogue context in a memory module and repeatedly search this context to select an adequate system response.
Instead of employing symbolic knowledge queries, \citet{dhingra-EtAl:2017:Long1} propose an induced ``soft'' posterior distribution over the knowledge base to search for matching entities. 
\citet{williamszweig2017Long} combine an RNN with domain-specific knowledge encoded as software and system action templates. 
The copying mechanism is shown to be effective as generative dialogue state tracking.
\citet{lei2018sequicity} propose an extendable framework to track dialogue states with a text span, including the constraints for a knowledge base query.
The limited amount of labeled data is a severe challenge for DST.
\citet{jin2018explicit} introduce a semi-supervised way to integrate a copy procedure with the dialogue state tracking.

While early studies on DST methods focused on a single-domain scenario, more recent studies have turned their attention to multi-domain DST with the release of a multi-domain DST benchmark dataset, MultiWoZ~\citep{budzianowski2018multiwoz}. 
\citet{ramadan2018large} jointly identify the domain and tracks the belief states corresponding to that domain to address the multi-domain DST problem.
\citet{zhou2019multi} formulate multi-domain DST as a question-answering task and used
reading comprehension techniques to generate the answers.
Similarly, \citet{gao2019dialog} also formulate DST as a reading comprehension task and propose an attention-based neural network to find the state answer as a span over tokens.
The DSTC challenges have provided a series of popular experimentation frameworks and dialogue datasets collected through human-machine interactions for benchmarking~\citep{henderson2014second,henderson2014third,williams2014dialog,williams2016dialog}. 

In real-world scenarios, it is often not practical to enumerate all possible slot value pairs and
perform scoring from a large, dynamically changing knowledge base~\citep{xu2018end}. 
\citet{wu2019transferable} propose a method to generate dialogue states from utterances using a copy mechanism, where tracking knowledge across domains is shared.
To alleviate data sparsity in DST, \citet{yin2020dialog} propose a reinforced data augmentation framework to increase both the amount and diversity of the training data.
\citet{chen2020schema} incorporate slot relations and model slot interactions in multi-domain dialogue state tracking to enhance the slot interrelation between disciplines. 
\citet{feng2022dynamic} extend this method by dynamically updating slot relations in the schema graph.
\citet{heck2020trippy} maintain two memories in DST: one for system inform slots and one for the previously seen slots.
\citet{li-etal-2021-generation} combine a generation and extraction method with hierarchical ontology integration for DST.
To tackle the understanding of ellipsis and reference expressions in open vocabulary-based methods, \citet{ouyang2020dialogue} propose a copy-augmented encoder-decoder model by connecting the target slot and its source slot explicitly.
\citet{liao2021multi} formulate multi-domain DST as a recursive inference mechanism to improve the generation performance.
Most DST models are trained offline, which requires a fixed dataset prepared in advance. 
Given a new domain in multi-domain DST, \citet{campagna2020zero} propose a zero-shot transfer learning method to handle new domains without incurring the high cost of data acquisition.
More recently, the granularity of dialogue history has been proposed to mitigate the sparseness in DST~\citep{yang2021comprehensive}.
\citet{guo2022beyond} propose a multi-perspective dialogue collaborative selector module to dynamically select the granularity of dialogue history in DST.

Pre-trained language models have been shown to be effective in dialogue state tracking.
\citet{lin2021zero} apply a pre-trained model for DST to exploit external knowledge from reading comprehension data.
Similarly, \citet{zhong2022dialoglm} verify that document summarization can provide helpful signals to improve DST. 
\citet{liu2021domain} introduce domain-lifelong learning into DST. The authors propose a knowledge preservation network that includes a multi-prototype enhanced retrospection component and a multi-strategy knowledge distillation component.
\citet{lin2021leveraging} successfully apply T5 to improve zero-shot cross-domain DST.
\citet{lee2021dialogue} introduce a solution for multi-domain DST by prompting knowledge from a large-scale pre-trained language model. 
\citet{lin2021knowledge} detail a hybrid method to integrate GPT-2 with graph attention networks to enhance the DST performance. 
To mitigate the problem of incorrect in DST, \citet{wang2022luna} design a BERT-based method by explicitly aligning each slot with its most relevant utterance.

Scalability, robustness, and efficiency in DST have also been addressed recently.
\citet{lei2018sequicity} formulate a two-stage copy-aware network demonstrating good scalability.
\citet{ren2019scalable} consider the DST task as a sequence generation problem and design a scalable hierarchical encoder-decoder neural network with constant inference time complexity. 
\citet{kumar2020ma} extend this to improve the encoding of dialogue context and slot semantics for DST to robustly capture critical dependencies between slots and the conversation history.
\citet{kim2020efficient} focus on an open vocabulary-based setting and consider the dialogue state as a memory that can be selectively overwritten to improve the efficiency in multi-domain DST.
\citet{zhu2020efficient} introduce an efficient multi-domain dialogue state tracker by jointly encoding the previous dialogue state, the current turn dialogue, and the schema graph by
internal and external attention mechanisms. 

The policy learning module in task-oriented dialogue systems is meant to generate the following available system action given the state generation result~\citep{cuayahuitl2015strategic}.
Traditional rule-based methods are first applied in the policy learning procedure~\citep{cuayahuitl2015strategic}.
Supervised and reinforcement learning have also proven to be effective in policy learning~\citep{su2016continuously,yan2017building}.
\citet{su2016continuously} propose a two-stage framework for policy learning, i.e., a supervised learning stage and a reinforcement learning stage.
\citet{chen2019agentgraph} propose a structured deep reinforcement learning approach for policy learning based on graph neural networks. 
The dialogue policy can be further trained in an end-to-end way with reinforcement learning to lead the system in making policies toward the final performance~\citep{yan2017building,chen2019agentgraph}.
In an e-commerce scenario, the policy learning component needs to trigger the ``recommendation'' or a concrete service provided by the customer service~\citep{sun2016conversational,sun2018conversational,zhao2021jddc}.
Most task-oriented dialogue datasets, including WoZ and MultiWoZ, focus on language understanding and dialogue state tracking. 
However, selecting actions in real life requires obeying user requests and following practical policy limitations.
Accordingly, \citet{chen2021action} present an action-based conversations dataset consisting of $10042$ conversations containing numerous actions with precise procedural requirements.
\citet{he2022galaxy} utilized semi-supervised pre-training to model explicit dialogue policy in task-oriented dialogue systems.

The natural language generation (NLG) component transfers a dialogue action into a natural language response~\citep{Chen2017A}. 
Neural network-based NLG approaches have been proposed for task-oriented dialogues~\citep{wen2015stochastic,wen2015semantically,tran2017semantic,zhou2016context}.
\citet{wen2015stochastic} apply an RNN-based generator module and a CNN-based module to rerank candidate utterances.
\citet{wen2015semantically} use an additional control cell to gate the dialogue act to address the slot information omitting and duplicating problems in surface realization.
\citet{tran2017semantic} extend this approach by gating the input token vector of an LSTM with the dialogue act.
A sequence-to-sequence approach is applied to produce natural language output and deep syntax dependency trees from input dialogue acts~\citep{duvsek2016sequence}.
\citet{zhou2016context} propose an encoder-decoder LSTM-based method to jointly incorporate the request information, semantic slot values, and dialogue act type to generate correct answers.
The copy mechanism~\citep{vinyals2015pointer,gu2016incorporating} has been successfully applied to the task-oriented dialogue systems to enhance the performance of NLG~\citep{eric2017copy,lei2018sequicity,jin2018explicit}.
Aiming to augment dialogue datasets through paraphrasing, \citet{gao2020paraphrase} jointly optimize dialogue paraphrasing and dialogue response generation via a paraphrase augmented response generation approach.
Pre-trained language models have shown supreme performance in text generation tasks~\citep{li2021pretrained}.
In recent years, more and more studies have applied pre-trained language models to enhance the performance of NLG in task-oriented dialogue systems~\citep{zhang2019dialogpt,peng-etal-2020-shot,wu2020tod,NEURIPS2020e9462095,zhong2022dialoglm}. 

End-to-end methods have been proposed for task-oriented dialogue systems.
\citet{wen2017network} propose an end-to-end trainable goal-oriented dialogue system with a new way of collecting dialogue data based on a pipeline framework toward end-to-end learning for DST and policy learning~\citep{zhao2016towards}.
The pipeline-aware method can also be implemented and trained end-to-end using the copy mechanism~\citep{lei2018sequicity,jin2018explicit,liao2021multi}.
A copy-augmented sequence-to-sequence architecture has been proposed to provide better performance in task-oriented dialogues~\citep{eric2017copy}, while~\citet{eric2017key} propose a key-value retrieval network for task-oriented dialogue response generation.
Using the copy mechanism, \citet{lei2018sequicity} formulate a theoretical framework that is end-to-end trainable using only one sequence-to-sequence model.
\citet{jin2018explicit} propose a semi-supervised copy flow neural network to train the end-to-end dialogue generation model.
\citet{madotto2018mem2seq} formulate a memory-to-sequence neural network that combines the multi-hop attention over memories with the idea of a pointer network.
\citet{xu2018end} also apply a pointer network to handle unknown slot values in the absence of a predefined ontology.
\citet{NEURIPS2020e9462095} enable modeling of the inherent dependencies between the sub-tasks of task-oriented dialogue by optimizing for all tasks in an end-to-end manner, recasting task-oriented dialogues as a simple and casual language modeling task.
\citet{liao2021multi} propose a recursive inference mechanism to resolve multi-domain DST in an end-to-end way. 
More recently, pre-trained language models have also been applied to end-to-end solutions for task-oriented dialogue systems~\citep{wu2020tod,lin2021knowledge,10.1145/3477495.3532069}.

In contrast to other types of dialogue systems, evaluation metrics in task-oriented dialogue systems need to consider specific metrics.
\emph{Entity match rate} evaluates task completion~\citep{wen2017network}; it determines if a system can generate all correct constraints to search the indicated entities of the user. This metric is either 0 or 1 for each dialogue.
The original \emph{success rate} metric measures if the system answered all the requested
information (e.g., address, phone number)~\citep{wen2015stochastic,mrkvsic2015multi}. However, this metric only evaluates recall. 
As a variant, \emph{Success F1} evaluates task completion and is modified from the success rate by balancing both recall and precision~\citep{lei2018sequicity}.
Automatic user satisfaction has received much attention in task-oriented dialogues.
User simulation is a promising approach to evaluate dialogue systems at scale in task-oriented dialogue scenarios~\citep{zhang2020evaluating}.
\citet{sunzhang2021simulator} formulate the task of simulating user satisfaction for evaluating task-oriented dialogue systems to enhance the evaluation of dialogue systems. The authors also share a dataset about user satisfaction simulation.
\citet{kim2022mismatch} propose the relative slot accuracy metric in DST evaluation, which is not affected by unseen slots in the current dialogue turn.

\subsection{Knowledge-grounded dialogue systems}
\label{ch6:sec3:kgc}

Although answering inquiries is essential for dialogue systems, especially for task-oriented dialogue systems, it is still far behind a natural knowledge-grounded dialogue system, which should be able to understand the facts involved in the current dialogue session (so-called fact matching) and diffuse them to other similar entities for knowledge-based dialogues  (i.e., entity diffusion): 
\begin{enumerate}[leftmargin=*,nosep,label=(\arabic*)]
\item \textit{Fact matching}: In dialogue systems, matching utterances to exact facts is much harder than answering explicit factoid inquiries. 
Though some utterances, whose subjects and relations can be easily recognized, are fact-related inquiries, the subjects and relations are often elusive, leading to challenges when matching exact facts.
Table \ref{tab6:intro4} shows an example, with items 1 and 2 talking about the film ``Titanic.'' Unlike item 1, which is a typical question-answering conversation, item 2 is a knowledge-related chit-chat without any explicit relation. It is difficult to define the exact fact match for item 2.
\item \textit{Entity diffusion}: Conversations usually drift from one entity to another. In Table \ref{tab6:intro4}, the utterances in items 3 and 4 are about the entity ``Titanic.'' However, the entities in the responses are other similar films. 
Current knowledge triplets rarely capture such entity diffusion relations.
The response in item 3 shows that the two entities, ``Titanic'' and ``Waterloo Bridge,'' are relevant through ``love stories.'' Item 4 suggests another similar shipwreck film ``Titanic.''
\end{enumerate}

\begin{table}[t!]
\centering
\begin{tabular}{ l  l }
\toprule
\bf ID & \bf Dialogue \\ 
\midrule
\multirow{4}{*}{1} & A: Who is the director of the \underline{Titanic}? \\
 & \begin{CJK}{UTF8}{gkai}\underline{泰坦尼克号}的导演是谁？\end{CJK} \\
 & B: \underline{James Cameron}. \\
 & \begin{CJK}{UTF8}{gkai}\underline{詹姆斯卡梅隆}。\end{CJK} \\ \hline
\multirow{4}{*}{2} & A: \underline{Titanic} is my favorite film! \\ 
 & \begin{CJK}{UTF8}{gkai}\underline{泰坦尼克号}是我最爱的电影！\end{CJK} \\
 & B: The love inside it is so touching. \\ 
 & \begin{CJK}{UTF8}{gkai}里面的爱情太感人了。\end{CJK} \\ \hline
\multirow{5}{*}{3} & A: Is there anything like the \underline{Titanic}? \\ 
 & \begin{CJK}{UTF8}{gkai}有什么像\underline{泰坦尼克号}一样的电影吗？\end{CJK} \\
 & B: I think the love story in film \underline{Waterloo} \\
 & \underline{Bridge} is beautiful, too. \\ 
 & \begin{CJK}{UTF8}{gkai}我觉得\underline{魂断蓝桥}中的爱情故事也很美。\end{CJK} \\ \hline
\multirow{4}{*}{4} & A: Is there anything like the \underline{Titanic}? \\ 
 & \begin{CJK}{UTF8}{gkai}有什么像\underline{泰坦尼克号}一样的电影吗？\end{CJK} \\
 & B: \underline{Poseidon} is also a classic marine film. \\ 
 & \begin{CJK}{UTF8}{gkai}\underline{海神号}也是一部经典的海难电影。\end{CJK} \\ 
\bottomrule
\end{tabular}
\caption{\label{example-table} Examples of knowledge grounded conversations. Knowledge entities are underlined. Image source: \citep{liu2018knowledge}.}
\label{tab6:intro4}
\end{table}

\noindent Knowledge-grounded dialogue systems address the aforementioned challenges.
Work on knowledge-grounded dialogue systems can be categorized into two groups.
Methods in the first group use \textit{structured knowledge}~(given knowledge graphs)~\citep{wu2020diverse, zhou-etal-2020-kdconv, wang2020improving, wu2020improving, xu2020knowledge, xu2020conversational, jung2020attnio, tuan2019dykgchat, wu2019proactive, moon2019opendialkg, zhou2018commonsense, liu2018knowledge}.
Methods in the second group focus on using \textit{unstructured knowledge}, such as \textit{document-based unstructured knowledge}~(given a whole document, e.g., a Wikipedia article)~\citep{chuanmeng2020refnet, ma2020survey, ma2020compare, tian-etal-2020-response, ren2019thinking,gopalakrishnan2019topical, zekangli2018incremental, qin-etal-2019-conversing, moghe2018towards, zhou2018dataset, parthasarathi2018extending} or \textit{piece-based unstructured knowledge}~(given some separate pieces of knowledge, e.g., Foursquare tips)~\citep{ghazvininejad2018knowledge, dinan2018wizard, meng2020dukenet, kim2020sequential, Lian2019Learning, zheng2020difference, chen2020bridging, zheng2020approximation, zhao2020knowledge, zheng2019enhancing, lin-etal-2020-generating}.
There are key research directions for both groups:
\begin{enumerate*}[label=(\roman*)]
\item improving knowledge selection~\citep{kim2020sequential};
\item improving knowledge-aware response generation~\citep{zhao2020knowledge} or response selection~\citep{young2018augmenting, zhao2019document, hua2020learning, sun2020history};
that is, given the chosen knowledge, how to better generate a response token by token or select a response from pre-defined response candidates;
\item using multiple knowledge modalities~\citep{liu2019knowledge, zhao2020multiple};
that is, how to use structured, unstructured, and even other types of knowledge simultaneously;
\item overcoming data scarcity~\citep{zhao2019low, li2020zero}.
\end{enumerate*}

The neural knowledge diffusion model introduces knowledge into the dialogue generation. This method can match the relevant facts for the input utterance and diffuse them to similar entities~\citep{liu2018knowledge}. 
Early studies on knowledge selection in knowledge-grounded dialogue systems calculate the weight of each piece of knowledge and obtain a weighted sum of their representations~\citep{ghazvininejad2018knowledge,zheng2019enhancing,lin-etal-2020-generating,zheng2020approximation}.
\citet{bordes2016learning} employ memory networks to address restaurant reservations, using a small number of keywords to handle entity types in a knowledge base (cuisine type, location, price range, party size, rating, phone number, and address).
\citet{ghazvininejad2018knowledge} adapt it to memorize relevant, grounded facts for a neural conversation model.
The hierarchical variational memory network (HVMN) adds  hierarchical structure and a variational memory network into a neural encoder-decoder network for non-task-oriented dialogue generation~\citep{ziwwsdm1}.

Several recent studies on knowledge selection focus on calculating a weight on each piece of knowledge and then directly sampling the amount of knowledge with the highest weight.
Specifically, \citet{dinan2018wizard} design the TMemNet model that uses context to predict a distribution over pieces of knowledge and then only sample one of them into a decoder.
They also introduce a knowledge selection loss to supervise knowledge selection during training.
\citet{Lian2019Learning} describe PostKS, which uses a context to predict a prior distribution over pieces of knowledge.
During training, the prior distribution is supervised by a posterior distribution predicted by the context and the corresponding response.
Similar to PostKS, \citet{zheng2020approximation} use a context and a piece of knowledge retrieved by the context to predict a distribution over fragments of knowledge, where the probability of the amount of knowledge retrieved by the corresponding response is maximized during training.
The former distills a context containing multiple utterances at different turns into a vector that is used to match with the representation of a piece of knowledge to get a score, while the latter matches every utterance in a context with a piece of knowledge to get matching features that are aggregated to get a score.
A piece of knowledge is chosen based on the score list for all pieces of knowledge.
\citet{kim2020sequential} propose a sequential knowledge transformer (SKT), which jointly uses previously selected knowledge and context to facilitate knowledge selection.
\citet{chen2020bridging} upgrade SKT by adding the knowledge distillation-based training strategy to improve knowledge selection.
\citet{zheng2020difference} detail a method that introduces the difference information between the previously selected knowledge and the current pieces of candidate knowledge to facilitate knowledge selection.
\citet{meng2020dukenet} design DukeNet, which regards tracking the previously selected knowledge and selecting the current knowledge as dual tasks within a dual learning paradigm~\citep{qin2020dual}.
\citet{zhao2020knowledge} describe a method, RLKS, where the selected knowledge is sent to a decoder to generate a response that would be compared with the ground truth response to give feedback to further supervised knowledge selection.
\citet{meng2021initiative} introduce a mixed-initiative knowledge selection method for knowledge-grounded conversations that explicitly distinguishes between user-initiative and system-initiative knowledge selection at each conversation turn to improve the performance of knowledge selection.
\citet{sun2021conversations} find that the amount of knowledge available in different languages is highly unbalanced. 
Hence, the authors address cross-lingual knowledge grounded conversations with a self-distillation knowledge selection and curriculum learning.

More recent years have witnessed the rapid development of pre-trained language models in open-domain dialogue systems.
Large pre-trained language models can store knowledge into their parameters during pre-training and can generate informative responses in conversations~\citep{Zhao2020ArePL}.
\citet{Petroni2019LanguageMA} have shown that pre-trained language models can serve as knowledge bases for downstream tasks (e.g., question-answering~\citep{Roberts2020HowMK}). 
On this basis, \citet{Zhao2020ArePL} have shown that pre-trained language models can ground open-domain dialogues using their implicit knowledge.
\citet{madotto2020learning} embed knowledge bases into model's parameters for end-to-end task-oriented dialogues.
\citet{Roller2021RecipesFB} fine-tune pre-trained language models on knowledge-grounded conversational data.
\citet{Cui2021KnowledgeEF} describe knowledge-enhanced fine-tuning methods to handle unseen entities.
\citet{Xu2021RetrievalFreeKD} propose a topic-aware adapter to adapt pre-trained language models in knowledge-grounded dialogues.
\citet{Liu2022MultiStagePF} introduce a multi-stage prompting approach for triggering knowledge in pre-trained language models.
\citet{Wu2022LexicalKI} design lexical knowledge internalization to integrate token-level knowledge into the model’s parameters.
The problem of hallucination is becoming more and more challenging.
\citet{sun2023contrastive} optimize an implicit knowledge eliciting process, i.e., they reduce hallucination of pre-trained language models in knowledge-grounded dialogues through a contrastive learning framework.

\subsection{Empathetic dialogue generation}
\label{ch6:sec3:emp}


Several approaches to data-driven open-domain dialogue generation generate emotional responses based on a manually specified label to control the dynamic content of the target output~\citep{zhou2018emotional,lis18, zhou2018mojitalk, huang2018automatic,wei2019emotion,colombo2019affect,shenf20}. 

Unlike emotional dialogue generation, the study of empathetic dialogue generation avoids an additional step of determining which emotion type to respond to explicitly~\citep{skowron2013affect}.
Several studies~\citep{Rashkin18, zhong2019affect, Shin19, rashkin2019towards, santhanam19, linmsxf19, lin2020caire, zhong2020, MajumderHPLGGMP20, li2020empdg} have attempted to make dialogue models more empathetic.
\citet{rashkin2019towards} combine models in different ways to produce empathetic responses.
\citet{linmsxf19} softly combine the possible emotional responses from several separate experts. 
\citet{MajumderHPLGGMP20} consider polarity-based emotion clusters and emotional mimicry.
\citet{li2020empdg} propose a multi-resolution adversarial framework that considers multi-granularity emotion factors and users' feedback.
\citet{li2020empathetic} investigate how to use external knowledge to explicitly improve the emotional understanding and expression in the task of empathetic dialogue generation.
\citet{sabour2022cem} focus on two aspects in empathetic dialogue generation: affection and cognition. 
The authors propose a method with various
commonsense reasoning to improve understanding of interlocutors' situations and feelings.

Besides advances in empathetic dialogue models, the emergence of new emotion-labeled dialogue corpora has also contributed to this research field~\citep{li2017dailydialog,hsuckhk18,rashkin2019towards}.
\citeauthor{rashkin2019towards} consider a rich and well-balanced set of emotions and release a dataset, \textsc{EmpatheticDialogues}, where a listener responds to a speaker in an emotional situation in an empathetic way.

\section{Emerging directions}
\label{ch6:sec4}

This section describes recent emerging directions on question-answering and dialogue systems in e-commerce. These emerging directions in QA and dialogue systems can be divided into five perspectives: safety, ethics, interpretability, privacy, and evaluation.

As discussed in Section~\ref{ch6:sec1}, e-commerce QA agents aim to answer questions based on large volumes of reviews.
However, reviews may not answer these questions as they may not contain any relevant answers for the question, or a query may be poorly phrased and therefore require additional clarification.
Moreover, untruthful comments and spam are widely observed in e-commerce reviews~\citep{carmel2018product}.
\citet{mihaylova2019semeval} investigate the fact-checking problem in a QA scenario with a system to classify the veracity of answers.
\citet{zhang2020answerfact} release a large-scale fact-checking dataset called AnswerFact for investigating the answer veracity in e-commerce QA. 
\citet{estes2022fact} develop a high-speed fact-verification system that has a very high false statement recall and very high true statement precision to product question-answering.
However, the authors still apply a rule-based method in their system, and find that pre-trained language models are unable to perform fact-checking well on structured catalog data.
As limitations such as poor generalization exist in rule-based methods, how to optimize pre-trained language models in fact-checking for product question-answering still needs more attention in future research. 

Ethical challenges in dialogue systems are attracting significant amounts of attention in recent years. Currently, most dialogue systems are developed from scratch with large corpora or fine-tuned through pre-trained language models. Large-scale datasets collected from the open internet have been applied during model training. However, offensive and malevolent content can be observed in the data~\citep{si2022so}. 
To avoid being unintentionally offensive or harming the user, studies have been performed to detect toxic speech around, e.g., religion, race, and violence~\citep{tripathi2019detecting,dinan2020queens,zhang2021human,kann2022open,si2022so}. 
\citet{zhang2021human} propose a human-machine collaborative evaluation framework for reliable toxic speech detection in dialogue systems.
\citet{si2022so} study toxic speech in open-domain dialogue systems to reveal that specific kinds of ``non-toxic'' queries are able to trigger an open-domain conversational assistant to output toxic responses. 
However, how to respond when these malevolence topics are being identified is still an open question~\citep{kann2022open}.
As more and more e-commerce dialogue systems have also been trained based on pre-trained language models, ethical challenges will need to be tackled in future research.

Poor explainability is a challenging problem for most e-commerce QA approaches.
Most e-commerce QA approaches apply end-to-end semantic matching methodologies, which tend to be black-box and directly output a matching score for each question answer pair. 
\citet{zhao2019riker} address the explainable QA problem through a hybrid retrieval-based framework.
The authors employ a bidirectional recurrent neural network in the internal word representation stage and apply a keyword-aware retrieval method during the second stage. In contrast, the tf-idf ranking function naturally exhibits much better interpretability owing to its transparency and intuitiveness.
Conversational recommendation is another typical application of e-commerce dialogue systems~\citep{sun2016conversational,mangili2020bayesian}.
Most approaches neglect explainability when learning recommendation actions. However, \citet{chen2020towards} propose an incremental multi-task learning framework using user feedback for the task of explainable conversational recommendation.
By considering user preferences as latent variables in a variational Bayesian manner, \citet{ren2022variational} employ a method to estimate explicit user preferences during the dialogue.

Privacy protection has received more and more attention in dialogue systems~\citep{papernot2016semi,henderson2018ethical}.
For many task-oriented dialogue systems, it is necessary to notice that we are using the same dialogue assistant. 
Recent studies on membership inference attacks have confirmed that privacy information in training data for sequence-to-sequence generative models and pre-trained language models can be attacked~\citep{hisamoto2020membership,liu2021encodermi}.
As we discussed in Section~\ref{ch6:sec2}, most e-commerce dialogue system models are designed based on sequence-to-sequence generative neural networks and pre-trained language models.  
By learning through interactions and communications, a dialogue assistant can inadvertently and implicitly store sensitive information.
Hence, consumers' privacy information may get obtained by attackers through membership inference attacks.
To address this problem, developing privacy-aware dialogue systems is likely to attract increased attention in the future.

The evaluation of e-commerce dialogue systems is a crucial part of the development process.
Recent studies on evaluating dialogue systems are either through offline evaluation or human evaluation~\citep{Lamel2000TheLA,Jurccek2011RealUE}.
Offline evaluation is often limited to single-turn assessments, while human evaluation is intrusive, time-intensive, and does not scale~\citep{Deriu2020SurveyOE}.
User simulators have been applied to exhaustively enumerate user goals to generate human-like conversations for simulated evaluation~\citep{Zhang2020RecentAA}.
However, user simulators as evaluation methods for e-commerce dialogue systems are still under-explored.
Today's simulators suffer from limited realism and evaluation capabilities~\citep{Balog2021Sim4IRTS}.
Moreover, evaluation metrics that specifically target e-commerce aspects are still underexplored and underexploited in today's e-commerce dialogue agents. 
Dedicated evaluation metrics in e-commerce search and recommendation scenarios will likely help to advance progress.


\chapter{Conclusion and outlook}
\label{chapter:conclusion}

We summarize the main topics presented in this survey in Section~\ref{sec:summarysurvey}. 
In Section~\ref{sec:concludingremarks}, we describe our outlook on future developments.

\section{Conclusion}
\label{sec:summarysurvey}

From the large number of user engagements on e-commerce platforms a large amount of information can be inferred.
The aim of this survey has been to give a broad overview of information discovery in e-commerce portals. 
Our overview has included methods about user behavior modeling, search, recommendation, question answering, and dialogue systems in e-commerce. 

Our strategy with this survey has been to provide a broad coverage of research directions about information discovery in e-commerce. 
Although we have tried our best to provide all key approaches in each direction as much as possible, the amount of technical details is limited. 
For areas that are broad enough to have their own survey, we have only focused on key publications and provide structure and pipelines for each direction.
Additionally, we only focus on areas that are relevant to information retrieval research; studies in other areas relevant to e-commerce, such as supply chains and computational advertising, are ignored in this survey.

In our introduction, we summarized the outline and topics covered in this survey, followed by a description of basic concepts and key definitions in Section~\ref{chapter:terms}.
Then we introduced preliminaries about e-commerce interfaces and users in Section~\ref{chapter:basic}. We formulated concepts of e-commerce infrastructures and summarized studies about information seeking via e-commerce interfaces.
We investigated e-commerce information components, i.e., titles, product descriptions, and reviews, and detailed characteristics of consumer behaviors in e-commerce portals.
We introduced studies into e-commerce user analyses concerning multiple behaviors, including clicks, purchases, engagements, and post-clicks, on e-commerce platforms.

The core of this survey is organized around five directions: e-commerce user modeling, e-commerce search, e-commerce recommendation, e-commerce QA, and e-commerce dialogue systems. We have detailed each of these in four sections (i.e., Section~\ref{chapter:user},~\ref{chapter:search},~\ref{chp:rec}, and~\ref{chapter:qa}).
Each section starts with an overview of the main direction discussed in the section, with characteristics and subtasks.
After that, key research studies of each subtasks were demonstrated with some level of detail. 
We discussed emerging research directions at the end of each key component.

In particular, in Section~\ref{chapter:user}, we introduced approaches to user modeling and profiling for e-commerce applications. 
We divided this section into two main components: user behavior modeling and user profiling.
We provided a summary of studies on modeling these e-commerce user behavior,  analyzed research on user profiling in e-commerce, and discussed emerging directions on user modeling in e-commerce.

In Section~\ref{chapter:search}, we focused on search technologies in e-commerce platforms. 
We provided the characteristics of e-commerce search and divided research studies based on matching strategies and ranking technologies for e-commerce search scenarios, respectively.
We presented approaches aiming for studying matching strategies for e-commerce search. 
And we  studied research approaches on ranking technologies for e-commerce search. 
Emerging research directions were discussed at the end of the section.

In Section~\ref{chp:rec}, we introduced the most prominent approaches to e-commerce recommendation methods.
We summarized the key characteristics of e-com\-merce recommendation, towards which a two-stage framework was developed that contains candidate retrieval and candidate ranking, forming the mainstream solution for e-commerce recommender systems. We reviewed models developed for the two stages and detailed mainstream learning methods for optimizing model parameters to provide a complete view of e-commerce recommender systems.

Section~\ref{chapter:qa} detailed research methods for question answering and dialogue systems in e-commerce. 
We addressed question answering and dialogue systems in a single section as most research background and approaches are shared between these two directions.
We reviewed previous work on question answering and then demonstrated the characteristics of e-commerce QA.
For e-commerce question answering (QA), we described studies both on extractive QA and generative QA.
For e-commerce dialogue systems, we demonstrated the patterns of e-commerce dialogue systems, especially about task-oriented dialogue systems, knowledge-grounded conversations, and empathetic dialogue systems.
We discussed emerging research directions around QA and dialogue agents in Section~\ref{ch6:sec4}.

\section{Outlook}
\label{sec:concludingremarks}

Information discovery is increasingly a mixed initiative scenario, where users and e-commerce platforms take turns. 
As described in the previous sections, the research presented on information discovery in e-commerce has been addressed from six angles: infrastructures, user modeling, search, recommendation, QA, and dialogue systems. 
As we summarized at the end of each section, a broad variety of emerging research has also been motivated following each angle. 
For user modeling, we consider three research topics as key emerging directions: graph learning for user behavior modeling, dynamic user behavior modeling and profiling, and multi-modal user profiling.
For emerging directions in e-commerce search, we focus on applications for multi-modal e-commerce search and ranking. 
Online learning to rank technologies also provide key insights. We also foresee the development of new learning theories that will improve e-commerce search and ranking performance in the future.
We divide emerging directions on e-commerce recommendation into three directions: 
\begin{enumerate*}[label=(\roman*)]
\item reasoning, recommendation, and explanations; 
\item conversational recommendation; and 
\item unifying recommendation and search.
\end{enumerate*}
In our view, future work on e-commerce language processing should include generating explainable reasons for search and recommendation, improving the robustness of e-commerce question answering, and improving conversational e-commerce search and recommendation.

\subsection{Four directions}

Among these emerging research approaches, we have identified potential directions of future work that are encountered across multiple angles.
In particular, we list future research directions in four bigger themes: conversational search, conversational recommendation, multi-modal information discovery, and generative information discovery.

\emph{Conversational search} refers to a novel search paradigm using multiple interactions between users and search engines. As we have discussed in Section~\ref{chapter:search}, conversational search is increasingly receiving more  attention in the IR community.
Different from the traditional query-aware search paradigm, conversational search allows users to express their information need by directly conducting conversations with search engines.
More recent studies have begun to apply conversational search to online shopping scenarios as it is able to provide a natural, adaptive and interactive shopping experience for consumers~\citep{xiao2021end}. In e-commerce, conversational search faces two  challenges: imperfect product attribute schemas and product knowledge.
The former exists as product attributes link lengthy multi-turn utterances with products in conversational search systems, whereas the latter derives from the lack of manually labelling in benchmark datasets.
Core tasks in conversational search, e.g., search intent detection, action prediction, query selection, passage selection, and response generation~\citep{ren2021wizard}, also provide  insights in future work about e-commerce conversational search systems.

\emph{Conversational recommendation} refers to recommendation systems that can elicit the dynamic preferences of users and take actions based on their current needs through real-time multi-turn interactions.
As we have discussed in Section~\ref{chp:rec}, conversational recommendation is an emerging direction in e-commerce recommendation.
Integrating more accurate domain-specific knowledge to promote the recommendation and conversation is a challenging problem in conversational recommendation~\citep{chen2020towards}.
Moreover, as with conversational search, current studies on conversational recommendation suffer from a lack of manually labelled data in benchmark datasets.
Recent studies on empathetic dialogue systems reveal that there exist some kind of dependency between commonsense knowledge and emotional preference~\citep{li2020empathetic,siro-2022-understanding}. 
Hence, conversational recommender systems that jointly combine emotion detection and knowledge exploration are worth studying in future.

\emph{Multi-modal information} can be widely observed in many e-commerce scenarios, e.g., user behavior, search, recommendation, and dialogue systems. 
Most previous e-commerce information discovery approaches are constructed only based on text understanding and retrieval; addressing multi-modal information, e.g., images and videos, still appears to be difficult.
In future work, it is more and more important to tackle challenges about multi-modality in e-commerce scenarios. 
Multimedia technologies focusing on integrating various types of modalities are expected to help to understand those multi-modal information for various types of e-commerce applications.
It is interesting to explore multi-modal generation through powerful generative deep neural networks in e-commerce review generation, question answering, and dialogue systems.

\emph{Large-scale generative models} have the potential to significantly enhance various e-commerce information discovery applications, such as search, recommendation, and conversational AI. Transformer-based pre-trained language models like BERT have already proven effective in both search and recommendation tasks in e-commerce. More recently, large language models (LLMs) based on auto-regressive mechanisms, such as T5 and GPT, have demonstrated promising capabilities in understanding and generating human-like information, making them valuable for e-commerce contexts.
Moreover, while traditional two-stage paradigms (i.e., retrieval followed by re-ranking) have been widely used in e-commerce search and recommendation scenarios. They face two limitations: 
\begin{enumerate*}[label=(\roman*)]
\item heterogeneous modules with different optimization objectives may lead to sub-optimal performance; and
\item a large document index is needed which may come with substantial memory and computational requirements.
\end{enumerate*}
This has motivated research into end-to-end solutions using generative models. Recent studies on generative retrieval, such as DSI~\citep{tay2022transformer}, have shown encouraging performance on several information retrieval benchmarks, suggesting that exploring generative models for end-to-end e-commerce search and recommendation could be a promising direction for future research.

\subsection{Beyond accuracy}
Besides the directions for future work listed above, we also consider the following important issues when it comes to information discovery tasks in e-commerce: \emph{fairness}, \emph{trustworthiness}, and \emph{explainability}~\citep{roegiest-2019-facts-ir,derijke-2023-beyond}.

Recently, the problem of bias has attracted considerable attention in the IR community, in multiple contexts, e.g., for user behavior modeling, profiling, ranking, and recommendation. 
To address the bias problem, \emph{fairness} is considered as a significant metric during the optimization procedure.
Early on, fairness was studied from the perspective of information exposure regarding sensitive attributes such as gender and race~\citep{singh2018fairness}.
Fairness in IR also focuses on how to let different items receive equal exposure, or exposure proportional to their utility or impact, depending on which exposure distribution is considered to be fair by the system~\citep{morik2020controlling,chen2020bias}.
In e-commerce, ranking-based interfaces are quite common in various scenarios; hence, fairness is a matter of great importance to information discovery in e-commerce.
Future work on interactive fairness-aware reranking can be helpful for debiasing user modeling, search, recommendation, and answer generation in e-commerce platforms~\citep{sarvi-2022-understanding}.
Also, knowledge-based and dynamic fairness-aware methods are able to address more real-world challenges.
The trade-off between accuracy and fairness is of importance in e-commerce search and recommendation scenarios, where equally treating different groups has been shown to sacrifice the performance~\citep{ariannezhad-2023-personalized}.
To address this problem, an important research direction is to understand the dimensions of causality and design fairness-aware algorithms.

Fake news and fake information are is increasingly widespread. 
It is now viewed as one of the greatest threats on the web~\citep{zhou2020survey}.
As we have discussed in Section~\ref{chapter:qa}, spam and fake reviews and answers are widely observed in many e-commerce platforms.
Therefore, pursuing \emph{trustworthiness} has become an important issue in e-commerce question answering and dialogue systems.
Several studies distinguish spam or fake reviews in online review systems via graph neural networks~\citep{kaghazgaran2018combating,rao2020xfraud,liu2020alleviating,dou2020enhancing}.
Textual generation methods based on deep neural networks have been applied to fake reviews or spam generation by camouflaged fraudsters. Thus, distinguishing the authenticity of e-commerce information is a challenging task.
Other patterns in the reviews, e.g., sentiments and emotions, can be applied to improve the detection.
Also, investigating inconsistency problems under multiple domains provides new avenues of research.

\emph{Explainability} in e-commerce aims to answer the question about why we receive a specific ranking, recommendation, or answer result. The task of explainability can be divided into explainability of the learning models and explainability of the results. 
The former aims to provide more transparent learning details for the proposed methods, whereas the latter focuses on provide more explainable results to various application scenarios, e.g., search, recommendation, and question answering, etc.
Explainability methods have been shown to be effective for enhancing the e-commerce search and recommendation~\citep{zhang2018towards,liu2020keywords}.
For future work, it is important to evaluate if and how users and other stakeholders are satisfied with the explanations generated from an e-commerce system, especially as these are increasingly conversational in nature~\citep{lucic-2021-multistakeholder}.
Generating coherent, faithful, and naturally-sounding explanations based on a sequence of reasoning steps (including search or recommendation system output) is still difficult.

As we have shown in this survey, the number of research studies in the area of information discovery for e-commerce is increasing rapidly.
We believe that this is only the beginning. 
The recent launch of a dedicated product search track at TREC seems to confirm this.\footnote{\url{https://trec-product-search.github.io}}
The volume of the work described in this survey and the steady pace of publications in the field, together with the arrival of open research challenges indicate a promising future ahead.
A lot remains to be done.

\begin{acknowledgements}
This survey grew out of a tutorial ``Information Discovery in e-Commerce'' taught at SIGIR 2018. 
We thank the audience for their feedback and questions.

We also thank our colleagues 
Mozhdeh Ariannezhad,
Hongshen Chen, 
Jiawei Chen,
Zhumin Chen, 
Songgaojun Deng,
Zhuoye Ding,
Yue Feng,
Stefan Grafberger,
Paul Groth,
Yulong Gu,
Shuyu Guo, 
Ziyi Guo, 
Maria Heuss,
Mariya Hendriksen,
Na Huang,
Sami Jullien,
Barrie Kersbergen,
Jiahuan Lei,
Dongdong Li,
Ming Li,
Xiang Li,
Xinyi Li, 
Zhenyang Li,
Xiaozhong Liu,
Hengliang Luo,
Si Luo, 
Yougang Lyu,
Jun Ma, 
Yao Ma, 
Pengjie Ren, 
Emma de Rijke,
Fatemeh Sarvi,
Sebastian Schelter,
Xinlei Shi,
Clemencia Siro,
Hongye Song,
Olivier Sprangers,
Changlong Sun,
Fei Sun,
Weiwei Sun,
Jiliang Tang, 
Bart Voorn,
Jingang Wang,
Shuaiqiang Wang, 
Xuepeng Wang,
Zihan Wang, 
Long Xia,
Xin Xin, 
Zhen Zhang, 
Jiashu Zhao, 
Xiangyu Zhao, 
and Yihong Zhao
for help, feedback, and inspiration.

We thank our editors Yiqun Liu and Mark Sanderson for support, patience, and valuable feedback.

This research was supported by
Alibaba DAMO Aca\-demy,
Baidu.com,
JD.com,
and
Meituan, 
as well as
AIRLab, a collaboration between Ahold Delhaize and the University of Amsterdam, 
the Hybrid Intelligence Center, a 10-year program funded by the Dutch Ministry of Education, Culture and Science through the Netherlands Organisation for Scientific Research, \url{https://hybrid-intelligence-centre.nl}, project nr.\ 024.004.022,
project LESSEN with project number NWA.1389.20.\-183 of the research program NWA ORC 2020/21, which is (partly) financed by the Dutch Research Council (NWO),
project ROBUST with project number KICH3.LTP.20.006, which is (partly) financed by the Dutch Research Council (NWO) and the Dutch Ministry of Economic Affairs and Climate Policy (EZK) under the program LTP KIC 2020-2023,
and
the FINDHR (Fairness and Intersectional Non-Discrimination in Human Recommendation) project that received funding from the European Union's Horizon Europe research and innovation program under grant agreement No 101070212.

All content represents the opinion of the authors, which is not necessarily shared or endorsed by their respective employers and/or sponsors.
\end{acknowledgements}

\appendix

\chapter{Datasets}
\label{chap:dataset}

In this appendix, we list benchmark datasets that are relevant for studying information discovery in e-commerce. 
We follow the topical organization of our sections, and divide the datasets into five types: e-commerce infrastructures, e-commerce user modeling, e-commerce search, e-commerce recommendation, and e-commerce QA \& dialogues.

\section{Datasets for e-commerce infrastructures}

To begin, we list benchmark datasets about e-commerce interfaces and users:

\begin{itemize}[leftmargin=*,nosep]
    \item \textbf{Taobao short title dataset}~\citep{sun2018multi}: This dataset contains 411,246 title-product pairs in 94 categories. Each item in the dataset is represented as a triple $\langle Q, K, S\rangle$, where $Q$ denotes the products' original titles, $K$ refers to the background knowledge about the products, and $S$ represents the human-written short titles.
    \item \textbf{eCOM-C2C dataset} about product categories and titles~\citep{wang2018multi}: This dataset takes advantage of realistic data from a well-known C2C website in China.
    The dataset contains 185,386 triplets in the Women’s Clothes category. Each item in the dataset is represented as a triple $\langle S, T, Q\rangle$, where $S$ refers to a product’s original title, $T$ denotes a handcrafted short title, and $Q$ is a successful transaction-leading search queries.
    \item \textbf{Walmart product summarization dataset}~\citep{mukherjee2020discriminative}: The dataset includes 40,445 top-selling Walmart grocery products during the calendar year 2018, together with their product titles and corresponding human-generated summaries. There are also descriptions, brand names, and category information of the products.
    \item \textbf{Taobao multi-modal title dataset}~\citep{miao2020multi}: The dataset contains 114,278 original titles with corresponding short titles and product images. The short titles are manually written by professional editors, whereas the images are selected by the seller.
    \item \textbf{Walmart e-commerce product dataset}~\citep{mukherjee2021unsupervised}: The dataset contains five parts: D-search includes the top 12 million product search queries on Walmart.com and their frequencies over a one year period. 
    D-product includes 250,000 top-selling Walmart products over a six month period. 
    D-com-human includes 40,445 human-generated title compressions from the Walmart catalog across eight different product categories. 
    D-meta-auto contains 40,000 meta-training examples. And D-meta-human is a dataset consisting of 16,000 human-generated 1-shot title compression examples.
    \item \textbf{LESD4EC dataset}~\citep{gong2019automatic}: The dataset consists of 6,481,623 pairs of original and short product titles in a module in Taobao named ``Youhua\-shuo.'' Each product in this dataset includes a long product title and a short title summary written by professional writers, along with a high-quality image and attributes tags.
\end{itemize}

Table~\ref{app:tb1} summarizes the key statistics of the datasets listed above.

\renewcommand\arraystretch{1.5}
\begin{sidewaystable}[!htbp]
\caption{Statistics of datasets about e-commerce infrastructures.} 
\centering 
\resizebox{18cm}{!}{
\begin{tabular}{l rr cc r l}  
\toprule 
\multirow{3}{*}{\textbf{Datasets}} 
& \multicolumn{5}{c}{\textbf{Statistics}} 
& \multirow{3}{*}{\textbf{References}}\\
\cmidrule(lr){2-6}
& \#Dataset size     
& \makecell[c]{\#Number \\of category }
& \makecell[c]{\#Avg.length \\of original titles}  
& \makecell[c]{\#Avg.length \\of short titles}   
& \makecell[c]{\#Avg.length \\of background knowledge}
\\ 
\midrule 
Taobao short title dataset
& 453,138   & 94  & 25.34 & 7.73  & 5.92
& \citep{sun2018multi}
\\
eCOM-C2C dataset
& 185,386   & 1  & 25.1 & 7.5  & 8.3
& \citep{wang2018multi} 
\\
Walmart product summarization dataset
& 40,445   &   & 4/10/35 & 1/2/5  &
& \citep{mukherjee2020discriminative} 
\\
Taobao multi-modal title dataset
& 114,278   &   &  &   & 
& \citep{miao2020multi}
\\  
Walmart e-commerce product dataset
& 40,000 + 16,000   & 4  &  &   & 
& \citep{mukherjee2021unsupervised}  
\\
LESD4EC dataset
& 6,481,623   &  & 12 & 5  &
& \citep{gong2019automatic}  
\\

\bottomrule 
\end{tabular}
}
\label{app:tb1}
\end{sidewaystable}

\section{Datasets for e-commerce user modeling}

Next, we list benchmark datasets about e-commerce user modeling:

\begin{itemize}[leftmargin=*,nosep]
    \item \textbf{Taobao Tianchi consumer dataset}~\citep{kim2021deep}: The dataset includes responses of users to advertisements of inventory in the user profile and advertising information. The time length of the data is eight days, and the dataset is divided into four tables: advertisement features, user profiles, past shopping behavior that users engaged in, and who received the advertisement with responses.\footnote{\url{https://tianchi.aliyun.com/dataset/dataDetail?dataId=56}}
    \item \textbf{Instacart.MB dataset}~\citep{sheng2021htda}: The Instacart Market Basket (Instacart.MB) dataset is anonymized and contains a sample of over 3 million grocery orders from more than 200,000 Instacart users. 
    For each user in the dataset, there are between 4 and 100 of their orders, with the sequence of products purchased in each order.\footnote{\url{https://www.kaggle.com/c/instacart-market-basket-analysis/data}}
    \item \textbf{Bing advertising service dataset}~\citep{lian2021multi}: The dataset contains user click logs within a two week period from the Bing Native Advertising service. It also includes users' online behavior history before their corresponding clicks. 
    The user behavior sequences are truncated to 100 in the dataset.
    \item \textbf{Feeds user dataset}~\citep{yi2021debiasedrec}: The feeds dataset is collected on Microsoft News App from August 1, 2020 to September 1, 2020. It contains 643,177 news items, over 10,000 users, 320,925 impressions, and 970,846 clicks.
    \item \textbf{JD user profiling dataset}~\citep{chen2019semi}: This dataset is collected from one of the largest e-commerce platforms in China.
    In this dataset, users, items, and attributes reflect real-world e-commerce consumers, products, and words in the titles of the products respectively. 
    The profiles of users are the age and gender labels. 
    \item \textbf{Twitter user behavior dataset}~\citep{al2012homophily}: Each attribute dataset consists of approximately 400 labeled Twitter users, 200 with one label (e.g., ``female'') and 200 with a second label (e.g., ``male''). In addition, all of the friends of these labeled users are identified; for each of these labeled and neighbor users, the most recent 1,000 tweets generated by the user were collected.
    \item \textbf{UCL social media user profiling dataset}~\citep{liang2017inferring}: This dataset was collected by UCL's Big Data Institute. The data set includes 1,375 active Twitter users chosen randomly and their tweets from the time they registered until May 31, 2015. The dataset has 3.78 million tweets in total. The length of a tweet is 12 words on average.
    \item \textbf{CALL dataset}~\citep{dong2014inferring}: The dataset is extracted from a collection of more than 1 billion (i.e., 1,000,229,603) call and text-message events from an anonymous country, which spans from August 2008 to September 2008. The data does not contain any communication content.
    \item \textbf{W-NUT dataset}~\citep{han2016twitter}: This is a user-level dataset of the geolocation prediction shared task released at the W-NUT workshop in 2016. The dataset consists of over 1 million training users, 10,000 development users, and 10,000 test users. The ground truth location of a user is decided by majority voting of the closest city center. 
    \item \textbf{Facebook user profiling dataset}~\citep{farnadi2018user}: This is a re-collected dataset based on Facebook's MyPersonality project dataset.\footnote{\url{http://www.mypersonality.org}}
    The dataset includes information about each user's demographics, friendship links, Facebook activities (e.g., number of group affiliations, page likes, education, and work history), status updates, profile picture, and Big Five Personality scores (ranging from 1 to 5).
\end{itemize}

Table~\ref{app:tb2} summarizes the key statistics of the datasets listed above.

\renewcommand\arraystretch{1.5}
\begin{sidewaystable}[!htbp]
\caption{Statistics of datasets about e-commerce user modeling.} 
\centering 
\resizebox{18cm}{!}{
\begin{tabular}{l rr rr c l}  
\toprule %
\multirow{2}{*}{\textbf{Datasets}} 
& \multicolumn{5}{c}{\textbf{Statistics}} 
& \multirow{2}{*}{\textbf{References}}\\
\cmidrule(lr){2-6}
& \#Users    &\#Items  &\#Interactions   &\#Avg.seq.len & TimeSpan
\\ 
\midrule %
Taobao Tianchi consumer dataset
& 1,140,000   &   &  & 26,000,000  & 20170506-20170513
& \citep{kim2021deep}  
\\
Instacart.MB dataset
& 11,464   & 42,207  & 7,764,043 &677.25   &
& \citep{sheng2021htda}
\\
Bing advertising service dataset
& 748,000   & 409,000  &  &74   &
& \citep{lian2021multi}
\\
Feeds user dataset
& 10,000   & 643,177  & 970,846 &   & 
& \citep{yi2021debiasedrec}
\\  
JD user profiling dataset
& 54,161   & 203,712  &  &   &
& \citep{chen2019semi}  
\\
Twitter user behavior dataset
& 400   & 400,000  &  & 1,000  & 
& \citep{al2012homophily}  
\\
UCL social media user profiling dataset
& 1,375   & 3,780,000  &  & 12  & time of registration-20150531
& \citep{liang2017inferring}
\\
CALL dataset
& 1,090,000  &  & &   & 200808-200809
& \citep{dong2014inferring}
\\   
W-NUT dataset
& 1,020,000   & 13,000,000  & &   & 
& \citep{han2016twitter}  
\\
Facebook user profiling dataset
& 5,670   & 49,372  &  &   & 
& \citep{farnadi2018user}  
\\

\bottomrule 
\end{tabular}
}
\label{app:tb2}
\end{sidewaystable}

\section{Datasets for e-commerce search}

We list benchmark datasets about e-commerce search as follows:

\begin{itemize}[leftmargin=*,nosep]
    \item \textbf{QUARTS e-commerce search dataset}~\citep{nguyen2020learning}: This is a human-labeled dataset of query-item pairs, obtained from an e-commerce search platform. There are in total 3.2 million pairs of which only a small fraction are mismatches. About 100,000 labeled pairs are used as a separate test set. Another 3 million query-item pairs are deemed ``matched'' by considering items that are purchased frequently in response to those queries from the search logs. 
    \item \textbf{SCEM product search dataset}~\citep{bi2019leverage}: The dataset contains three category-specific datasets, namely, ``Toys \& Games,'' ``Garden \& Outdoor,'' and ``Cell Phones \& Accessories,'' from the logs of a commercial product search engine spanning ten months between years 2017 and 2018. The datasets include up to a few million query sessions containing several hundred thousand unique queries.
    \item \textbf{Walmart product search dataset}~\citep{karmaker2017application}:  This is a subset obtained from Walmart's online product catalog. The dataset consists of more than 2,800 randomly selected product search queries and a catalog of around 5 million products. For each query, the top 120 products are retrieved.
    \item \textbf{Walmart query log dataset}~\citep{magnani2019neural}: This is a large query log dataset on shoe segments during a six-month window from May 2018 to October 2018 on Walmart.com. Historical data of the extra features such as clicks and orders are collected from the query log six months before May 2018. The dataset is composed of more than 100 million query and product pairs, of which there are more than 1 million unique queries and more than 1 million unique item titles.
    \item \textbf{Bestbuy dataset}~\citep{duan2013supporting}: The dataset consists of a full crawl of the ``Laptop \& Netbook Computers'' category of Bestbuy.com. In total, there are 864 laptops in the database, each entity has 44 specifications on average. And 260 laptops have user reviews. The annotated datasets contain 40 queries, on average, there are 2.8 keywords per query and 3.8 keywords per query for the hard queries.
    \item \textbf{Amazon product dataset}~\citep{bi2020learning,mcauley2015image}: The Amazon product dataset is a well-known benchmark for product search and recommendation. It contains information for millions of customers, products and associated metadata, including descriptions, reviews, brands, and categories.\footnote{\url{http://jmcauley.ucsd.edu/data/amazon}}
    \item \textbf{Etsy product search dataset}~\citep{wu2018turning}: The dataset contains 4 weeks worth of search log data with clicks and purchases from Etsy.\footnote{\url{https://www.etsy.com}} In total, there are 334,931 search sessions with 239,928 queries and $6,347,251$ items. In total, 270,239 buyers and 550,025 sellers are involved in the transactions, whereas 631,778 keywords are used by sellers to describe their items.
\end{itemize}

Table~\ref{app:tb3} summarizes the key statistics of the datasets listed above.
\renewcommand\arraystretch{1.5}
\begin{sidewaystable}[!htbp]
\caption{Statistics of datasets about e-commerce search.} 
\centering 
\resizebox{18cm}{!}{
\begin{tabular}{l rrr rr l}  
\toprule 
\multirow{2}{*}{\textbf{Datasets}} 
& \multicolumn{5}{c}{\textbf{Statistics}} 
& \multirow{2}{*}{\textbf{References}}\\
\cmidrule(lr){2-6}
& \#Queries     &\#Products   &\#Pairs & Product title length & Vocabulary size
\\ 
\midrule 
QUARTS e-commerce search dataset
&    &   & 3,200,000 &   & 
& \citep{nguyen2020learning}  
\\ 
\midrule 
SCEM product search dataset
&   &   &  &   & 
& \citep{bi2019leverage} 
\\
SCEM-Toys\&Games
&   &  &  &13.14$\pm$6.46   & 381,620
\\
SCEM-Garden\&Outdoor
&   &  &  &16.39$\pm$7.38   & 1,054,980
\\
SCEM-CellPhones\&Accessories
&   &  &  &22.02$\pm$7.34   & 194,022
\\
\midrule 
Walmart product search dataset
& 2,800   & 5,000,000  &  &   &
& \citep{karmaker2017application} 
\\
Walmart query log dataset
& 1,000,000+   & 1,000,000+  & 100,000,000+ &   & 
& \citep{magnani2019neural} 
\\  
Bestbuy dataset
& 40   & 864  &  &   & 
& \citep{duan2013supporting} 
\\
Amazon product dataset
&    &   &  &   & 
& \citep{bi2020learning,mcauley2015image}  
\\
Etsy product search dataset
& 239,928   & 6,347,251  &  &26.5   &
& \citep{wu2018turning}  
\\

\bottomrule 
\end{tabular}
}
\label{app:tb3}
\end{sidewaystable}

\section{Datasets for e-commerce recommendations}

Next, we list benchmark datasets about e-commerce recommender systems:

\begin{itemize}[leftmargin=*,nosep]
    \item \textbf{Amazon product dataset}~\citep{he2016ups,mcauley2015image}: For e-commerce recommendations, the Amazon product dataset is split by top-level product categories in amazon and is notable for its high sparsity and variability. This dataset contains product reviews and metadata from Amazon, including 142.8 million reviews spanning May 1996--July 2014. This dataset includes reviews (i.e., ratings, text, helpfulness votes), product metadata (i.e., descriptions, category information, price, brand, and image features), and links (i.e., substitutive/complementary relations).
    \item \textbf{Amazon soc dataset}~\citep{mcauley2015image}: A large-scale database of 230,000 users; each data sample includes a user’s profile, user feedback on a product, and social relationship among users. More specifically, the user’s profile includes gender, income, age, and hobby. User feedback includes the user’s comments and browsing history.
    \item \textbf{AliExpress dataset}~\citep{ahmed2021deep}: This dataset is collected from an online retailer service owned by the Alibaba group. There are about 2,260,923 records from AliExpress, the data for about fourteen months from January 1, 2019 to February 23, 2020. The dataset contains 1,506,850 users that submitted reviews against 49,221 items in 205 different categories, such as electronics, entertainment, education, house, and garden, etc., and the items are rated from 1 to 5 scale.
    \item \textbf{Instacart orders dataset}: This is an anonymized dataset collected from the Instacart site.\footnote{\url{https://www.instacart.com}} It contains a sample of over 3 million grocery orders from more than 200,000 Instacart users. For each user, 4 and 100 of his/her orders are provided, with the sequence of products purchased in each order. There are also the week and hour of the day the order was placed and a relative measure of time between orders.\footnote{\url{https://www.instacart.com/datasets/grocery-shopping-2017}}
    \item \textbf{Movielens dataset}~\citep{harper2015movielens}: This is a widely used benchmark dataset collected from \url{https://movielens.org}.
    The dataset contains user ratings and timestamps for the movie. There is side-info of users and movies. According to the year and the size of the dataset, there are multiple specific versions.\footnote{\url{https://grouplens.org/datasets/movielens/}}
    \item \textbf{Yoochoose dataset}~\citep{ben2015recsys}: This dataset is collected from the 2015 recommender systems challenge (RecSys Challenge 2015).
    The dataset includes six months of user activities for a large European e-commerce business that sells various consumer goods, including garden tools, toys, clothes, electronics, and more. There are 33,040,175 records in the click file and 1,177,769 records in the buys file. 
    The training set consists of 9,512,786 unique sessions, and the test file consists of 2,312,432 click sessions.
    \item \textbf{Alibaba Cloud/TIANCHI dataset}~\citep{zhu2018learning}: The dataset was randomly selected from Taobao; it contains about 1 million users with their behavior, which includes clicks, purchases, adding items to the shopping cart, and item favoring from November 25 to December 3, 2017. The dataset is organized in a very similar form to MovieLens-20M, i.e., each line represents a specific user-item interaction, which consists of user ID, item ID, item's category ID, behavior type, and timestamp, separated by commas.\footnote{\url{https://tianchi.aliyun.com/dataset/dataDetail?dataId=649\&userId=1\&lang=en-us}}
\end{itemize}

Table~\ref{app:tb4} summarizes the key statistics of the datasets listed above.

\begin{sidewaystable}[!htbp]
\caption{Statistics of datasets about e-commerce recommendations.} 

\resizebox{18cm}{!}{
\begin{tabular}{l rrrr c l}  
\toprule 
\multirow{2}{*}{\textbf{Datasets}} 
& \multicolumn{5}{c}{\textbf{Statistics}} 
& \multirow{2}{*}{\textbf{References}}\\
\cmidrule(lr){2-6}
& \#Users   & \#Items  & \#Records &\#Categories   & TimeSpan
\\ 
\midrule 
Amazon product dataset
& 20,980,320   & 5,933,184  & 143,663,229 & 11  & 
& \citep{he2016ups,mcauley2015image}
\\
Amazon soc dataset
& 230,000   &   &  &   & 
& \citep{mcauley2015image}
\\
AliExpress dataset
& 1,506,850   & 49,221  & 2,260,923 & 205  & 20190101-20200223
& \citep{ahmed2021deep}
\\
Instacart orders dataset
& 200,000+   & 3,000,000  &  &   & 
& Instacart dataset\footnote{\url{https://www.instacart.com/datasets/grocery-shopping-2017}}
\\  
\midrule 
Movielens-ML100K
& 943   & 1,682  & 100,000 &   & 199709-199804
\\
Movielens-ML1M
& 6,040   & 3,706  & 1,000,209 &   & 200004-200302
& \citep{harper2015movielens}
\\
Movielens-ML10M
& 69,878   & 10,681  & 10,000,054 &  & 199501-200901
\\
Movielens-ML20M
& 138,493   & 27,278  & 20,000,263 &   & 199501-201503

\\
\midrule
Yoochoose dataset
& 9,249,729   & 52,739  & 34,154,697 &   &
& \citep{ben2015recsys}
\\
Alibaba Cloud/TIANCHI dataset
& 1,000,000   & 4,023,451  & 100,934,102 & 9,378  & 20171125-20171203
& \citep{zhu2018learning}
\\

\bottomrule 
\end{tabular}
}
\label{app:tb4}
\end{sidewaystable}

\section{Datasets for e-commerce QA and dialogues}

Next, we list benchmark datasets about e-commerce question answering and dialogue systems:

\begin{itemize}[leftmargin=*,nosep]
    \item \textbf{JD product question answering}~\citep{gao2019product}: This dataset consists of online product-aware QA pairs. Each QA pair is associated with the reviews and attributes of the corresponding product. The corpus covers 469,953 products and 38 product categories. The average length of the question is 9.03 words, and the ground truth answer is 10.3 words. The average number of attributes is 9.0 key-value pairs. 
    \item \textbf{Taobao question answering dataset}~\citep{chen2019driven}: This dataset is collected on Taobao. The dataset includes 4,457 and 47,979 products under the category Cellphone and Household Electrics, respectively. For each product, the associated question-answering pairs and user reviews are included. After pre-processing, Cellphone/Household Electrics products have 356,842 and 798,688 QA-pairs in two subsets, respectively.
    \item \textbf{Amazon complex question/answer dataset}~\citep{mcauley2016addressing}: This dataset was collected from Amazon, including reviews and descriptions of products and QA data. This dataset contains 1.4 million answered questions on 191 thousand products and 13 million related reviews. 
    \item \textbf{Hierarchical product review corpus}~\citep{yu2011domain}: This corpus contains consumer reviews on 11 popular products in four domains. These reviews were crawled from several prevalent forum websites, including cnet.com, viewpoints.com, reevoo.com, and gsmarena.com. All of the reviews were posted between June 2009 and September 2010. The aspects of the reviews, as well as the opinions on the aspects, were manually annotated. 
    \item \textbf{Amazon question answering dataset}~\citep{deng2020opinion}: This dataset is constructed by combining Amazon Question Answering Dataset~\citep{mcauley2016addressing} and Amazon Product Review Dataset~\citep{he2016ups} by matching the product ID.
    In this dataset, each QA sample contains a question, a reference answer, the answer opinion type label, and a set of relevant review snippets with corresponding ratings. 
    After collecting the final dataset, each QA sample contains a question, a reference answer, the answer opinion type label, and a set of relevant review snippets with corresponding ratings. There are three categories, namely Electronics, Home \& Kitchen, and Sports \& Outdoors, with 193,960 (Electronics), 90,269 (Home \& Kitchen), and 50,020 pairs (Sports \& Outdoors).
    \item \textbf{JDDC e-commerce dialogue dataset}~\citep{chen2020jddc}: JDDC is a large-scale real scenario Chinese E-commerce conversation corpus, with more than one million multi-turn dialogues, 20 million utterances, and 150 million words, which contains conversations about after-sales topics between users and customer service staffs in an e-commerce scenario. JDDC was updated with multi-modal customer service information in 2021~\citep{zhao2021jddc}.  
    \item \textbf{E-commerce dialogue corpus dataset}~\citep{zhang2018modeling}: The dataset is collected from the real-world conversations between customers and customer service staff on Taobao. It contains over five types of conversations (i.e., commodity consultation, logistics express, recommendation, negotiation, and chitchat) based on over 20 commodities.\footnote{\url{https://drive.google.com/file/d/154J-neBo20ABtSmJDvm7DK0eTuieAuvw/view?usp=sharing}}
\end{itemize}

Table~\ref{app:tb5} summarizes the key statistics for the datasets listed above.

\begin{sidewaystable}[!htbp]
\caption{Statistics of datasets about e-commerce question answering and dialogues.} 
\centering 
\resizebox{18cm}{!}{
\begin{tabular}{l rrrrr l}  
\toprule 
\multirow{2}{*}{\textbf{Datasets}} 
& \multicolumn{5}{c}{\textbf{Statistics}} 
& \multirow{2}{*}{\textbf{References}}\\
\cmidrule(lr){2-6}
&\#Products     &\#Q-A pairs   &\#Categories  
&\makecell[c]{\#Avg.length \\of questions }
&\makecell[c]{\#Avg.length \\of ground truth}
\\ 
\midrule 
JD product question answering
& 469,953   &   & 38 & 9.03  & 10.3
& \citep{gao2019product} 
\\
Taobao question answering dataset
& 4,457/47,979   & 356,842/798,688  & 2 & 9/8  & 13/13
& \citep{chen2019driven}
\\
Amazon complex question/answer dataset
& 191,185   & 1,447,173  & 8 &   & 
& \citep{mcauley2016addressing}  
\\
Hierarchical product review corpus
& 11   &  & 4 &   &
& \citep{yu2011domain}
\\  
Amazon question answering dataset
&    & 334,249  & 3 &  &
& \citep{deng2020opinion}
\\
\toprule %

&\#Dialogues     &\#Utterance   &\#Total words  
&\makecell[c]{\#Avg.length \\of utterances }
&\makecell[c]{\#Avg.length \\of dialogues}
\\
\midrule %
JDDC e-commerce dialogue dataset
& 1,024,196   & 20,451,337  & 150,716,172 & 7.4  & 20
& \citep{chen2020jddc} 
\\
E-commerce Dialogue Corpus dataset
&\makecell[c]{1,000,000 (Train)\\10,000 (Valid)\\10,000 (Test)}    
&   
& 
&\makecell[c]{7.02 (Train)\\6.99 (Valid)\\7.11 (Test)}  
&\makecell[c]{5.51 (Train)\\5.48 (Valid)\\5.64 (Test)} 
& \citep{zhang2018modeling} 
\\

\bottomrule %
\end{tabular}
}
\label{app:tb5}
\end{sidewaystable}

\backmatter  
\printbibliography

@inproceedings{derijke-2023-beyond,
	author = {{de Rijke}, Maarten},
	booktitle = {Proceedings of WSDM},
	pages = {2-3},
	publisher = {ACM},
	title = {Beyond Accuracy Goals, Again},
	year = {2023}}

@inproceedings{siro-2022-understanding,
	author = {Siro, Clemencia and Aliannejadi, Mohammad and de Rijke, Maarten},
	booktitle = {Proceedings of SIGIR},
	pages = {2018--2023},
	publisher = {ACM},
	title = {Understanding User Satisfaction with Task-Oriented Dialogue Systems},
	year = {2022}}

@inproceedings{Balog2021Sim4IRTS,
	author = {Krisztian Balog and David Maxwell and Paul Thomas and Shuo Zhang},
	booktitle = {Proceedings of SIGIR},
	pages = {2697--2698},
	title = {Sim4IR: The SIGIR 2021 Workshop on Simulation for Information Retrieval Evaluation},
	year = {2021}}

@article{Zhang2020RecentAA,
	author = {Zheng Zhang and Ryuichi Takanobu and Minlie Huang and Xiaoyan Zhu},
	journal = {ArXiv},
	title = {Recent Advances and Challenges in Task-oriented Dialog System},
	volume = {abs/2003.07490},
	year = {2020}}

@article{Deriu2020SurveyOE,
	author = {Jan Deriu and {\'A}lvaro Rodrigo and Arantxa Otegi and Guillermo Echegoyen and Sophie Rosset and Eneko Agirre and Mark Cieliebak},
	journal = {Artificial Intelligence Review},
	pages = {755--810},
	title = {Survey on Evaluation Methods for Dialogue Systems},
	volume = {54},
	year = {2020}}

@inproceedings{Jurccek2011RealUE,
	author = {Filip Jurcicek and Simon Keizer and Milica Gasic and Francois Mairesse and Blaise Thomson and Kai Yu and Steve J. Young},
	booktitle = {Proceedings of INTERSPEECH},
	title = {Real User Evaluation of Spoken Dialogue Systems Using Amazon Mechanical Turk},
	year = {2011}}

@inproceedings{liu2021encodermi,
	author = {Liu, Hongbin and Jia, Jinyuan and Qu, Wenjie and Gong, Neil Zhenqiang},
	booktitle = {Proceedings of CCS},
	pages = {2081--2095},
	title = {EncoderMI: Membership inference against pre-trained encoders in contrastive learning},
	year = {2021}}

@article{hisamoto2020membership,
	author = {Hisamoto, Sorami and Post, Matt and Duh, Kevin},
	journal = {Transactions of the Association for Computational Linguistics},
	number = {1},
	pages = {49--63},
	title = {Membership Inference Attacks on Sequence-to-Sequence Models: Is My Data In Your Machine Translation System?},
	volume = {8},
	year = {2020}}

@inproceedings{zhang2021human,
	author = {Zhang, Yangjun and Ren, Pengjie and de Rijke, Maarten},
	booktitle = {Proceedings of ACL-IJCNLP},
	pages = {5612--5623},
	title = {A Human-machine Collaborative Framework for Evaluating Malevolence in Dialogues},
	year = {2021}}

@inproceedings{dinan2020queens,
	author = {Dinan, Emily and Fan, Angela and Williams, Adina and Urbanek, Jack and Kiela, Douwe and Weston, Jason},
	booktitle = {Proceedings of EMNLP},
	pages = {8173--8188},
	title = {Queens are Powerful too: Mitigating Gender Bias in Dialogue Generation},
	year = {2020}}

@inproceedings{tripathi2019detecting,
	author = {Tripathi, Rahul and Dhamodharaswamy, Balaji and Jagannathan, Srinivasan and Nandi, Abhishek},
	booktitle = {Proceedings of DSAA},
	pages = {374--381},
	title = {Detecting Sensitive Content in Spoken Language},
	year = {2019}}

@inproceedings{kann2022open,
	author = {Kann, Katharina and Ebrahimi, Abteen and Koh, Joewie and Dudy, Shiran and Roncone, Alessandro},
	booktitle = {Proceedings of ConvAI},
	pages = {148--165},
	title = {Open-domain Dialogue Generation: What We Can Do, Cannot Do, And Should Do Next},
	year = {2022}}

@inproceedings{si2022so,
	author = {Si, Wai Man and Backes, Michael and Blackburn, Jeremy and De Cristofaro, Emiliano and Stringhini, Gianluca and Zannettou, Savvas and Zhang, Yang},
	booktitle = {Proceedings of CCS},
	pages = {2659--2673},
	title = {Why So Toxic? Measuring and Triggering Toxic Behavior in Open-Domain Chatbots},
	year = {2022}}

@inproceedings{estes2022fact,
	author = {Estes, Alex and Vedula, Nikhita and Collins, Marcus D and Cecil, Matthew and Rokhlenko, Oleg},
	booktitle = {Proceedings of EMNLP},
	title = {Fact Checking Machine Generated Text with Dependency Trees},
	year = {2022}}

@inproceedings{sun2023contrastive,
	author = {Sun, Weiwei and Shi, Zhengliang and Gao, Shen and Ren, Pengjie and de Rijke, Maarten and Ren, Zhaochun},
	booktitle = {Proceedings of AAAI},
	title = {Contrastive Learning Reduces Hallucination in Conversations},
    pages = {13618--13626},
	year = {2023}}

@inproceedings{Wu2022LexicalKI,
	author = {Zhiyong Wu and Wei Bi and Xiang Li and Lingpeng Kong and Benjamin C.M. Kao},
	booktitle = {Proceedings of ACL},
	pages = {7945--7958},
	title = {Lexical Knowledge Internalization for Neural Dialog Generation},
	year = {2022}}

@inproceedings{Liu2022MultiStagePF,
	author = {Zihan Liu and Mostofa Ali Patwary and Ryan J. Prenger and Shrimai Prabhumoye and Wei Ping and Mohammad Shoeybi and Bryan Catanzaro},
	booktitle = {Proceedings of ACL},
	pages = {1317--1337},
	title = {Multi-Stage Prompting for Knowledgeable Dialogue Generation},
	year = {2022}}

@inproceedings{Xu2021RetrievalFreeKD,
	author = {Yan Xu and Etsuko Ishii and Zihan Liu and Genta Indra Winata and Dan Su and Andrea Madotto and Pascale Fung},
	booktitle = {Proceedings of ACL},
	pages = {93--107},
	title = {Retrieval-Free Knowledge-Grounded Dialogue Response Generation with Adapters},
	year = {2022}}

@inproceedings{Cui2021KnowledgeEF,
	author = {Leyang Cui and Yu Wu and Shujie Liu and Yue Zhang},
	booktitle = {Proceedings of EMNLP},
	pages = {2328--2337},
	title = {Knowledge Enhanced Fine-Tuning for Better Handling Unseen Entities in Dialogue Generation},
	year = {2021}}

@inproceedings{Roller2021RecipesFB,
	author = {Stephen Roller and Emily Dinan and Naman Goyal and Da Ju and Mary Williamson and Yinhan Liu and Jing Xu and Myle Ott and Kurt Shuster and Eric Michael Smith and Y.-Lan Boureau and Jason Weston},
	booktitle = {Proceedings of EACL},
	pages = {300--325},
	title = {Recipes for Building an Open-Domain Chatbot},
	year = {2021}}

@inproceedings{Roberts2020HowMK,
	author = {Adam Roberts and Colin Raffel and Noam M. Shazeer},
	booktitle = {Proceedings of EMNLP},
	pages = {5418--5426},
	title = {How Much Knowledge Can You Pack into the Parameters of a Language Model?},
	year = {2020}}

@inproceedings{Petroni2019LanguageMA,
	author = {Fabio Petroni and Tim Rockt{\"a}schel and Patrick Lewis and Anton Bakhtin and Yuxiang Wu and Alexander H. Miller and Sebastian Riedel},
	booktitle = {Proceedings of EMNLP},
	pages = {2463--2473},
	title = {Language Models as Knowledge Bases?},
	year = {2019}}

@article{Zhao2020ArePL,
	author = {Yufan Zhao and Wei Wu and Can Xu},
	journal = {arXiv preprint arXiv:2011.09708},
	title = {Are Pre-trained Language Models Knowledgeable to Ground Open Domain Dialogues?},
	year = {2020}}

@inproceedings{peng-etal-2020-shot,
	author = {Peng, Baolin and Zhu, Chenguang and Li, Chunyuan and Li, Xiujun and Li, Jinchao and Zeng, Michael and Gao, Jianfeng},
	booktitle = {Proceedings of EMNLP},
	pages = {172--182},
	title = {Few-shot Natural Language Generation for Task-oriented Dialog},
	year = {2020}}

@inproceedings{zhang2019dialogpt,
	author = {Zhang, Yizhe and Sun, Siqi and Galley, Michel and Chen, Yen-Chun and Brockett, Chris and Gao, Xiang and Gao, Jianfeng and Liu, Jingjing and Dolan, Bill},
	booktitle = {Proceedings of ACL},
	pages = {270--278},
	title = {DialoGPT: Large-scale Generative Pre-training for Conversational Response Generation},
	year = {2020}}

@inproceedings{li2021pretrained,
	author = {Li, Junyi and Tang, Tianyi and Zhao, Wayne Xin and Wen, Ji-Rong},
	booktitle = {Proceedings of IJCAI},
	pages = {4492--4499},
	title = {Pretrained Language Models for Text Generation: A Survey},
	year = {2021}}

@inproceedings{kim2022mismatch,
	author = {Kim, Takyoung and Yoon, Hoonsang and Lee, Yukyung and Kang, Pilsung and Kim, Misuk},
	booktitle = {Proceedings of ACL},
	pages = {297--309},
	title = {Mismatch between Multi-turn Dialogue and its Evaluation Metric in Dialogue State Tracking},
	year = {2022}}

@inproceedings{guo2022beyond,
	author = {Guo, Jinyu and Shuang, Kai and Li, Jijie and Wang, Zihan and Liu, Yixuan},
	booktitle = {Proceedings of ACL},
	pages = {2320--2332},
	title = {Beyond the Granularity: Multi-Perspective Dialogue Collaborative Selection for Dialogue State Tracking},
	year = {2022}}

@inproceedings{yang2021comprehensive,
	author = {Yang, Puhai and Huang, He-Yan and Mao, Xian-Ling},
	booktitle = {Proceedings of ACL-IJCNLP},
	pages = {2481--2491},
	title = {Comprehensive Study: How the Context Information of Different Granularity Affects Dialogue State Tracking?},
	year = {2021}}

@inproceedings{wang2022luna,
	author = {Wang, Yifan and Zhao, Jing and Bao, Junwei and Duan, Chaoqun and Wu, Youzheng and He, Xiaodong},
	booktitle = {Proceedings of NAACL},
	pages = {3319--3328},
	title = {LUNA: Learning Slot-Turn Alignment for Dialogue State Tracking},
	year = {2022}}

@inproceedings{lin2021knowledge,
	author = {Lin, Weizhe and Tseng, Bo-Hsiang and Byrne, Bill},
	booktitle = {Proceedings of EMNLP},
	pages = {7871--7881},
	title = {Knowledge-Aware Graph-Enhanced GPT-2 for Dialogue State Tracking},
	year = {2021}}

@inproceedings{lin2021leveraging,
	author = {Lin, Zhaojiang and Liu, Bing and Moon, Seungwhan and Crook, Paul A and Zhou, Zhenpeng and Wang, Zhiguang and Yu, Zhou and Madotto, Andrea and Cho, Eunjoon and Subba, Rajen},
	booktitle = {Proceedings of NAACL-HLT},
	pages = {5640--5648},
	title = {Leveraging Slot Descriptions for Zero-Shot Cross-Domain Dialogue StateTracking},
	year = {2021}}

@inproceedings{lee2021dialogue,
	author = {Lee, Chia-Hsuan and Cheng, Hao and Ostendorf, Mari},
	booktitle = {Proceedings of EMNLP},
	pages = {4937--4949},
	title = {Dialogue State Tracking with a Language Model using Schema-Driven Prompting},
	year = {2021}}

@inproceedings{zhu2020efficient,
	author = {Zhu, Su and Li, Jieyu and Chen, Lu and Yu, Kai},
	booktitle = {Proceedings of EMNLP},
	pages = {766--781},
	title = {Efficient Context and Schema Fusion Networks for Multi-Domain Dialogue State Tracking},
	year = {2020}}

@inproceedings{lin2021zero,
	title = {Zero-shot Dialogue State Tracking via Cross-task Transfer},
	booktitle = {Proceedings of EMNLP},
	pages = {7890--7900},
	author = {Lin, Zhaojiang and Liu, Bing and Madotto, Andrea and Moon, Seungwhan and Crook, Paul and Zhou, Zhenpeng and Wang, Zhiguang and Yu, Zhou and Cho, Eunjoon and Subba, Rajen and others},
	year = {2021}}

@inproceedings{feng2022dynamic,
	author = {Feng, Yue and Lipani, Aldo and Ye, Fanghua and Zhang, Qiang and Yilmaz, Emine},
	booktitle = {Proceedings of ACL},
	pages = {115--126},
	title = {Dynamic Schema Graph Fusion Network for Multi-Domain Dialogue State Tracking},
	year = {2022}}

@article{kim2022ask,
	author = {Kim, Su Young and Park, Hyeonjin and Shin, Kyuyong and Kim, Kyung-Min},
	journal = {arXiv preprint arXiv:2207.02516},
	title = {Ask Me What You Need: Product Retrieval Using Knowledge from GPT-3},
	year = {2022}}

@article{brown2020language,
	author = {Brown, Tom B},
	journal = {arXiv preprint arXiv:2005.14165},
	title = {Language Models are Few-shot Learners},
	year = {2020}}

@inproceedings{zhong2022dialoglm,
	author = {Zhong, Ming and Liu, Yang and Xu, Yichong and Zhu, Chenguang and Zeng, Michael},
	booktitle = {Proceedings of AAAI},
	pages = {11765--11773},
	title = {DialogLM: Pre-trained Model for Long Dialogue Understanding and Summarization},
	year = {2022}}

@inproceedings{10.1145/3477495.3532069,
	author = {He, Wanwei and Dai, Yinpei and Yang, Min and Sun, Jian and Huang, Fei and Si, Luo and Li, Yongbin},
	booktitle = {Proceedings of SIGIR},
	pages = {187--200},
	title = {Unified Dialog Model Pre-Training for Task-Oriented Dialog Understanding and Generation},
	year = {2022}}

@inproceedings{zhang2021effectiveness,
	author = {Zhang, Haode and Zhang, Yuwei and Zhan, Li-Ming and Chen, Jiaxin and Shi, Guangyuan and Wu, Xiao-Ming and Lam, Albert},
	booktitle = {Findings of EMNLP},
	pages = {1114--1120},
	title = {Effectiveness of Pre-training for Few-shot Intent Classification},
	year = {2021}}

@inproceedings{he2022galaxy,
	author = {He, Wanwei and Dai, Yinpei and Zheng, Yinhe and Wu, Yuchuan and Cao, Zheng and Liu, Dermot and Jiang, Peng and Yang, Min and Huang, Fei and Si, Luo and others},
	booktitle = {Proceedings of AAAI},
	pages = {10749--10757},
	title = {Galaxy: A Generative Pre-trained Model for Task-oriented Dialog with Semi-supervised Learning and Explicit Policy Injection},
	year = {2022}}

@inproceedings{yuan2022mcic,
	author = {Yuan, Shaozu and Shen, Xin and Zhao, Yuming and Liu, Hang and Yan, Zhiling and Liu, Ruixue and Chen, Meng},
	booktitle = {Proceedings of NLPCC},
	pages = {749--761},
	title = {MCIC: Multimodal Conversational Intent Classification for E-commerce Customer Service},
	year = {2022}}

@inproceedings{li2022unsupervised,
	author = {Li, Jia and Tao, Chongyang and Hu, Huang and Xu, Can and Chen, Yining and Jiang, Daxin},
	booktitle = {Proceedings of WSDM},
	pages = {562--570},
	title = {Unsupervised Cross-Domain Adaptation for Response Selection Using Self-Supervised and Adversarial Training},
	year = {2022}}

@inproceedings{feng2022reciprocal,
	author = {Feng, Jiazhan and Tao, Chongyang and Li, Zhen and Liu, Chang and Shen, Tao and Zhao, Dongyan},
	booktitle = {Proceedings of COLING},
	pages = {389--399},
	title = {Reciprocal Learning of Knowledge Retriever and Response Ranker for Knowledge-Grounded Conversations},
	year = {2022}}

@inproceedings{tao2021pre,
	author = {Tao, Chongyang and Chen, Changyu and Feng, Jiazhan and Wen, Ji-Rong and Yan, Rui},
	booktitle = {Proceedings of ACL-IJCNLP},
	pages = {4446--4457},
	title = {A Pre-training Strategy for Zero-Resource Response Selection in Knowledge-Grounded Conversations},
	year = {2021}}

@inproceedings{han2021fine,
	author = {Han, Janghoon and Hong, Taesuk and Kim, Byoungjae and Ko, Youngjoong and Seo, Jungyun},
	booktitle = {Proceedings of NAACL},
	pages = {1549--1558},
	title = {Fine-grained Post-training for Improving Retrieval-based Dialogue Systems},
	year = {2021}}

@inproceedings{lin2022task,
	author = {Lin, Tzu-Hsiang and Chi, Ta-Chung and Rumshisky, Anna},
	booktitle = {Proceedings of ACL-IJCNLP},
	pages = {665--669},
	title = {Domain-Adaptive Pretraining Methods for Dialogue Understanding},
	year = {2021}}

@inproceedings{xu-etal-2021-discovering,
	author = {Xu, Jun and Lei, Zeyang and Wang, Haifeng and Niu, Zheng-Yu and Wu, Hua and Che, Wanxiang},
	booktitle = {Proceedings of ACL-IJCNLP},
	pages = {1726--1739},
	title = {Discovering Dialog Structure Graph for Coherent Dialog Generation},
	year = {2021}}

@inproceedings{yu2022xdai,
	author = {Yu, Jifan and Zhang, Xiaohan and Xu, Yifan and Lei, Xuanyu and Guan, Xinyu and Zhang, Jing and Hou, Lei and Li, Juanzi and Tang, Jie},
	booktitle = {Proceedings of KDD},
	pages = {4422--4432},
	title = {XDAI: A Tuning-free Framework for Exploiting Pre-trained Language Models in Knowledge Grounded Dialogue Generation},
	year = {2022}}

@inproceedings{li2022knowledge,
	author = {Li, Qintong and Li, Piji and Ren, Zhaochun and Ren, Pengjie and Chen, Zhumin},
	booktitle = {Proceedings of AAAI},
	pages = {10993--11001},
	title = {Knowledge Bridging for Empathetic Dialogue Generation},
	year = {2022}}

@inproceedings{lewis2020question,
	author = {Lewis, Patrick and Stenetorp, Pontus and Riedel, Sebastian},
	booktitle = {Proceedings of ACL},
	pages = {1000--1008},
	title = {Question and Answer Test-train Overlap in Open-domain Question Answering Datasets},
	year = {2021}}

@inproceedings{NEURIPS2021_da3fde15,
	author = {Singh, Devendra and Reddy, Siva and Hamilton, Will and Dyer, Chris and Yogatama, Dani},
	booktitle = {Proceedings of NIPS},
	pages = {25968--25981},
	title = {End-to-End Training of Multi-Document Reader and Retriever for Open-Domain Question Answering},
	year = {2021}}

@inproceedings{sachan2021end,
	author = {Sachan, Devendra Singh and Patwary, Mostofa and Shoeybi, Mohammad and Kant, Neel and Ping, Wei and Hamilton, William L and Catanzaro, Bryan},
	booktitle = {Proceedings of ACL},
	pages = {6648--6662},
	title = {End-to-end Training of Neural Retrievers for Open-domain Question Answering},
	year = {2021}}

@inproceedings{mao2020generation,
	author = {Mao, Yuning and He, Pengcheng and Liu, Xiaodong and Shen, Yelong and Gao, Jianfeng and Han, Jiawei and Chen, Weizhu},
	booktitle = {Proceedings of ACL},
	pages = {4089--4100},
	title = {Generation-augmented Retrieval for Open-domain Question Answering},
	year = {2021}}

@inproceedings{yu-etal-2022-kg,
	author = {Yu, Donghan and Zhu, Chenguang and Fang, Yuwei and Yu, Wenhao and Wang, Shuohang and Xu, Yichong and Ren, Xiang and Yang, Yiming and Zeng, Michael},
	booktitle = {Proceedings of ACL},
	pages = {4961--4974},
	title = {KG-FiD: Infusing Knowledge Graph in Fusion-in-decoder for Open-domain Question Answering},
	year = {2022}}

@inproceedings{zhao2020data,
	author = {Zhao, Mingxiao},
	booktitle = {Proceedings of BDEIM},
	pages = {61--65},
	title = {Data-driven Scene Marketing Based on Consumer Insight},
	year = {2020}}

@inproceedings{lin2022automatic,
	author = {Lin, Peng and Zou, Yanyan and Wu, Lingfei and Ma, Mian and Ding, Zhuoye and Long, Bo},
	booktitle = {Proceedings of EMNLP},
	title = {Automatic Scene-based Topic Channel Construction System for E-Commerce},
	year = {2022}}

@inproceedings{Hao2021wwwsentiment,
	author = {Fei, Hao and Ren, Yafeng and Wu, Shengqiong and Li, Bobo and Ji, Donghong},
	booktitle = {Proceedings of Web Conference},
	pages = {553--564},
	title = {Latent Target-Opinion as Prior for Document-Level Sentiment Classification: A Variational Approach from Fine-Grained Perspective},
	year = {2021}}

@inproceedings{nguyen2022adaptive,
	author = {Nguyen, Thong and Wu, Xiaobao and Luu, Anh-Tuan and Nguyen, Cong-Duy and Hai, Zhen and Bing, Lidong},
	booktitle = {Proceedings of EMNLP},
	title = {Adaptive Contrastive Learning on Multimodal Transformer for Review Helpfulness Predictions},
	year = {2022}}

@inproceedings{han2022sancl,
	author = {Han, Wei and Chen, Hui and Hai, Zhen and Poria, Soujanya and Bing, Lidong},
	booktitle = {Proceedings of COLING},
	pages = {5666--5677},
	title = {SANCL: Multimodal Review Helpfulness Prediction with Selective Attention and Natural Contrastive Learning},
	year = {2022}}

@inproceedings{hu-2020-modeling,
	author = {Hu, Haoji and He, Xiangnan and Gao, Jinyang and Zhang, Zhi-Li},
	booktitle = {Proceedings of SIGIR},
	pages = {1071--1080},
	title = {Modeling Personalized Item Frequency Information for Next-basket Recommendation},
	year = {2020}}

@inproceedings{yu-2020-predicting,
	author = {Yu, Le and Sun, Leilei and Du, Bowen and Liu, Chuanren and Xiong, Hui and Lv, Weifeng},
	booktitle = {Proceedings of KDD},
	pages = {1083--1091},
	title = {Predicting Temporal Sets with Deep Neural Networks},
	year = {2020}}

@article{sprangers-2023-parameter,
	author = {Sprangers, Olivier and Schelter, Sebastian and de Rijke, Maarten},
	journal = {International Journal of Forecasting},
	number = {1},
	pages = {332--345},
	title = {Parameter Efficient Deep Probabilistic Forecasting},
	volume = {39},
	year = {2023}}

@inproceedings{liu-2019-characterizing,
	author = {Yue Liu and Helena Lee and Palakorn Achananuparp and Ee-Peng Lim and Tzu-Ling Cheng and Shou-De Lin},
	booktitle = {Proceedings of DPH},
	pages = {11--20},
	publisher = {{ACM}},
	title = {Characterizing and Predicting Repeat Food Consumption Behavior for Just-in-Time Interventions},
	year = {2019},
}

@article{li-2023-next,
	author = {Li, Ming and Jullien, Sami and Ariannezhad, Mozhdeh and de Rijke, Maarten},
	journal = {ACM Transactions on Information Systems},
	title = {A Next Basket Recommendation Reality Check},
	year = {2023},
    volume =41,
    number = {4},
    pages = {Article 116}}

@article{li-2023-who,
	author = {Li, Ming and Ariannezhad, Mozhdeh and Yates, Andrew and de Rijke, Maarten},
	journal = {ACM Transactions on Recommender Systems},
	note = {Accepted subject to major revisions},
	title = {Who Will Purchase this Item Next? {Reverse} Next Period Recommendation in Grocery Shopping},
	year = {2023}}

@inproceedings{ariannezhad-2023-personalized,
	author = {Ariannezhad, Mozhdeh and Li, Ming and Schelter, Sebastian and de Rijke, Maarten},
	booktitle = {Proceedings of WSDM},
	publisher = {ACM},
	title = {A Personalized Neighborhood-based Model for Within-basket Recommendation in Grocery Shopping},
	year = {2023}}

@inproceedings{ariannezhad-2022-recanet,
	author = {Ariannezhad, Mozhdeh and Jullien, Sami and Li, Ming and Fang, Min and Schelter, Sebastian and de Rijke, Maarten},
	booktitle = {Proceedings of SIGIR},
	pages = {1240--1250},
	publisher = {ACM},
	title = {ReCANet: A Repeat Consumption-Aware Neural Network for Next Basket Recommendation in Grocery Shopping},
	year = {2022}}

@inproceedings{ariannezhad-2020-demand,
	author = {Ariannezhad, Mozhdeh and Schelter, Sebastian and de Rijke, Maarten},
	booktitle = {Proceedings of AALTD},
	pages = {46-62},
	publisher = {Springer},
	series = {LNCS 12588},
	title = {Demand Forecasting in the Presence of Privileged Information},
	year = {2020}}

@inproceedings{huang2022clickrec,
	author = {Huang, Jin and Oosterhuis, Harrie and de Rijke, Maarten},
	booktitle = {Proceedings of WSDM},
	pages = {381--389},
	title = {It Is Different When Items Are Older: Debiasing Recommendations When Selection Bias and User Preferences Are Dynamic},
	year = {2022}}

@inproceedings{xu2021transformer,
	author = {Xu, Dongkuan and Liang, Junjie and Cheng, Wei and Wei, Hua and Chen, Haifeng and Zhang, Xiang},
	booktitle = {Proceedings of AAAI},
	pages = {4546--4554},
	title = {Trans\-former-style Relational Reasoning with Dynamic Memory Updating for Temporal Network Modeling},
	year = {2021}}

@inproceedings{zhao2021sigirwgcn,
	author = {Zhao, Yunxiang and Qi, Jianzhong and Liu, Qingwei and Zhang, Rui},
	booktitle = {Proceedings of SIGIR},
	pages = {624--633},
	title = {WGCN: Graph Convolutional Networks with Weighted Structural Features},
	year = {2021}}

@inproceedings{liu2022sigirgraph,
	author = {Liu, Yonghao and Li, Mengyu and Li, Ximing and Giunchiglia, Fausto and Feng, Xiaoyue and Guan, Renchu},
	booktitle = {Proceedings of SIGIR},
	pages = {471--481},
	title = {Few-Shot Node Classification on Attributed Networks with Graph Meta-Learning},
	year = {2022}}

@inproceedings{gaowsdm2022rec,
	author = {Gao, Chen and Wang, Xiang and He, Xiangnan and Li, Yong},
	booktitle = {Proceedings of WSDM},
	pages = {1623--1625},
	title = {Graph Neural Networks for Recommender System},
	year = {2022}}

@inproceedings{chen2022kddasymptotically,
	author = {Chen, Yu and Jin, Jiaqi and Zhao, Hui and Wang, Pengjie and Liu, Guojun and Xu, Jian and Zheng, Bo},
	booktitle = {Proceedings of Web Conference},
	pages = {369--379},
	title = {Asymptotically Unbiased Estimation for Delayed Feedback Modeling via Label Correction},
	year = {2022}}

@article{yoshikawa2018nonparametric,
	author = {Yoshikawa, Yuya and Imai, Yusaku},
	journal = {arXiv preprint arXiv:1802.00255},
	title = {A Nonparametric Delayed Feedback Model for Conversion Rate Prediction},
	year = {2018}}

@inproceedings{chapelle2014modeling,
	author = {Chapelle, Olivier},
	booktitle = {Proceedings of KDD},
	pages = {1097--1105},
	title = {Modeling Delayed Feedback in Display Advertising},
	year = {2014}}

@inproceedings{guo2021enhanced,
	author = {Guo, Siyuan and Zou, Lixin and Liu, Yiding and Ye, Wenwen and Cheng, Suqi and Wang, Shuaiqiang and Chen, Hechang and Yin, Dawei and Chang, Yi},
	booktitle = {Proceedings of SIGIR},
	pages = {275--284},
	title = {Enhanced Doubly Robust Learning for Debiasing Post-click Conversion Rate Estimation},
	year = {2021}}

@inproceedings{zhang2020large,
	author = {Zhang, Wenhao and Bao, Wentian and Liu, Xiao-Yang and Yang, Keping and Lin, Quan and Wen, Hong and Ramezani, Ramin},
	booktitle = {Proceedings of Web Conference},
	pages = {2775--2781},
	title = {Large-scale Causal Approaches to Debiasing Post-click Conversion Rate Estimation with Multi-task Learning},
	year = {2020}}

@inproceedings{ma2018entire,
	author = {Ma, Xiao and Zhao, Liqin and Huang, Guan and Wang, Zhi and Hu, Zelin and Zhu, Xiaoqiang and Gai, Kun},
	booktitle = {Proceedings of SIGIR},
	pages = {1137--1140},
	title = {Entire Space Multi-task Model: An Effective Approach for Estimating Post-click Conversion Rate},
	year = {2018}}

@inproceedings{dai2022kddcvr,
	author = {Dai, Quanyu and Li, Haoxuan and Wu, Peng and Dong, Zhenhua and Zhou, Xiao-Hua and Zhang, Rui and Zhang, Rui and Sun, Jie},
	booktitle = {Proceedings of KDD},
	pages = {252--262},
	title = {A Generalized Doubly Robust Learning Framework for Debiasing Post-Click Conversion Rate Prediction},
	year = {2022}}

@inproceedings{zhao2022personaot,
	author = {Zhao, Mengxue and Yang, Yang and Li, Miao and Wang, Jingang and Wu, Wei and Ren, Pengjie and de Rijke, Maarten and Ren, Zhaochun},
	booktitle = {Proceedings of SIGIR},
	pages = {1066--1076},
	title = {Personalized Abstractive Opinion Tagging},
	year = {2022}}

@inproceedings{jianhao2022session,
	author = {Yuan, Jiahao and Ji, Wendi and Zhang, Dell and Pan, Jinwei and Wang, Xiaoling},
	booktitle = {Proceedings ICDE},
	title = {Micro-Behavior Encoding for Session-based Recommendation},
	year = {2022}}

@inproceedings{li2021conversionpred,
	author = {Li, Haoming and Pan, Feiyang and Ao, Xiang and Yang, Zhao and Lu, Min and Pan, Junwei and Liu, Dapeng and Xiao, Lei and He, Qing},
	booktitle = {Proceedings of SIGIR},
	pages = {1915--1919},
	title = {Follow the Prophet: Accurate Online Conversion Rate Prediction in the Face of Delayed Feedback},
	year = {2021}}

@inproceedings{hou2021conversion,
	author = {Hou, Yilin and Zhao, Guangming and Liu, Chuanren and Zu, Zhonglin and Zhu, Xiaoqiang},
	booktitle = {Proceedings ICDM},
	pages = {191--199},
	title = {Conversion Prediction with Delayed Feedback: A Multi-task Learning Approach},
	year = {2021}}

@inproceedings{yang2021capturing,
	author = {Yang, Jia-Qi and Li, Xiang and Han, Shuguang and Zhuang, Tao and Zhan, De-Chuan and Zeng, Xiaoyi and Tong, Bin},
	booktitle = {Proceedings of AAAI},
	pages = {4582--4589},
	title = {Capturing Delayed Feedback in Conversion Rate Prediction via Elapsed-time Sampling},
	year = {2021}}

@inproceedings{wang2021clicks,
	author = {Wang, Wenjie and Feng, Fuli and He, Xiangnan and Zhang, Hanwang and Chua, Tat-Seng},
	booktitle = {Proceedings of SIGIR},
	pages = {1288--1297},
	title = {Clicks Can Be Cheating: Counterfactual Recommendation for Mitigating Clickbait Issue},
	year = {2021}}

@inproceedings{wen2019leveraging,
	author = {Wen, Hongyi and Yang, Longqi and Estrin, Deborah},
	booktitle = {Proceedings of RecSys},
	pages = {278--286},
	title = {Leveraging Post-click Feedback for Content Recommendations},
	year = {2019}}

@inproceedings{meng2020sigirclick,
	author = {Meng, Wenjing and Yang, Deqing and Xiao, Yanghua},
	booktitle = {Proceedings of SIGIR},
	pages = {1091--1100},
	title = {ncorporating User Micro-Behaviors and Item Knowledge into Multi-Task Learning for Session-Based Recommendation},
	year = {2020}}

@inproceedings{lu2018between,
	author = {Lu, Hongyu and Zhang, Min and Ma, Shaoping},
	booktitle = {Proceedings of SIGIR},
	pages = {435--444},
	title = {Between Clicks and Satisfaction: Study on Multi-phase User Preferences and Satisfaction for Online News Reading},
	year = {2018}
}

@inproceedings{cheng2022dynamic,
	author = {Cheng, Yuan},
	booktitle = {Proceedings of CIKM},
	pages = {3888--3892},
	title = {Dynamic Explicit Embedding Representation for Numerical Features in Deep CTR Prediction},
	year = {2022}}

@inproceedings{guo2022icdectr,
	author = {Guo, Wei and Zhang, Can and He, Zhicheng and Qin, Jiarui and Guo, Huifeng and Chen, Bo and Tang, Ruiming and He, Xiuqiang and Zhang, Rui},
	booktitle = {Proceedings of ICDE},
	pages = {727-740},
	title = {MISS: Multi-Interest Self-Supervised Learning Framework for Click-Through Rate Prediction},
	year = {2022}}

@article{zhu2022tkdectr,
	author = {Zhu, Chenxu and Chen, Bo and Zhang, Weinan and Lai, Jincai and Tang, Ruiming and He, Xiuqiang and Li, Zhenguo and Yu, Yong},
	journal = {IEEE Transactions on Knowledge and Data Engineering},
	title = {AIM: Automatic Interaction Machine for Click-Through Rate Prediction},
	year = {2021}}

@inproceedings{chen2021deepctr,
	author = {Chen, Bo and Wang, Yichao and Liu, Zhirong and Tang, Ruiming and Guo, Wei and Zheng, Hongkun and Yao, Weiwei and Zhang, Muyu and He, Xiuqiang},
	booktitle = {Proceedings of CIKM},
	pages = {3757--3766},
	title = {Enhancing Explicit and Implicit Feature Interactions via Information Sharing for Parallel Deep CTR Models},
	year = {2021}}

@inproceedings{wen2021hierarchy,
	author = {Wen, Hong and Zhang, Jing and Lv, Fuyu and Bao, Wentian and Wang, Tianyi and Chen, Zulong},
	booktitle = {Proceedings of SIGIR},
	pages = {2187--2191},
	title = {Hierarchically Modeling Micro and Macro Behaviors via Multi-Task Learning for Conversion Rate Prediction},
	year = {2021}}

@inproceedings{bian2021contra,
	author = {Bian, Shuqing and Zhao, Wayne Xin and Zhou, Kun and Cai, Jing and He, Yancheng and Yin, Cunxiang and Wen, Ji-Rong},
	booktitle = {Proceedings of CIKM},
	pages = {3737--3746},
	title = {Contrastive Curriculum Learning for Sequential User Behavior Modeling via Data Augmentation},
	year = {2021}}

@inproceedings{gong2020edgerec,
	author = {Gong, Yu and Jiang, Ziwen and Feng, Yufei and Hu, Binbin and Zhao, Kaiqi and Liu, Qingwen and Ou, Wenwu},
	booktitle = {Proceedings of CIKM},
	pages = {2477--2484},
	title = {EdgeRec: Recommender System on Edge in Mobile Taobao},
	year = {2020}}

@inproceedings{huang2019ecompred,
	author = {Huang, Chao and Wu, Xian and Zhang, Xuchao and Zhang, Chuxu and Zhao, Jiashu and Yin, Dawei and Chawla, Nitesh V.},
	booktitle = {Proceedings of KDD},
	pages = {2613--2622},
	title = {Online Purchase Prediction via Multi-Scale Modeling of Behavior Dynamics},
	year = {2019}}

@article{vakulenko2021large,
	author = {Vakulenko, Svitlana and Kanoulas, Evangelos and de Rijke, Maarten},
	journal = {ACM Transactions on Information Systems},
	number = {4},
	pages = {1--32},
	title = {A Large-scale Analysis of Mixed Initiative in Information-seeking Dialogues for Conversational Search},
	volume = {39},
	year = {2021}}

@inproceedings{ye2022structured,
	author = {Ye, Chenchen and Liao, Lizi and Feng, Fuli and Ji, Wei and Chua, Tat-Seng},
	booktitle = {Proceedings of SIGIR},
	pages = {155--164},
	title = {Structured and Natural Responses Co-generation for Conversational Search},
	year = {2022}}

@article{zamani2022conversational,
	author = {Zamani, Hamed and Trippas, Johanne R and Dalton, Jeff and Radlinski, Filip},
	journal = {arXiv preprint arXiv:2201.08808},
	title = {Conversational Information Seeking},
	year = {2022}}

@article{zou2022learning,
	author = {Zou, Jie and Huang, Jimmy Xiangji and Ren, Zhaochun and Kanoulas, Evangelos},
	journal = {ACM Transactions on Information Systems},
	title = {Learning to Ask: Conversational Product Search via Representation Learning},
	year = {2022}}

@inproceedings{gao2022search,
	author = {Gao, Chang and Lam, Wai},
	booktitle = {Proceedings of ECIR},
	pages = {230--243},
	title = {Search Clarification Selection via Query-Intent-Clarification Graph Attention},
	year = {2022}}

@inproceedings{ghanem2022question,
	author = {Ghanem, Bilal and Coleman, Lauren Lutz and Dexter, Julia Rivard and von der Ohe, Spencer and Fyshe, Alona},
	booktitle = {Findings of ACL},
	pages = {2131--2146},
	title = {Question Generation for Reading Comprehension Assessment by Modeling How and What to Ask},
	year = {2022}}

@inproceedings{liu2021learning,
	author = {Liu, Zhongkun and Ren, Pengjie and Chen, Zhumin and Ren, Zhaochun and de Rijke, Maarten and Zhou, Ming},
	booktitle = {Proceedings of ACL},
	pages = {5638--5650},
	title = {Learning to Ask Conversational Questions by Optimizing Levenshtein Distance},
	year = {2021}}

@inproceedings{kaiser2021reinforcement,
	author = {Kaiser, Magdalena and Saha Roy, Rishiraj and Weikum, Gerhard},
	booktitle = {Proceedings of SIGIR},
	pages = {459--469},
	title = {Reinforcement learning from reformulations in conversational question answering over knowledge graphs},
	year = {2021}}

@inproceedings{vakulenko2021question,
	author = {Vakulenko, Svitlana and Longpre, Shayne and Tu, Zhucheng and Anantha, Raviteja},
	booktitle = {Proceedings of SIGIR},
	pages = {355--363},
	title = {Question rewriting for conversational question answering},
	year = {2021}}

@inproceedings{qu2020open,
	author = {Qu, Chen and Yang, Liu and Chen, Cen and Qiu, Minghui and Croft, W Bruce and Iyyer, Mohit},
	booktitle = {Proceedings of SIGIR},
	pages = {539--548},
	title = {Open-retrieval conversational question answering},
	year = {2020}}

@inproceedings{dalton2022conversational,
	author = {Dalton, Jeffrey and Fischer, Sophie and Owoicho, Paul and Radlinski, Filip and Rossetto, Federico and Trippas, Johanne R and Zamani, Hamed},
	booktitle = {Proceedings of SIGIR},
	pages = {3455--3458},
	title = {Conversational Information Seeking: Theory and Application},
	year = {2022}}

@inproceedings{azzopardi2022towards,
	author = {Azzopardi, Leif and Aliannejadi, Mohammad and Kanoulas, Evangelos},
	booktitle = {Proceedings ECIR},
	pages = {31--38},
	title = {Towards Building Economic Models of Conversational Search},
	year = {2022}}

@inproceedings{aliannejadi2021analysing,
	author = {Aliannejadi, Mohammad and Azzopardi, Leif and Zamani, Hamed and Kanoulas, Evangelos and Thomas, Paul and Craswell, Nick},
	booktitle = {Proceedings of CIKM},
	pages = {16--26},
	title = {Analysing mixed initiatives and search strategies during conversational search},
	year = {2021}}

@article{keyvan2022approach,
	author = {Keyvan, Kimiya and Huang, Jimmy Xiangji},
	journal = {ACM Computing Surveys},
	title = {How to Approach Ambiguous Queries in Conversational Search? A Survey of Techniques, Approaches, Tools and Challenges},
	year = {2022}}

@inproceedings{liu2022pretraining,
	author = {Liu, Xinyi and Guan, Wanxian and Li, Lianyun and Li, Hui and Lin, Chen and Li, Xubin and Chen, Si and Xu, Jian and Deng, Hongbo and Zheng, Bo},
	booktitle = {Proceedings of KDD},
	pages = {3429--3437},
	title = {Pretraining Representations of Multi-modal Multi-query E-commerce Search},
	year = {2022}}

@article{wang2022siamese,
	author = {Wang, Yunxiao and Liu, Meng and Wei, Yinwei and Cheng, Zhiyong and Wang, Yinglong and Nie, Liqiang},
	journal = {IEEE Transactions on Multimedia},
	number = {8},
	pages = {1--13},
	title = {Siamese Alignment Network for Weakly Supervised Video Moment Retrieval},
	volume = {14},
	year = {2022}}

@article{wang2020metasearch,
	author = {Wang, Qi and Liu, Xinchen and Liu, Wu and Liu, An-An and Liu, Wenyin and Mei, Tao},
	journal = {IEEE Transactions on Image Processing},
	number = {1},
	pages = {7549--7564},
	title = {Metasearch: Incremental Product Search via Deep Meta-learning},
	volume = {29},
	year = {2020}}

@inproceedings{zhu2022cross,
	author = {Zhu, Rui and Zhao, Yiming and Qu, Wei and Liu, Zhongyi and Li, Chenliang},
	booktitle = {Proceedings of CIKM},
	pages = {3746--3755},
	title = {Cross-Domain Product Search with Knowledge Graph},
	year = {2022}}

@inproceedings{tan2022bit,
	author = {Tan, Wentao and Zhu, Lei and Guan, Weili and Li, Jingjing and Cheng, Zhiyong},
	booktitle = {Proceedings of SIGIR},
	pages = {982--991},
	title = {Bit-aware Semantic Transformer Hashing for Multi-modal Retrieval},
	year = {2022}}

@article{wei2021universal,
	author = {Wei, Jiwei and Yang, Yang and Xu, Xing and Zhu, Xiaofeng and Shen, Heng Tao},
	journal = {IEEE Transactions on Pattern Analysis and Machine Intelligence},
	number = {10},
	pages = {6534-6545},
	title = {Universal weighting metric learning for cross-modal retrieval},
	volume = {44},
	year = {2022}}

@inproceedings{qu2021dynamic,
	author = {Qu, Leigang and Liu, Meng and Wu, Jianlong and Gao, Zan and Nie, Liqiang},
	booktitle = {Proceedings of SIGIR},
	pages = {1104--1113},
	title = {Dynamic Modality Interaction Modeling for Image-text Retrieval},
	year = {2021}}

@inproceedings{zou2021pre,
	author = {Zou, Lixin and Zhang, Shengqiang and Cai, Hengyi and Ma, Dehong and Cheng, Suqi and Wang, Shuaiqiang and Shi, Daiting and Cheng, Zhicong and Yin, Dawei},
	booktitle = {Proceedings of KDD},
	pages = {4014--4022},
	title = {Pre-trained Language Model based Ranking in Baidu Search},
	year = {2021}}

@inproceedings{chu2022h,
	author = {Chu, Xiaokai and Zhao, Jiashu and Zou, Lixin and Yin, Dawei},
	booktitle = {Proceedings of SIGIR},
	pages = {1478--1489},
	title = {H-ERNIE: A Multi-Granularity Pre-Trained Language Model for Web Search},
	year = {2022}}

@inproceedings{wu2022multi,
	author = {Wu, Xuyang and Magnani, Alessandro and Chaidaroon, Suthee and Puthenputhussery, Ajit and Liao, Ciya and Fang, Yi},
	booktitle = {Proceedings of Web Conference},
	pages = {493--501},
	title = {A Multi-task Learning Framework for Product Ranking with BERT},
	year = {2022}}

@inproceedings{fan2022modeling,
	author = {Fan, Lu and Li, Qimai and Liu, Bo and Wu, Xiao-Ming and Zhang, Xiaotong and Lv, Fuyu and Lin, Guli and Li, Sen and Jin, Taiwei and Yang, Keping},
	booktitle = {Proceedings of Web Conference},
	title = {Modeling User Behavior with Graph Convolution for Personalized Product Search},
	year = {2022}}

@inproceedings{guo2016deep,
	author = {Guo, Jiafeng and Fan, Yixing and Ai, Qingyao and Croft, W Bruce},
	booktitle = {Proceedings of CIKM},
	pages = {55--64},
	title = {A Deep Relevance Matching Model for Ad-hoc Retrieval},
	year = {2016}}

@inproceedings{yao2022reprbert,
	author = {Yao, Shaowei and Tan, Jiwei and Chen, Xi and Zhang, Juhao and Zeng, Xiaoyi and Yang, Keping},
	booktitle = {Proceedings of KDD},
	pages = {4363--4371},
	title = {ReprBERT: Distilling BERT to an Efficient Representation-Based Relevance Model for E-Commerce},
	year = {2022}}

@inproceedings{chen-top-n-2017,
	author = {Chen, Yifan and Zhao, Xiang and de Rijke, Maarten},
	booktitle = {Proceedings of SIGIR},
	publisher = {ACM},
	title = {Top-N Recommendation with High-dimensional Side Information via Locality Preserving Projection},
    pages = {985--988},
	year = {2017}}

@inproceedings{kersbergen-2021-learnings,
	author = {Kersbergen, Barrie and Schelter, Sebastian},
	booktitle = {Proceedings of ICDE},
	pages = {2447-2452},
	title = {Learnings from a Retail Recommendation System on Billions of Interactions at bol.com},
	year = {2021}}

@article{malone-1987-intelligent,
	author = {Malone, Thomas W. and Grant, Kenneth R. and Turbak, Franklyn A. and Brobst, Stephen A. and Cohen, Michael D.},
	journal = {Communications of the ACM},
	number = {5},
	pages = {390--402},
	title = {Intelligent Information Sharing Systems},
	volume = {30},
	year = {1987}}

@article{gao-2021-advances,
	author = {Gao, Chongming and Lei, Wenqiang and He, Xiangnan and de Rijke, Maarten and Chua, Tat-Seng},
	journal = {AI Open},
	pages = {100--126},
	title = {Advances and Challenges in Conversational Recommender Systems: A Survey},
	volume = {2},
	year = {2021}}

@article{lin-2021-pretrained,
	author = {Lin, Jimmy and Nogueira, Rodrigo and Yates, Andrew},
	journal = {Synthesis Lectures on Human Language Technologies},
	number = {4},
	pages = {1--325},
	publisher = {Morgan \& Claypool Publishers},
	title = {Pretrained Transformers for Text Ranking: {BERT} and Beyond},
	volume = {14},
	year = {2021}}

@article{onal-neural-2018,
	author = {Onal, Kezban Dilek and Zhang, Ye and Altingovde, Ismail Sengor and Rahman, Md Mustafizur and Karagoz, Pinar and Braylan, Alex and Dang, Brandon and Chang, Heng-Lu and Kim, Henna and McNamara, Quinten and Angert, Aaron and Banner, Edward and Khetan, Vivek and McDonnell, Tyler and Nguyen, An Thanh and Xu, Dan and Wallace, Byron C. and de Rijke, Maarten and Lease, Matthew},
	journal = {Information Retrieval Journal},
	number = {2--3},
	pages = {111--182},
	title = {Neural Information Retrieval: At the End of the Early Years},
	volume = {21},
	year = {2018}}

@inproceedings{mishne-deriving-2006,
	author = {Mishne, Gilad and de Rijke, Maarten},
	booktitle = {Proceedings of Web Conference},
	publisher = {ACM},
	title = {Deriving Wishlists from Blogs: Show us your Blog, and We'll Tell you What Books to Buy},
	year = {2006}}

@inproceedings{sarvi-2020-comparison,
	author = {Sarvi, Fatemeh and Voskarides, Nikos and Mooiman, Lois and Schelter, Sebastian and de Rijke, Maarten},
	booktitle = {Proceedings of SIGIR Workshop on eCommerce},
	publisher = {ACM},
	title = {A Comparison of Supervised Learning to Match Methods for Product Search},
	year = {2020}}

@inproceedings{vardasbi-2022-probabilistic,
	author = {Vardasbi, Ali and Sarvi, Fatemeh and de Rijke, Maarten},
	booktitle = {Proceedings of SIGIR},
	publisher = {ACM},
	title = {Probabilistic Permutation Graph Search: Black-Box Optimization for Fairness in Ranking},
	year = {2022},
    pages = {715--725},
}

@inproceedings{sarvi-2022-understanding,
	author = {Sarvi, Fatemeh and Heuss, Maria and Aliannejadi, Mohammad and Schelter, Sebastian and de Rijke, Maarten},
	booktitle = {Proceedings of WSDM},
	pages = {861--869},
	publisher = {ACM},
	title = {Understanding and Mitigating the Effect of Outliers in Fair Ranking},
	year = {2022}}

@article{reddy-etal-2019-coqa,
	author = {Reddy, Siva and Chen, Danqi and Manning, Christopher D.},
	journal = {Transactions of the Association for Computational Linguistics},
	pages = {249--266},
	publisher = {MIT Press},
	title = {{C}o{QA}: A Conversational Question Answering Challenge},
	volume = {7},
	year = {2019}}

@inproceedings{radlinski-2017-theoretical,
	author = {Radlinski, Filip and Craswell, Nick},
	booktitle = {Proceedings of CHIIR},
	pages = {117--126},
	title = {A Theoretical Framework for Conversational Search},
	year = {2017}}

@inproceedings{azzopardi2018conceptualizing,
	author = {Azzopardi, Leif and Dubiel, Mateusz and Halvey, Martin and Dalton, Jeffery},
	booktitle = {Proceedings of CAIR},
	publisher = {ACM},
	title = {Conceptualizing Agent-human Interactions During the Conversational Search Process},
	year = {2018}}

@inproceedings{hendriksen-2020-analyzing,
	author = {Hendriksen, Mariya and Kuiper, Ernst and Nauts, Pim and Schelter, Sebastian and {de Rijke}, Maarten},
	booktitle = {Proceedings of SIGIR Workshop on eCommerce},
	publisher = {ACM},
	title = {Analyzing and Predicting Purchase Intent in E-commerce: Anonymous vs. Identified Customers},
	year = {2020}}

@inproceedings{ariannezhad-2021-understanding,
	author = {Ariannezhad, Mozhdeh and Jullien, Sami and Nauts, Pim and Fang, Min and Schelter, Sebastian and {de Rijke}, Maarten},
	booktitle = {Proceedings of CIKM},
	publisher = {ACM},
	title = {Understanding Multi-channel Customer Behavior in Retail},
	year = {2021}}

@inproceedings{mishne-2006-deriving,
	author = {Mishne, Gilad and {de Rijke}, Maarten},
	booktitle = {Proceedings of Web Conference},
	pages = {925--926},
	title = {Deriving Wishlists from Blogs: Show us your Blog, and We'll Tell you What Books to Buy},
	year = {2006}}

@book{hearst-2009-search,
	author = {Hearst, Marti},
	publisher = {Cambridge University Press},
	title = {Search User Interfaces},
	year = {2009}}

@article{tsagkias-2020-challenges,
	author = {Tsagkias, Manos and King, Tracy Holloway and Kallumadi, Surya and Murdock, Vanessa and {de Rijke}, Maarten},
	journal = {SIGIR Forum},
	number = {1},
	title = {Challenges and Research Opportunities in eCommerce Search and Recommendations},
	volume = {54},
	year = {2020}}

@inproceedings{jin2018explicit,
	author = {Jin, Xisen and Lei, Wenqiang and Ren, Zhaochun and Chen, Hongshen and Liang, Shangsong and Zhao, Yihong and Yin, Dawei},
	booktitle = {Proceedings of CIKM},
	pages = {1403--1412},
	title = {Explicit State Tracking with Semi-Supervisionfor Neural Dialogue Generation},
	year = {2018}}

@inproceedings{tran2017semantic,
	author = {Tran, Van-Khanh and Nguyen, Le-Minh},
	booktitle = {Proceedings of PACLING},
	pages = {63--75},
	title = {Semantic Refinement GRU-Based Neural Language Generation for Spoken Dialogue Systems},
	year = {2017}}

@inproceedings{duvsek2016sequence,
	author = {Du{\v{s}}ek, Ondrej and Jurc{\i}cek, Filip},
	booktitle = {Proceedings of ACL},
	pages = {45},
	title = {Sequence-to-Sequence Generation for Spoken Dialogue via Deep Syntax Trees and Strings},
	year = {2016}}

@inproceedings{zhou2016context,
	author = {Zhou, Hao and Huang, Minlie and others},
	booktitle = {Proceedings of COLING},
	pages = {2032--2041},
	title = {Context-aware Natural Language Generation for Spoken Dialogue Systems},
	year = {2016}}

@inproceedings{wen2015semantically,
	author = {Wen, Tsung-Hsien and Gasic, Milica and Mrk{\v{s}}i{\'c}, Nikola and Su, Pei-Hao and Vandyke, David and Young, Steve},
	booktitle = {Proceedings of EMNLP},
	pages = {1711--1721},
	title = {Semantically Conditioned LSTM-based Natural Language Generation for Spoken Dialogue Systems},
	year = {2015}}

@inproceedings{eric2017copy,
	author = {Eric, Mihail and Manning, Christopher D},
	booktitle = {Proceedings of EACL},
	pages = {468--473},
	title = {A Copy-Augmented Sequence-to-Sequence Architecture Gives Good Performance on Task-Oriented Dialogue},
	year = {2017}}

@article{Young2013POMDP,
	author = {Young, Steve and Ga{\v{s}}i{\'c}, Milica and Thomson, Blaise and Williams, Jason D},
	journal = {Proceedings of IEEE},
	number = {5},
	pages = {1160--1179},
	publisher = {IEEE},
	title = {POMDP-Based Statistical Spoken Dialog Systems: A Review},
	volume = {101},
	year = {2013}}

@inproceedings{Ritter2011Data,
	author = {Ritter, Alan and Cherry, Colin and Dolan, William B.},
	booktitle = {Proceedings of EMNLP},
	pages = {583--593},
	title = {Data-Driven Response Generation in Social Media},
	year = {2011}}

@inproceedings{Banchs2013IRIS,
	author = {Banchs, Rafael E. and Li, Haizhou},
	booktitle = {Proceedings of ACL},
	pages = {37--42},
	title = {IRIS: A Chat-oriented Dialogue System based on the Vector Space Model},
	year = {2013}}

@inproceedings{Ameixa2014Luke,
	author = {Ameixa, David and Coheur, Luisa and Fialho, Pedro and Quaresma, Paulo},
	booktitle = {Proceedings of IVA},
	pages = {13--21},
	publisher = {Springer International Publishing},
	title = {Luke, I am Your Father: Dealing with Out-of-Domain Requests by Using Movies Subtitles},
	year = {2014}}

@inproceedings{lei2018sequicity,
	author = {Lei, Wenqiang and Jin, Xisen and Kan, Min-Yen and Ren, Zhaochun and He, Xiangnan and Yin, Dawei},
	booktitle = {Proceedings of ACL},
	pages = {1437--1447},
	title = {{S}equicity: Simplifying Task-oriented Dialogue Systems with Single Sequence-to-Sequence Architectures},
	year = {2018}}

@inproceedings{Kingma2014Auto,
	author = {Kingma, Diederik P and Welling, Max},
	booktitle = {Proceedings of ICLR},
	title = {Auto-Encoding Variational Bayes},
	year = {2014}}

@article{kim2021deep,
	author = {Kim, Kyungwon and Kwon, Eun and Park, Jaram},
	journal = {IEEE Access},
	pages = {9812--9821},
	publisher = {IEEE},
	title = {Deep User Segment Interest Network Modeling for Click-through Rate Prediction of Online Advertising},
	volume = {9},
	year = {2021}}

@inproceedings{ren2018information,
	author = {Ren, Zhaochun and He, Xiangnan and Yin, Dawei and de Rijke, Maarten},
	booktitle = {Proceedings of SIGIR},
	pages = {1379--1382},
	title = {Information Discovery in E-commerce: Half-day SIGIR 2018 Tutorial},
	year = {2018}}

@inproceedings{han2016twitter,
	author = {Han, Bo and Rahimi, Afshin and Derczynski, Leon and Baldwin, Timothy},
	booktitle = {Proceedings of Workshop on Noisy User-generated Text},
	pages = {213--217},
	title = {Twitter Geolocation Prediction Shared Task of the 2016 Workshop on Noisy User-generated Text},
	year = {2016}}

@inproceedings{gao2020paraphrase,
	author = {Gao, Silin and Zhang, Yichi and Ou, Zhijian and Yu, Zhou},
	booktitle = {Proceedings of ACL},
	title = {Paraphrase Augmented Task-Oriented Dialog Generation},
	year = {2020}}

@inproceedings{liu2018knowledge,
	author = {Liu, Shuman and Chen, Hongshen and Ren, Zhaochun and Feng, Yang and Liu, Qun and Yin, Dawei},
	booktitle = {Proceedings of ACL},
	pages = {1489--1498},
	title = {Knowledge Diffusion for Neural Dialogue Generation},
	year = {2018}}

@book{chuklin-click-2015,
	author = {Chuklin, Aleksandr and Markov, Ilya and {de Rijke}, Maarten},
	publisher = {Morgan \& Claypool Publishers},
	series = {Synthesis Lectures on Information Concepts, Retrieval, and Services},
	title = {Click Models for Web Search},
	year = {2015}}

@inproceedings{LiRCRLM17,
	author = {Jing Li and Pengjie Ren and Zhumin Chen and Zhaochun Ren and Tao Lian and Jun Ma},
	booktitle = {Proceedings of CIKM},
	pages = {1419--1428},
	title = {Neural Attentive Session-based Recommendation},
	year = {2017}}

@inproceedings{He2017NCF,
	author = {He, Xiangnan and Liao, Lizi and Zhang, Hanwang and Nie, Liqiang and Hu, Xia and Chua, Tat-Seng},
	booktitle = {Proceedings of Web Conference},
	pages = {173--182},
	publisher = {ACM},
	title = {Neural Collaborative Filtering},
	year = {2017}}

@article{zhang2019deep,
	author = {Zhang, Shuai and Yao, Lina and Sun, Aixin and Tay, Yi},
	journal = {ACM Computing Surveys},
	number = {1},
	pages = {1--38},
	publisher = {ACM},
	title = {Deep Learning Based Recommender System: A Survey and New Perspectives},
	volume = {52},
	year = {2019}}

@inproceedings{Li2017,
	author = {Li, Piji and Wang, Zihao and Ren, Zhaochun and Bing, Lidong and Lam, Wai},
	booktitle = {Proceedings of SIGIR},
	pages = {345--354},
	title = {Neural Rating Regression with Abstractive Tips Generation for Recommendation},
	year = {2017}}

@article{Koren2009Matrix,
	author = {Koren, Yehuda and Bell, Robert and Volinsky, Chris},
	journal = {Computer},
	number = {8},
	pages = {30--37},
	publisher = {IEEE},
	title = {Matrix Factorization Techniques for Recommender Systems},
	volume = {42},
	year = {2009}}

@inproceedings{Kingma2014Adam,
	author = {Diederik P. Kingma and Jimmy Ba},
	booktitle = {Proceedings of ICLR},
	title = {Adam: A Method for Stochastic Optimization},
	year = {2015}}

@inproceedings{Wang2018Improving,
	author = {Wang, Zihan and Jiang, Ziheng and Ren, Zhaochun and Tang, Jiliang and Yin, Dawei},
	booktitle = {Proceedings of WSDM},
	pages = {619--627},
	title = {A Path-constrained Framework for Discriminating Substitutable and Complementary Products in E-commerce},
	year = {2018}}

@inproceedings{ziwwsdm1,
	author = {Chen, Hongshen and Ren, Zhaochun and Tang, Jiliang and Zhao, Yihong E. and Yin, Dawei},
	booktitle = {Proceedings of Web Conference},
	title = {Hierarchical Variational Memory Network for Dialogue Generation},
	year = {2018}}

@inproceedings{papernot2016semi,
	author = {Papernot, Nicolas and Abadi, Mart{\'\i}n and Erlingsson, {\'U}lfar and Goodfellow, Ian and Talwar, Kunal},
	booktitle = {Proceedings of ICLR},
	title = {Semi-supervised Knowledge Transfer for Deep Learning from Private Training Data},
	year = {2016}}

@article{Chen2017A,
	author = {Chen, Hongshen and Liu, Xiaorui and Yin, Dawei and Tang, Jiliang},
	journal = {ACM SIGKDD Explorations Newsletter},
	number = {2},
	publisher = {ACM},
	title = {A Survey on Dialogue Systems: Recent Advances and New Frontiers},
	volume = {19},
	year = {2017}}

@inproceedings{zhouwsdm2018,
	author = {Zhou, Meizi and Ding, Zhuoye and Jiang, Ziheng and Yin, Dawei},
	booktitle = {Proceedings of WSDM},
	pages = {727--735},
	title = {Micro Behaviors: A New Perspective in E-commerce Recommender Systems},
	year = {2018}}

@inproceedings{he2016fast,
	author = {He, Xiangnan and Zhang, Hanwang and Kan, Min-Yen and Chua, Tat-Seng},
	booktitle = {Proceedings of SIGIR},
	pages = {549--558},
	title = {Fast Matrix Factorization for Online Recommendation with Implicit Feedback},
	year = {2016}}

@inproceedings{degenhardt-ecom-2017,
	author = {Degenhardt, Jon and Kallumadi, Surya and {de Rijke}, Maarten and Si, Luo and Trotman, Andrew and Yinghui, Xu},
	booktitle = {Proceedings of SIGIR},
	title = {{eCom}: The {SIGIR} 2017 Workshop on eCommerce},
	year = {2017}}

@inproceedings{kenter-neural-2017,
	author = {Kenter, Tom and Borisov, Alexey and {Van Gysel}, Christophe and Dehghani, Mostafa and {de Rijke}, Maarten and Mitra, Bhaskar},
	booktitle = {Proceedings of SIGIR},
	title = {Neural Networks for Information Retrieval ({NN4IR})},
	year = {2017}}

@inproceedings{zoghi-copeland-2015,
	author = {Zoghi, Masrour and Whiteson, Shimon and Karnin, Zohar and {de Rijke}, Maarten},
	booktitle = {Proceedings of NIPS},
	pages = {307--315},
	title = {Copeland Dueling Bandits},
	year = {2015}}

@inproceedings{oosterhuis-balancing-2017,
	author = {Oosterhuis, Harrie and {de Rijke}, Maarten},
	booktitle = {Proceedings of CIKM},
	pages = {277--286},
	title = {Balancing Speed and Quality in Online Learning to Rank for Information Retrieval},
	year = {2017}}

@inproceedings{schuth-multileave-2016,
	author = {Schuth, Anne and Oosterhuis, Harrie and Whiteson, Shimon and {de Rijke}, Maarten},
	booktitle = {Proceedings of WSDM},
	pages = {457--466},
	title = {Multileave Gradient Descent for Fast Online Learning to Rank},
	year = {2016}}

@inproceedings{ai2017learning,
	author = {Ai, Qingyao and Zhang, Yongfeng and Bi, Keping and Chen, Xu and Croft, W Bruce},
	booktitle = {Proceedings of SIGIR},
	pages = {645--654},
	title = {Learning a Hierarchical Embedding Model for Personalized Product Search},
	year = {2017}}

@inproceedings{yi2014beyond,
	author = {Yi, Xing and Hong, Liangjie and Zhong, Erheng and Liu, Nanthan Nan and Rajan, Suju},
	booktitle = {Proceedings of RecSys},
	pages = {113--120},
	title = {Beyond Clicks: Dwell Time for Personalization},
	year = {2014}}

@inproceedings{duan2013probabilistic,
	author = {Duan, Huizhong and Zhai, ChengXiang and Cheng, Jinxing and Gattani, Abhishek},
	booktitle = {Proceedings of CIKM},
	pages = {2179--2188},
	title = {A Probabilistic Mixture Model for Mining and Analyzing Product Search Log},
	year = {2013}}

@inproceedings{he2016ups,
	author = {He, Ruining and McAuley, Julian},
	booktitle = {Proceedings of Web Conference},
	pages = {507--517},
	title = {Ups and Downs: Modeling the Visual Evolution of Fashion Trends with One-class Collaborative Filtering},
	year = {2016}}

@inproceedings{yu2011domain,
	author = {Yu, Jianxing and Zha, Zheng-Jun and Wang, Meng and Wang, Kai and Chua, Tat-Seng},
	booktitle = {Proceedings of EMNLP},
	pages = {140--150},
	title = {Domain-Assisted Product Aspect Hierarchy Generation: Towards Hierarchical Organization of Unstructured Consumer Reviews},
	year = {2011}}

@inproceedings{mcauley2015image,
	author = {McAuley, Julian and Targett, Christopher and Shi, Qinfeng and Van Den Hengel, Anton},
	booktitle = {Proceedings of SIGIR},
	pages = {43--52},
	title = {Image-Based Recommendations on Styles and Substitutes},
	year = {2015}}

@article{duan2013supporting,
	author = {Duan, Huizhong and Zhai, ChengXiang and Cheng, Jinxing and Gattani, Abhishek},
	journal = {Proceedings of VLDB},
	number = {14},
	pages = {1786--1797},
	publisher = {VLDB Endowment},
	title = {Supporting Keyword Search in Product Database: A Probabilistic Approach},
	volume = {6},
	year = {2013}}

@article{jansen2006effectiveness,
	author = {Jansen, Bernard J and Molina, Paulo R},
	journal = {Information Processing \& Management},
	number = {4},
	pages = {1075--1098},
	publisher = {Elsevier},
	title = {The Effectiveness of Web Search Engines for Retrieving Relevant Ecommerce Links},
	volume = {42},
	year = {2006}}

@inproceedings{nurmi2008product,
	author = {Nurmi, Petteri and Lagerspetz, Eemil and Buntine, Wray and Flor{\'e}en, Patrik and Kukkonen, Joonas},
	booktitle = {Proceedings of SIGIR},
	pages = {781--782},
	title = {Product Retrieval for Grocery Stores},
	year = {2008}}

@article{rowley2000product,
	author = {Rowley, Jennifer},
	journal = {Journal of Consumer Marketing},
	number = {1},
	pages = {20--35},
	publisher = {MCB UP Ltd},
	title = {Product Search in E-shopping: A Review and Research Propositions},
	volume = {17},
	year = {2000}}

@inproceedings{van2016learning,
	author = {{Van Gysel}, Christophe and {de Rijke}, Maarten and Kanoulas, Evangelos},
	booktitle = {Proceedings of CIKM},
	pages = {165--174},
	title = {Learning Latent Vector Spaces for Product Search},
	year = {2016}}

@inproceedings{lo2016understanding,
	author = {Lo, Caroline and Frankowski, Dan and Leskovec, Jure},
	booktitle = {Proceedings of KDD},
	pages = {531--540},
	title = {Understanding Behaviors That Lead to Purchasing: A Case Study of Pinterest},
	year = {2016}}

@inproceedings{zhang2014explicit,
	author = {Zhang, Yongfeng and Lai, Guokun and Zhang, Min and Zhang, Yi and Liu, Yiqun and Ma, Shaoping},
	booktitle = {Proceedings of SIGIR},
	pages = {83--92},
	title = {Explicit Factor Models for Explainable Recommendation Based on Phrase-level Sentiment Analysis},
	year = {2014}}

@article{mitra2017introduction,
	author = {Mitra, Bhaskar and Craswell, Nick},
	journal = {Foundations and Trends in Information Retrieval},
	title = {An Introduction to Neural Information Retrieval},
	year = {2017},
    pages = {1--126},
    volume = 13,
    numer=1}

@phdthesis{InfNeed4RS,
	address = {Minneapolis, MN, USA},
	advisor = {Konstan, Joseph A.},
	author = {Mcnee, Sean Michael},
	publisher = {University of Minnesota},
	school = {University of Minnesot},
	title = {Meeting User Information Needs in Recommender Systems},
	year = {2006}}

@inproceedings{purchaserate1,
	author = {Wang, Jian and Sarwar, Badrul and Sundaresan, Neel},
	booktitle = {Proceedings of RecSys},
	pages = {329--332},
	title = {Utilizing Related Products for Post-Purchase Recommendation in E-Commerce},
	year = {2011}}

@article{kim2005development,
	author = {Kim, Yong Soo and Yum, Bong-Jin and Song, Junehwa and Kim, Su Myeon},
	journal = {Expert Systems with Applications},
	number = {2},
	pages = {381--393},
	title = {Development of a Recommender System based on Navigational and Behavioral Patterns of Customers in E-Commerce Sites},
	volume = {28},
	year = {2005}}

@inproceedings{kim2007impact,
	author = {Kim, Young and Srivastava, Jaideep},
	booktitle = {Proceedings of EC},
	pages = {293--302},
	title = {Impact of Social Influence in E-Commerce Decision Making},
	year = {2007}}

@inproceedings{jiang2015life,
	author = {Jiang, Peng and Zhu, Yadong and Zhang, Yi and Yuan, Quan},
	booktitle = {Proceedings of KDD},
	pages = {1879--1888},
	title = {Life-Stage Prediction for Product Recommendation in E-Commerce},
	year = {2015}}

@inproceedings{RS4EC,
	author = {Sarwar, Badrul and Karypis, George and Konstan, Joseph and Riedl, John},
	booktitle = {Proceedings of EC},
	pages = {158--167},
	title = {Analysis of Recommendation Algorithms for E-commerce},
	year = {2000}}

@inproceedings{yin2011exploiting,
	author = {Yin, Dawei and Hong, Liangjie and Davison, Brian D},
	booktitle = {Proceedings of Web Conference},
	pages = {167--168},
	title = {Exploiting Session-like Behaviors in Tag Prediction},
	year = {2011}}

@inproceedings{ren2019lifelong,
	author = {Ren, Kan and Qin, Jiarui and Fang, Yuchen and Zhang, Weinan and Zheng, Lei and Bian, Weijie and Zhou, Guorui and Xu, Jian and Yu, Yong and Zhu, Xiaoqiang and others},
	booktitle = {Proceedings of SIGIR},
	pages = {565--574},
	title = {Lifelong Sequential Modeling with Personalized Memorization for User Response Prediction},
	year = {2019}}

@inproceedings{pi2019practice,
	author = {Pi, Qi and Bian, Weijie and Zhou, Guorui and Zhu, Xiaoqiang and Gai, Kun},
	booktitle = {Proceedings of KDD},
	pages = {2671--2679},
	title = {Practice on Long Sequential User Behavior Modeling for Click-through Rate Prediction},
	year = {2019}}

@article{liu2023enhancing,
	author = {Liu, Xin and Li, Zheng and Gao, Yifan and Yang, Jingfeng and Cao, Tianyu and Wang, Zhengyang and Yin, Bing and Song, Yangqiu},
	journal = {arXiv preprint arXiv:2312.16199},
	title = {Enhancing User Intent Capture in Session-Based Recommendation with Attribute Patterns},
	year = {2023}}

@inproceedings{cen2020controllable,
	author = {Cen, Yukuo and Zhang, Jianwei and Zou, Xu and Zhou, Chang and Yang, Hongxia and Tang, Jie},
	booktitle = {Proceedings of KDD},
	pages = {2942--2951},
	title = {Controllable Multi-interest Framework for Recommendation},
	year = {2020}}

@inproceedings{zhu2022bars,
	author = {Zhu, Jieming and Dai, Quanyu and Su, Liangcai and Ma, Rong and Liu, Jinyang and Cai, Guohao and Xiao, Xi and Zhang, Rui},
	booktitle = {Proceedings of SIGIR},
	pages = {2912--2923},
	title = {Bars: Towards Open Benchmarking for Recommender Systems},
	year = {2022}}

@inproceedings{zhu2021open,
	author = {Zhu, Jieming and Liu, Jinyang and Yang, Shuai and Zhang, Qi and He, Xiuqiang},
	booktitle = {Proceedings of CIKM},
	pages = {2759--2769},
	title = {Open Benchmarking for Click-through Rate Prediction},
	year = {2021}}

@inproceedings{xiao2020deep,
	author = {Xiao, Zhibo and Yang, Luwei and Jiang, Wen and Wei, Yi and Hu, Yi and Wang, Hao},
	booktitle = {Proceedings of CIKM},
	pages = {2265--2268},
	title = {Deep Multi-interest Network for Click-through Rate Prediction},
	year = {2020}}

@inproceedings{chang2023twin,
	author = {Chang, Jianxin and Zhang, Chenbin and Fu, Zhiyi and Zang, Xiaoxue and Guan, Lin and Lu, Jing and Hui, Yiqun and Leng, Dewei and Niu, Yanan and Song, Yang and others},
	booktitle = {Proceedings of KDD},
	pages = {3785--3794},
	title = {TWIN: Two-stage Interest Network for Lifelong User Behavior Modeling in CTR Prediction at Kuaishou},
	year = {2023}}

@inproceedings{song2017summarizing,
	author = {Song, Hongya and Ren, Zhaochun and Liang, Shangsong and Li, Piji and Ma, Jun and {de Rijke}, Maarten},
	booktitle = {Proceedings of WSDM},
	pages = {405--414},
	title = {Summarizing Answers in Non-factoid Community Question-answering},
	year = {2017}}

@inproceedings{he2014practical,
	author = {He, Xinran and Pan, Junfeng and Jin, Ou and Xu, Tianbing and Liu, Bo and Xu, Tao and Shi, Yanxin and Atallah, Antoine and Herbrich, Ralf and Bowers, Stuart and others},
	booktitle = {Proceedings of ADKDD},
	pages = {1--9},
	title = {Practical Lessons from Predicting Clicks on Ads at Facebook},
	year = {2014}}

@inproceedings{deng2012use,
	author = {Deng, Li and Tur, Gokhan and He, Xiaodong and Hakkani-Tur, Dilek},
	booktitle = {Proceedings of SLT},
	pages = {210--215},
	title = {Use of Kernel Deep Convex Networks and End-to-end Learning for Spoken Language Understanding},
	year = {2012}}

@inproceedings{sarikaya2011deep,
	author = {Sarikaya, Ruhi and Hinton, Geoffrey E and Ramabhadran, Bhuvana},
	booktitle = {Proceedings of ICASSP},
	pages = {5680--5683},
	title = {Deep Belief Nets for Natural Language Call-routing},
	year = {2011}}

@inproceedings{yao2013recurrent,
	author = {Yao, Kaisheng and Zweig, Geoffrey and Hwang, Mei-Yuh and Shi, Yangyang and Yu, Dong},
	booktitle = {Proceedings of Interspeech},
	pages = {2524--2528},
	title = {Recurrent Neural Networks for Language Understanding},
	year = {2013}}

@inproceedings{yao2014spoken,
	author = {Yao, Kaisheng and Peng, Baolin and Zhang, Yu and Yu, Dong and Zweig, Geoffrey and Shi, Yangyang},
	booktitle = {Proceedings of SLT},
	pages = {189--194},
	title = {Spoken Language Understanding using Long Short-term Memory Neural Networks},
	year = {2014}}

@inproceedings{wang2013simple,
	author = {Wang, Zhuoran and Lemon, Oliver},
	booktitle = {Proceedings of SIGDIAL},
	pages = {423--432},
	title = {A Simple and Generic Belief Tracking Mechanism for the Dialog State Tracking Challenge: On the believability of observed information},
	year = {2013}}

@article{young2010hidden,
	author = {Young, Steve and Gasic, Milica and Keizer, Simon and Mairesse, Fran{\c{c}}ois and Schatzmann, Jost and Thomson, Blaise and Yu, Kai},
	journal = {Computer Speech \& Language},
	number = {2},
	pages = {150--174},
	publisher = {Elsevier},
	title = {The Hidden Information State model: A Practical Framework for POMDP-based Spoken Dialogue Management},
	volume = {24},
	year = {2010}}

@inproceedings{williams2012belief,
	author = {Williams, Jason},
	booktitle = {Proceedings of NAACL},
	pages = {23--24},
	title = {A Belief Tracking Challenge Task for Spoken Dialog Systems},
	year = {2012}}

@inproceedings{lee2013structured,
	author = {Lee, Sungjin},
	booktitle = {Proceedings of SIGDIAL},
	pages = {442--451},
	title = {Structured Discriminative Model For Dialog State Tracking},
	year = {2013}}

@inproceedings{lee2013recipe,
	author = {Lee, Sungjin and Eskenazi, Maxine},
	booktitle = {Proceedings of SIGDIAL},
	pages = {414--422},
	title = {Recipe For Building Robust Spoken Dialog State Trackers: Dialog State Tracking Challenge System Description},
	year = {2013}}

@inproceedings{ren2013dialog,
	author = {Ren, Hang and Xu, Weiqun and Zhang, Yan and Yan, Yonghong},
	booktitle = {Proceedings of SIGDIAL},
	pages = {457--461},
	title = {Dialog State Tracking using Conditional Random Fields},
	year = {2013}}

@inproceedings{mesnil2013investigation,
	author = {Mesnil, Gr{\'e}goire and He, Xiaodong and Deng, Li and Bengio, Yoshua},
	booktitle = {Proceedings of Interspeech},
	pages = {3771--3775},
	title = {Investigation of Recurrent-Neural-Network Architectures and Learning Methods for Spoken Language Understanding},
	year = {2013}}

@inproceedings{williams2013multi,
	author = {Williams, Jason},
	booktitle = {Proceedings of SIGDIAL},
	pages = {433--441},
	title = {Multi-domain Learning and Generalization in Dialog State Tracking},
	year = {2013}}

@inproceedings{deoras2013deep,
	author = {Deoras, Anoop and Sarikaya, Ruhi},
	booktitle = {Proceedings of Interspeech},
	pages = {2713--2717},
	title = {Deep Belief Network based Semantic Taggers for Spoken Language Understanding},
	year = {2013}}

@inproceedings{henderson2013deep,
	author = {Henderson, Matthew and Thomson, Blaise and Young, Steve},
	booktitle = {Proceedings of SIGDIAL},
	pages = {467--471},
	title = {Deep Neural Network Approach for the Dialog State Tracking Challenge},
	year = {2013}}

@inproceedings{tur2012towards,
	author = {Tur, Gokhan and Deng, Li and Hakkani-T{\"u}r, Dilek and He, Xiaodong},
	booktitle = {Proceedings of ICASSP},
	pages = {5045--5048},
	title = {Towards Deeper Understanding: Deep Convex Networks for Semantic Utterance Classification},
	year = {2012}}

@inproceedings{bordes2016learning,
	author = {Bordes, Antoine and Weston, Jason},
	booktitle = {Proceedings of ICLR},
	title = {Learning End-to-End Goal-Oriented Dialog},
	year = {2017}}

@inproceedings{mrkvsic2017neural,
	author = {Mrk{\v{s}}i{\'c}, Nikola and S{\'e}aghdha, Diarmuid {\'O} and Wen, Tsung-Hsien and Thomson, Blaise and Young, Steve},
	booktitle = {Proceedings of ACL},
	pages = {1777--1788},
	title = {Neural Belief Tracker: Data-Driven Dialogue State Tracking},
	year = {2017}}

@article{cuayahuitl2015strategic,
	author = {Cuay{\'a}huitl, Heriberto and Keizer, Simon and Lemon, Oliver},
	journal = {arXiv preprint arXiv:1511.08099},
	title = {Strategic Dialogue Management via Deep Reinforcement Learning},
	year = {2015}}

@inproceedings{yan2017building,
	author = {Yan, Zhao and Duan, Nan and Chen, Peng and Zhou, Ming and Zhou, Jianshe and Li, Zhoujun},
	booktitle = {Proceedings of AAAI},
	pages = {4618--4625},
	title = {Building Task-oriented Dialogue Systems for Online Shopping},
	year = {2017}}

@inproceedings{zhao2016towards,
	author = {Zhao, Tiancheng and Eskenazi, Maxine},
	booktitle = {Proceedings of Annual Meeting of the Special Interest Group on Discourse and Dialogue},
	pages = {1},
	title = {Towards End-to-End Learning for Dialog State Tracking and Management using Deep Reinforcement Learning},
	year = {2016}}

@inproceedings{eric2017key,
	author = {Eric, Mihail and Krishnan, Lakshmi and Charette, Francois and Manning, Christopher D},
	booktitle = {Proceedings of SIGDIAL},
	pages = {37--49},
	title = {Key-Value Retrieval Networks for Task-Oriented Dialogue},
	year = {2017}}

@inproceedings{geigle2016scaling,
	author = {Geigle, Chase and Zhai, ChengXiang},
	booktitle = {Proceedings of ACM Conference on Learning @ Scale},
	pages = {257--260},
	title = {Scaling up Online Question Answering via Similar Question Retrieval},
	year = {2016}}

@inproceedings{liu2016retrieving,
	author = {Liu, Mengwen and Fang, Yi and Park, Dae Hoon and Hu, Xiaohua and Yu, Zhengtao},
	booktitle = {Proceedings of SIGIR},
	pages = {385--394},
	title = {Retrieving Non-Redundant Questions to Summarize a Product Review},
	year = {2016}}

@inproceedings{lalmas2018tutorial,
	author = {Lalmas, Mounia and Hong, Liangjie},
	booktitle = {Proceedings of WSDM},
	pages = {781--782},
	title = {Tutorial on Metrics of User Engagement: Applications to News, Search and E-Commerce},
	year = {2018}}

@inproceedings{perozzi2014deepwalk,
	author = {Perozzi, Bryan and Al-Rfou, Rami and Skiena, Steven},
	booktitle = {Proceedings of KDD},
	pages = {701--710},
	title = {DeepWalk: Online Learning of Social Representations},
	year = {2014}}

@inproceedings{chen2016xgboost,
	author = {Chen, Tianqi and Guestrin, Carlos},
	booktitle = {Proceedings of KDD},
	pages = {785--794},
	title = {XGBoost: A Scalable Tree Boosting System},
	year = {2016}}

@inproceedings{cheng2016wide,
	author = {Cheng, Heng-Tze and Koc, Levent and Harmsen, Jeremiah and Shaked, Tal and Chandra, Tushar and Aradhye, Hrishi and Anderson, Glen and Corrado, Greg and Chai, Wei and Ispir, Mustafa and others},
	booktitle = {Proceedings of Workshop on Deep Learning for Recommender Systems},
	pages = {7--10},
	title = {Wide \& Deep Learning for Recommender Systems},
	year = {2016}}

@inproceedings{wang2017factorization,
	author = {Wang, Huazheng and Wu, Qingyun and Wang, Hongning},
	booktitle = {Proceedings of AAAI},
	pages = {2695--2702},
	title = {Factorization Bandits for Interactive Recommendation.},
	year = {2017}}

@inproceedings{yu2017modelling,
	author = {Yu, Jianfei and Qiu, Minghui and Jiang, Jing and Huang, Jun and Song, Shuangyong and Chu, Wei and Chen, Haiqing},
	booktitle = {Proceedings of WSDM},
	pages = {682--690},
	title = {Modelling Domain Relationships for Transfer Learning on Retrieval-based Question Answering Systems in E-commerce},
	year = {2018}}

@inproceedings{yu2012answering,
	author = {Yu, Jianxing and Zha, Zheng-Jun and Chua, Tat-Seng},
	booktitle = {Proceedings of EMNLP-CoNLL},
	pages = {391--401},
	title = {Answering Opinion Questions on Products by Exploiting Hierarchical Organization of Consumer Reviews},
	year = {2012}}

@inproceedings{mcauley2016addressing,
	author = {McAuley, Julian and Yang, Alex},
	booktitle = {Proceedings of Web Conference},
	pages = {625--635},
	title = {Addressing Complex and Subjective Product-Related Queries with Customer Reviews},
	year = {2016}}

@article{tapeh2008knowledge,
	author = {Tapeh, Ali Ghobadi and Rahgozar, Maseud},
	journal = {Knowledge-Based Systems},
	number = {8},
	pages = {946--950},
	publisher = {Elsevier},
	title = {A Knowledge-based Question Answering System for B2C eCommerce},
	volume = {21},
	year = {2008}}

@inproceedings{yin2015neural,
	author = {Yin, Jun and Jiang, Xin and Lu, Zhengdong and Shang, Lifeng and Li, Hang and Li, Xiaoming},
	booktitle = {Proceedings of IJCAI},
	pages = {2972--2978},
	title = {Neural Generative Question Answering},
	year = {2016}}

@inproceedings{wang2018multi,
	author = {Wang, Jingang and Tian, Junfeng and Qiu, Long and Li, Sheng and Lang, Jun and Si, Luo and Lan, Man},
	booktitle = {Proceedings of AAAI},
	pages = {451--458},
	title = {A Multi-task Learning Approach for Improving Product Title Compression with User Search Log Data},
	year = {2018}}

@inproceedings{vanderveld2016engagement,
	author = {Vanderveld, Ali and Pandey, Addhyan and Han, Angela and Parekh, Rajesh},
	booktitle = {Proceedings of KDD},
	pages = {293--302},
	title = {An Engagement-Based Customer Lifetime Value System for E-Commerce},
	year = {2016}}

@inproceedings{Huangwww2018,
	author = {Huang, Hong and Zhao, Bo and Zhao, Hao and Zhuang, Zhou and Wang, Zhenxuan and Yao, Xiaoming and Wang, Xinggang and Jin, Hai and Fu, Xiaoming},
	booktitle = {Proceedings of Web Conference},
	pages = {1785--1794},
	title = {A Cross-Platform Consumer Behavior Analysis of Large-Scale Mobile Shopping Data},
	year = {2018}}

@article{kanoje2015user,
	author = {Kanoje, Sumitkumar and Girase, Sheetal and Mukhopadhyay, Debajyoti},
	journal = {arXiv preprint arXiv:1503.07474},
	title = {User Profiling Trends, Techniques and Applications},
	year = {2015}}

@inproceedings{zou2018drlunderreview,
	author = {Zou, Lixin and Xia, Long and Ding, Zhuoye and Liu, Weidong and Zhao, Yihong and Yin, Dawei},
	booktitle = {Proceedings of KDD},
	pages = {95--103},
	title = {Reinforcement Learning to Optimize Long-term User Engagement in Recommender Systems},
	year = {2020}}

@inproceedings{guoziyi2018,
	author = {Liu, Yiding and Gu, Yulong and Ding, Zhuoye and Gao, Junchao and Guo, Ziyi and Bao, Yongjun and Yan, Weipeng},
	booktitle = {Proceedings of CIKM},
	pages = {2621--2628},
	title = {Decoupled Graph Convolution Network for Inferring Substitutable and Complementary Items},
	year = {2020}}

@inproceedings{fawcett1996combining,
	author = {Fawcett, Tom and Provost, Foster J},
	booktitle = {Proceedings of KDD},
	pages = {8--13},
	title = {Combining Data Mining and Machine Learning for Effective User Profiling.},
	year = {1996}}

@inproceedings{adomavicius1999user,
	author = {Adomavicius, Gediminas and Tuzhilin, Alexander},
	booktitle = {Proceedings of KDD},
	pages = {377--381},
	title = {User Profiling in Personalization Applications through Rule Discovery and Validation},
	year = {1999}}

@inproceedings{zhang2013predicting,
	author = {Zhang, Yongzheng and Pennacchiotti, Marco},
	booktitle = {Proceedings of Web Conference},
	pages = {1521--1532},
	title = {Predicting Purchase Behaviors from Social Media},
	year = {2013}}

@inproceedings{gupta2014identifying,
	author = {Gupta, Vineet and Varshney, Devesh and Jhamtani, Harsh and Kedia, Deepam and Karwa, Shweta},
	booktitle = {Proceedings of ICWSM},
	pages = {180--186},
	title = {Identifying Purchase Intent from Social Posts},
	year = {2014}}

@inproceedings{rahdari2017analysis,
	author = {Rahdari, Behnam and Arabghalizi, Tahereh and Brambilla, Marco},
	booktitle = {Proceedings of International Cross-Domain Conference},
	pages = {219--236},
	title = {Analysis of Online User Behaviour for Art and Culture Events},
	year = {2017}}

@inproceedings{hollerit2013towards,
	author = {Hollerit, Bernd and Kr{\"o}ll, Mark and Strohmaier, Markus},
	booktitle = {Proceedings of Web Conference},
	pages = {629--632},
	title = {Towards Linking Buyers and Sellers: Detecting Commercial Intent on Twitter},
	year = {2013}}

@article{solomon1994buying,
	author = {Solomon, Michael R and Behavior, Consumer},
	journal = {London: Prenticle Hall},
	title = {Consumer Buying, Having and Being},
	year = {1994}}

@article{braynov2003personalization,
	author = {Braynov, Sviatoslav},
	journal = {The Internet Encyclopedia},
	publisher = {Wiley Online Library},
	title = {Personalization and Customization Technologies},
	year = {2003}}

@inproceedings{kuflik2000generation,
	author = {Kuflik, Tsvi and Shoval, Peretz},
	booktitle = {Proceedings of SIGIR},
	pages = {313--315},
	title = {Generation of User Profiles for Information Filtering --- Research Agenda},
	year = {2000}}

@article{cufoglu2014user,
	author = {Cufoglu, Ayse},
	journal = {International Journal of Computer Applications},
	number = {3},
	publisher = {Foundation of Computer Science},
	title = {User Profiling-A Short Review},
	volume = {108},
	year = {2014}}

@inproceedings{liu2010iui,
	author = {Liu, Jiahui and Dolan, Peter and Pedersen, Elin R{\o}nby},
	booktitle = {Proceedings of IUI},
	pages = {31--40},
	title = {Personalized News Recommendation Based on Click Behavior},
	year = {2010}}

@inproceedings{ding2015mining,
	author = {Ding, Xiao and Liu, Ting and Duan, Junwen and Nie, Jian-Yun},
	booktitle = {Proceedings of AAAI},
	pages = {2389--2395},
	title = {Mining User Consumption Intention from Social Media Using Domain Adaptive Convolutional Neural Network.},
	year = {2015}}

@inproceedings{gelli2017personality,
	author = {Gelli, Francesco and He, Xiangnan and Chen, Tao and Chua, Tat-Seng},
	booktitle = {Proceedings of MM},
	pages = {1828--1837},
	title = {How Personality Affects our Likes: Towards a Better Understanding of Actionable Images},
	year = {2017}}

@article{tang2010combination,
	author = {Tang, Jie and Yao, Limin and Zhang, Duo and Zhang, Jing},
	journal = {ACM Transactions on Knowledge Discovery from Data (TKDD)},
	number = {1},
	pages = {2},
	publisher = {ACM},
	title = {A Combination Approach to Web User Profiling},
	volume = {5},
	year = {2010}}

@article{godoy2005user,
	author = {Godoy, Daniela and Amandi, Analia},
	journal = {The Knowledge Engineering Review},
	number = {4},
	pages = {329--361},
	publisher = {Cambridge University Press},
	title = {User Profiling in Personal Information Agents: A Survey},
	volume = {20},
	year = {2005}}

@article{adomavicius2005toward,
	author = {Adomavicius, Gediminas and Tuzhilin, Alexander},
	journal = {IEEE Transactions on Knowledge \& Data Engineering},
	number = {6},
	pages = {734--749},
	publisher = {IEEE},
	title = {Toward the Next Generation of Recommender Systems: A Survey of the State-of-the-Art and Possible Extensions},
	volume = {17},
	year = {2005}}

@article{su2009survey,
	author = {Su, Xiaoyuan and Khoshgoftaar, Taghi M},
	journal = {Advances in Artificial Intelligence},
	publisher = {Hindawi},
	title = {A Survey of Collaborative Filtering Techniques},
	volume = {2009},
	year = {2009}}

@inproceedings{mao2014estimating,
	author = {Mao, Jiaxin and Liu, Yiqun and Zhang, Min and Ma, Shaoping},
	booktitle = {Proceedings of SIGIR},
	pages = {263--274},
	title = {Estimating Credibility of User Clicks with Mouse Movement and Eye-Tracking Information},
	year = {2014}}

@inproceedings{su2018detecting,
	author = {Su, Ning and Liu, Yiqun and Li, Zhao and Liu, Yuli and Zhang, Min and Ma, Shaoping},
	booktitle = {Proceedings of Web Conference},
	pages = {1673--1682},
	title = {Detecting Crowdturfing ``Add to Favorites'' Activities in Online Shopping},
	year = {2018}}

@article{bellman1999predictors,
	author = {Bellman, Steven and Lohse, Gerald L and Johnson, Eric J},
	journal = {Communications of the ACM},
	number = {12},
	pages = {32--38},
	publisher = {ACM},
	title = {Predictors of Online Buying Behavior},
	volume = {42},
	year = {1999}}

@inproceedings{mikolov2013efficient,
	author = {Mikolov, Tomas and Chen, Kai and Corrado, Greg and Dean, Jeffrey},
	booktitle = {Proceedings of ICLR},
	title = {Efficient Estimation of Word Representations in Vector Space},
	year = {2013}}

@article{brown2003buying,
	author = {Brown, Mark and Pope, Nigel and Voges, Kevin},
	journal = {European Journal of Marketing},
	number = {11/12},
	pages = {1666--1684},
	publisher = {MCB UP Ltd},
	title = {Buying or Browsing? An Exploration of Shopping Orientations and Online Purchase Intention},
	volume = {37},
	year = {2003}}

@article{ferro2019boosting,
	author = {Ferro, Nicola and Lucchese, Claudio and Maistro, Maria and Perego, Raffaele},
	journal = {Information Retrieval Journal},
	number = {6},
	pages = {1--27},
	publisher = {Springer},
	title = {Boosting Learning to Rank with User dynamics and Continuation Methods},
	volume = {23},
	year = {2019}}

@article{ahmed2021deep,
	author = {Ahmed, Adeel and Saleem, Khalid and Khalid, Osman and Rashid, Umer},
	journal = {Expert Systems with Applications},
	pages = {114757},
	publisher = {Elsevier},
	title = {On Deep Neural Network for Trust Aware Cross Domain Recommendations in E-commerce},
	volume = {174},
	year = {2021}}

@inproceedings{agichtein2006improving,
	author = {Agichtein, Eugene and Brill, Eric and Dumais, Susan},
	booktitle = {Proceedings of SIGIR},
	pages = {19--26},
	title = {Improving Web Search Ranking by Incorporating User Behavior Information},
	year = {2006}}

@inproceedings{wu2018turning,
	author = {Wu, Liang and Hu, Diane and Hong, Liangjie and Liu, Huan},
	booktitle = {Proceedings of SIGIR},
	pages = {365--374},
	title = {Turning Clicks into Purchases: Revenue Optimization for Product Search in E-Commerce},
	year = {2018}}

@article{lu2006clustering,
	author = {Lu, Yiyao and He, Hai and Peng, Qian and Meng, Weiyi and Yu, Clement},
	journal = {Data \& Knowledge Engineering},
	number = {2},
	pages = {231--246},
	publisher = {Elsevier},
	title = {Clustering E-commerce Search Engines based on their Search Interface Pages using WISE-Cluster},
	volume = {59},
	year = {2006}}

@inproceedings{aryafar2017ensemble,
	author = {Aryafar, Kamelia and Guillory, Devin and Hong, Liangjie},
	booktitle = {Proceedings of ADKDD},
	pages = {10},
	title = {An Ensemble-based Approach to Click-Through Rate Prediction for Promoted Listings at Etsy},
	year = {2017}}

@inproceedings{regelson2006predicting,
	author = {Regelson, Moira and Fain, D},
	booktitle = {Proceedings of Second Workshop on Sponsored Search Auctions},
	pages = {1--6},
	title = {Predicting Click-Through Rate Using Keyword Clusters},
	volume = {9623},
	year = {2006}}

@article{anderson2002new,
	author = {Anderson, Philip and Anderson, Erin},
	journal = {MIT Sloan Management Review},
	number = {4},
	pages = {53},
	publisher = {Massachusetts Institute of Technology, Cambridge, MA},
	title = {The New E-commerce Intermediaries},
	volume = {43},
	year = {2002}}

@article{lee2001visualization,
	author = {Lee, Juhnyoung and Podlaseck, Mark and Schonberg, Edith and Hoch, Robert},
	journal = {Data Mining and Knowledge Discovery},
	number = {1-2},
	pages = {59--84},
	publisher = {Springer},
	title = {Visualization and Analysis of Clickstream Data of Online Stores for Understanding Web Merchandising},
	volume = {5},
	year = {2001}}

@inproceedings{rosales2012post,
	author = {Rosales, R{\'o}mer and Cheng, Haibin and Manavoglu, Eren},
	booktitle = {Proceedings of WSDM},
	pages = {293--302},
	title = {Post-Click Conversion Modeling and Analysis for Non-Guaranteed Delivery Display Advertising},
	year = {2012}}

@article{swinyard2004activities,
	author = {Swinyard, William R and Smith, Scott M},
	journal = {International Business and Economics Research Journal},
	pages = {37--48},
	title = {Activities, Interests, and Opinions of Online Shoppers and Non-Shoppers},
	volume = {3},
	year = {2004}}

@inproceedings{li2011towards,
	author = {Li, Beibei and Ghose, Anindya and Ipeirotis, Panagiotis G},
	booktitle = {Proceedings of Web Conference},
	pages = {327--336},
	title = {Towards a Theory Model for Product Search},
	year = {2011}}

@article{kim2003combination,
	author = {Kim, Eunju and Kim, Wooju and Lee, Yillbyung},
	journal = {Decision Support Systems},
	number = {2},
	pages = {167--175},
	publisher = {Elsevier},
	title = {Combination of Multiple Classifiers for the Customer's Purchase Behavior Prediction},
	volume = {34},
	year = {2003}}

@article{sismeiro2004modeling,
	author = {Sismeiro, Catarina and Bucklin, Randolph E},
	journal = {Journal of marketing research},
	number = {3},
	pages = {306--323},
	publisher = {American Marketing Association},
	title = {Modeling Purchase Behavior at an E-Commerce Web Site: A Task-Completion Approach},
	volume = {41},
	year = {2004}}

@article{suh2004prediction,
	author = {Suh, Euiho and Lim, Seungjae and Hwang, Hyunseok and Kim, Suyeon},
	journal = {Expert Systems with Applications},
	number = {2},
	pages = {245--255},
	publisher = {Elsevier},
	title = {A Prediction Model for the Purchase Probability of Anonymous Customers to Support Real Time Web Marketing: A Case Study},
	volume = {27},
	year = {2004}}

@article{van2005predicting,
	author = {Van den Poel, Dirk and Buckinx, Wouter},
	journal = {European Journal of Operational Research},
	number = {2},
	pages = {557--575},
	publisher = {Elsevier},
	title = {Predicting Online-purchasing Behaviour},
	volume = {166},
	year = {2005}}

@article{young2004predicting,
	author = {Young Kim, Eun and Kim, Youn-Kyung},
	journal = {European Journal of Marketing},
	number = {7},
	pages = {883--897},
	publisher = {Emerald Group Publishing Limited},
	title = {Predicting Online Purchase Intentions for Clothing Products},
	volume = {38},
	year = {2004}}

@article{mcduff2015predicting,
	author = {McDuff, Daniel and El Kaliouby, Rana and Cohn, Jeffrey F and Picard, Rosalind W},
	journal = {IEEE Transactions on Affective Computing},
	number = {3},
	pages = {223--235},
	publisher = {IEEE},
	title = {Predicting Ad Liking and Purchase Intent: Large-Scale Analysis of Facial Responses to Ads},
	volume = {6},
	year = {2015}}

@inproceedings{guo2011role,
	author = {Guo, Stephen and Wang, Mengqiu and Leskovec, Jure},
	booktitle = {Proceedings of EC},
	pages = {157--166},
	title = {The Role of Social Networks in Online Shopping: Information Passing, Price of Trust, and Consumer Choice},
	year = {2011}}

@article{gunawan2015viral,
	author = {Gunawan, Dedy Darsono and Huarng, Kun-Huang},
	journal = {Journal of Business Research},
	number = {11},
	pages = {2237--2241},
	publisher = {Elsevier},
	title = {Viral Effects of Social Network and Media on Consumers' Purchase Intention},
	volume = {68},
	year = {2015}}

@article{hajli2017social,
	author = {Hajli, Nick and Sims, Julian and Zadeh, Arash H and Richard, Marie-Odile},
	journal = {Journal of Business Research},
	pages = {133--141},
	publisher = {Elsevier},
	title = {A Social Commerce Investigation of the Role of Trust in a Social Networking Site on Purchase Intentions},
	volume = {71},
	year = {2017}}

@article{testa2018social,
	author = {Testa, Francesco and Russo, Michael V and Cornwell, T Bettina and McDonald, Aaron and Reich, Brandon},
	journal = {Journal of Public Policy \& Marketing},
	number = {1},
	pages = {152--166},
	publisher = {American Marketing Association},
	title = {Social Sustainability as Buying Local: Effects of Soft Policy, Meso-Level Actors, and Social Influences on Purchase Intentions},
	volume = {37},
	year = {2018}}

@inproceedings{bhatt2010predicting,
	author = {Bhatt, Rushi and Chaoji, Vineet and Parekh, Rajesh},
	booktitle = {Proceedings of CIKM},
	pages = {1039--1048},
	title = {Predicting Product Adoption in Large-Scale Social Networks},
	year = {2010}}

@inproceedings{mathur2016engagement,
	author = {Mathur, Akhil and Lane, Nicholas D and Kawsar, Fahim},
	booktitle = {Proceedings of UbiComp},
	pages = {622--633},
	title = {Engagement-Aware Computing: Modelling User Engagement from Mobile Contexts},
	year = {2016}}

@article{o2010development,
	author = {O'Brien, Heather L and Toms, Elaine G},
	journal = {Journal of the American Society for Information Science and Technology},
	number = {1},
	pages = {50--69},
	publisher = {Wiley Online Library},
	title = {The Development and Evaluation of a Survey to Measure User Engagement},
	volume = {61},
	year = {2010}}

@inproceedings{lehmann2012models,
	author = {Lehmann, Janette and Lalmas, Mounia and Yom-Tov, Elad and Dupret, Georges},
	booktitle = {Proceedings of UMAP},
	pages = {164--175},
	title = {Models of User Engagement},
	year = {2012}}

@inproceedings{wu2017returning,
	author = {Wu, Qingyun and Wang, Hongning and Hong, Liangjie and Shi, Yue},
	booktitle = {Proceedings of CIKM},
	pages = {1927--1936},
	title = {Returning is Believing: Optimizing Long-Term User Engagement in Recommender Systems},
	year = {2017}}

@inproceedings{lalmas2015promoting,
	author = {Lalmas, Mounia and Lehmann, Janette and Shaked, Guy and Silvestri, Fabrizio and Tolomei, Gabriele},
	booktitle = {Proceedings of KDD},
	pages = {1929--1938},
	title = {Promoting Positive Post-Click Experience for In-Stream Yahoo Gemini Users},
	year = {2015}}

@inproceedings{wan2018item,
	author = {Wan, Mengting and McAuley, Julian},
	booktitle = {Proceedings of RecSys},
	pages = {86--94},
	title = {Item Recommendation on Monotonic Behavior Chains},
	year = {2018}}
\end{document}